\documentclass[longauth]{aa} 

\usepackage{multirow} 
\usepackage{booktabs} 
\usepackage{amsmath} 

\usepackage{graphicx}
\usepackage{keyval}
\usepackage{subfigure}
\usepackage{rotating}
\usepackage{adjustbox}
\usepackage{orcidlink}
\usepackage{natbib}
\usepackage{txfonts}
\usepackage{xcolor} 
\usepackage{hyperref}
\hypersetup{colorlinks=true, linkcolor=blue, filecolor=magenta, urlcolor=cyan, citecolor=blue}
\begin{document}

 \title{Supermassive Black Hole Winds in X-rays: SUBWAYS}

 \subtitle{IV. Tracing radio emission and unveiling the role of winds}

 \author{E. Amenta
 \inst{1,2}\orcidlink{0009-0006-9487-0358}
 \and
 M. Brienza\inst{2}\orcidlink{0000-0003-4120-9970}\and
 G. Bruni\inst{3}\orcidlink{0000-0002-5182-6289}\and
 M. Brusa\inst{1,4}\orcidlink{0000-0002-5059-6848}\and
 R. Morganti\inst{5,6}\orcidlink{0000-0002-9482-6844}\and
 F. Panessa\inst{3}\orcidlink{0000-0003-0543-3617}\and 
 R. D. Baldi\inst{2}\orcidlink{0000-0002-1824-0411}\and
 E. Behar\inst{7}\orcidlink{0000-0001-9735-4873}\and
 G. Lanzuisi\inst{4}\orcidlink{0000-0001-9094-0984}\and
 T. Shimwell\inst{5}\orcidlink{0000-0001-5648-9069}\and
 F. Tombesi\inst{8,9,10}\orcidlink{0000-0002-6562-8654}\and 
 S. Bianchi\inst{11}\orcidlink{0000-0002-4622-4240}\and
 G. Chartas\inst{12}\orcidlink{0000-0003-1697-6596}\and
 A. Comastri\inst{4}\orcidlink{0000-0003-3451-9970}\and
 G. Cresci\inst{13}\orcidlink{0000-0002-5281-1417}\and
 B. De Marco\inst{14}\orcidlink{0000-0003-2743-6632}\and
 F. Fiore\inst{15}\orcidlink{0000-0002-4031-4157}\and
 M. Gaspari\inst{16}\orcidlink{0000-0003-2754-9258}\and
 V. E. Gianolli\inst{17}\orcidlink{0000-0002-9719-8740}\and
 R. Gilli\inst{4}\orcidlink{0000-0001-8121-6177}\and
 S. B. Kraemer\inst{18}\and
 G. Kriss\inst{19}\orcidlink{0000-0002-2180-8266}\and
 Y. Krongold\inst{20}\orcidlink{0000-0001-6291-5239}\and
 F. La Franca\inst{11}\orcidlink{0000-0002-1239-2721}\and
 A. L. Longinotti\inst{20}\orcidlink{0000-0001-8825-3624}\and
 M. Mehdipour\inst{19,21}\and
 E. Nardini\inst{13}\orcidlink{0000-0001-9226-8992}\and
 M. Perna\inst{22}\orcidlink{0000-0002-0362-5941}\and
 P. Petrucci\inst{23}\orcidlink{0000-0001-6061-3480}\and
 E. Piconcelli\inst{9}\orcidlink{0000-0001-9095-2782}\and
 G. Ponti\inst{24,25,26}\orcidlink{0000-0003-0293-3608}\and
 F. Ricci\inst{11}\orcidlink{0000-0001-5742-5980}\and
 L. Zappacosta\inst{9}\orcidlink{0000-0002-4205-6884}
}

 \institute{
 Dipartimento di Fisica e Astronomia, Università di Bologna, Via P. Gobetti 93/2, 40129 Bologna, Italy\\
 \email{elisa.amenta2@unibo.it}
 \and 
 INAF - Istituto di Radioastronomia, Via P. Gobetti 101, 40129 Bologna, Italy
 \and
 INAF – Istituto di Astrofisica e Planetologia Spaziali, Via Fosso del Cavaliere, 00133 Roma, Italy
 \and
 INAF - Osservatorio di Astrofisica e Scienza dello Spazio di Bologna, Via Gobetti, 93/3, 40129 Bologna, Italy
 \and
 ASTRON, the Netherlands Institute for Radio Astronomy, Postbus 2, 7990 AA Dwingeloo, The Netherlands
 \and
 Kapteyn Astronomical Institute, University of Groningen, P.O. Box 800, 9700 AV Groningen, The Netherlands
 \and
 Department of Physics, Technion, Haifa 32000, Israel
 \and
 Physics Department, Tor Vergata University of Rome, Via della Ricerca Scientifica 1, 00133 Rome, Italy
 \and
 INAF – Astronomical Observatory of Rome, Via Frascati 33, 00040 Monte Porzio Catone, Italy
 \and
 INFN - Rome Tor Vergata, Via della Ricerca Scientifica 1, 00133 Rome, Italy
 \and 
 Dipartimento di Matematica e Fisica Università degli Studi Roma Tre, Via della Vasca Navale 84, 00146, Rome, Italy
 \and
 Department of Physics and Astronomy, College of Charleston, Charleston, SC 29424, USA
 \and
 INAF – Osservatorio Astrofisco di Arcetri, Largo E. Fermi 5, 50127 Firenze, Italy
 \and
 Departament de F\'{i}sica, EEBE, Universitat Polit\'ecnica de Catalunya, Av. Eduard Maristany 16, S-08019 Barcelona, Spain 
 \and
 INAF - Osservatorio Astronomico di Trieste, via G.B. Tiepolo 11, 34143, Trieste
 \and
 Department of Physics, Informatics and Mathematics, University of Modena and Reggio Emilia, 41125 Modena, Italy
 \and
 Department of Physics and Astronomy, Clemson University, Kinard Lab of Physics, Clemson, SC 29634, USA
 \and
 Department of Physics, Institute for Astrophysics and Computational Sciences, The Catholic University of America, Washington, DC 20064, USA 
 \and
 Space Telescope Science Institute, 3700 San Martin Drive, Baltimore, MD 21218, USA
 \and 
 Instituto de Astronom\'ia, Universidad Nacional Aut\'onoma de M\'exico, Circuito Exterior, Ciudad Universitaria, Ciudad de M\'exico 04510, M\'exico 
 \and
 Department of Astronomy, University of Michigan, 1085 South University Avenue, Ann Arbor, MI 48109, USA
 \and 
 Centro de Astrobiolog\'ia (CAB), CSIC--INTA, Cra. de Ajalvir Km.~4, 28850 -- Torrej\'on de Ardoz, Madrid, Spain 
 \and
 Univ. Grenoble Alpes, CNRS, IPAG, 38000 Grenoble, France
 \and
 INAF – Osservatorio Astronomico di Brera, Via Bianchi 46, 23807 Merate (LC), Italy
 \and
 Max-Planck-Institut für extraterrestrische Physik (MPE), Gießenbachstraße 1, D-85748 Garching bei München, Germany
 \and
 Como Lake Center for Astrophysics (CLAP), DiSAT, Università degli Studi dell’Insubria, via Valleggio 11, 22100 Como, Italy\\
 }

 \date{Received December 17, 2025; accepted April 29, 2026}

 \abstract
 {Most active galactic nuclei (AGNs) are radio quiet (RQ), with radio emission that may arise from star-formation activity, AGN-driven winds, weak jets, and coronal activity. Disentangling these mechanisms is challenging and requires detailed multi-wavelength investigations, but it is crucial for quantifying AGN feedback in galaxy evolution.}
 {We present a detailed radio investigation of 22 X-ray-selected AGNs in the Supermassive Black Hole Winds in X-Rays (SUBWAYS) sample ($L_{bol} \simeq 10^{44.9-46.3}$ erg/s, z=0.1-0.5), selected to systematically search for ultra-fast outflows (UFOs). The UFOs are detected in $\sim30\%$ of the targets, with measured velocities and kinetic luminosities, making the sample particularly well-suited for investigating the role and signatures of multi-scale outflows at different frequencies.}
 {We built the radio spectral energy distribution of the sources complementing our proprietary data, collected with the JVLA at 1.5 and 6 GHz, with images from LoTSS (145 MHz) and other publicly available radio surveys between 150 and 1400 MHz.
 
 We investigated the role and occurrence of the aforementioned mechanisms, with particular interest in outflows and their possible relation to UFOs. To achieve this, we combined information on spectral indices, luminosities, and morphologies of radio emission with properties derived in other wavebands, such as the star formation rate, the X-ray luminosity, the Eddington ratio, or the UFO kinetic luminosity.}
 {All sources are detected and are mostly consistent with RQ AGNs. For $\sim80\%$ of the sources, the data suggest the presence of an outflow (wind or weak jet). Interestingly, our results indicate that AGNs with UFOs tend to have larger radio extension and a steep radio spectrum consistent with outflows. Moreover, the radio emission of the six UFO hosts is consistent with predictions from wind-driven shock models, possibly indicating a direct connection between the two phases. Alternatively, this may simply reflect physical conditions that favour the production of both phenomena.} 
{}

 \maketitle
%
\section{Introduction}

Accretion onto supermassive black holes (SMBHs) can light up the nuclear regions of galaxies, giving rise to the so-called active galactic nuclei (AGNs), which are among the most energetic objects in the Universe (see \citealt{KormendyHo_2013} and \citealt{Alexander_2025} for reviews). Active galactic nuclei 
 have been shown to inject significant amounts of energy and momentum into their surroundings from parsec to Mpc scales. This occurs via multiple channels, including direct radiation (e.g. \citealt{Ciotti_2010}), jets of relativistic plasma (e.g. \citealt{Fabian_2012}), and massive multi-phase gas winds (e.g. \citealt{Crenshaw_2003}). 
 In particular, through the interaction with the interstellar medium (ISM), powerful AGN-driven outflows, in the form of winds or jets, have proven to be capable of reproducing the scaling relations between SMBH mass and galaxy properties (\citealt{Cattaneo_1999}; \citealt{Kauffman_2000}; \citealt{King_2015}; \citealt{King_2005}; \citealt{Gaspari_2020}; \citealt{King_2015}; \citealt{Fabian_2012}).
While synchrotron emission from powerful relativistic jets dominates the radio luminosity only in a minority of AGNs, classified as radio loud (RL, $L_{6GHz}=10^{23-27}\ W/Hz$), the majority falls into the radio quiet (RQ; $L_{6GHz}=10^{21-23}\ W/Hz$) category, where the origin of radio emission is still debated. In RQ AGNs, radio emission can arise from a plethora of processes, possibly overlapping, including AGN-driven winds, weak jets, coronal activity, and star-formation activity (see \citealt{Padovani_2017}; \citealt{Panessa_2019} for extensive reviews). Disentangling these mechanisms and understanding how the energy they release couples with the ISM is challenging and requires detailed spectral and morphological analysis. However, obtaining observational constraints is crucial for modelling AGN feedback and galaxy evolution.

The AGN-driven winds are indeed detected in any gas phase, at different ionisation states, velocities, and distances from the inner SMBH. These range from sub-relativistic ($0.1-0.25\ c$), highly ionised, ultra-fast outflows (UFOs), arising at few gravitational radii (e.g. \citealt{Chartas_2002}; \citealt{Tombesi_2010}; \citealt{Pounds_2014}; \citealt{Tombesi_2015}; \citealt{Matzeu_2023}; \citealt{Gianolli_2024}), 
down to the slower ($\sim 500 - 2000$ km/s) outflows detected on kpc scales, in the atomic and molecular gas phases (e.g. \citealt{Zakamska_Greene_2014}; \citealt{Morganti_2016}, and see \citealt{Cicone_2018} for a review). In luminous RQ AGNs,
 it has been proposed that the shocks generated by radiatively driven winds can accelerate electrons, producing synchrotron radio emission on scales larger than 100 pc with $\nu$L$_{\nu}$ $\sim$ 10$^{-5}$ L$_{\rm AGN}$ (\citealt{Zakamska_Greene_2014}; \citealt{Nims_2015}). However, how the different wind phases are connected to each other is still poorly understood. 
 
Recent observations also showed that low-luminosity (log $P_{1.4 GHz}<10^{24}$ W/Hz) galaxy-scale radio jets, which are much more common than their high-power counterparts (\citealt{Mauch_2007}, \citealt{Sabater_2019}; \citealt{Venturi_2021}; \citealt{Speranza_2022}; \citealt{Murthy_2022}; \citealt{Cresci_2023}; \citealt{Ulivi_2024}), 
remain trapped in the ISM for longer, affecting the surrounding medium over a large volume (e.g. \citealt{Bicknell_1994}; \citealt{Jarvis_2019}; \citealt{Mukherjee_2025}). Such low-power jets can still account for the radio emission observed in RQ AGNs. However, AGN winds and jets can coexist (e.g. \citealt{Tombesi_2012}; \citealt{Mestici_2024}), making it complicated to understand their potential causal connection. In addition, recent work 
discusses the possibility that UFOs can accelerate ultra-high-energy cosmic rays. If such associations are confirmed, they would provide further motivation to test whether X-ray UFOs correlate with extended non-thermal radio emission (\citealt{Karwin_2023}; \citealt{Ehlert_2025}).
\begin{figure}[tp]
 \centering
 {\includegraphics[width=0.48\textwidth]{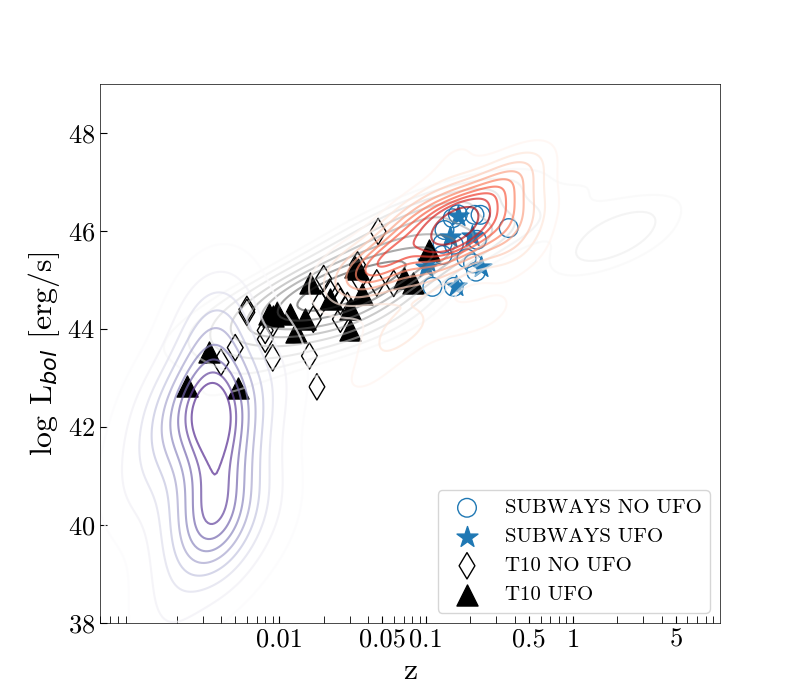}}
                \caption{Bolometric luminosity as a function of redshift. The grey contours highlight the AGNs of \citet[][Y24]{Yamada_2024}, the red ones are the data from \citet[][LB08]{Laor_Behar_2008}, and those in purple the data from \citet[][PG13]{Panessa_Giroletti_2013}. We choose LB08 and PG13 as a comparison for the study of the X-ray-radio luminosity correlation. Then we plot the objects in the SUBWAYS sample (blue empty dots) and a comparison sample from \citet[][T10, black empty diamonds]{Tombesi_2010}. UFO hosts are marked by stars and triangles in SUBWAYS and T10 respectively.}
        \label{fig:matzeu23}
\end{figure}

The current literature providing a detailed investigation of the radio emission in UFO hosts has been largely limited to single-object analyses or archival studies (e. g. \citealt{Falcke_1998}; \citealt{Tombesi_2010}, \citealt{Tombesi_2012}; \citealt{Tombesi_2014}; \citealt{Morganti_2017}; \citealt{Giroletti_2017}; \citealt{Wang_2021}; \citealt{Longinotti_2023}; \citealt{Maksym_2023}; \citealt{Zanchettin_2023}; \citealt{yamada2024}).

In this paper we present the analysis of new Jansky Very Large Array (JVLA) observations, obtained at 1.5 and 6 GHz of 22 quasars belonging to the Supermassive Black Holes Winds in X-rays (SUBWAYS) sample.
By combining JVLA data with radio observations from additional surveys, together with the wealth of multi-wavelength data available for the SUBWAYS sample, we investigate the origin of the radio emission (star formation, coronae, or outflows) and characterise the contribution of AGN-driven outflows\footnote{Throughout the paper,  when we do not distinguish between winds and jets we refer to them more generally as outflows.}, with particular emphasis on UFO hosts.
This work constitutes the first systematic analysis that exploits dedicated follow-up observations for a sample of RQ AGNs with detected UFOs. 

 This paper is organised as follows: in Sect.~\ref{sec:sample} and ~\ref{sec:data} we describe the sample and how the data have been collected and reduced; in Sect.~\ref{sec:Results} and ~\ref{sec:discussion} we discuss the results and consider the potential origins of the radio emission, investigating how the radio properties of the sample relate to those derived in other wavebands. In Sect.~\ref{sec:Summary} we draw our conclusions.

 Throughout this work, a standard Lambda cold dark matter ($\Lambda$CDM) cosmology has been adopted, with $H_0 = 71$ kms$^{-1}$ Mpc$^{-1}$, $\Omega_{\Lambda}=0.73$, and $\Omega_m = 0.27$. The uncertainties are always reported at 1$\sigma$.

\begin{table*}[htp!]
\begin{minipage}[b]{0.9\textwidth}
\vfill
\caption{General parameters of the sources in the SUBWAYS sample.}
\label{tab:images_general}
\small
\centering
\begin{tabular}{llcccccccccc}
\hline \hline
ID&Target Name & R.A. & Dec. & $z$ & $D_L$ & $\log L_{\mathrm{bol}}$ & $\log L_{2-10\,\mathrm{keV}}$ & $\log M_{\mathrm{BH}}$ & $\log \lambda_{\mathrm{Edd}}$ & $v_{\mathrm{out}}$ \\
 && J2000 & J2000 & & (Gpc) & (erg/s) & (erg/s) & ($M_\odot$) & & ($c$) \\
 (1) & (2) & (3) & (4) & (5) & (6) & (7) & (8) & (9) & (10) &(11)\\
\hline
 5&PG0804+761 & 08:10:58.60 & +76:02:43.00 & 0.100 & 0.46 & 45.27 & 44.46$\pm$0.01 & $8.31^{+0.04}_{-0.04}$ & -1.07 & $0.13\pm 0.01$ \\
 6&PG0947+396 & 09:50:48.42 & +39:26:50.64 & 0.205 & 1.01 & 45.89 & 44.37$\pm$0.01 & $8.68^{+0.08}_{-0.10}$ & -1.24 & $0.31^{+0.02}_{-0.04}$ \\
 11&2MASXJ1051+35 & 10:51:44.24 & +35:39:30.76 & 0.159 & 0.76 & 44.88 & 43.70$\pm$0.01 & $8.40^{+0.30}_{-0.30}$ & -1.92 & $0.24\pm 0.01$ \\
 10&PG1114+445 & 11:17:06.40 & +44:13:33.31 & 0.144 & 0.68 & 45.87 & 44.09$\pm$0.01 & $8.59^{+0.09}_{-0.09}$ & -1.26 & $0.07\pm 0.02$ \\
 22&PG1202+281 & 12:04:42.12 & +27:54:12.11 & 0.165 & 0.79 & 46.30 & 44.40$\pm$0.01 & $8.61^{+0.08}_{-0.10}$ & -0.45 & 0.11$\pm$0.01 \\
 18&LBQS1338$-$0038 & 13:41:13.94 & $-$00:53:14.97 & 0.237 & 1.19 & 45.27 & 44.52$\pm$0.01 & $7.74^{+0.03}_{-0.03}$ & -0.52 & $0.15\pm 0.02$ \\
 2&MASXJ1653+23 & 16:53:15.06 & +23:49:42.96 & 0.103 & 0.48 & 45.37 & 43.80$\pm$0.01 & $6.98^{+0.30}_{-0.30}$ & 0.35 & $0.11^{+0.02}_{-0.01}$ \\
\hline
 9&PG0052+251 & 00:54:52.10 & +25:25:37.99 & 0.154 & 0.74 & 45.72 & 44.61$\pm$0.01 & $8.41^{+0.09}_{-0.09}$ & -0.83 & \\
 7&2MASXJ0220-07 & 02:20:14.58 & $-$07:28:59.23 & 0.213 & 1.06 & 46.33 & 44.21$\pm$0.01 & $8.87^{+0.30}_{-0.30}$ & -1.89 & \\
 8&WISEJ0537-02 & 05:37:56.30 & $-$02:45:13.27 & 0.110 & 0.51 & 44.86 & 43.69$\pm$0.01 & $7.73^{+0.30}_{-0.30}$ & -0.60 & \\
 12&PG0953+414 & 09:56:52.40 & +41:15:22.00 & 0.234 & 1.17 & 46.33 & 44.60$\pm$0.01 & $8.24^{+0.06}_{-0.09}$ & -0.05 & \\
 3&PG1307+085 & 13:09:47.00 & +08:19:48.22 & 0.154 & 0.74 & 44.86 & 44.31$\pm$0.01 & $7.90^{+0.12}_{-0.12}$ & -1.18 & \\
 4&PG1352+183 & 13:54:35.72 & +18:05:18.05 & 0.151 & 0.72 & 46.26 & 43.89$\pm$0.01 & $8.42^{+0.08}_{-0.10}$ & -0.30 & \\
 13&2MASXJ1402+26 & 14:02:51.19 & +26:31:17.63 & 0.188 & 0.92 & 45.44 & $< 44.23$ & $< 8.55$ & -1.21 & \\
 15&PG1402+261 & 14:05:16.21 & +25:55:34.13 & 0.164 & 0.79 & 46.34 & 44.03$\pm$0.01 & $7.94^{+0.08}_{-0.10}$ & -0.99 & \\
 17&PG1416$-$129 & 14:19:03.82 & $-$13:10:44.78 & 0.129 & 0.59 & 45.74 & 44.17$\pm$0.01 & $8.12^{+0.08}_{-0.10}$ & 0.00 & \\
 14&PG1425+267 & 14:27:35.61 & +26:32:14.63 & 0.364 & 1.95 & 46.06 & 44.82$\pm$0.01 & $9.22^{+0.30}_{-0.30}$ & -1.30 & \\
 19&PG1427+480 & 14:29:43.10 & +47:47:26.02 & 0.221 & 1.10 & 45.82 & $< 44.20$ & $< 8.09$ & -0.41 & \\
 16&PG1435$-$067 & 14:38:16.10 & $-$06:58:21.00 & 0.129 & 0.59 & 45.51 & 43.68$\pm$0.01 & $7.77^{+0.08}_{-0.10}$ & -0.40 & \\
 20&SDSSJ1444+06 & 14:44:14.67 & +06:33:06.77 & 0.208 & 1.02 & 45.34 & 44.47$\pm$0.01 & $8.33^{+0.30}_{-0.30}$ & -0.94 & \\
 21&HB891529+050 & 15:32:28.79 & +04:53:58.46 & 0.218 & 1.08 & 45.17 & 44.22$\pm$0.01 & $7.46^{+0.30}_{-0.30}$ & -1.47 & \\
 1&PG1626+554 & 16:27:56.10 & +55:22:32.02 & 0.133 & 0.63 & 46.02 & 44.08$\pm$0.01 & $8.54^{+0.08}_{-0.10}$ & -0.66 & \\
\hline
\end{tabular}
\end{minipage}
\tablefoot{
Column 1: Reference label. Column 2: Target name. Column 3 and 4: Coordinates. Column 5: Redshift. Column 6: Luminosity distance. Column 7: Bolometric luminosity. Column 8: X-ray luminosity (derived from \textit{XMM-Newton} analysis, Paper I). Column 9: SMBH mass in units of solar masses. Column 10: Eddington ratio ( Paper I). Column 11: Outflow velocity in units of c ( Paper III) for the UFO detected with confidence interval $\geq 99\%$ ( Paper I). PG1202+281 was not observed by JVLA. For completeness we list its properties here, even though it was not analysed in this work. The UFO subsample is shown before the horizontal line.
}
\end{table*}

\section{The sample}\label{sec:sample}
SUBWAYS was born with the aim of increasing the number of known UFO hosts with multi-wavelength observations. The whole project is based on a large XMM-Newton campaign ($\sim1.5$ Ms of observing time in 2019-2020; \citealt{Brusa2022}) of a sample of 17 (plus five collected from the XMM archive with comparable spectral quality) X-ray selected AGNs (mostly Type 1 quasar (QSO),
with log$(M_{BH}/M_{odot}) = 7.46-9.22$ and $-1.92<$log$\lambda_{Edd}<0.35$), chosen at redshift z=0.1-0.5 and with bolometric luminosity $L_{bol} \simeq 10^{44.9-46.3}$ erg/s. In Table~\ref{tab:images_general} we list the 22 sources belonging to SUBWAYS with their basic properties. Sources with a UFO detection in the confidence interval $\geq 99\%$ are placed above the horizontal line. The unprecedented quality reached by XMM-Newton spectra allowed a detailed investigation of under-explored ranges of redshift and luminosity in UFO studies. From the main analysis of the XMM-Newton observations, \citet[Paper I]{Matzeu_2023} reported, in seven of the 22 AGNs ($\sim30\%$), significant detection of absorption lines corresponding to high column density, highly ionised iron with outflow velocities in the range $-0.3\leq v_{out}/c\leq -0.05$ 
, consistent with the expectations for UFOs. 
Complementary UV imaging data have been obtained with \textit{Hubble} Space Telescope (HST; PI: G. Kriss), but no counterpart for the UFO was found (\citealt[ Paper II]{Mehdipour_2023}). These results could be attributed either to an excessively high ionisation state or to a low covering fraction of the UV-emitting source. Moreover, correlations of SUBWAYS UFO parameters with X-ray luminosity are found to be much stronger than those with the UV luminosity (\citealt[ Paper III]{Gianolli_2024}), suggesting that the innermost regions of the AGNs are more significantly involved than the accretion disk in powering sub-relativistic winds. In fact, though for four objects with UFO detection thermal pressure could in theory launch the outflow, it cannot explain the outflow velocity of UFOs, suggesting that a combination of magnetically and radiatively driven processes is needed (see Fig. 11b of Paper II).

Figure~\ref{fig:matzeu23} shows the bolometric luminosity ($L_{bol}$) redshift plane of the SUBWAYS sample in the context of literature samples of detected X-ray outflows: the one compiled by \citet[Y24]{Yamada_2024}, which includes 132 AGNs at z=0-4, and encompasses 93 UFO hosts known so far (including the SUBWAYS sample) and the one from \citet[T10, included in Y24 as well]{Tombesi_2010}, which represents the low-redshift ($z\leq 0.1$) comparison sample used in Paper III and, for consistency, also adopted in this work. The red and purple contours instead show the location in the $L_{bol}-z$ plane of the samples of \citet{Laor_Behar_2008} and \citet{Panessa_Giroletti_2013}, which we chose as a comparison for the study of the correlation of X-ray and radio luminosity. 

\section{Data}\label{sec:data}

\subsection{JVLA data reduction}\label{ss:JVLAobs}

We made dedicated observations of the sample with the JVLA interferometer between August 2020, May 2023 (PI: F. Panessa), and January 2026 (PI: E. Amenta) in B configuration ($b_{min}\sim 0.21 \text{ km}$, $b_{max}\sim 11.1 \text{ km}$). This set-up has, for the purposes of this work, the right trade-off in terms of maximum and minimum baseline length, providing indeed both good resolution and good sensitivity, necessary for a meaningful morphological analysis. 

The data were collected using the L- and C-band receivers (1.5 and 6 GHz respectively), with 30 minute and 5 minute observations of each target, respectively. Taking into account a $50\%$ overhead for each observing block, which includes telescope initial set-up, phase and amplitude calibration, and slewing between sources, the total observing time is equal to 18.5 observing hours in the L band and 2.88 in the C band. 
The predicted resolution is $4.3''$ at 1.5 GHz and $1''$ at 6 GHz, with $1''$ being $\sim2-4$ kpc at the redshift of our sources.

For all observing blocks, we downloaded the datasets calibrated by the observatory pipeline\footnote{\footnotesize{VLA CASA Calibration Pipeline 5.6.2}}, then checked the goodness of automatic calibration and, if necessary, improved target flagging with \texttt{AOFlagger}\footnote{\href{https://aoflagger.readthedocs.io/en/latest/}{AOFlagger documentation}} (\citealt{Offringa_2010}). 

The imaging was performed with the Common Astronomy Software Applications package (\citealt{CASATeam_2022} \texttt{CASA} 6.6.3.22 version\footnote{\href{https://casadocs.readthedocs.io/en/latest/notebooks/introduction.html}{CASA documentation}}) task \texttt{tclean}, weighting the visibilities with the Briggs scheme (\citealt{Briggs_1995}), with weighting parameter as listed in Table~\ref{tab:image_params}. We based the flux-density scale on measurements of VLA primary amplitude calibrators by \cite{Perley_Butler_2013}. Whenever the noise displayed a non-uniform pattern, before attempting the self-calibration, we tried imaging the data with \texttt{w-stacking clean} (\texttt{WSClean})\footnote{\href{https://wsclean.readthedocs.io/en/latest/}{WSC version 3.1}} (\citealt{Offringa_2014}) and this often improved the results. The images produced with this software are labelled $^*$ in Table~\ref{tab:image_params}.
This was not sufficient for two of the targets, PG1425+267 and HB891529+050, for which we performed one round of phase self-calibration within \texttt{CASA}.

All parameters relative to the final images are listed in Table~\ref{tab:image_params}. The typical resolutions are $\theta_{1.5GHz}\sim 4''.5$ and $\theta_{6GHz}\sim 1''$, with a median noise equal to $\sigma_{rms}$=0.02 mJy beam$^{-1}$ and 0.012 mJy beam$^{-1}$ at 1.5 and 6 GHz respectively. 

\begin{figure*}[htp!]
        \centering      
 \begin{subfigure}{\includegraphics[width=0.32\textwidth]{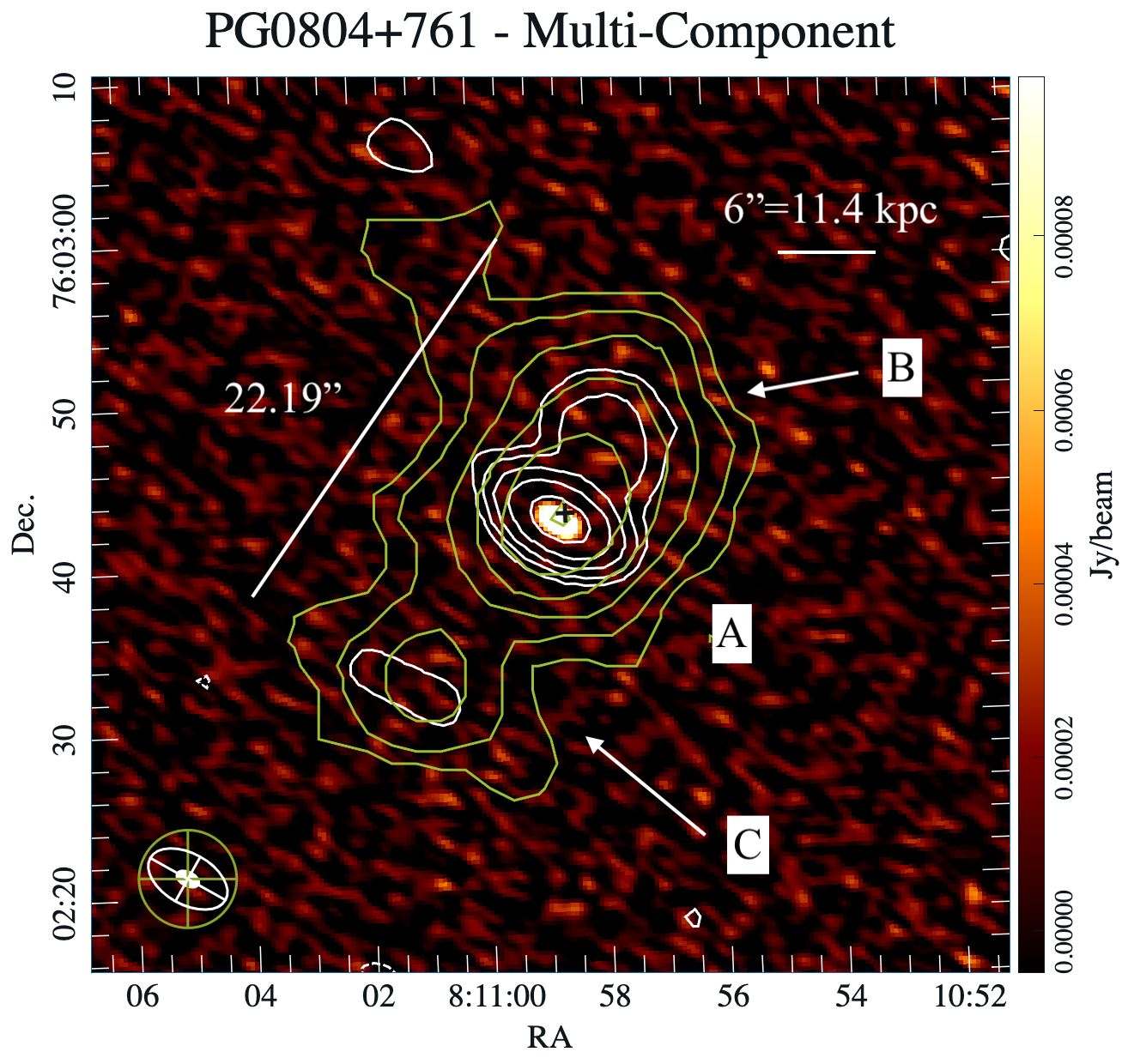}}
        \end{subfigure}
 \hfill
 \begin{subfigure}{\includegraphics[width=0.32\textwidth]{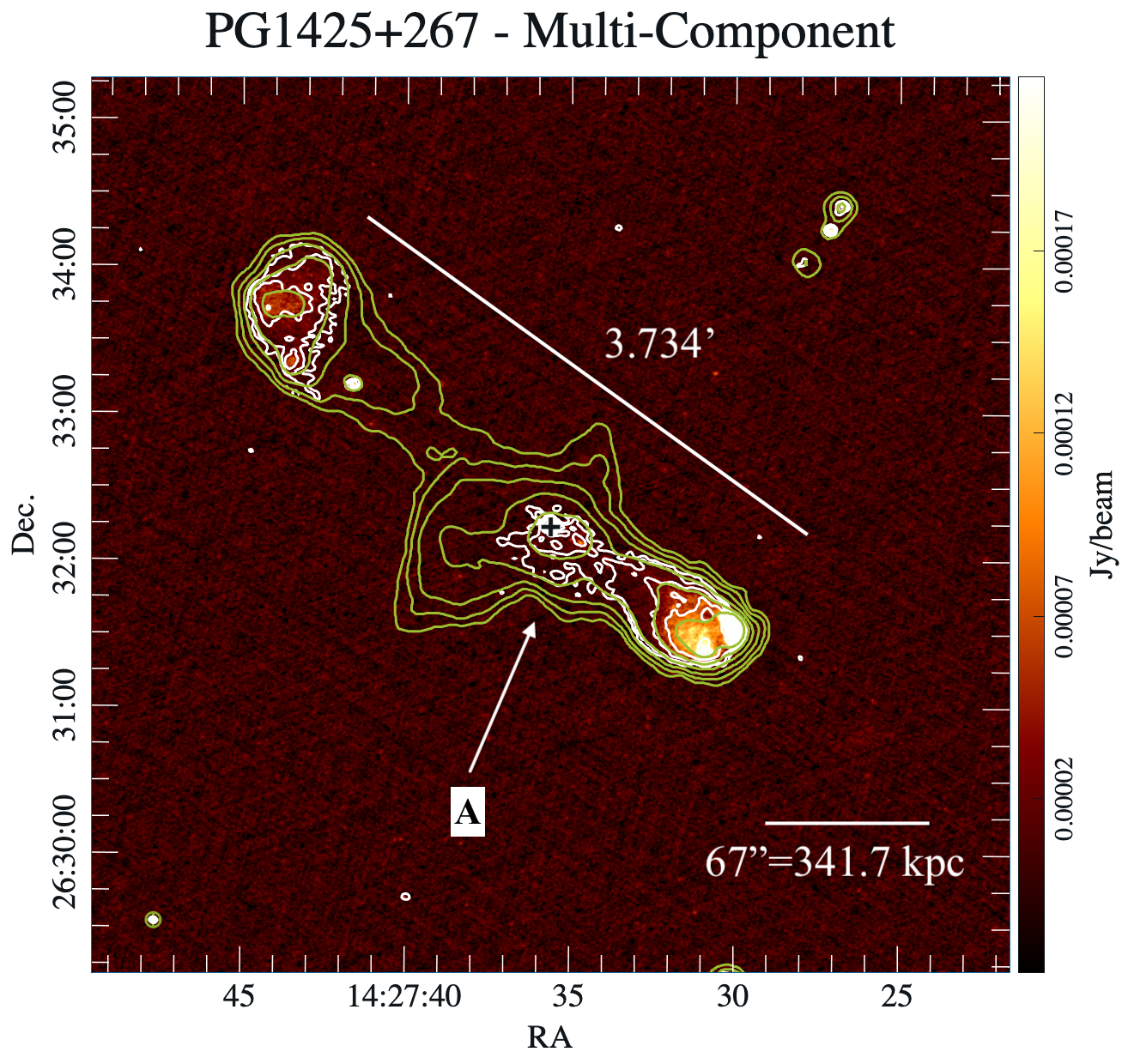}}
 \end{subfigure}
 \hfill
 \begin{subfigure}
 {\includegraphics[width=0.32\textwidth]{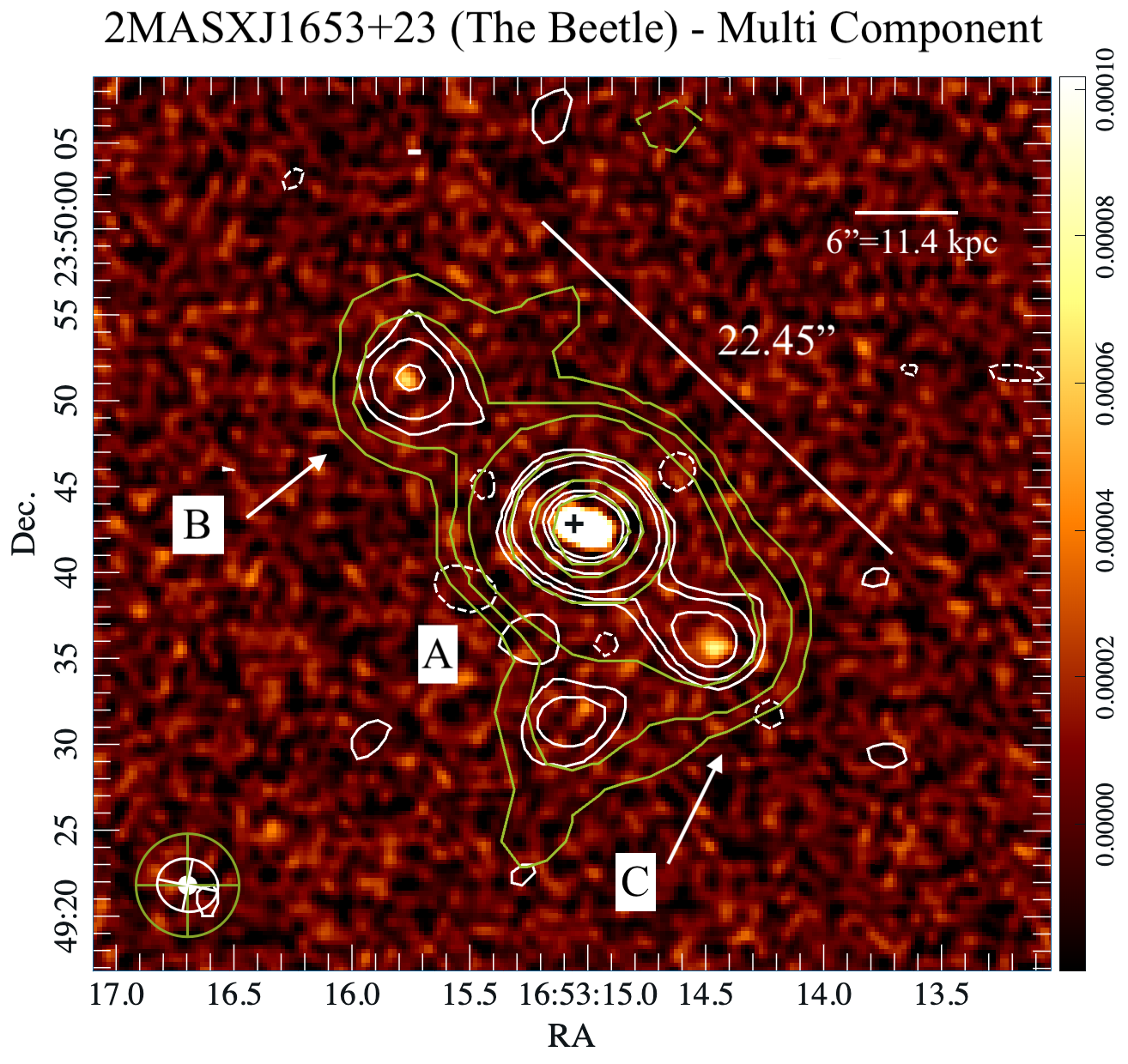}}
\end{subfigure}
        \caption{Images of the 'multi-component' sources. Colours at 6 GHz with $\sigma_{rms}\cdot [-3,3,6,12,24,48,96]$ contours at 1.5 GHz (white) and 145 MHz (green). The beams are shown in the bottom-left corner of each panel. Labels A, B, and C mark distinct radio components: A denotes the nuclear (core) component spatially coincident with the optical AGN position, while B and C denote spatially-separated extranuclear components (i.e. features that are not coincident with the optical/IR nucleus and that extend beyond the host’s optical light profile) likely related to AGN outflows.}
        \label{img:MC}
\end{figure*}

\subsection{Complementary data}
To investigate the morphology of the sources at different frequencies and their broadband radio spectral shape, we combined the aforementioned JVLA 1.5 and 6 GHz simultaneous data with the LOw Frequency ARray Two-metre Sky Survey (LoTSS; \citealt{Shimwell_2017}; \citealt{Shimwell_2022}) data. 
The aim of LoTSS 
is to survey the whole northern sky at 145 MHz with a resolution of $\sim6''$, comparable to that of the proprietary data, and a sensitivity $\sim0.10$ mJy beam$^{-1}$. There are LoTSS mosaics available for 18 of 22 targets (the remaining four are too far south to be observed with LOFAR); in particular, 11 of the mosaics are currently released within the LoTSS collaboration in DR3 (\citealt{Shimwell_2026}). 

To further sample the 145 MHz-6 GHz frequency range and test temporal variability, we also considered images from other surveys, which we list here together with their main properties:

- TGSS (TIFR Giant Metrewave Radio Telescope) at about 150 MHz (ADR1; \citealt{Intema_2017})\footnote{\href{https://vo.astron.nl/tgssadr/q$\_$fits/cutout/form}{ADR1}} with a $2''$ resolution and a median noise of 3.5 mJy beam$^{-1}$ (\citealt{Intema_2017}). We rely on TGSS for the four sources located too far south to fall into the LoTSS footprint.

- RACS - low Data Release 1 (Rapid ASKAP Continuum Survey DR1) images at 888 MHz have a common resolution of $25''$, with a uniform sensitivity of about 0.25 mJy beam$^{-1}$ (\citealp{McConnell_2020}; \citealp{Hale_2021}). The RACS cutouts are available for 15 targets. 

- FIRST (Faint Images of the Radio Sky at Twenty-cm) Survey is a survey done with VLA in B configuration at 1.4 GHz, with a $5''$ resolution and median noise of 0.15 mJy beam$^{-1}$ \citep{Becker_1995}. Among the objects belonging to the SUBWAYS sample, four are not covered by FIRST.

- VLASS (Very Large Array Sky Survey) is conducted by the VLA in B configuration, with multi-epoch observations at 3 GHz, separated by approximately 32 months. Observations are available for three epochs, with a sensitivity $\sim0.16$ mJy beam$^{-1}$ and a resolution $\sim2.4''$ (\citealt{Lacy_2016}; \citealt{Lacy_2020}; \citealt{Gordon_2021}). Cutouts of the three epochs are available for all objects, but, unless otherwise specified, we refer only to the third because it is the most recent one.\\

A more in-depth discussion of the use of these data can be found in the following section. The final images are displayed in \ref{A:SEDs} with L-band white contours superimposed on the C-band colours. The green contours belong to the LoTSS images.

\section{Results}\label{sec:Results}

\subsection{Detection rates and morphologies}\label{lab:morphology}

All sources ( 22/22, $100\%$) are detected at 1.5 and 6 GHz, and 17/18 ($\sim94\%$) are detected at 150 MHz.
Due to the worse resolution and/or sensitivity, in the remaining complementary data the fraction of detected objects decreases resulting in 11/15 ($\sim73\%$), 13/18 ($\sim72\%$) and 13/22 ($\sim59\%$) at 888 MHz, 1.4, and 3 GHz, respectively. This shows the importance and need for dedicated observations for such faint objects.

A compact emitting region, spatially coincident with the optical centre of the AGN, is detected for all of the sources at both 1.5 and 6 GHz, with a signal-to-noise ratio larger or equal to 6 in all cases (see Appendix \ref{tabs}). An exception is the 1.5 GHz image of PG1352+183, where the emission is detected at 3$\sigma_{rms}$ only (see \ref{A:SEDs}). For the four targets not covered by LoTSS we derive upper limits from TGSS (Col. 3 of Table~\ref{tab:cutouts_tab}, more comments in Sect.~\ref{lab:SEDs}).

We provide a morphological classification of the sources by comparing the highest (6 GHz) frequency proprietary images and the lowest (145 MHz) frequency LoTSS one, which may be sensitive to different components. 
To do so, for individual components, we fit a Gaussian within the 3$\sigma_{rms}$ contours in \texttt{CASAviewer} at both frequencies. A deconvolved value is given by \texttt{gaussfit}, to which the algorithm already associated uncertainty. If the source is found to be resolved, this approach provides the least instrumentally biased measurements, which is important for compact emitting regions, as in the case (of the majority) of this sample.

\begin{table*}[htp!]
\caption{Morphologies and sizes of the targets from JVLA (6 GHz) and LoTSS images.}
\label{tab:elongations_combined}
\centering
\begin{tabular}{l | c c c | c c c}
\hline\hline
&&JVLA (6 GHz)&&&LoTSS&\\
\hline
 Target Name & Morphology & Angular Size & Linear Size & Morphology & Angular Size& Linear Size \\
 & &(arcsec) & (kpc) & & (arcsec) & (kpc) \\
(1) & (2) & (3) & (4) & (5) & (6) & (7) \\
PG0804+761 & Extended & $21.36 \pm 0.53$ & $40.58 \pm 1.00$ & Extended & $22.00 \pm 0.60$ & $41.80 \pm 1.10$ \\
PG0947+396 & Unresolved & $<1.26$ & $<4.28$ & Unresolved & $<5.93$ & $<20.16$ \\
2MASXJ1051+35 & Extended & $0.46 \pm 0.06$ & $1.29 \pm 0.17$ & Extended & $3.11 \pm 0.08$ & $8.70 \pm 0.20$ \\
PG1114+445 & Extended & $1.04 \pm 0.13$ & $2.60 \pm 0.32$ & Extended & $4.94 \pm 0.78$ & $22.35 \pm 1.95$ \\
 PG1202+281 & Extended & $ 0.39\pm0.06$& $ 1.12\pm0.17$ & Extended&$ 3.0 \pm0.3$&$ 8.7\pm 0.9$\\
LBQS1338$-$0038 & Extended & $0.70 \pm 0.20$ & $2.66 \pm 0.76$ & Extended & $6.87 \pm 0.76$ & $26.10 \pm 2.90$ \\
2MASXJ1653+23 & Extended & $23.20 \pm 0.37$ & $44.08 \pm 0.70$ & Extended & $24.00 \pm 0.60$ & $45.60 \pm 1.10$ \\
\hline
PG0052+251 & Extended & $1.19 \pm 0.05$ & $3.21 \pm 0.14$ & Extended & $4.24 \pm 0.47$ & $11.40 \pm 1.10$ \\
2MASXJ0220-07 & Unresolved & $<1.05$ & $<3.66$ & - & - & - \\
WISEJ0537-02 & Unresolved & $<0.90$ & $<1.80$ & - & - & - \\
PG0953+414 & Unresolved & $<1.04$ & $<3.85$ & Unresolved & $<6.16$ & $<22.79$ \\
PG1307+085 & Extended & $1.99 \pm 0.43$ & $5.37 \pm 1.16$ & Extended & $13.00 \pm 3.00$ & $35.10 \pm 8.10$ \\
PG1352+183 & Unresolved & $<0.94$ & $<2.44$ & Unresolved & - & - \\
2MASXJ1402+26 & Unresolved & $<1.38$ & $<4.42$ & Unresolved & $<5.96$ & $<19.07$ \\
PG1402+261 & Unresolved & $<2.59$ & $<7.25$ & Unresolved & $<6.06$ & $<16.97$ \\
PG1416-129 & Unresolved & $<0.87$ & $<2.00$ & - & - & -\\
PG1425+267 & Extended & $242 \pm 0.34$ & $1235 \pm 1.70$ & Extended & $245 \pm 0.60$ & $1250 \pm 3$ \\
PG1427+480 & Unresolved & $<0.84$ & $<3.02$ & Extended & $10.41\pm2.14$ & $37.48\pm7.70$ \\
PG1435-067 & Unresolved & $<1.23$ & $<2.83$ & - & - & - \\
SDSSJ1444+06 & Unresolved & $<0.87$ & $<2.96$ & Unresolved & $<9.16$ & $<31.14$ \\
HB891529+050 & Extended & $0.57 \pm 0.02$ & $1.99 \pm 0.07$ & Extended & $4.10 \pm 0.26$ & $14.35 \pm 0.91$ \\
PG1626+554 & Unresolved & $<0.87$ & $<2.09$ & Unresolved & $<5.71$ & $<13.07$ \\
\hline
\end{tabular}
\tablefoot{Column 1: Target name. Columns 2--4: Morphological classification, deconvolved angular size, and linear size from JVLA. Columns 5--7: Same parameters from LoTSS. The UFO subsample is shown before the horizontal line.}
\end{table*}

\begin{figure}
\centering
\includegraphics[width=0.35\textwidth]{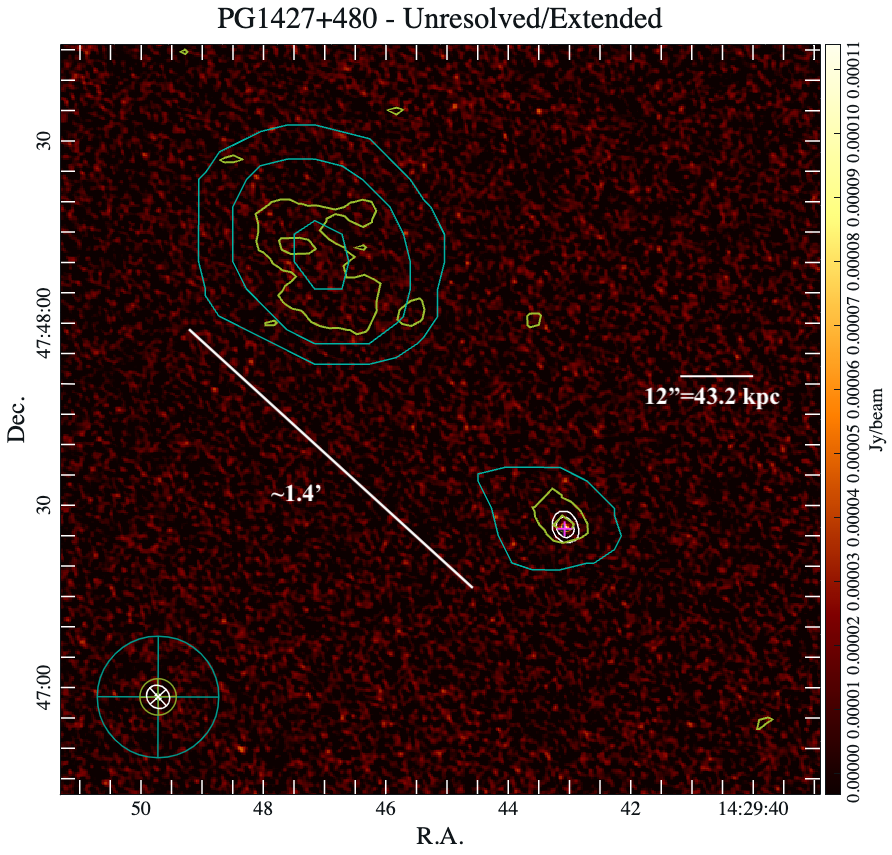}
 \caption{Image of PG1427+480, showing the extended emission detected in the north-east direction of the target, which is located in the lower-right region of the image. Colours at 6 GHz with $\sigma_{rms} \cdot [-3, 3, 6, 12, 24, 48, 96]$ contours at 1.5 GHz (white) and 145 MHz (6'' in green, 20'' in cyan). The beams are shown in the bottom-left corner of the panel. The black cross marks the optical position of the AGN.}
 \label{fig:Pg1427+480bis}
\end{figure}

With this approach 11/22 ($50\%$) sources are classified as unresolved at all frequencies. In these cases, we associate with their extension an upper limit given by the major axis of the restoring beam. Only one source (PG1427+480), appears to be unresolved at 1.5 GHz (L band) and 6 GHz (C band), but shows extensions at 150 MHz, thus we classify it overall as resolved (Fig.~\ref{fig:Pg1427+480bis}). Moreover, at a distance $\sim 1.4'$ in the north-east direction of PG1427+480, some very faint extended emission is detected in the LoTSS image. In Fig.~\ref{fig:Pg1427+480bis} we plot in green the contours of the low resolution LoTSS image ($20\times20''$) where the detection of such features reaches 9$\sigma_{rms}$. There is no clear morphological connection with the target nuclear emission, and overlays with the optical/IR surveys reveal some background objects in the same region. However, there is no clear optical/IR identification, leaving open the possibility that the emission is associated with PG1427+480.

All objects with an angular size larger than the beam major axis are resolved and classified as extended. In all the targets that belong to this class, the compact central emitting source is surrounded by extended emission. Among these, three sources (PG0804+761, 2MASXJ1653+23, and PG1425+267) exhibit multiple components (Fig.~\ref{img:MC}), where some features are well distinguished from the compact central structure, unambiguously extended for several arcseconds, and are not associated with neighbouring sources (see also \ref{A:SEDs}). We measure the maximum extension of these three targets with the ruler tool in \texttt{CASAviewer}, taking as a reference the 3$\sigma_{rms}$ contours.

At both 1.5 and 6 GHz 10/22 ($\sim45\%$) targets belong to the class of extended sources, with three of them being multi-components ($\sim14\%$ of the total sample). In LoTSS images 11 objects are resolved with the same number (3) of multi-components seen by JVLA. The results are summarised in Table~\ref{tab:elongations_combined}.

We also note that there are two targets that show adjacent radio emission, namely PG0947+396 and PG0953+414 (see \ref{A:SEDs}). However, further multi-wavelength investigation reveals that the emission is unrelated to the targets but is instead associated with an edge-on spiral galaxy in the first case, and a blazar, SDSSJ095651+411558 at $z\sim 0.6$, in the second.

\subsection{Flux densities}\label{sec:FD}
For unresolved sources, we extracted the integrated flux densities within the $3\sigma_{rms}$ contours from the images at their original resolution, at all frequencies, using the task \texttt{imstat}.

\begin{table*}[htp!]
 \caption{Integrated flux densities used in this work and their associated uncertainties.}
 \label{tab:cutouts_tab}
 \centering
 \small
 \begin{tabular}{l c c c c c c c}
 \hline\hline
 Target Name & TGSS & LoTSS & RACS & FIRST & JVLA L & VLASS & JVLA C \\
 & 150 MHz & 145 MHz & 888 MHz & 1.4 GHz & 1.5 GHz & 3 GHz & 6 GHz\\
 & (mJy) & (mJy) & (mJy) & (mJy) & (mJy) & (mJy) & (mJy) \\
 \hline
 PG0804+761 & - & $12.5 \pm 1.9$ & - & - & $2.31 \pm 0.06$ & $0.87 \pm 0.13$ & $0.89 \pm 0.19$ \\
 PG0804+761 (A) & - &$5.6 \pm 1.1$ & -&-& $1.9\pm0.3$ &-&$0.89 \pm 0.19$\\
 PG0804+761 (B) & - & $3.0 \pm 0.5$ & -& - & $0.29 \pm 0.03$ & - & - \\
 PG0804+761 (C) & - & $1.00 \pm0.18$ & -& -& $0.11 \pm 0.03$ & - & -\\
 PG0947+396 & - & $0.59 \pm 0.11$ & - & $<0.44$ & $0.13 \pm 0.04$ & $<0.38$ & $0.099 \pm 0.015$ \\
 2MASXJ1051+35 & - & $16 \pm 2$ & - & $11.3 \pm 0.7$ & $11.2 \pm 0.6$ & $7.0 \pm 0.4$ & $4.4 \pm 0.2$ \\
 PG1114+445 & - & $3.1 \pm 0.6$ & - & $0.42 \pm 0.15$ & $0.81 \pm 0.04$ & $0.14 \pm 0.13$ & $0.32 \pm 0.03$ \\
 PG1202+281 & - &$ 4.6 \pm0.7$& $ 2.3\pm0.3$&$ 0.65\pm0.15$&$ 1.2\pm0.6$ &$ 0.85\pm0.13$&$ 0.62\pm 0.03$\\
 LBQS1338-0038 & - & $22 \pm 4$ & $9.8 \pm 1.1$ & $4.8 \pm 0.5$ & $5.6 \pm 0.3$ & $2.9\pm 0.2$ & $2.36 \pm 0.17$ \\
 2MASXJ1653+23 & - & $42 \pm 6$ & $9.9 \pm 1.4$ & $7.0 \pm 0.6$ & $8.20 \pm 0.14$ & $3.8 \pm 0.3$ & $2.7 \pm 0.4$ \\
 2MASXJ1653+23 (A)& - &$32 \pm 6$ & -&-& $7.3\pm0.4$ &-&$2.53\pm 0.13$\\
 2MASXJ1653+23 (B) & - & $2.2 \pm 0.4$ & -& -& $0.37 \pm 0.03$ & -& $0.06 \pm 0.01$ \\
 2MASXJ1653+23 (C) & - & $3.4 \pm 0.5$ & -& -& $0.53 \pm 0.03$ & -& $0.07 \pm 0.01$ \\
 \hline
 PG0052+251 & - & $8 \pm 1$ & $2.1 \pm 0.3$ & - & $1.61 \pm 0.09$ & $0.73 \pm 0.13$ & $0.89 \pm 0.08$ \\
 2MASXJ0220-07 & $<7.8$ & - & $0.3 \pm 0.3$ & $33.9 \pm 1.7$ & $1.49 \pm 0.08$ & $1.10 \pm 0.16$ & $1.22 \pm 0.06$ \\
 WISEJ0537-02 & $<17.7$ & - & $<2.16$ & - & $0.41 \pm 0.04$ & $<0.49$ & $0.22 \pm 0.02$ \\
 PG0953+414 & - & $1.02 \pm 0.16$ & - & $<0.42$ & $0.31 \pm 0.02$ & $0.07 \pm 0.13$ & $0.210 \pm 0.016$ \\
 PG1307+085 & - & $2.4 \pm 0.4$ & $0.6 \pm 0.4$ & $0.22 \pm 0.15$ & $0.59 \pm 0.04$ & $<0.45$ & $0.300 \pm 0.019$ \\
 PG1352+183 & - & $<0.25$ & $<1.02$ & $<0.45$ & $0.07 \pm 0.03$ & $<0.41$ & $0.089 \pm 0.014$ \\
 2MASXJ1402+26 & - & $0.55 \pm 0.11$ & $<0.81$ & $<0.47$ & $0.35 \pm 0.03$ & $<0.36$ & $0.170 \pm 0.016$ \\
 PG1402+261 & - & $2.8 \pm 0.4$ & $1.1 \pm 0.2$ & $0.44 \pm 0.15$ & $1.04 \pm 0.06$ & $0.41 \pm 0.12$ & $0.42 \pm 0.02$ \\
 PG1416-129 & $<26.4$ & - & $0.30 \pm 0.18$ & - & $2.12 \pm 0.09$ & $1.00 \pm 0.15$ & $1.41 \pm 0.08$ \\
 PG1425+267 (total)& - & $2770 \pm 1001$ & $366 \pm 22$ & $210 \pm 19$ & $286 \pm 4$ & $311 \pm 399$ & $72 \pm 14$ \\
 PG1425+267 (core)& - &$131\pm20$        & $65\pm 4$     &$42\pm 2$      &$29.2\pm 1.5$   &$24.2\pm 1.2$          &$27.3\pm 1.4$\\
 PG1427+480 & - & $0.45 \pm 0.09$ & - & $0.30 \pm 0.15$ & $0.162 \pm 0.017$ & $<0.39$ & $0.120 \pm 0.012$ \\
 PG1435-067 & $<11.7$ & - & $0.1 \pm 0.3$ & $0.36 \pm 0.15$ & $0.19 \pm 0.03$ & $<0.42$ & $0.11 \pm 0.02$ \\
 SDSSJ1444+06 & - & $0.7 \pm 0.9$ & $<1.56$ & $<0.45$ & $0.35 \pm 0.02$ & $<0.47$ & $0.230 \pm 0.018$ \\
 HB891529+050 & - & $41 \pm 6$ & $14.9 \pm 1.3$& $11.5 \pm 0.7$ & $11.4 \pm 0.6$ & $5.0 \pm 0.5$ & $4.8 \pm 0.3$ \\
 PG1626+554 & - & $0.88 \pm 0.15$ & - & $<0.44$ & $0.23 \pm 0.03$ & $<0.4$ & $0.166 \pm 0.013$ \\
 \hline
 \end{tabular}
 \tablefoot{Column 1: Target name. Columns 2--8: Integrated flux densities from TGSS (150 MHz), LoTSS (145 MHz), RACS (888 MHz), FIRST (1.4 GHz), JVLA L-band (1.5 GHz), VLASS (3 GHz), and JVLA C-band (6 GHz). The UFO subsample is shown before the horizontal line.}
\end{table*}

For extended sources instead, we first match the image resolution of the proprietary data to derive consistent measurements that can be used to compute spectral indices reliably between 145 MHz and 1.5 GHz and between 1.5 and 6 GHz. We note that all sources, even when resolved, except for PG1425+267, have sizes smaller than the largest C-band angular scale recoverable by the instrument ($29''$), ensuring no flux loss. Hence, only for PG1425+267 we set a cut in baseline length during imaging ($b_{min}\sim2585 \lambda$), so that all data are sensitive to the same scales of emission at both frequencies (see Table~\ref{tab:image_params}).

For all the extended targets, we first created new images at 6 GHz with a natural weighting, then smoothed the outcomes with the task \texttt{imsmooth} to match the beam at 1.5 GHz. The final images at 6 GHz have $\theta_{6GHz}\sim 4''-5''$, for the targets labelled with $\dagger$ in Table~\ref{tab:image_params}. The flux densities are then measured from these images following the 3$\sigma_{rms}$ contours. The same approach is used to extract flux densities from complementary surveys. We note that the resolution of the archival images from FIRST and VLASS is similar enough to LoTSS and the proprietary data to allow for a qualitative comparison. Lastly, despite the much worse resolution, RACS data can be useful in reconstructing the overall SED (spectral energy distribution) shape. Further discussion on this can be found in Sect.~\ref{lab:SEDs}.

The uncertainty associated with the integrated flux density $F_{\nu}$ is computed as
\begin{equation}
        \sigma_{F_{\nu}}=\sqrt{(\delta_{F_{\nu}} \cdot F_{\nu})^2+\big(\langle \sigma_{rms}\rangle_{\nu}\cdot \sqrt{n_{beams}}\big)^2},
        \label{eq:fluxerr}
\end{equation}

 \noindent where $\delta_{F_{\nu}}$ is the fraction of uncertainty associated with the flux density calibration, dependent on instrument and frequency, $\langle \sigma_{rms}\rangle_{\nu}$ is the average noise at frequency $\nu$, and $n_{beams}$ the area encompassed by the source in the beam unit. We used the standard JVLA flux density scale calibration uncertainty ($5\%$) for the proprietary data, FIRST and VLASS, while for LoTSS images we used the standard flux density calibration uncertainty for the DR2 $\delta_{F_{\nu}}=15\%$.
For RACS images we assumed a flux density scale calibration uncertainty of $5\%$. All measurements are reported in Table~\ref{tab:cutouts_tab}. The tabulated VLASS flux density is measured from the third and most recent epoch.

To compute the $\sigma_{F_{\nu}}$ of unresolved objects, $n_{beams}$ is equal to unity, while in all the other cases it is the area used for measurement in units of beam.
Whenever the target is undetected upper limits are derived as 3$\sigma_{rms}$.

\subsection{Spectral indices}\label{sec:spixes}

To study the radio spectral shape, we computed the slopes of the spectra between 145 MHz and 1.5 GHz ($\alpha_{1500}^{145}$) and between 1.5 and 6 GHz ($\alpha_{6000}^{1500}$). 

The spectral index is computed as
\begin{equation}
        \alpha_R=- \frac{ln(F_{\nu_{2}}/F_{\nu_{1}})}{ln(\nu_2/\nu_1)} \ \ \text{where}\ \ F_{\nu}\propto\nu^{-\alpha_R}
        \label{eq:spix}
\end{equation}
with associated uncertainty 
\begin{equation}
        \sigma_{\alpha_R}=\frac{1}{ln(\nu_2/\nu_1)}\sqrt{\Big(\frac{\sigma_{F\nu_{1}}}{F_{\nu_{1}}}\Big)^2+\Big(\frac{\sigma_{F\nu_{2}}}{F_{\nu_{2}}}\Big)^2},
        \label{eq:spixerr}
\end{equation}
where $\nu_1$ and $\nu_2$ are, respectively, the lower and higher frequency between which the spectral slope is computed.

\begin{figure*}[htp!]
\centering
 \includegraphics[width=0.30\textwidth]{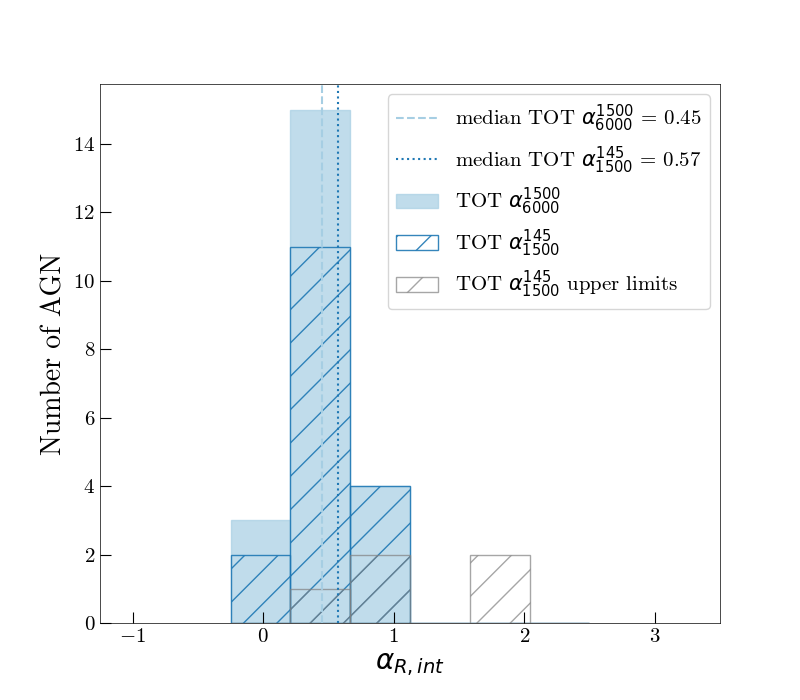}
\hfill
 \includegraphics[width=0.38\textwidth]{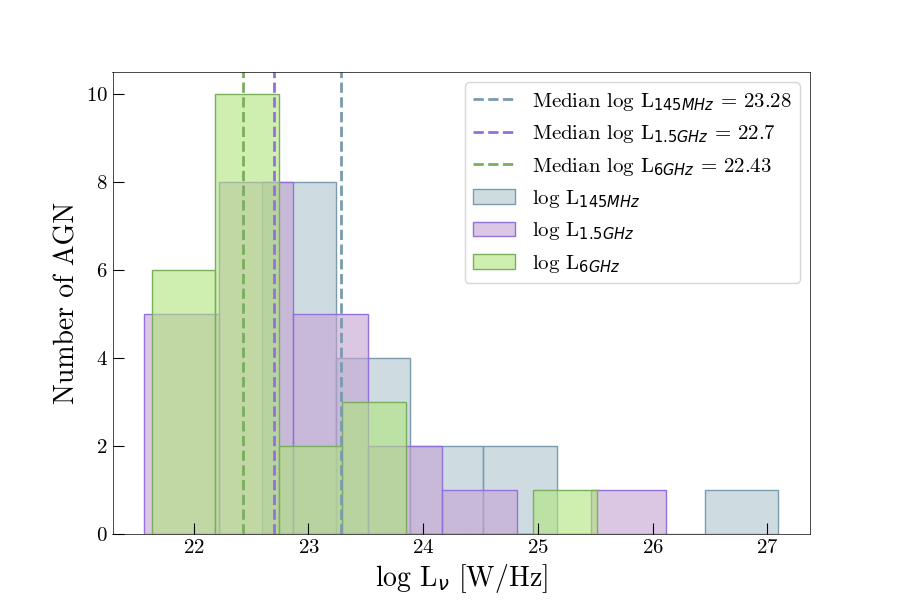}
\hfill
 \includegraphics[width=0.30\textwidth]{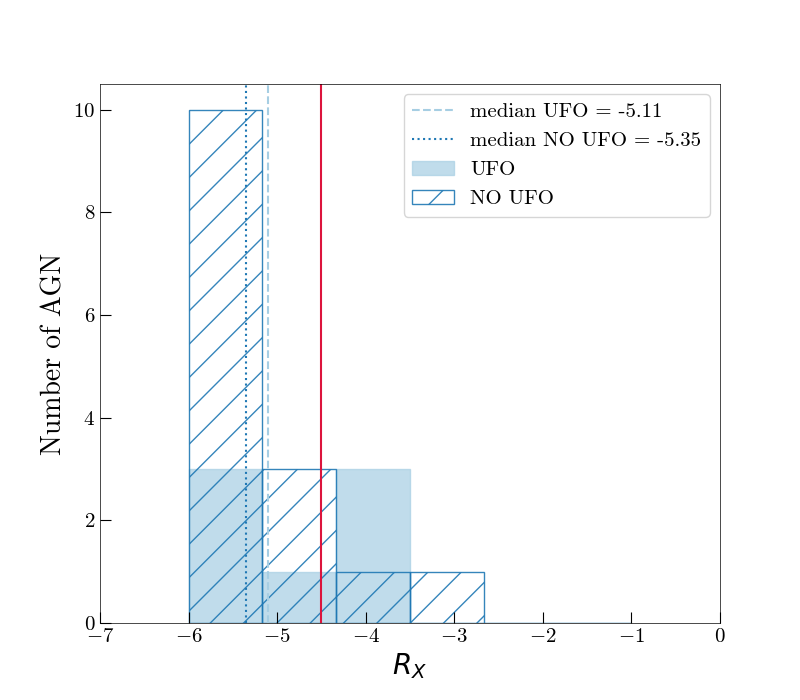}
\caption{Left: Spectral index distribution of the sample. The filled histogram represents the values computed between 1.5 and 6 GHz, while the barred one those computed between 145 MHz and 1.5 GHz. Middle: Histogram comparing the total radio luminosities of the sample. Blue: 145 MHz. Violet: 1.5 GHz. Green: 6 GHz. The upper limits at 145 MHz have been excluded.
Right: Radio loudness parameter distribution of the sample. The convention followed is $R_X=\text{log}\ (L_R(6cm)/L_X(2-10KeV))$ (\citealt{TW_2003}). The filled histogram represents the values of $R_X$ computed for the UFO hosts, and the barred histogram those derived for the non-UFO hosts. The vertical red line marks the -4.5 limit set by \citet{TW_2003}. RQ AGNs tend to dominate the sample, while RL AGNs represent only a small fraction, specifically 6/21 with the original limit $R_X$ = -4.5.}
\label{fig:GHzspixes}
\end{figure*}

In the left panel of Fig.~\ref{fig:GHzspixes}, we show the distribution of the spectral index; more specifically, the filled histogram represents the distribution of the values computed between 1.5 and 6 GHz, while the barred one represents the distribution between 145 MHz and 1.5 GHz. The upper limits are shown in grey. We note that the values of $\alpha_{6000}^{1500}$ derive from simultaneous observations, thus they constitute a robust view of the spectral slope at those frequencies, avoiding potential artificial steepening and/or flattening due to strong variability.

The spectral index of the nuclear and extranuclear components of 2MASXJ1653+23 and PG0804+761 are reported separately in Table~\ref{tab:combined_spix_lum}. 
Overall the spectral index distributions span from 0.15 up to 0.68 at lower frequencies, and from -0.17 up to 0.99 at higher ones, with median values 0.57$\pm 0.05$ and 0.45$\pm 0.08$ for $\alpha_{1500}^{145}$ and $\alpha_{6000}^{1500}$ respectively, computed taking into account the upper limits through a Kaplan-Maier (KM) survival analysis\footnote{
The Kaplan-Meier estimator is a non-parametric method adapted to estimate the cumulative distribution function (CDF) for data containing non-detects (left-censored data). 
The median survival time is estimated as the point where the CDF drops to 0.5.}.\\

\subsection{Radio luminosities and radio loudness}\label{lums_res}
We calculate the characteristic monochromatic radio powers at 145 MHz, 1.5, and 6 GHz, using the following expression
\begin{equation}
        L_{\nu}=4\pi D_L^2(1+z)^{(\alpha_{R}-1)}F_{\nu},
        \label{eq:monochlum}
\end{equation}

\noindent where $F_{\nu}$ is the flux density, $D_L$ the luminosity distance at redshift z (see Table~\ref{tab:images_general}), and $\alpha_{R}$ the radio spectral index, which is substituted with $\alpha_{1500}^{145}$ when computing $L_{145 MHz}$, while $\alpha_{6000}^{1500}$ is preferred for estimating both $L_{1.5 GHz}$ and $L_{6 GHz}$.
An uncertainty is associated according to the following
\begin{equation}
        \sigma_{L_{\nu}}=4\pi D_L^2(1+z)^{(\alpha_{R}-1)}\sqrt{(F_{\nu}\text{ln}(1+z)\sigma_{\alpha_{R}})^2+\sigma_{F_{\nu}}^2}.
        \label{eq:err_lum}
\end{equation}

\noindent The values are reported in Table~\ref{tab:combined_spix_lum} and shown in the histograms in the middle panel of Fig.~\ref{fig:GHzspixes}. The characteristic spectral luminosities range between $22.60<\text{log} L_{145MHz} < 27.06$ W/Hz at the lowest frequency, and between $21.63<\text{log} L_{6GHz} < 25.51$ W/Hz at the highest, with median values 23.28, 22.70, 22.43 W/Hz at 145 MHz, 1.5 GHz, and 6 GHz, respectively.

In addition, taking into account the entire radio emission, we compute the radio loudness parameter $R_X = \text{log}\ (L_{R}(6cm)/L_{X}(2-10 keV))$ (\citealt{TW_2003}) and report the values in Table~\ref{tab:WINDS_SFR_ordered}. Assuming $R_X=-4.5$ as the transition between the RQ and RL regime, following \citet{TW_2003}, we find that RQ AGNs tend to dominate the sample, while RL AGNs represent only a small fraction (six out of 22). 
However, the sample overall spans the region around the traditional boundary value (red line in Fig.~\ref{fig:GHzspixes}, right). This supports the growing difficulty in setting a well-defined difference between RQ and RL AGNs, since the probability distribution of $R_X$ typically does not show strong signs of bi-modality (e.g. \citealt{LaFranca_2010}).

\begin{table*}[ht!]
\centering
\caption{Spectral indices and radio luminosities of the targets. }
\label{tab:combined_spix_lum}
\begin{tabular}{lccccc ccc}
\hline\hline
Target Name & z & $\alpha_{1500}^{145}$ & $\alpha_{6000}^{1500}$ &
$\alpha_{6000}^{1500} - \alpha_{1500}^{145}$ &
$\log L_{145\,\mathrm{MHz}}$ & $\log L_{1.5\,\mathrm{GHz}}$ & $\log L_{6\,\mathrm{GHz}}$ \\
 & & & & & (W\,Hz$^{-1}$) & (W\,Hz$^{-1}$) & (W\,Hz$^{-1}$) \\
 (1) & (2) & (3) &(4) & (5) & (6) & (7) & (8) \\
\hline
PG0804+761 &0.100 & 0.72 $\pm$ 0.07 & 0.69 $\pm$ 0.16 & -0.02 $\pm$ 0.17 & 23.49 $\pm$ 0.07 & 22.76 $\pm$ 0.01 & 22.34 $\pm$ 0.09 \\
PG0804+761 (A) &0.100 & 0.48 $\pm$ 0.07 & 0.56 $\pm$ 0.05 & & & & \\
PG0804+761 (B) &0.100 & 0.99 $\pm$ 0.10 & >0.87 & & & & \\
PG0804+761 (C) &0.100 & 0.94 $\pm$ 0.14 & >0.80 & & & & \\
PG0947+396 &0.205 & 0.65 $\pm$ 0.15 & 0.2 $\pm$ 0.3 & -0.45 $\pm$ 0.29 & 22.83 $\pm$ 0.08 & 22.14 $\pm$ 0.15 & 22.02 $\pm$ 0.06 \\
2MASXJ1051+35 &0.159 & 0.15 $\pm$ 0.07 & 0.68 $\pm$ 0.05 & 0.52 $\pm$ 0.08 & 23.99 $\pm$ 0.07 & 23.87 $\pm$ 0.03 & 23.46 $\pm$ 0.02 \\
PG1114+445 &0.144 & 0.58 $\pm$ 0.09 & 0.66 $\pm$ 0.08 & 0.08 $\pm$ 0.11 & 23.22 $\pm$ 0.09 & 22.64 $\pm$ 0.02 & 22.24 $\pm$ 0.04 \\
PG1202+281&$ 0.165$&$ 0.58\pm0.22 $&$  0.5\pm0.4$&$ -0.1 \pm 0.4$&$ 23.50\pm0.07$&$ 22.9\pm0.2$&$ 22.63\pm0.03$\\
LBQS1338$-$0038 &0.237 & 0.59 $\pm$ 0.08 & 0.62 $\pm$ 0.06 & 0.03 $\pm$ 0.10 & 24.54 $\pm$ 0.07 & 23.94 $\pm$ 0.03 & 23.57 $\pm$ 0.03 \\
2MASXJ1653+23 & 0.103 & 0.70 $\pm$ 0.07 & 0.81 $\pm$ 0.11 & 0.17 $\pm$ 0.13 & 24.04 $\pm$ 0.07 & 23.34 $\pm$ 0.01 & 22.85 $\pm$ 0.06 \\
2MASXJ1653+23 (A) &0.103&0.62 $\pm$ 0.07 & 1.03 $\pm$ 0.05 & & & & \\
2MASXJ1653+23 (B) &0.103&0.88 $\pm$ 0.10 & 1.0 $\pm$ 0.2 & & & & \\
2MASXJ1653+23 (C) &0.103&0.89 $\pm$ 0.09 & 1.24 $\pm$ 0.18 & & & & \\
\hline
PG0052+251 &0.154 & 0.68 $\pm$ 0.07 & 0.43 $\pm$ 0.08 & -0.25 $\pm$ 0.11 & 23.70 $\pm$ 0.07 & 22.99 $\pm$ 0.03 & 22.73 $\pm$ 0.04 \\
2MASXJ0220$-$07 &0.213 & <0.71 & 0.14 $\pm$ 0.05 & -0.56 $\pm$ 0.17 & - & 23.23 $\pm$ 0.03 & 23.14 $\pm$ 0.02 \\
WISEJ0537$-$02 &0.110 & <1.61 & 0.45 $\pm$ 0.10 & -1.16 $\pm$ 0.19 & - & 22.08 $\pm$ 0.05 & 21.81 $\pm$ 0.04 \\
PG0953+414 &0.234 & 0.51 $\pm$ 0.07 & 0.28 $\pm$ 0.07 & -0.23 $\pm$ 0.10 & 23.18 $\pm$ 0.07 & 22.64 $\pm$ 0.03 & 22.47 $\pm$ 0.03 \\
PG1307+085 &0.154 & 0.60 $\pm$ 0.07 & 0.49 $\pm$ 0.07 & 0.03 $\pm$ 0.10 & 23.16 $\pm$ 0.07 & 22.55 $\pm$ 0.03 & 22.26 $\pm$ 0.02 \\
PG1352+183 &0.151 & <0.54 & -0.2 $\pm$ 0.3 & -1.64 $\pm$ 0.41 & - & 21.6 $\pm$ 0.2 & 21.67 $\pm$ 0.06 \\
2MASXJ1402+26 &0.188 & 0.19 $\pm$ 0.09 & 0.52 $\pm$ 0.09 & 0.24 $\pm$ 0.13 & 22.68 $\pm$ 0.09 & 22.51 $\pm$ 0.04 & 22.20 $\pm$ 0.04 \\
PG1402+261 &0.164 & 0.42 $\pm$ 0.07 & 0.65 $\pm$ 0.05 & 0.25 $\pm$ 0.09 & 23.28 $\pm$ 0.07 & 22.86 $\pm$ 0.03 & 22.47 $\pm$ 0.02 \\
PG1416$-$129 &0.129 & <1.08 & 0.29 $\pm$ 0.05 & -0.78 $\pm$ 0.16 & - & 22.91 $\pm$ 0.02 & 22.73 $\pm$ 0.02 \\
PG1425+267 (total) &0.364 & 1.0 $\pm$ 0.4 & 0.99 $\pm$ 0.14 & 0.05 $\pm$ 0.22 & 27.1 $\pm$ 0.4 & 26.11 $\pm$ 0.02 & 25.51 $\pm$ 0.07 \\
PG1425+267 (core) &0.364 & 0.64$\pm$0.07        & 0.05 $\pm$ 0.05 & && & \\
PG1427+480 &0.221 & 0.44 $\pm$ 0.10 & 0.22 $\pm$ 0.10 & -0.22 $\pm$ 0.14 & 22.76 $\pm$ 0.09 & 22.30 $\pm$ 0.05 & 22.17 $\pm$ 0.04 \\
PG1435$-$067 &0.129 & <1.77 & 0.39 $\pm$ 0.17 & -1.38 $\pm$ 0.24 & - & 21.86 $\pm$ 0.08 & 21.63 $\pm$ 0.07 \\
SDSSJ1444+06 &0.208 & 0.30 $\pm$ 0.5 & 0.30 $\pm$ 0.07 & -0.16 $\pm$ 0.13 & 22.9 $\pm$ 0.5 & 22.58 $\pm$ 0.03 & 22.40 $\pm$ 0.03 \\
HB89 1529+050 &0.218 & 0.55 $\pm$ 0.07 & 0.62 $\pm$ 0.06 & 0.01 $\pm$ 0.09 & 24.72 $\pm$ 0.07 & 24.17 $\pm$ 0.03 & 23.79 $\pm$ 0.02 \\
PG1626+554 & 0.133 & 0.57 $\pm$ 0.09 & 0.24 $\pm$ 0.11 & -0.34 $\pm$ 0.14 & 22.60 $\pm$ 0.07 & 21.99 $\pm$ 0.06 & 21.85 $\pm$ 0.03 \\
\hline
\end{tabular}
\tablefoot{Spectral indices of the single extranuclear components of multi-component sources are reported as well. Column 1: Target name. Column 2: Redshift. Columns 3--4: Spectral index between 145 MHz and 1.5 GHz and between 1.5 and 6 GHz, respectively. Column 5: Spectral curvature. Columns 6--8: Logarithmic radio luminosities at 145 MHz, 1.5 GHz, and 6 GHz. The UFO subsample is shown before the horizontal line.}
\end{table*}

\subsection{Radio spectral energy distributions (SED)}\label{lab:SEDs}
 To investigate the broad band SED we combined the values reported in Sect.~\ref{sec:FD} with measurements from TGSS, RACS, FIRST, and VLASS (see \ref{A:SEDs}). We plotted as blue dots the data used for computing the spectral indices (the ones from LoTSS and the proprietary data) and connected them with a solid line when representing actual detections. The grey dots represent all the remaining archival data from RACS, FIRST, and VLASS. In case no LoTSS mosaics are available, we add the TGSS upper limit as a grey dot at 145 MHz. The VLASS flux densities, as mentioned above, are from epoch 3, unless they show significant offset with respect to the computed $\alpha_{6000}^{1500}$; in such cases, we also plot all available epochs as crosses.

We stress that the newly added flux densities are derived from images with resolutions similar to the proprietary data, with the exception of RACS ($25''$). 
However, we verified from high-resolution images at both frequencies higher and lower than that of RACS that a $25''$ does not encompass unwanted extended emission from nearby sources.
We also verify that the images are sensitive to the maximum source spatial scale.
We also report in the legend of each SED in \ref{A:SEDs} the computed values of $\alpha_{1500}^{145}$ and $\alpha_{6000}^{1500}$, and when consistent with each other within the uncertainties, we fit all the SED points with a single power-law of index $\alpha_{fit}$.
For PG0804+761 and 2MASXJ1653+23 we also show the SED of each component separately (labelled as A, B, and C in Fig.~\ref{img:MC}).

The flux densities are integrated, at all frequencies, on the whole extension of the targets, including as a consequence all the extended features and the resolved components, so that the physical interpretation of the spectral signatures presented in the following section takes into account the potential superposition of different contributions from different regions. Moreover, only the proprietary data are simultaneous; therefore, we cannot exclude that variability may affect some of our conclusions.

Indeed, in some objects, variability is suggested by the observed discrepancy between the flux densities measured between FIRST and JVLA proprietary data (e.g. PG1427+480 and PG1435-067 with FIRST measurements being a factor of $\sim 3.3$ and $\sim 2.8$ higher, respectively; see \ref{A:SEDs}). In such cases, as a further test, we plot all three VLASS epochs.

\section{Discussion}\label{sec:discussion}

\begin{figure*}[hpt!]
\sidecaption
\includegraphics[width=12cm]{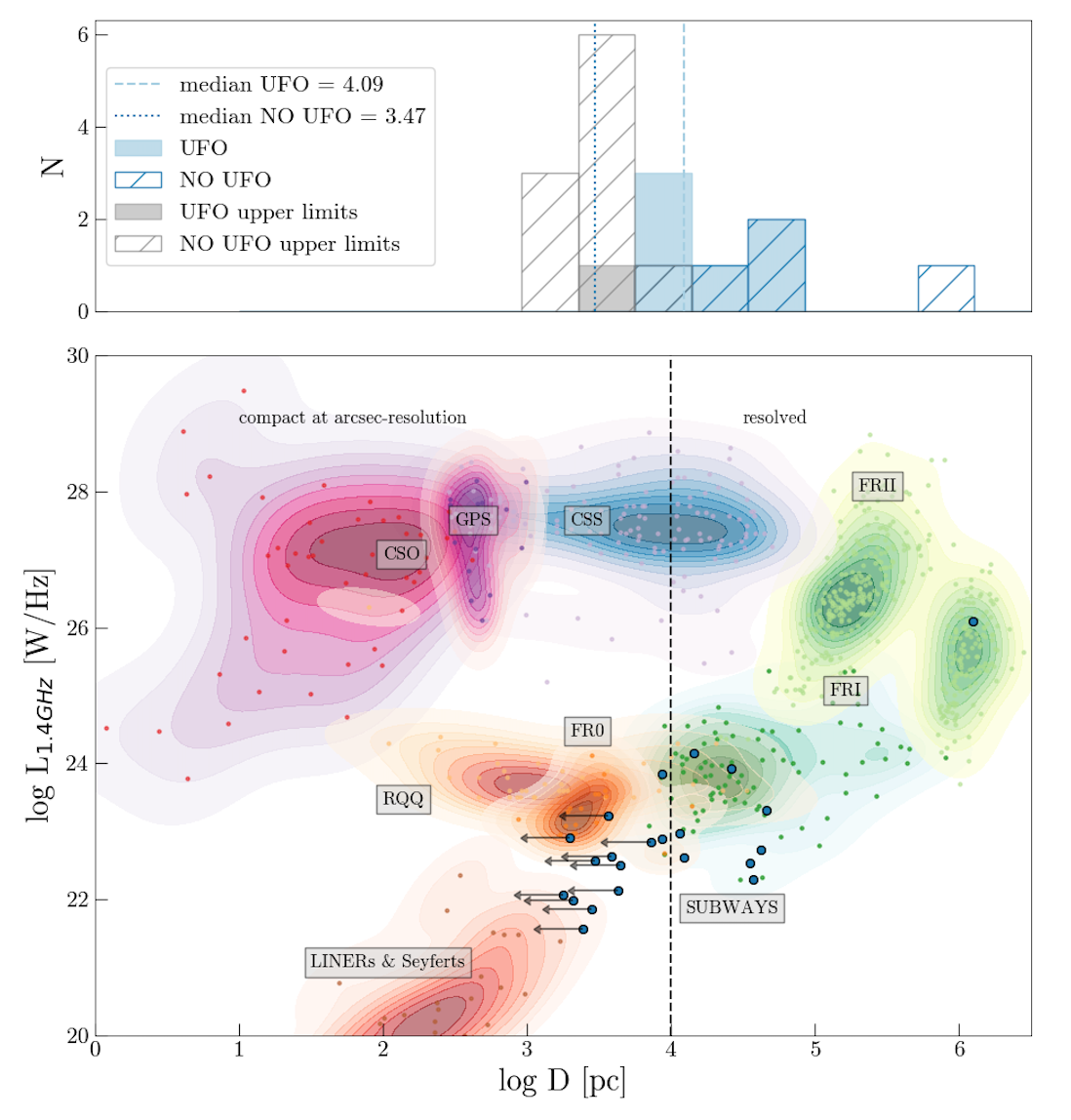}
        \caption{Radio power at 1.4 GHz versus linear size plot (P-D diagram) for different types of RL and RQ AGNs, adapted from \citet{Baldi_2023}. Points show individual objects and coloured contours represent a smoothed estimator of source density. The different categories of source shown are: CSO (compact symmetric objects), GPS (GHz-peaked spectrum sources), CSS (compact 
    steep spectrum sources), FRI, FRII, RQ quasars, Seyferts and LINERS, and FR0s. The vertical dashed line roughly marks the separation between resolved and unresolved compact sources based on arcsecond angular resolution. The blue dots represent how SUBWAYS AGNs overlap to LINERS $\&$ Seyferts, FRIs, FR0s and RQQ. PG1425+267 is a strong outlier, falling on the FRIIs region. Such variety is consistent with the sample being selected based on its X-ray properties. In the upper panel we show the linear size distribution of the sample, with the UFO-subsample (filled histogram) separated from the non-UFO one (barred histogram). The median physical size measured for the UFO subsample is higher ($12.35\pm5$ kpc) with respect to the non-UFO one ($2.96\pm 1$ kpc), potentially supporting multi-phase outflow models (e.g. \citealt{Crenshaw_2003}). 
    Medians are computed taking into account the upper limits through KM survival analysis. Catalogue references: LINERS $\&$ Seyferts (\citealt{Baldi_2018b}), FR0 (\citealt{Baldi_2019}; \citealt{Cheng_2018}); RQQ (\citealt{Jarvis_2021}); GPS (\citealt{Liu_2007}); CSS (\citealt{KB_2010}); CSO (\citealt{AAnB_2012}); FRI (\citealt{Dabhade_2020}; \citealt{JG_2019}); FRII (\citealt{Lao_2024}; \citealt{Parma_1987}).}
        \label{fig:PD}
\end{figure*}
In this section, we investigate the main mechanisms at the origin of the radio emission and integrate our results into the relevant scientific framework. We note that, in more than one case, the SED are best interpreted by a superposition of different processes. However, here we aim at distinguishing the dominant one rather than doing an SED fitting decomposition. In the following, we focus mainly on the general view; for details on single objects we refer to \ref{A:SEDs}.

\subsection{Morphology and linear sizes}\label{sec:sizes}

The morphology and linear size are powerful tools to distinguish the origin of the radio emission: coronal emission coming from sub-pc scales is expected to be compact and unresolved, and star-forming regions can extend up to the same scales as the host galaxy. Only outflows are capable of going beyond the stellar body of the host. We convert the angular sizes (see Cols. 3 and 6 of Table~\ref{tab:elongations_combined}) into linear ones (Cols. 4 and 7).

 {While for unresolved targets the highest resolution of the JVLA at 6 GHz constrains the sizes to linear scales $<1$ kpc, all ten targets resolved at all frequencies – six of which have UFOs 
- display clear extensions on scales ranging from 11 kpc to 1250 kpc (Col. 4 of Table~\ref{tab:elongations_combined}). These extensions suggest that diffuse star formation (SF) and winds and/or jets are present (\citealt{Behar_2015}; \citealt{Panessa_2019}; \citealt{Kawamuro_2022}; \citealt{del_Palacio_2025}; \citealt{Ricci_2023}), although an unresolved coronal component cannot be ruled out (e.g. \citealt{Chen_2024}). 

Interestingly, among these, a giant radio quasar (GRQ; e.g. \citealt{Dabhade_2020}, \citealt{Kuzmicz_2021}), PG1425+267, is identified. In the original selection, RL AGNs with prominent jets were excluded on purpose; however, the quality of the archival information on PG1425+267 was evidently insufficient to classify it as such. An in-depth discussion on the properties of PG1425+267 is beyond the aims of this work. 

Both the other two multi-component sources, 2MASXJ1653+23 (which is a well-studied AGN; \citealt{Villar_Martin_2017}) and PG0804+761 host a 
UFO (with $v_{out}\simeq0.128c$ and $0.108c$, 
respectively, Table~\ref{tab:images_general}), and show resolved extranuclear components separated by $45.6\pm1.1$ kpc and $41.81\pm1.1$ kpc, respectively (Table~\ref{tab:elongations_combined}). 

 Interestingly, all but one of the seven UFO hosts (PG0947+396) are resolved, suggesting that the radio emission is related to shocks driven by a large-scale outflow, potentially linked to the nuclear wind. A similar connection between extended radio emission and X-ray outflow has been proposed for other RQ objects such as Mrk 34 (\citealt{Falcke_1998}; \citealt{Maksym_2023}), IRAS17020+4544 (\citealt{Longinotti_2023}), and NGC 2992 (\citealt{Zanchettin_2023}). Although with the present sample size the result should be considered tentative \footnote{A two-sided Fisher exact test on the 2×2 contingency table ($[6,1],[4,11]$) yields $p \approx 0.020$ (the association is marginally significant at the $\sim2\% $ level)}, the higher incidence of extended radio morphologies in AGNs hosting a UFO (UFO subsample) with respect to those that do not (non-UFO subsample), 
 suggests a physical connection between extended radio emission and X-ray winds, and favours a scenario in which only a subset of AGNs launch powerful outflows (\citealt{Tombesi_2010};\citealt{Tombesi_2012}; Paper I; Paper II).

 To quantify this, we verified that, using the most representative sizes, measured at 6 GHz for the unresolved objects and at 145 MHz for the resolved ones, the median physical size of the radio emission measured for the UFO subsample (the filled histogram in Fig.~\ref{fig:PD}) is higher ($12.35\pm5$ kpc) with respect to that of the non-UFO one ($2.96\pm1$ kpc, barred histogram in Fig.~\ref{fig:PD}). However, a log-rank test yields a p-value of $ \sim0.52$, indicating that there is no significant difference between the two distributions. Again, larger statistics are needed to confirm or reject this tentative result. Similarly, the two subsamples do not show any clear bimodality in $R_X$, with median values equal to -5.11 and -5.35 for the UFO and non-UFO subsamples, respectively, and the p-value $p\sim0.44$.\vspace{-2mm}
\begin{figure*}[htp!]
\sidecaption
\includegraphics[width=12cm]{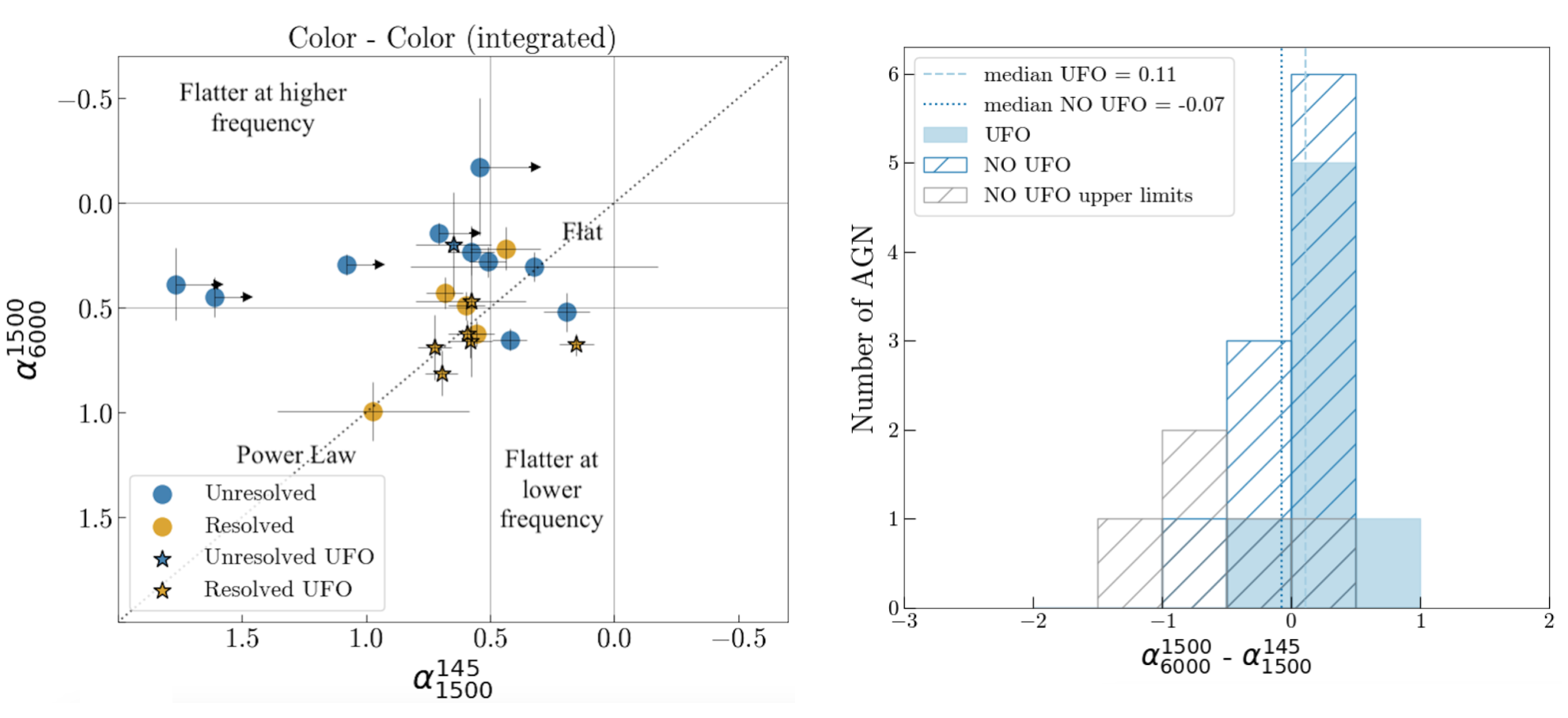}
\caption{Left: radio colour-colour plots with the 1.5 - 6 GHz versus 145 MHz - 1.5 GHz spectral slopes. The yellow dots are the resolved objects while the blue ones are the unresolved ones. Dots are substituted with stars for UFO hosts. Upper limits are plotted with an arrow. Right: Histogram showing the distribution of the spectral curvature, defined as $\alpha_{6000}^{1500}-\alpha_{1500}^{145}$, for the UFO (filled) and NO UFO subsamples (barred). The SEDs of the UFO subsample (almost coincident with the extended class) typically steepen at GHz frequencies, consistently with the observed larger median extension, supporting the hypothesis that nuclear outflows evolve into large-scale winds, resulting in elongated radio morphologies and radio SEDs dominated by extended optically thin emitting regions.}
        \label{fig:colorcolor}
        \vspace{-2mm}
\end{figure*}
\vspace{-2mm}

By locating the sample in a radio luminosity-linear size (P-D) diagram, as shown in Fig.~\ref{fig:PD}, it is immediately noticeable that SUBWAYS (blue dots) occupies a region wide enough to encompass at least four different classes of objects: 
 classical RQ quasars (RQQ, yellow), low-ionisation nuclear emission-line region galaxies (LINERs; \citealt{Singh_2013}) $\&$ Seyferts (brown; \citealt{Seyfert_1943}), low-power Fanaroff-Riley Is (FRI, dark green; \citealt{FR_1974}) and Fanaroff-Riley 0 (FR0, orange; \citealt{Baldi_2016}). The only point located in the region occupied by Fanaroff-Riley IIs (FRII; \citealt{FR_1974}) is the GRQ PG1425+267, which represents a clear outlier. The sample shows how an X-ray selection picks up a variety of radio-emitting AGNs; therefore, we expect a non-homogeneous classification of the mechanisms responsible for the radio emission. This variety is consistent with the radio intermediate (e.g. \citealt{Falcke_1996}; \citealt{Yuan_2008}) values of $R_X$ in SUBWAYS. 
Interestingly, the points straddle the line between resolved and unresolved compact sources (vertical dashed black line) for telescopes with arcsecond resolving powers (e.g. \citealt{Baldi_2023}). Sub-arcsecond resolution observations may reveal compact jet structures or unresolved winds (\citealt{Baldi_2015}; \citealt{Baldi_2023}; \citealt{Chen_2024}).

\subsection{Radio spectral characteristics}\label{sec:cc}

 Using the present observations, spanning the 145 MHz – 6 GHz interval, we compare the SED slopes in the 145 MHz–1.5 GHz and 1.5–6 GHz ranges (see Table~\ref{tab:combined_spix_lum}) plotting them in Fig.~\ref{fig:colorcolor} (left, also called colour-colour plot). 
To further quantify these spectral behaviours, we also compute the spectral curvature (SPC), defined here as the difference between the 1.5 - 6 GHz and 145 MHz - 1.5 GHz spectral indices ($\alpha_{6000}^{1500}-\alpha_{1500}^{145}$; Col. 5 of Table~\ref{tab:combined_spix_lum}).
In the left panel of Fig.~\ref{fig:colorcolor}, horizontal and vertical lines at 0 and 0.5 define the region where flat spectra are, while we classify as steep slopes above 0.5 and as inverted those below 0 (e.g. \citealt{Panessa_2022}). If the value of SPC is null, the spectrum can be fitted by a single power law across the entire available frequency range, and the points are located along the bisector. Those located below exhibit instead a high frequency curvature and a positive SPC, consistent with spectral ageing of optically thin plasma. The SEDs of the objects above the bisector flatten towards higher frequencies (negative SPC), suggesting a combination of steep and flat components.
\vspace{-0.5mm}
If the upper limits and uncertainties are properly considered, all but one of the unresolved targets (PG1402+261) fall in the region where the spectral slope is flat in both regimes, indicative of single or multiple self-absorbed components consistent with either a corona or self-absorbed jet. In contrast, the resolved ones are all either steep or curved. This trend, coupled with the considerations on the sizes of Sect.~\ref{sec:sizes}, and the UFO detection in six of the resolved targets, suggests that compact and unresolved ones tend to exhibit a flat spectrum, typical of nuclear dense regions, whilst, coherently with the elongation of extended targets, star-formation activity or an outflow must be responsible for a steep or curved spectrum. This can be further quantified with a KM survival analysis, which shows that the UFO subsample has a median SPC around 0.11, while it is -0.07 for the other targets.
However, also in this case
if a log-rank test is performed, the p-value $p\sim 0.15$ is insufficient to state whether the UFO subsample belongs to a different distribution. 

\subsection{SFR contribution to radio emission}\label{sec:SF}
                
First of all, we verify whether SF can be considered the main mechanism for the production of radio emission. We derive the SFR from the 1.4 GHz luminosity  following \citet{Condon_1992}, and taking care of extrapolating the 1.5 GHz luminosity to 1.4 GHz using the spectral indices $\alpha_{6000}^{1500}$ listed in Table \ref{tab:combined_spix_lum}.  
For the sources exhibiting multiple components we exclude the extranuclear ones, whose morphology clearly suggests an origin not attributable to the AGN host galaxy. The results are in Table~\ref{tab:WINDS_SFR_ordered} (Col. 3). 

        \begin{figure}[hpt!]
                \centering
                \includegraphics[width=0.45\textwidth]{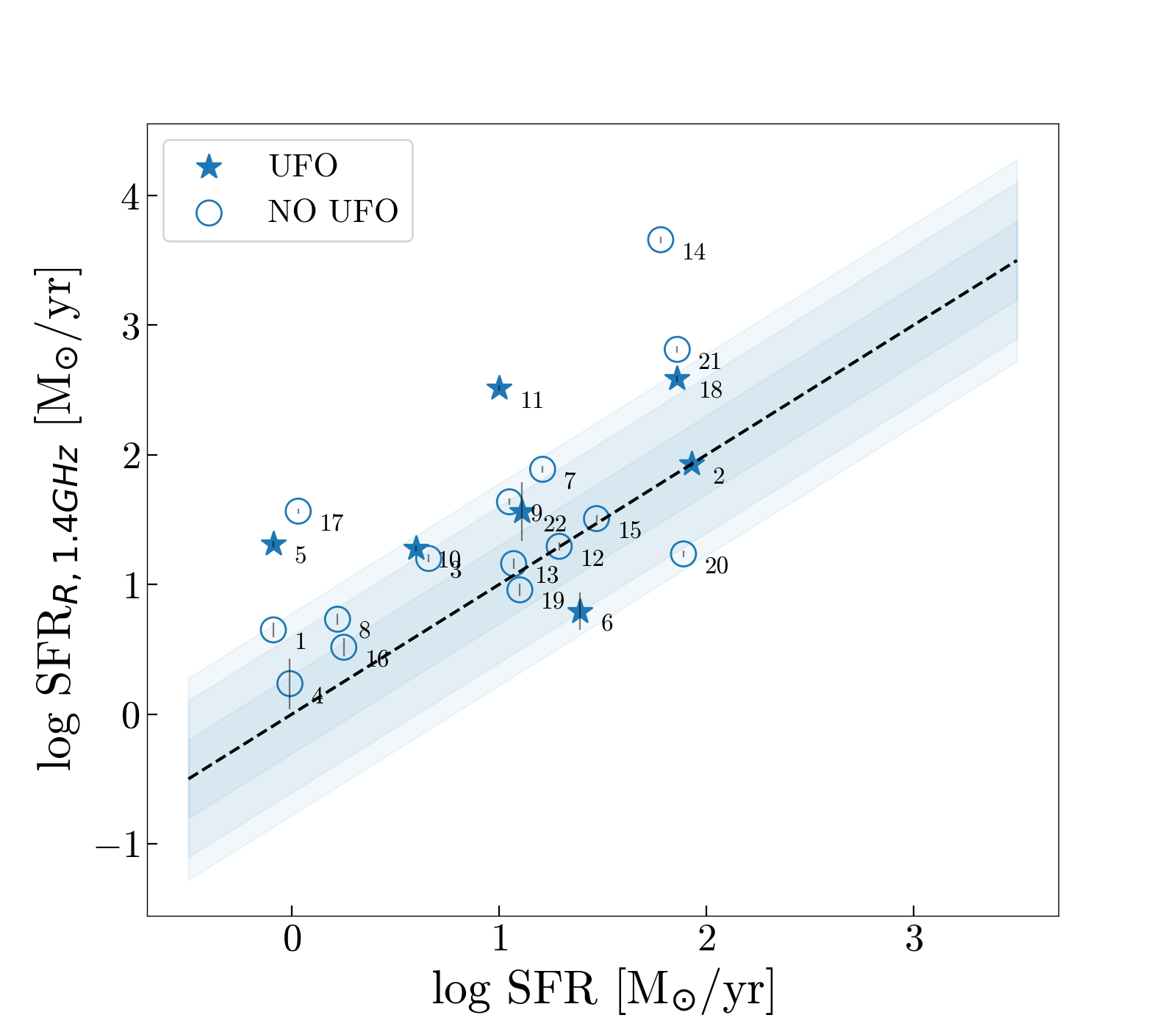}
                \caption{logSFR derived from the radio luminosity following \citet{Condon_1992} against literature values of logSFR derived from IR diagnostics (\citealt{Zhang_2016}). The blue shaded region represents the 1$\sigma$, 2$\sigma$, 3$\sigma$ uncertainty on the relation. Some targets, labelled with ($^*$) in Table~\ref{tab:WINDS_SFR_ordered}, are not studied in \citet{Zhang_2016}, thus their SFR has been computed from WISE flux densities according to the computations of \citet{Cluver_2017}. Numbers refer to single targets as in Table~\ref{tab:WINDS_SFR_ordered}. Apart from five targets exceeding the $3\sigma$ uncertainty region, SUBWAYS AGNs follow the relation quite well. However, only PG1402+261 (15) and 2MASXJ1402+26 (13) exhibit SEDs consistent with being dominated by SF.}
                \label{fig:sfrvssfrradio}
        \end{figure}

In the same table (Col. 4) we list the SFR values either based on near-infrared spectra and far-infrared photometry (\citealt{Zhang_2016}), or derived from Wide-field Infrared Survey Explorer (WISE; \citealt{Wright_2010}) flux densities (using Eq. 5 of \citealt{Cluver_2017}). The latter are labelled with $^*$. 

\begin{table*}[ht!]
\vfill
\caption{Combined properties of the sample, including radio loudness parameters.}
\label{tab:WINDS_SFR_ordered}
\begin{minipage}[b]{0.8\textwidth}
\small
\begin{tabular}{l c c c c c c c c c}
\hline\hline
Target & Label & $\log \mathrm{SFR}_R$ & $\log \mathrm{SFR}$ &
$\log L_{5\,\mathrm{GHz}}$ & $R_X$ &
$\log L_{k,i}$ & $\log L_{\mathrm{exp,en}}$ &
$\log L_{\mathrm{exp,mom}}$ & $\log \nu L_{\nu,\mathrm{obs}}$ \\
 & & (M$_{\odot}$/yr) & (M$_{\odot}$/yr) &
(erg\,s$^{-1}$\,Hz$^{-1}$) & & 
(erg\,s$^{-1}$) & (erg\,s$^{-1}$) & (erg\,s$^{-1}$) &
(erg\,s$^{-1}$) \\
(1) & (2) & (3) & (4) & (5) & (6) & (7) & (8) & (9) & (10) \\
\hline
PG0804+761 & 5 & 1.31 $\pm$ 0.02 & -0.09 & 29.29 $\pm$ 0.10 & -5.3 $\pm$ 0.10 & $<$46.21 & $<$41.2 & $<$39.3 $\pm$ 1.0 & 39.12 $\pm$ 0.09 \\
PG0947+396 & 6 & 0.79 $\pm$ 0.14 & 1.39 & 29.00 $\pm$ 0.07 & -5.5 $\pm$ 0.07 & 46.16 $\pm$ 0.45 & 41.2 $\pm$ 0.2& 39.2 $\pm$ 1.0 & 38.80 $\pm$ 0.06 \\
2MASXJ1051+35 & 11 & 2.51 $\pm$ 0.02 & 1.00 & 30.41 $\pm$ 0.02 & -3.6 $\pm$ 0.02 & 46.53 $\pm$ 0.35& 41.6 $\pm$ 0.2& 39.6 $\pm$ 1.0 & 40.24 $\pm$ 0.02 \\
PG1114+445 & 10 & 1.28 $\pm$ 0.02 & 0.60 & 29.19 $\pm$ 0.04 & -5.2 $\pm$ 0.04 & 45.26 $\pm$ 0.36& 40.3 $\pm$ 0.2& 38.3 $\pm$ 1.0 & 39.02 $\pm$ 0.04 \\
 PG1202+281&$ 22$&$ 1.6\pm0.2$&$ 1.11$&$ 22.40\pm0.03$&$ -4.99\pm0.03$&$ <45.39$&$ <40.4$&$ <38.4$&$ 39.41\pm0.03$\\
LBQS1338-0038 & 18$^*$ & 2.59 $\pm$ 0.02 & 1.86 & 30.52 $\pm$ 0.03 & -4.3 $\pm$ 0.03 & 45.98 $\pm$ 0.66 & 41.0 $\pm$ 0.2& 39.0 $\pm$ 1.0 & 40.34 $\pm$ 0.03 \\
2MASXJ1653+23 & 2 & 1.93 $\pm$ 0.02 & 1.93 & 29.79 $\pm$ 0.07 & -4.3 $\pm$ 0.07 & 44.98 $\pm$ 0.62 & 40.0 $\pm$ 0.2 & 38.0 $\pm$ 1.0 & 39.63 $\pm$ 0.06 \\
\hline
PG0052+251 & 9 & 1.64 $\pm$ 0.02 & 1.05 & 29.70 $\pm$ 0.04 & -5.2 $\pm$ 0.04 & - & - & - & 39.51 $\pm$ 0.03 \\
2MASXJ0220-07 & 7$^*$ & 1.89 $\pm$ 0.02 & 1.21 & 30.13 $\pm$ 0.02 & -4.4 $\pm$ 0.02 & - & - & - & 39.92 $\pm$ 0.02 \\
WISEJ0537-02 & 8$^*$ & 0.73 $\pm$ 0.04 & 0.22 & 28.78 $\pm$ 0.04 & -5.2 $\pm$ 0.04 & - & - & - & 38.59 $\pm$ 0.04 \\
PG0953+414 & 12 & 1.29 $\pm$ 0.03 & 1.29 & 29.45 $\pm$ 0.03 & -5.4 $\pm$ 0.03 & - & - & - & 39.25 $\pm$ 0.03 \\
PG1307+085 & 3 & 1.20 $\pm$ 0.03 & 0.66 & 29.22 $\pm$ 0.03 & -5.4 $\pm$ 0.03 & - & - & - & 39.03 $\pm$ 0.02 \\
PG1352+183 & 4 & 0.23 $\pm$ 0.19 & -0.01 & 28.68 $\pm$ 0.07 & -5.5 $\pm$ 0.07 & - & - & - & 38.45 $\pm$ 0.06 \\
2MASXJ1402+26 & 13 & 1.16 $\pm$ 0.04 & 1.07 & 29.16 $\pm$ 0.04 & -5.4 $\pm$ 0.04 & - & - & - & 38.98 $\pm$ 0.04 \\
PG1402+261 & 15 & 1.51 $\pm$ 0.03 & 1.47 & 29.42 $\pm$ 0.02 & -4.9 $\pm$ 0.02 & - & - & - & 39.25 $\pm$ 0.02 \\
PG1416-129 & 17 & 1.56 $\pm$ 0.02 & 0.03 & 29.71 $\pm$ 0.03 & -4.8 $\pm$ 0.03 & - & - & - & 39.52 $\pm$ 0.02 \\
PG1425+267 & 14 & 3.66 $\pm$ 0.03 & 1.78 & 32.44 $\pm$ 0.08 & -2.7 $\pm$ 0.08 & - & - & - & 42.29 $\pm$ 0.07 \\
PG1427+480 & 19 & 0.96 $\pm$ 0.05 & 1.10 & 29.15 $\pm$ 0.04 & -5.3 $\pm$ 0.04 & - & - & - & 38.95 $\pm$ 0.04 \\
PG1435-067 & 16 & 0.51 $\pm$ 0.07 & 0.25 & 28.60 $\pm$ 0.08 & -5.4 $\pm$ 0.08 & - & - & - & 38.41 $\pm$ 0.07 \\
SDSSJ1444+06 & 20$^*$ & 1.23 $\pm$ 0.03 & 1.89 & 29.38 $\pm$ 0.03 & -5.4 $\pm$ 0.03 & - & - & - & 39.18 $\pm$ 0.03 \\
HB891529+050 & 21$^*$ & 2.81 $\pm$ 0.02 & 1.86 & 30.74 $\pm$ 0.03 & -3.8 $\pm$ 0.03 & - & - & - & 40.57 $\pm$ 0.02 \\
PG1626+554 & 1 & 0.65 $\pm$ 0.06 & -0.09 & 28.83 $\pm$ 0.03 & -5.5 $\pm$ 0.03 & - & - & - & 38.63 $\pm$ 0.03 \\
\hline
\end{tabular}
\end{minipage}
\tablefoot{Column (1): Target name. 
Column (2): Reference label as reported in the figures. 
Column (3): Logarithm of the star-formation rate derived from core radio luminosity (following \citealt{Condon_1992}). 
Column (4): Literature values for the logarithm of the star-formation rate derived from IR data in \citet{Zhang_2016}, or computed from WISE flux densities following \citet{Cluver_2017} for objects marked with ($^*$). 
Column (5): Radio luminosity at 5 GHz extrapolated from the measured 6 GHz (core) luminosity. 
Column (6): Radio loudness parameter $R_X$. 
Column (7): Logarithm of the wind kinetic luminosity from Paper III. 
Column (8): Logarithm of the radio luminosity expected from wind shocks according to Eq.~\ref{eq:winds}, energy-conserving case. 
Column (9): Same as (8), momentum-conserving case. 
Column (10): Logarithm of the observed 6\,GHz radio luminosity.
The UFO subsample is shown before the horizontal line.}
\end{table*}

Radio versus infrared (IR) values are plotted in Fig.~\ref{fig:sfrvssfrradio}. 
Five targets exceed the 3$\sigma$ uncertainty region, implying that different processes beyond SF are required to explain their radio emission (e.g. \citealt{White_2017}). 
In all other cases, the targets sit within 3$\sigma$ of the relation. 
Despite this, either the radio emission extension, larger than the host galaxy as observed in optical images (see \ref{A:SEDs}), or the flat SED led us to exclude SF as main responsible for the radio emission. Only PG1402+261 and 2MASXJ1402+26 are unresolved (see \ref{A:SEDs}) and exhibit values of $\alpha_{6000}^{1500}$ consistent with being dominated by an optically thin component ($\alpha\sim0.6$) at GHz frequencies, making SF a viable source for the production of most radio emission.
The overall observed trend suggests that some contribution from nuclear SF cannot be ruled out, though not dominating the emission, and consequently, the SED shape. This confirms the challenges in identifying the origin of radio emission in RQQ and emphasises the importance of employing complementary indicators.

\subsection{X-ray - radio luminosity correlation}\label{sec:radioxraycorr}

In works by \citet{Panessa_2007}, \citet{Laor_Behar_2008}, and \citet{Panessa_Giroletti_2013}, it has been observed that the 5 GHz radio luminosity and the 2-10 KeV X-ray one in RQ AGNs follow the empirical relation $L_{R}/L_{X}\sim10^{-5.5}$ (\citealt{Laor_Behar_2008}), which is remarkably similar to the Güdel–Benz relation observed for coronally active stars (\citealt{Guedel_1993}). The latter is considered suggestive of a magnetically heated corona: magnetic reconnection releases energy, with a nearly constant fraction converted into heat (seen as X-ray emission) and particle acceleration (seen as synchrotron radio emission). 

To test this scenario, we extrapolate the 6 GHz radio luminosities to 5 GHz, using as $\alpha_{6000}^{1500}$ those reported in Table~\ref{tab:combined_spix_lum}. 
In Fig.~\ref{fig:LRvsLX}, we plot the 5 GHz radio luminosity against the X-ray one derived in Paper I, and the empirical relation $L_{R}/L_{X}\sim10^{-5.5}$ (dashed black line, with associated 1,2,3$\sigma$ uncertainties). For the three objects that exhibit extranuclear components, we consider only the luminosity of the nuclear one (A in Fig.~\ref{img:MC}). 
\begin{figure}[hpt!]
        \centering
 \includegraphics[width=0.45\textwidth]{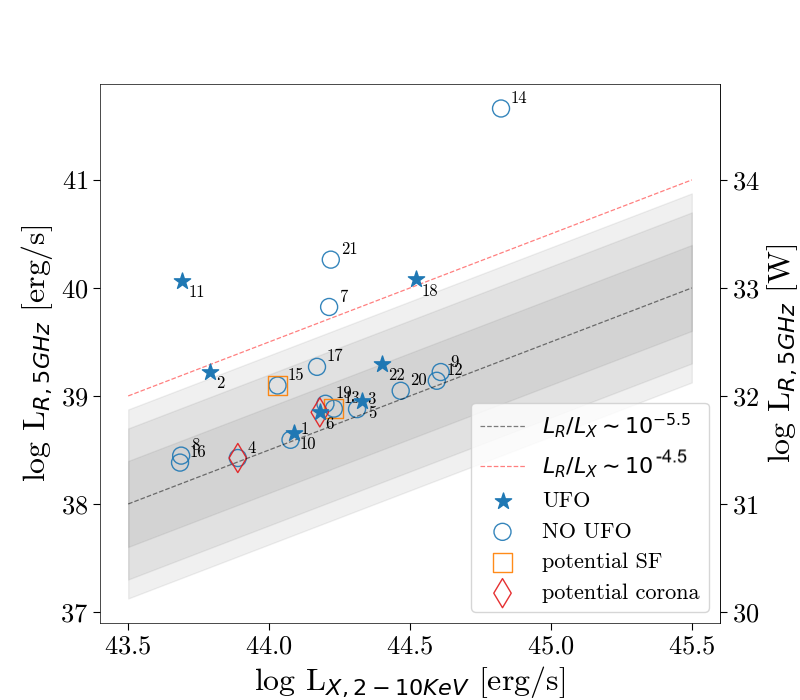}
        \caption{Correlation between X-ray and radio luminosity. The dashed black line is the relationship found by \citet{Laor_Behar_2008} and the grey regions represent the 1$\sigma$, 2$\sigma$, 3$\sigma$ uncertainty. The AGNs where the bulk of radio emission is due to SF are highlighted with orange squares while those where the emission is connected to the corona are surrounded by a red diamond. The dashed red line is the $R_X$=-4.5 traditional threshold between RL and RQ AGNs (\citealt{TW_2003}). Six RL objects exceed the 3$\sigma$ uncertainty region, but all other SUBWAYS targets follow the $L_R/L_X\sim10^{-5.5}$ relation. PG0947+396 (6) and PG1352+183 (4) (red diamonds in the left panel), are unresolved and satisfy the expectations for a corona-dominated spectrum at GHz frequencies. The overall trend confirms that RQQ follow the Güdel–Benz relation almost independently on their radio spectral shape (e.g. \citealt{Laor_Behar_2008}; \citealt{Panessa_Giroletti_2013}).
 }
        \label{fig:LRvsLX}
\end{figure}

First, it is possible to appreciate that the six objects exceeding the 3$\sigma$ uncertainty region are coherently catalogued as RL according to \citet{TW_2003}.
Specifically, the fraction of UFO hosts deviating from the relation is 3/7, much larger than that of non-UFO hosts (3/15). This supports the interpretation that in AGNs with nuclear winds shocks from winds or small jets dominate radio emission at 5 GHz, and can significantly increase $L_R $ relative to $L_X$ (e.g. \citealt{Zakamska_2016}; \citealt{Mancuso_2017}; \citealt{Richards_2021}; \citealt{Baldi_2022}). 

All the remaining objects closely follow the relation. However, as already mentioned in Sec.~\ref{sec:sizes}, current models indicate that corona-dominated emission, which is confined to sub-pc scales, should appear unresolved at arcsecond resolution and exhibit a self-absorbed radio spectrum at GHz frequencies, with spectral indices $\alpha_R \leq 0.2$ (\citealt{Raginski_Laor_2016}; \citealt{del_Palacio_2025}). Only PG0947+396 and PG1352+183 (surrounded by red diamonds in Fig.~\ref{fig:LRvsLX}) match these predictions and are therefore good candidates for hosting corona-dominated radio emission.

In conclusion, similarly to the discussion related to SF in Sec.~\ref{sec:SF}, although the Güdel–Benz relation seems to hold for RQQ almost independently on their spectral shape, a more detailed analysis using complementary diagnostics reveals that the origin of the radio emission cannot be uniquely attributed to coronal processes, highlighting the importance of a careful and multi-approach analysis.

In general, indeed, coronal emission can be more efficiently traced at millimeter wavelengths (100s GHz), where the millimeter excess eventually dominates and an unbiased and direct view of the corona is possible (\citealt{Behar_2015}; \citealt{Panessa_2019}; \citealt{Kawamuro_2022}; \citealt{Ricci_2023}; \citealt{del_Palacio_2025}). Therefore, our current results would be strengthened by a future analysis of the millimeter emission, which could provide proper evidence for a significant coronal contribution to the overall emission.

\subsection{Expected radio emission from winds}\label{sec:winds}

Based on the results presented so far, while for two sources the radio emission could be explained with SF and for other two with the corona, for all the other objects all the diagnostics suggest the culprit should be an outflow, either in the form of wind or low-power jet. The observational signatures for the two mechanisms are very similar (i.e. steep GHz spectrum and extended emission), making it particularly complicated to disentangle (\citealt{Wang_2023}) unless higher resolution observations are available. 

Focusing on the UFO subsample, in this section we explore the interesting possibility that the observed radio emission originates from large-scale winds, which are the galactic-scale continuation of the X-ray nuclear outflows (\citealt{King_2015}; \citealt{Zubovas_2020}; \citealt{Longinotti_2023}; \citealt{Zanchettin_2023}). We attempt a connection between the kinetic luminosity of the nuclear winds to the properties of the radio emission, assuming that we observe radio synchrotron rising where the nuclear wind shocks the ISM.

As noted in Sect.~\ref{sec:sizes}, all UFO hosts except PG0947+396 are extended over tens of kpc and show steep SEDs, consistent with large-scale outflows and possibly supporting multi-phase wind models. In order to place more quantitative constraints, we follow \citet{Nims_2015}. Starting from the energy conserving wind model of \citet{FGQ_2012}, the authors propose an estimate for the expected radio luminosity at a frequency $\nu \geq \nu_{cool}$ (defined as the frequency after which $\nu L_{\nu}$ becomes flat)
\begin{equation}
                L_{R,exp}=\nu L_{\nu}\approx 10^{-5}\ \xi_{-2}\ L_{AGN}\ \Big(\frac{L_{k,o}}{0.05L_{AGN}}\Big) \ \text{erg/s},
                \label{eq:winds}
\end{equation}
 where $\xi_{-2}$ is the energy conversion efficiency in units of $10^{-2}$, $L_{AGN}$ the AGN bolometric luminosity, and $L_{k,o}$ the kinetic luminosity of the radio wind. Under energy-conserving assumptions, we obtain $L_{R,exp}$ substituting $L_{k,o}$ with the value of the X-ray wind ($L_{k,i}$) derived for the SUBWAYS UFO subsample ( Paper III). In Col. 9 of Table~\ref{tab:WINDS_SFR_ordered} we report for each UFO host the $5\%$ efficiency estimates, with symmetric uncertainty computed as the distance from the $1\%$ and $10\%$ efficiency values, which is a reasonable efficiency range reported in the literature (e.g. \citealt{Zubovas_King_2012}; \citealt{King_2015}; \citealt{Zubovas_2020}). 
In Fig.~\ref{fig:winds_lklbol} the blue stars represent the comparison between such values and the observed 6 GHz radio luminosity for the UFO subsample (computed on the whole extension of the targets). However, it is unlikely for the winds to maintain the same kinetic luminosity from UFO scales to kpc ones, where some energy dissipation should occur. For this reason, the blue stars should be interpreted as upper limits.

 In order to find lower limits as well, we also consider the momentum-conserving scenario (following, e.g. \citealt{FGQ_2012} and \citealt{Zubovas_King_2012}). In this case, the kinetic luminosity of the radio outflow $L_{k,o}$ differs from $L_{k,i}$ and can be estimated by comparing the momentum rate of the inner and outer wind: $\dot{P_i}=\dot{P_o}$. 
We get $L_{k,o}=L_{k,i}\cdot(v_{o}/v_{i})$. The ratio between the typical velocity of a large-scale wind (ionised, HI or cold molecular) of $\sim500-2000$ km/s (\citealt{Zakamska_Greene_2014};\citealt{Morganti_2016}; \citealt{Fiore_2017}; \citealt{Bischetti_2019}) and that of a UFO ($0.1-0.25c$ \citealt{Chartas_2002}; \citealt{Tombesi_2012}; Paper I) is around 0.01. Therefore, the radio wind roughly preserves only $1\%$ of the nuclear wind power. The results are reported in Col. 9 of Table~\ref{tab:WINDS_SFR_ordered} and are plotted as red stars in Fig.~\ref{fig:winds_lklbol}. 

 The observed radio luminosity of our sample is included between the upper and lower limits, coherently with recent observations of multi-phase outflows on single sources (e.g. \citealt{Tozzi_2021}; \citealt{Bonanomi_2023}; \citealt{Zanchettin_2023}; \citealt{Longinotti_2023}, \citealt{Baldini2024}) that suggest reality actually lies in between the two limiting cases. Therefore, our result is consistent with radio emission being dominated by outflows.
 In particular, in terms of radio and X-ray properties, our UFO hosts share interesting similarities with IRAS17020+4544 (\citealt{Longinotti_2023}) and NGC 2992 (\citealt{Zanchettin_2023}), both hosting a multi-phase outflow.
Moreover, the work by \citet{Villar_Martin_2017} on 2MASXJ1653+23, strongly supports an outflow-related scenario. 
\begin{figure}[hpt!]
                {\includegraphics[width=0.49\textwidth]{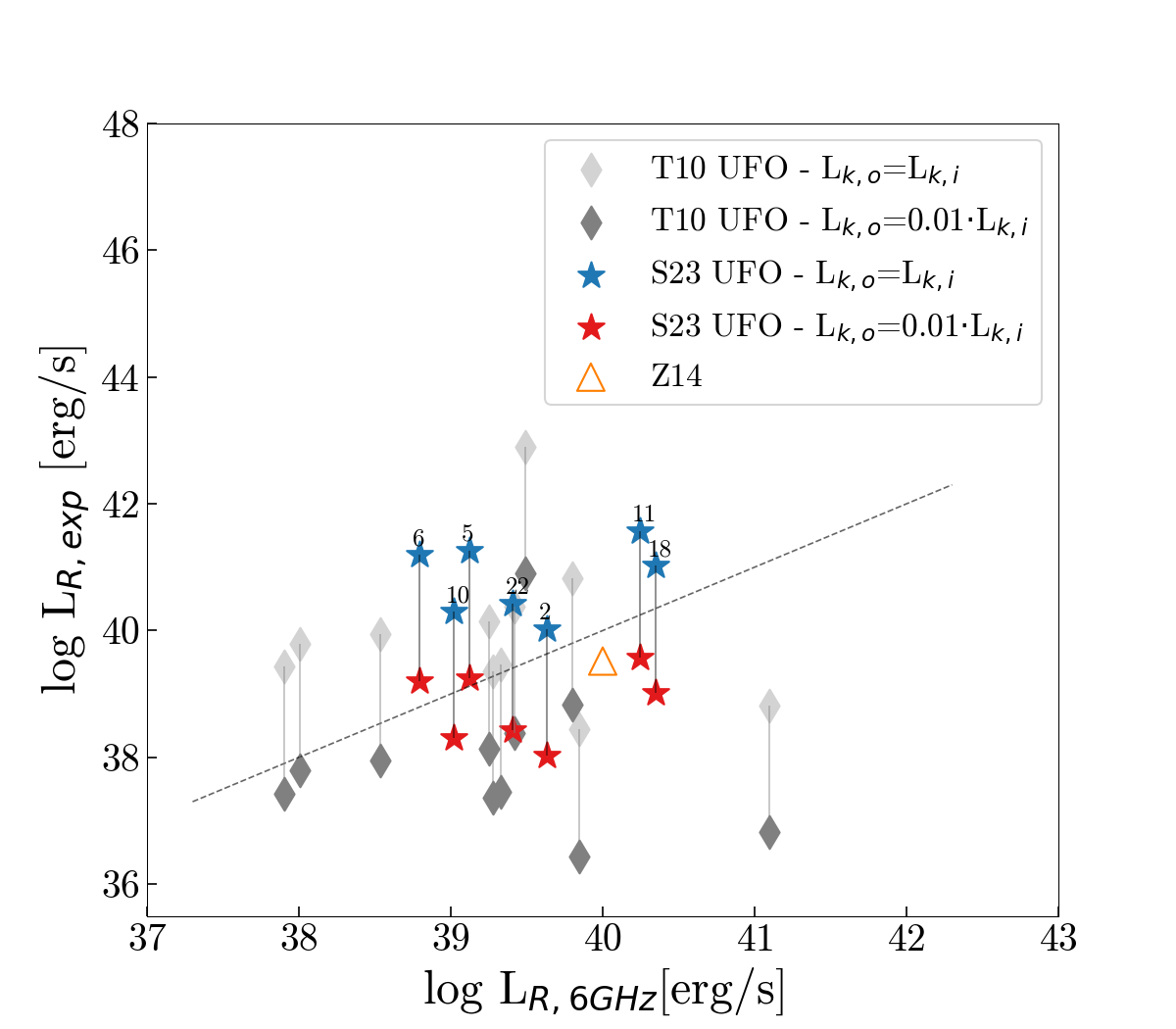}}
        \caption{Expected radio luminosity versus observed one at 6 GHz for the SUBWAYS UFO subsample (M23, blue and red stars) and the comparison samples from \citet[T10, grey diamonds]{Tombesi_2010} and \citet[Z14, orange empty triangle]{Zakamska_Greene_2014}. For the latter only the median values are plotted. $L_{R,exp}$ is computed as from Eq.~\ref{eq:winds} with $L_{k}$ from Paper III in both the energy-conserving (blue stars and light grey diamonds) and momentum-conserving (red stars and dark grey diamonds) scenario. The observed radio luminosity of our sample is included between upper and lower limits and is coherent with recent observations of multi-phase outflows (e.g. \citealt{Zanchettin_2023}; \citealt{Longinotti_2023}). The hypothesis of outflow-driven radio emission is further supported by the extended radio morphology and the steep radio spectral indices of UFO hosts.}
        \label{fig:winds_lklbol}
\end{figure}

Following the same procedure described above, we use the $L_k$ values derived in Paper III for the local RQ AGNs belonging to \citet{Tombesi_2010} sample to compute the expected radio luminosity. The observed radio luminosity of this set of AGNs is the result of extrapolations based on NASA/IPAC Extragalactic Database (NED) archival data\footnote{https://ned.ipac.caltech.edu}. In Fig.~\ref{fig:winds_lklbol} we plot the upper limits as light grey diamonds and the lower limits as dark grey ones. Once again, with the exception of two cases, the one-to-one line is encompassed by the two threshold values, thus, a reasoning similar to that used for our sample applies here. With an orange triangle, we represent the median values for the sample of 568 luminous obscured QSO studied by \citet[][Z14]{Zakamska_Greene_2014}. The authors derive $L_k$ through a detailed analysis of gas kinematics and not from the UFO parameters, thus substituting its value in Eq. \ref{eq:winds} we get a direct estimate of how much kinetic energy is converted into radiation. 

In general, further modelling is needed to fully explain the complex interplay between the outflow and the ISM. 
We note that large-scale ($\sim kpc$) radio outflows develop on $\sim Myr$ timescales, much longer than the typical lifetimes of nuclear UFOs (\citealt{Zubovas_King_2012}; \citealt{King_2015}; \citealt{Zubovas_2020}). This mismatch implies that, if the radio outflow represents the large-scale extension of a nuclear wind, it is more likely associated with a past episode of nuclear activity of comparable intensity, rather than with the currently observed UFO. Alternatively, the observed coexistence of the two outflows may simply reflect physical conditions that favour the production of both phenomena.

To conclude, the above exercise shows that the bulk of radio emission in UFO hosts is consistent with being driven by an evolution of the nuclear wind. In addition we note that within our UFO subsample we detect steep-spectrum features extending beyond the stellar body of the galaxy, matching with outflow expectations. Although all previous observations fit well within a wind scenario, we stress that we cannot rule out the jet scenario based on this analysis. Higher resolution imaging and an estimate of the polarisation fraction may help to distinguish between the two (\citealt{Wang_2023}; \citealt{Chen_2023}; \citealt{Meenakshi_2024}; \citealt{Chen_2024}).

\section{Summary and conclusions}\label{sec:Summary}

We have analysed radio data of the AGNs in the SUBWAYS sample (z=0.1-0.5, $L_{bol}=10^{44.9-46.3}$ erg/s) and investigated the origin of their radio emission, with a focus on radio-UFO connection, exploiting the availability of a unique combination of multi-band observations, as for RQQ the most likely origins considered are star-formation activity, AGN corona, wind, or jets. The radio proprietary data, simultaneously collected with JVLA at 1.5 and 6 GHz, have been complemented with archival images from LoTSS at 145 MHz, and from additional radio surveys (GMRT, RACS, FIRST, and VLASS) to cover a larger frequency range.

Below we summarise our most important results: 
 
- The detection rate in the proprietary data is $100\%$. This decreases to $94\%$ in LoTSS and even lower values in the other archival data, due to sensitivity limitations. 

- The derived radio luminosities (with median values 23.28, 22.70, 22.43 W/Hz at 145 MHz, 1.5 GHz, and 6 GHz respectively; see Fig.~\ref{fig:GHzspixes}), together with the radio loudness parameter computed by exploiting the available X-ray data, suggest that the sources can be classified as RQ or intermediate.

- The morphological classifications based on JVLA 6-GHz and LoTSS images are consistent, revealing twelve unresolved sources and ten resolved ones. For the unresolved sources, the upper limits of the emission extension range from $\sim1$ kpc to $<10$ kpc, therefore winds or low-power jets at sub-kpc and/or kpc scales, if present, could not be resolved. The resolved sources have linear sizes from $\sim10$ kpc to 1000s kpc. One source, PG1425+267, is a Giant Radio QSO, extending for more than 1000 kpc (Fig.~\ref{img:MC}). This previously unidentified object is a serendipitous discovery of this work, which may be worth a detailed investigation in the future (see also Fig.~\ref{fig:PD}). 

- The radio SEDs, the radio colour-colour plots, and the spectral curvature (Fig.~\ref{fig:GHzspixes}, left and middle) highlight a variety of spectral shapes, from simple power-laws, suggestive of an optically thin emitting plasma, to inverted and broken spectra, indicating the presence of an optically thick component dominating either at lower or higher frequencies (see Fig.~\ref{fig:colorcolor}). This is a hint of different mechanisms at the origin of the radio emission.

- Interestingly, six out of the seven UFO hosts show extended radio emission with steep spectra, consistent with outflowing optically thin plasma. While more robust conclusions require larger statistics and in-depth analysis, this may hint at a connection between the two phases, or simply reflect a more suitable environment for producing both phenomena.

- For all sources, we tested whether the observed radio emission is consistent with SF through the radio-IR correlation (Fig.~\ref{fig:sfrvssfrradio}). Similarly, we placed our targets in the $L_R - L_X$ plane to verify their location with respect to the Güdel–Benz relation ($L_{R}/L_{X}\sim10^{-5.5}$; Fig.~\ref{fig:LRvsLX}). 

Overall, by combining the information from the correlations with the spectral and morphological properties, we find that only in two sources the radio emission may be dominated by SF and in the other two by the corona. 
With the currently available data, a dominant radio emission mechanism can be identified in just four sources ($ 18\%$), highlighting the difficulty of analysing radio emission in RQQs, and underscoring the importance of employing complementary diagnostics.

- For all the rest it is likely that an outflow, either in the form of wind or jet, is mainly responsible for the radio emission. In particular, for the seven QSO showing UFOs we computed both upper (energy-conserving) and lower (momentum-conserving) limits for the expected wind radio emission. The observed radio luminosities of our sample lie between the upper and lower predicted limits, suggesting that winds might be a viable mechanism for explaining radio emission (Fig.~\ref{fig:winds_lklbol}). However, because of the huge degeneracy between the observables, we do not exclude the compact jet scenario.

\subsection*{Future perspectives}
To further explore and confirm the results presented in this work, it would be interesting to verify the presence and properties of outflows in other gas phases and test their relation to the observed radio outflows and UFO. As an example, a uniform analysis of the already available optical SDSS spectra would allow us to test whether or not outflows as traced by [OIII] are detected (e.g. \citealt{Zakamska_Greene_2014}).

Also higher resolution observations of unresolved objects could give stronger constraints on their linear sizes, while for the resolved ones with flux densities of the order of mJy (e.g. with JVLA A configuration, e-MERLIN, and VLBI), those may reveal collimated sub-kpc-scale structures, helping to distinguish jets from winds, as noted in \citet{Panessa_Giroletti_2013} and \citet{Patil_2020}. Furthermore, the nuclear radio emission, measured from sub-arcsecond resolution observations, is expected to better correlate with the X-ray one, further supporting the corona scenario in some cases (\citealt{Chen_2024} ).
We stress that future observations at 100-300 GHz with ALMA will be of great importance for more robust estimates of the corona contribution, not only to investigate spectral shapes (because of the lower frequencies sampled by our data a flat spectrum may be hidden) but also correlated mm-X-ray variability.

Finally, both the Next Generation VLA (ngVLA) and Square Kilometer Array (SKA), offering sub-arcsecond resolution (down to mas with ngVLA) and unprecedented sensitivities, will open new avenues for studying compact objects, even at higher redshift, providing the ideal opportunity to extend this kind of radio analysis to larger samples of sources hosting UFOs.

\section*{Data availability}
Data associated with this article, including supplementary figures (e.g. the full SEDs and radio images), are also available in the Zenodo repository \href{https://doi.org/10.5281/zenodo.19921061}{Appendix B}, under the Creative Commons License.

\begin{acknowledgements}
 We acknowledge support from PRINMIUR2017PH3WAT (‘Black hole winds and the baryon life cycle of galaxies’). All the italian co-authors from the SUBWAYS collaboration acknowledge support and fundings from Accordo Attuativo ASI-INAF n. 2017-14-H.0. 
 MBrienza and EAmenta acknowledge financial support from INAF under the Mini Grant 2023 funding scheme (project ‘Low radio frequencies as a probe of AGN jetfeedback at low and high redshift’).
 MBrienza acknowledges financial support from Next Generation EU funds within the National Recovery and Resilience Plan (PNRR), Mission 4 - Education and Research, Component 2 - From Research to Business (M4C2), Investment Line 3.1 - Strengthening and creation of Research Infrastructures, Project IR0000034 – “STILES - Strengthening the Italian Leadership in ELT and SKA”.

LOFAR is the Low Frequency Array designed and constructed by ASTRON. It has observing, data processing, and data storage facilities in several countries, which are owned by various parties (each with their own funding sources), and which are collectively operated by the LOFAR ERIC under a joint scientific policy. The LOFAR resources have benefited from the following recent major funding sources: CNRS-INSU, Observatoire de Paris and Universit\'e d'Orl\'eans, France; BMFTR, MKW-NRW, MPG, Germany; Science Foundation Ireland (SFI), Department of Business, Enterprise and Innovation (DBEI), Ireland; NWO, The Netherlands; The Science and Technology Facilities Council, UK; Ministry of Science and Higher Education, Poland; The Istituto Nazionale di Astrofisica (INAF), Italy.
This research made use of the Dutch national e-infrastructure with support of the SURF Cooperative (e-infra 180169) and the LOFAR e-infra group. The J\"ulich LOFAR Long Term Archive and the German LOFAR network are both coordinated and operated by the J\"ulich Supercomputing Centre (JSC), and computing resources on the supercomputer JUWELS at JSC were provided by the Gauss Centre for Supercomputing e.V. (grant CHTB00) through the John von Neumann Institute for Computing (NIC).
This research made use of the University of Hertfordshire high-performance computing facility and the LOFAR-UK computing facility located at the University of Hertfordshire and supported by STFC [ST/P000096/1], and of the Italian LOFAR-IT computing infrastructure supported and operated by INAF, including the resources within the PLEIADI special "LOFAR" project by USC-C of INAF, and by the Physics Department of Turin university (under an agreement with Consorzio Interuniversitario per la Fisica Spaziale) at the C3S Supercomputing Centre, Italy.
This research is part of the project LOFAR Data Valorization (LDV) [project numbers 2020.031, 2022.033, and 2024.047] of the research programme Computing Time on National Computer Facilities using SPIDER that is (co-)funded by the Dutch Research Council (NWO), hosted by SURF through the call for proposals of Computing Time on National Computer Facilities.

VEG acknowledges funding under NASA contract 80NSSC24K1403.

This research used data from HST program number 15890, sponsored by the Space Telescope Science Institute, which is operated by the Association of Universities for Research in Astronomy, Incorporated, under NASA contract NAS5-26555.

BDM acknowledges support via the Spanish MINECO grants PID2023-148661NB-I00, PID2022-136828NB-C44, and the AGAUR/Generalitat de Catalunya grant SGR-386/2021.

MG acknowledges support from the ERC Consolidator Grant BlackHoleWeather
(101086804)

MP acknowledges support through the grants PID2021-127718NB-I00, PID2024-159902NA-I00, and RYC2023-044853-I, funded by the Spain Ministry of Science and Innovation/State Agency of Research MCIN/AEI/10.13039/501100011033 and El Fondo Social Europeo Plus FSE+.

ALL acknowledges support from DGAPA-PAPIIT IA103625. 

YK acknowledges support form UNAM-PAPIIT grant IN 102023. 

EB was supported by The Israel Science Foundation (grant No. 2617/25).

POP acknowledges support from the french spatial agency CNES and from the « Action Thématique Phénomènes Extrêmes et Multimessager » of the Astronomy-Astrophysics National Programme from INSU/CNRS.

GP acknowledges support from the European Research Council (ERC) under the European Union’s Horizon 2020 research and innovation program HotMilk (grant agreement No. 865637) and from the Framework per l’Attrazione e il Rafforzamento delle Eccellenze (FARE) per la ricerca in Italia (R20L5S39T9). 

LZ acknowledges financial support from the Bando Ricerca Fondamentale INAF 2022 Large Grant “Toward an holistic view of the Titans: multi-band observations of z > 6 QSOs powered by greedy supermassive black holes”, Bando Ricerca Fondamentale INAF 2024 Large Grant “The DEepest study of LUminous QSOs in X-ray at z=2-7 (DELUX)” and from the European Union – Next Generation EU, PRIN/MUR 2022 2022TKPB2P – BIG-z. FP acknowledges financial support from the Bando Ricerca Fondamentale INAF and "Programma di Ricerca Fondamentale INAF 2023 and 2024.

\end{acknowledgements}

%
 \bibliographystyle{aa_url} 
 \bibliography{refs} 

@ARTICLE{Baldini2024,
       author = {{Baldini}, P. and {Lanzuisi}, G. and {Brusa}, M. and {Merloni}, A. and {Gkimisi}, K. and {Perna}, M. and {L{\'o}pez}, I.~E. and {Bertola}, E. and {Igo}, Z. and {Waddell}, S. and {Musiimenta}, B. and {Aydar}, C. and {Arcodia}, R. and {Matzeu}, G.~A. and {Luminari}, A. and {Buchner}, J. and {Vignali}, C. and {Dadina}, M. and {Comastri}, A. and {Cresci}, G. and {Marchesi}, S. and {Gilli}, R. and {Tombesi}, F. and {Serafinelli}, R.},
        title = "{Winds of change: The nuclear and galaxy-scale outflows and the X-ray variability of 2MASS 0918+2117}",
      journal = {\aap},
         year = 2024,
        month = jun,
       volume = {686},
        pages = {A217},
          doi = {10.1051/0004-6361/202349071},
archivePrefix = {arXiv},
       eprint = {2402.16966},
 primaryClass = {astro-ph.HE},
       adsurl = {https://ui.adsabs.harvard.edu/abs/2024A&A...686A.217B}
}

@INPROCEEDINGS{Brusa2022,
       author = {{Brusa}, M. and {Matzeu}, G. and {Bianchi}, S. and {Piconcelli}, E. and {Dadina}, M. and {Lanzuisi}, G. and {Bongiorno}, A. and {Cappi}, M. and {Comastri}, A. and {Costanzo}, D. and {Cresci}, G. and {Duras}, F. and {Feruglio}, C. and {Gilli}, R. and {La Franca}, F. and {Luminari}, A. and {Matt}, G. and {Middei}, R. and {Nardini}, E. and {Ponti}, G. and {Tombesi}, F. and {Torresi}, E. and {Vignali}, C. and {Zaino}, A. and {Zappacosta}, L.},
        title = "{Supermassive Black Hole Winds in X-rays}",
    booktitle = {Memorie della Societa Astronomica Italiana},
         year = 2022,
       volume = {93},
        month = nov,
        pages = {48},
          doi = {10.36116/MEMSAIT_93N2_3.2022.6},
       adsurl = {https://ui.adsabs.harvard.edu/abs/2022MmSAI..93b..48B}
}

@article{Matzeu_2023,
		author = {{Matzeu}, G. A. and {Brusa}, M. and {Lanzuisi}, G. and {Dadina}, M. and {Bianchi}, S. and {Kriss}, G. and {Mehdipour}, M. and {Nardini}, E. and {Chartas}, G. and {Middei}, R. and {Piconcelli}, E. and {Gianolli}, V. and {Comastri}, A. and {Longinotti}, A. L. and {Krongold}, Y. and {Ricci}, F. and {Petrucci}, P. O. and {Tombesi}, F. and {Luminari}, A. and {Zappacosta}, L. and {Miniutti}, G. and {Gaspari}, M. and {Behar}, E. and {Bischetti}, M. and {Mathur}, S. and {Perna}, M. and {Giustini}, M. and {Grandi}, P. and {Torresi}, E. and {Vignali}, C. and {Bruni}, G. and {Cappi}, M. and {Costantini}, E. and {Cresci}, G. and {De Marco}, B. and {De Rosa}, A. and {Gilli}, R. and {Guainazzi}, M. and {Kaastra}, J. and {Kraemer}, S. and {La Franca}, F. and {Marconi}, A. and {Panessa}, F. and {Ponti}, G. and {Proga}, D. and {Ursini}, F. and {Baldini}, P. and {Fiore}, F. and {King}, A.R. and {Maiolino}, R. and {Matt}, G. and {Merloni}, A.},
		title = {Supermassive Black Hole Winds in X-rays: SUBWAYS. I. Ultra-fast outflows in quasars beyond the local Universe},
		journal = {\aap},
		year = {2023},
		month = {02},
		volume = {670},
		pages = {A182},
		doi = {10.1051/0004-6361/202245036},
		archivePrefix = {arXiv},
		eprint = {2212.02960},
		primaryClass = {astro-ph.HE},
		adsurl = {https://ui.adsabs.harvard.edu/abs/2023AA&AA...670A.182M}
	}

@article{Gianolli_2024,
		author = {{Gianolli}, V. E. and {Bianchi}, S. and {Petrucci}, P. -O. and {Brusa}, M. and {Chartas}, G. and {Lanzuisi}, G. and {Matzeu}, G. A. and {Parra}, M. and {Ursini}, F. and {Behar}, E. and {Bischetti}, M. and {Comastri}, A. and {Costantini}, E. and {Cresci}, G. and {Dadina}, M. and {De Marco}, B. and {De Rosa}, A. and {Fiore}, F. and {Gaspari}, M. and {Gilli}, R. and {Giustini}, M. and {Guainazzi}, M. and {King}, A.R. and {Kraemer}, S. and {Kriss}, G. and {Krongold}, Y. and {La Franca}, F. and {Longinotti}, A. L. and {Luminari}, A. and {Maiolino}, R. and {Marconi}, A. and {Mathur}, S. and {Matt}, G. and {Mehdipour}, M. and {Merloni}, A. and {Middei}, R. and {Miniutti}, G. and {Nardini}, E. and {Panessa}, F. and {Perna}, M. and {Piconcelli}, E. and {Ponti}, G. and {Ricci}, F. and {Serafinelli}, R. and {Tombesi}, F. and {Vignali}, C. and {Zappacosta}, L.},
		title = {{Supermassive Black Hole Winds in X-rays: SUBWAYS. III. A population study on ultra-fast outflows}},
		journal = {\aap},
		year = {2024},
		month = {07},
		volume = {687},
		pages = {A235},
		doi = {10.1051/0004-6361/202348908},
		adsurl = {https://ui.adsabs.harvard.edu/abs/2024AA&AA...687A.235G}
	}

@article{Mehdipour_2023,
	author = {{Mehdipour}, M. and {Kriss}, G. A. and {Brusa}, M. and {Matzeu}, G. A. and {Gaspari}, M. and {Kraemer}, S. B. and {Mathur}, S. and {Behar}, E. and {Bianchi}, S. and {Cappi}, M. and {Chartas}, G. and {Costantini}, E. and {Cresci}, G. and {Dadina}, M. and {De Marco}, B. and {De Rosa}, A. and {Dunn}, J. P. and {Gianolli}, V. E. and {Giustini}, M. and {Kaastra}, J. S. and {King}, A.R. and {Krongold}, Y. and {La Franca}, F. and {Lanzuisi}, G. and {Longinotti}, A. L. and {Luminari}, A. and {Middei}, R. and {Miniutti}, G. and {Nardini}, E. and {Perna}, M. and {Petrucci}, P. -O. and {Piconcelli}, E. and {Ponti}, G. and {Ricci}, F. and {Tombesi}, F. and {Ursini}, F. and {Vignali}, C. and {Zappacosta}, L.},
	title = {{Supermassive Black Hole Winds in X-rays: SUBWAYS. II. HST UV spectroscopy of winds at intermediate redshifts}},
	journal = {\aap},
	year = {2023},
	month = {02},
	volume = {670},
	pages = {A183},
	doi = {10.1051/0004-6361/202245047},
	archivePrefix = {arXiv},
	eprint = {2212.02961},
	primaryClass = {astro-ph.HE},
	adsurl = {https://ui.adsabs.harvard.edu/abs/2023AA&AA...670A.183M}
}

@article{Shimwell_2022,
				author = {{Shimwell}, T. W. and {Hardcastle}, M. and {Tasse}, C. and {Best}, P. N. and {Röttgering}, H. J. A. and {Williams}, W. L. and {Botteon}, A. and {Drabent}, A. and {Mechev} A. and {Shulevski}, A. and {van Weeren}, R. J. and {Bester}, L. and {Brüggen}, M. and {Brunetti}, G. and {Callingham}, J. R. and {Chyży}, K. T. and {Conway}, J. E. and {Dijkema}, T. J. and {Duncan}, K. and {de Gasperin}, F. and {Hale}, C. L. and {Haverkorn}, M. and {Hugo}, B. and {Jackson}, N. and {Mevius}, M. and {Miley}, G. K. and {Morabito}, L. K. and {Morganti}, R. and {Offringa}, A. and {Oonk}, J. B. R. and {Rafferty}, D. and {Sabater}, J. and {Smith}, D. J. B. and {Schwarz}, D. J. and {Smirnov}, O. and {O’Sullivan}, S. P. and {Vedantham}, H. and {White}, G. J. and {Albert}, J. G. and {Alegre}, L. and {Asabere}, B. and {Bacon}, D. J. and {Bonafede}, A. and {Bonnassieux}, E. and {Brienza}, M. and {Bilicki}, M. and {Bonato}, M. and {Calistro Rivera}, G. and {Cassano}, R. and {Cochrane}, R. and {Croston}, J. H. and {Cuciti}, V. and {Dallacasa}, D. and {Danezi}, A. and {Dettmar}, R. J. and {Di Gennaro}, G. and {Edler}, H. W. and {Enßlin}, T. A. and {Emig}, K. L. and {Franzen}, T. M. O. and {García-Vergara}, C. and {Grange}, Y. G. and {Gürkan}, G. and {Hajduk}, M. and {Heald}, G. and {Heesen}, V. and {Hoang}, D. N. and {Hoeft}, M. and {Horellou}, C. and {Iacobelli}, M. and {Jamrozy}, M. and {Jelić}, V. and {Kondapally}, R. and {Kukreti}, P. and {Kunert-Bajraszewska}, M. and {Magliocchetti}, M. and {Mahatma}, V. and {Małek}, K. and {Mandal}, S. and {Massaro}, F. and {Meyer-Zhao}, Z. and {Mingo}, B. and {Mostert}, R. I. J. and {Nair}, D. G. and {Nakoneczny}, S. J. and {Nikiel-Wroczyński}, B. and {Orrú}, E. and {Pajdosz-Śmierciak}, U. and {Pasini}, T. and {Prandoni}, I. and {van Piggelen}, H. E. and {Rajpurohit}, K. and {Retana-Montenegro}, E. and {Riseley}, C. J. and {Rowlinson}, A. and {Saxena}, A. and {Schrijvers}, C. and {Sweijen}, F. and {Siewert}, T. M. and {Timmerman}, R. and {Vaccari}, M. and {Vink}, J. and {West}, J. L. and {Wołowska}, A. and {Zhang}, X. and {Zheng}, J.},
			title = {The LOFAR Two-metre Sky Survey - V. Second data release},
				DOI= {10.1051/0004-6361/202142484},
				url= {https://doi.org/10.1051/0004-6361/202142484},
				journal = {\aap},
				year = {2022},
				volume = {659},
				pages = {A1}}

@article{Shimwell_2017,
	author = {{Shimwell} and {R{\"o}ttgering}, H. J. A. and {Best}, P. N. and {Williams}, W. L. and {Dijkema}, T. J. and {de Gasperin}, F. and {Hardcastle}, M. J. and {Heald}, G. H. and {Hoang}, D. N. and {Horneffer}, A. and {Intema}, H. and {Mahony}, E. K. and {Mandal}, S. and {Mechev}, A. P. and {Morabito}, L. and {Oonk}, J. B. R. and {Rafferty}, D. and {Retana-Montenegro}, E. and {Sabater}, J. and {Tasse}, C. and {van Weeren}, R. J. and {Br{\"u}ggen}, M. and {Brunetti}, G. and {Chy{\.z}y}, K. T. and {Conway}, J. E. and {Haverkorn}, M. and {Jackson}, N. and {Jarvis}, M. J. and {McKean}, J. P. and {Miley}, G. K. and {Morganti}, R. and {White}, G. J. and {Wise}, M. W. and {van Bemmel}, I. M. and {Beck}, R. and {Brienza}, M. and {Bonafede}, A. and {Calistro Rivera}, G. and {Cassano}, R. and {Clarke}, A. O. and {Cseh}, D. and {Deller}, A. and {Drabent}, A. and {van Driel}, W. and {Engels}, D. and {Falcke}, H. and {Ferrari}, C. and {Fr{\"o}hlich}, S. and {Garrett}, M. A. and {Harwood}, J. J. and {Heesen}, V. and {Hoeft}, M. and {Horellou}, C. and {Israel}, F. P. and {Kapi{\'n}ska}, A. D. and {Kunert-Bajraszewska}, M. and {McKay}, D. J. and {Mohan}, N. R. and {Orr{\'u}}, E. and {Pizzo}, R. F. and {Prandoni}, I. and {Schwarz}, D. J. and {Shulevski}, A. and {Sipior}, M. and {Smith}, D. J. B. and {Sridhar}, S. S. and {Steinmetz}, M. and {Stroe}, A. and {Varenius}, E. and {van der Werf}, P. P. and {Zensus}, J. A. and {Zwart}, J. T. L.},
	title = {The LOFAR Two-metre Sky Survey. I. Survey description and preliminary data release},
	journal = {\aap},
	year = {2017},
	month = {02},
	volume = {598},
	pages = {A104},
	doi = {10.1051/0004-6361/201629313},
	archivePrefix = {arXiv},
	eprint = {1611.02700},
	primaryClass = {astro-ph.IM},
	adsurl = {https://ui.adsabs.harvard.edu/abs/2017AA&AA...598A.104S}
}

@article{Lacy_2020,
	author ={{Lacy}, M. and {Baum}, S. A. and {Chandler}, C. J. and {Chatterjee}, S. and {Clarke}, T. E. and {Deustua}, S. and {English}, J. and {Farnes}, J. and {Gaensler}, B. M. and {Gugliucci}, N. and {Hallinan}, G. and {Kent}, B. R. and {Kimball}, A. and {Law}, C. J. and {Lazio}, T. J. W. and {Marvil}, J. and {Mao}, S. A. and {Medlin}, D. and {Mooley}, K. and {Murphy}, E. J. and {Myers}, S. and {Osten}, R. and {Richards}, G. T. and {Rosolowsky}, E. and {Rudnick}, L. and {Schinzel}, F. and {Sivakoff}, G. R. and {Sjouwerman}, L. O. and {Taylor}, R. and {White}, R. L. and {Wrobel}, J. and {Andernach}, H. and {Beasley}, A. J. and {Berger}, E. and {Bhatnager}, S. and {Birkinshaw}, M. and {Bower}, G. C. and {Brandt}, W. N. and {Brown}, S. and {Burke-Spolaor}, S. and {Butler}, B. J. and {Comerford}, J. and {Demorest}, P. B. and {Fu}, H. and {Giacintucci}, S. and {Golap}, K. and {G{\"u}th}, T. and {Hales}, C. A. and {Hiriart}, R. and {Hodge}, J. and {Horesh}, A. and {Ivezi{\'c}}, {\v{Z}}. and {Jarvis}, M. J. and {Kamble}, A. and {Kassim}, N. and {Liu}, X. and {Loinard}, L. and {Lyons}, D. K. and {Masters}, J. and {Mezcua}, M. and {Moellenbrock}, G. A. and {Mroczkowski}, T. and {Nyland}, K. and {O'Dea}, C. P. and {O'Sullivan}, S. P. and {Peters}, W. M. and {Radford}, K. and {Rao}, U. and {Robnett}, J. and {Salcido}, J. and {Shen}, Y. and {Sobotka}, A. and {Witz}, S. and {Vaccari}, M. and {van Weeren}, R. J. and {Vargas}, A. and {Williams}, P. K. G. and {Yoon}, I.},
	title = {The Karl G. Jansky Very Large Array Sky Survey (VLASS). Science Case and Survey Design},
	journal = {\pasp},
	year = {2020},
	month = {03},
	volume = {132},
	number = {1009},
	pages = {035001},
	doi = {10.1088/1538-3873/ab63eb},
	archivePrefix = {arXiv},
	eprint = {1907.01981},
	primaryClass = {astro-ph.IM},
	adsurl = {https://ui.adsabs.harvard.edu/abs/2020PASP..132c5001L}
}

@article{Chartas_2002,
       author = {{Chartas}, G. and {Brandt}, W.~N. and {Gallagher}, S.~C. and {Garmire}, G.~P.},
        title = "{CHANDRA Detects Relativistic Broad Absorption Lines from APM 08279+5255}",
      journal = {\apj},
         year = 2002,
        month = nov,
       volume = {579},
       number = {1},
        pages = {169-175},
          doi = {10.1086/342744},
archivePrefix = {arXiv},
       eprint = {astro-ph/0207196},
 primaryClass = {astro-ph},
       adsurl = {https://ui.adsabs.harvard.edu/abs/2002ApJ...579..169C}
}

@article{Lacy_2016,
	author = {{Lacy}, Mark and {Baum}, Stefi Alison and {Chandler}, Claire J. and {Chatterjee}, Shami and {Murphy}, Eric J. and {Myers}, Steven T. and {VLASS Survey Science Group}},
	title = {The VLA Sky Survey (VLASS): Description and Science Goals},
	booktitle = {American Astronomical Society Meeting Abstracts \#227},
	year = {2016},
	journal = {BAAS},
	volume = {227},
	month = {01},
	pages = {324.09},
	adsurl = {https://ui.adsabs.harvard.edu/abs/2016AAS...22732409L}
}

@article{McConnell_2020,
	author={{McConnell}, D. and Hale, C. L. and Lenc, E. and Banfield, J. K. and Heald, George and Hotan, A. W. and Leung, James K. and Moss, Vanessa A. and Murphy, Tara and O’Brien, Andrew and et al.},
	title={The Rapid ASKAP Continuum Survey I: Design and first results}, 
	volume={37},
    DOI={10.1017/pasa.2020.41},
	journal={\pasa},  
	year={2020}, 
    pages={e048}
	  }

@article{Hale_2021, 
	title={The Rapid ASKAP Continuum Survey Paper II: First Stokes I Source Catalogue Data Release}, 
	volume={38}, 
	DOI={10.1017/pasa.2021.47}, 
	journal={\pasa}, 
	author={{Hale}, Catherine L. and {McConnell}, D. and {Thomson}, A. J. M. and {Lenc}, E. and {Heald}, G. H. and {Hotan}, A. W. and {Leung}, J. K. and {Moss}, V. A. and {Murphy}, T. and {Pritchard}, J. and et al.}, 
	year={2021}, 
	pages={e058}
	}

@article{Becker_1995,
	author = {{Becker}, Robert H. and {White}, Richard L. and {Helfand}, David J.},
	title = {{The FIRST Survey: Faint Images of the Radio Sky at Twenty Centimeters}},
	journal = {\apj},
	year = {1995},
	month = {09},
	volume = {450},
	pages = {559},
	doi = {10.1086/176166},
	adsurl = {https://ui.adsabs.harvard.edu/abs/1995ApJ...450..559B}
}

@article{Gordon_2021,
			author = {{Gordon}, Yjan A. and {Boyce}, Michelle M. and {O'Dea}, Christopher P. and {Rudnick}, Lawrence and {Andernach}, Heinz and {Vantyghem}, Adrian N. and {Baum}, Stefi A. and {Bui}, Jean-Paul and {Dionyssiou}, Mathew and {Safi-Harb}, Samar and {Sander}, Isabel},
			title = {A Quick Look at the 3 GHz Radio Sky. I. Source Statistics from the Very Large Array Sky Survey},
			journal = {\apjs},
			year = {2021},
			month = {08},
			volume = {255},
			number = {2},
			pages = {30},
			doi = {10.3847/1538-4365/ac05c0},
			archivePrefix = {arXiv},
			eprint = {2102.11753},
			primaryClass = {astro-ph.GA},
			adsurl = {https://ui.adsabs.harvard.edu/abs/2021ApJS..255...30G}}

@article{CASATeam_2022,
	year = {2022},
	month = {11},
	publisher = {The Astronomical Society of the Pacific},
	volume = {134},
	number = {1041},
	pages = {114501},
	author ={{The CASA Team} and Bean, Ben and Bhatnagar, Sanjay and Castro, Sandra and Meyer, Jennifer Donovan and Emonts, Bjorn and Garcia, Enrique and Garwood, Robert and Golap, Kumar and Villalba, Justo Gonzalez and Harris, Pamela and Hayashi, Yohei and Hoskins, Josh and Hsieh, Mingyu and Jagannathan, Preshanth and Kawasaki, Wataru and Keimpema, Aard and Kettenis, Mark and Lopez, Jorge and Marvil, Joshua and Masters, Joseph and McNichols, Andrew and Mehringer, David and Miel, Renaud and Moellenbrock, George and Montesino, Federico and Nakazato, Takeshi and Ott, Juergen and Petry, Dirk and Pokorny, Martin and Raba, Ryan and Rau, Urvashi and Schiebel, Darrell and Schweighart, Neal and Sekhar, Srikrishna and Shimada, Kazuhiko and Small, Des and Steeb, Jan-Willem and Sugimoto, Kanako and Suoranta, Ville and Tsutsumi, Takahiro and van Bemmel, Ilse M. and Verkouter, Marjolein and Wells, Akeem and Xiong, Wei and Szomoru, Arpad and Griffith, Morgan and Glendenning, Brian and Kern, Jeff},
	title = {CASA, the Common Astronomy Software Applications for Radio Astronomy},
	journal = {\pasp}
}

@misc{Offringa_2010,
	author = {{Offringa}, A. R.},
	title = {{AOFlagger: RFI Software}},
	howpublished = {Astrophysics Source Code Library, record ascl:1010.017},
	year = {2010},
	month = {10},
	eid = {ascl:1010.017},
}

@article{Offringa_2014,
	author = {{Offringa}, A. R. and {McKinley}, B. and {Hurley-Walker} et al.},
	title = {WSClean: an implementation of a fast, generic wide-field imager for radio astronomy},
	volume = {444},
	number = {1},
	pages = {606--619},
	year = {2014},
	doi = {10.1093/mnras/stu1368},
	journal = {\mnras}
}

@article{Intema_2017,
	author = {{Intema}, H. T. and {Jagannathan}, P. and {Mooley}, K. P. and {Frail}, D. A.},
	title = {The GMRT 150 MHz all-sky radio survey. First alternative data release TGSS ADR1},
	journal = {\aap},
	year = {2017},
	month = {02},
	volume = {598},
	pages = {A78},
	doi = {10.1051/0004-6361/201628536},
	archivePrefix = {arXiv},
	eprint = {1603.04368},
	primaryClass = {astro-ph.CO},
	adsurl = {https://ui.adsabs.harvard.edu/abs/2017AA&AA...598A..78I}
}

@article{Zhang_2016,
	doi = {10.3847/2041-8205/819/2/L27},
	adsurl = {https://dx.doi.org/10.3847/2041-8205/819/2/L27},
	year = {2016},
	month ={03},
	publisher = {The American Astronomical Society},
	volume = {819},
	number = {2},
	pages = {L27},
	author = {{Zhang}, Z. and {Shi}, Y. and {Rieke}, G. H. and {Xia}, X. and {Wang}, Y. and {Sun}, B. and {Wan}, L.},
	title = {DISTRIBUTION OF QUASAR HOSTS ON THE GALAXY MAIN SEQUENCE PLANE},
	journal = {\apjl}
}

@article{Tombesi_2010,
	author = {{Tombesi}, F. and {Cappi}, M. and {Reeves}, J. N. and {Palumbo}, G. G. C. and {Yaqoob}, T. and {Braito}, V. and {Dadina}, M.},
	title = {Evidence for ultra-fast outflows in radio-quiet AGNs. I. Detection and statistical incidence of Fe K-shell absorption lines},
	journal = {\aap},
	year = {2010},
	month = {10},
	volume = {521},
	pages = {A57},
	doi = {10.1051/0004-6361/200913440},
	archivePrefix = {arXiv},
	eprint = {1006.2858},
	primaryClass = {astro-ph.HE},
	adsurl = {https://ui.adsabs.harvard.edu/abs/2010AA&AA...521A..57T}
}

@article{Wang_2021,
   title={The obstructed jet in Mrk 231},
   volume={504},
   ISSN={1365-2966},
   url={http://dx.doi.org/10.1093/mnras/stab587},
   DOI={10.1093/mnras/stab587},
   number={3},
   journal={\mnras},
   publisher={Oxford University Press (OUP)},
   author={Wang, Ailing and An, Tao and Jaiswal, Sumit and Mohan, Prashanth and Wang, Yuchan and Baan, Willem A and Zhang, Yingkang and Yang, Xiaolong},
   year={2021},
   month=mar, pages={3823–3830} }

@article{Maksym_2023,
doi = {10.3847/1538-4357/acd7f1},
adsurl = {https://doi.org/10.3847/1538-4357/acd7f1},
year = {2023},
month = {jul},
publisher = {The American Astronomical Society},
volume = {951},
number = {2},
pages = {146},
author = {Maksym, W. Peter and Elvis, Martin and Fabbiano, Giuseppina and Trindade Falcão, Anna and Kraemer, Steven B. and Fischer, Travis C. and Crenshaw, D. Michael and Storchi-Bergmann, Thaisa},
title = {A UFO Seen Edge-on? Resolving Ultrafast Outflow Emission on ∼200 pc Scales with Chandra in the Active Nucleus of Mrk 34},
journal = {\apj}}

@article{Falcke_1998,
       author = {{Falcke}, Heino and {Wilson}, Andrew S. and {Simpson}, Chris},
        title = "{Hubble Space Telescope and VLA Observations of Seyfert 2 Galaxies: The Relationship between Radio Ejecta and the Narrow-Line Region}",
      journal = {\apj},
         year = 1998,
        month = jul,
       volume = {502},
       number = {1},
        pages = {199-217},
          doi = {10.1086/305886},
archivePrefix = {arXiv},
       eprint = {astro-ph/9801086},
 primaryClass = {astro-ph},
       adsurl = {https://ui.adsabs.harvard.edu/abs/1998ApJ...502..199F}
}

@article{Gaspari_2020,
       author = {{Gaspari}, Massimo and {Tombesi}, Francesco and {Cappi}, Massimo},
        title = "{Linking macro-, meso- and microscales in multiphase AGN feeding and feedback}",
      journal = {Nature Astronomy},
         year = 2020,
        month = jan,
       volume = {4},
        pages = {10-13},
          doi = {10.1038/s41550-019-0970-1},
archivePrefix = {arXiv},
       eprint = {2001.04985},
 primaryClass = {astro-ph.GA},
       adsurl = {https://ui.adsabs.harvard.edu/abs/2020NatAs...4...10G}
}

@article{Zanchettin_2023,
       author = {{Zanchettin}, M.~V. and {Feruglio}, C. and {Massardi}, M. and {Lapi}, A. and {Bischetti}, M. and {Cantalupo}, S. and {Fiore}, F. and {Bongiorno}, A. and {Malizia}, A. and {Marinucci}, A. and {Molina}, M. and {Piconcelli}, E. and {Tombesi}, F. and {Travascio}, A. and {Tozzi}, G. and {Tripodi}, R.},
        title = "{NGC 2992: Interplay between the multiphase disc, wind, and radio bubbles}",
      journal = {\aap},
         year = 2023,
        month = nov,
       volume = {679},
        pages = {A88},
          doi = {10.1051/0004-6361/202245729},
archivePrefix = {arXiv},
       eprint = {2308.04108},
 primaryClass = {astro-ph.GA},
       adsurl = {https://ui.adsabs.harvard.edu/abs/2023A&A...679A..88Z}
}

@article{Longinotti_2023,
   title={NOEMA spatially resolved view of the multiphase outflow in IRAS17020+4544: a shocked wind in action?},
   volume={521},
   ISSN={1365-2966},
   url={http://dx.doi.org/10.1093/mnras/stad540},
   DOI={10.1093/mnras/stad540},
   number={2},
   journal={\mnras},
   publisher={Oxford University Press (OUP)},
   author={Longinotti, Anna Lia and Salomé, Q and Feruglio, C and Krongold, Y and García-Burillo, S and Giroletti, M and Panessa, F and Stanghellini, C and Vega, O and Patiño-Álvarez, V M and Chavushyan, V and Elías-Chavez, M and Robleto-Orús, A},
   year={2023},
   month=feb, pages={2134–2148} }

@article{yamada2024,
      author = {{Yamada}, Tomoya and {Sakai}, Nobuyuki and {Inoue}, Yoshiyuki and {Michiyama}, Tomonari},
        title = "{Deciphering Radio Emissions from Accretion Disk Winds in Radio-quiet Active Galactic Nuclei}",
      journal = {\apj},
         year = 2024,
        month = jun,
       volume = {968},
       number = {2},
        pages = {116},
          doi = {10.3847/1538-4357/ad3a63},
archivePrefix = {arXiv},
       eprint = {2404.04632},
 primaryClass = {astro-ph.HE},
       adsurl = {https://ui.adsabs.harvard.edu/abs/2024ApJ...968..116Y}
}

@article{Cresci_2023,
       author = {{Cresci}, G. and {Tozzi}, G. and {Perna}, M. and {Brusa}, M. and {Marconcini}, C. and {Marconi}, A. and {Carniani}, S. and {Brienza}, M. and {Giroletti}, M. and {Belfiore}, F. and {Ginolfi}, M. and {Mannucci}, F. and {Ulivi}, L. and {Scholtz}, J. and {Venturi}, G. and {Arribas}, S. and {{\"U}bler}, H. and {D'Eugenio}, F. and {Mingozzi}, M. and {Balmaverde}, B. and {Capetti}, A. and {Parlanti}, E. and {Zana}, T.},
        title = "{Bubbles and outflows: The novel JWST/NIRSpec view of the z = 1.59 obscured quasar XID2028}",
      journal = {\aap},
         year = 2023,
        month = apr,
       volume = {672},
        pages = {A128},
          doi = {10.1051/0004-6361/202346001},
archivePrefix = {arXiv},
       eprint = {2301.11060},
 primaryClass = {astro-ph.GA},
       adsurl = {https://ui.adsabs.harvard.edu/abs/2023A&A...672A.128C}
}

@article{Seyfert_1943,
	author = {{Seyfert}, C. K.},
	title = {Nuclear Emission in Spiral Nebulae.},
	journal = {\apj},
	year = {1943},
	month = {01},
	volume = {97},
	pages = {28},
	doi = {10.1086/144488},
	adsurl = {https://ui.adsabs.harvard.edu/abs/1943ApJ....97...28S}
}

@article{Singh_2013,
       author = {{Singh}, R. and {van de Ven}, G. and {Jahnke}, K. and {Lyubenova}, M. and {Falc{\'o}n-Barroso}, J. and {Alves}, J. and {Cid Fernandes}, R. and {Galbany}, L. and {Garc{\'\i}a-Benito}, R. and {Husemann}, B. and {Kennicutt}, R. C. and {Marino}, R. A. and {M{\'a}rquez}, I. and {Masegosa}, J. and {Mast}, D. and {Pasquali}, A. and {S{\'a}nchez}, S. F. and {Walcher}, J. and {Wild}, V. and {Wisotzki}, L. and {Ziegler}, B.},
        title = {The nature of LINER galaxies:. Ubiquitous hot old stars and rare accreting black holes},
      journal = {\aap},
         year = {2013},
        month = {10},
       volume = {558},
        pages = {A43},
          doi = {10.1051/0004-6361/201322062},
archivePrefix = {arXiv},
       eprint = {1308.4271},
 primaryClass = {astro-ph.GA},
       adsurl = {https://ui.adsabs.harvard.edu/abs/2013AA&AA...558A..43S}
}

@article{Baldi_2023,
	author={{Baldi}, R. D.},
	 title = "{The nature of compact radio sources: the case of FR 0 radio galaxies}",
      journal = {\aapr},
         year = 2023,
        month = dec,
       volume = {31},
       number = {1},
        pages = {3},
          doi = {10.1007/s00159-023-00148-3},
archivePrefix = {arXiv},
       eprint = {2307.08379},
 primaryClass = {astro-ph.GA},
       adsurl = {https://ui.adsabs.harvard.edu/abs/2023A&ARv..31....3B}}

@article{Baldi_2015,
       author = {{Baldi}, Ranieri D. and {Capetti}, Alessandro and {Giovannini}, Gabriele},
        title = "{Pilot study of the radio-emitting AGN population: the emerging new class of FR 0 radio-galaxies}",
      journal = {\aap},
         year = 2015,
        month = apr,
       volume = {576},
        pages = {A38},
          doi = {10.1051/0004-6361/201425426},
archivePrefix = {arXiv},
       eprint = {1502.00427},
 primaryClass = {astro-ph.GA},
       adsurl = {https://ui.adsabs.harvard.edu/abs/2015A&A...576A..38B}
}

@article{Wright_2010,
	author = {{Wright}, Edward L. and {Eisenhardt}, Peter R. M. and {Mainzer}, Amy K. and {Ressler}, Michael E. and {Cutri}, Roc M. and {Jarrett}, Thomas and {Kirkpatrick}, J. Davy and {Padgett}, Deborah and {McMillan}, Robert S. and {Skrutskie}, Michael and {Stanford}, S. A. and {Cohen}, Martin and {Walker}, Russell G. and {Mather}, John C. and {Leisawitz}, David and {Gautier}, III, Thomas N. and {McLean}, Ian and {Benford}, Dominic and {Lonsdale}, Carol J. and {Blain}, Andrew and {Mendez}, Bryan and {Irace}, William R. and {Duval}, Valerie and {Liu}, Fengchuan and {Royer}, Don and {Heinrichsen}, Ingolf and {Howard}, Joan and {Shannon}, Mark and {Kendall}, Martha and {Walsh}, Amy L. and {Larsen}, Mark and {Cardon}, Joel G. and {Schick}, Scott and {Schwalm}, Mark and {Abid}, Mohamed and {Fabinsky}, Beth and {Naes}, Larry and {Tsai}, Chao-Wei},
	title = {The Wide-field Infrared Survey Explorer (WISE): Mission Description and Initial On-orbit Performance},
	journal = {\aj},
	year = {2010},
	month = {12},
	volume = {140},
	number = {6},
	pages = {1868--1881},
	doi = {10.1088/0004-6256/140/6/1868},
	archivePrefix = {arXiv},
	eprint = {1008.0031},
	primaryClass = {astro-ph.IM},
	adsurl = {https://ui.adsabs.harvard.edu/abs/2010AJ....140.1868W}
}

@article{Baldi_2016,
	title={The new class of FR 0 radio galaxies},
	volume={337},
	ISSN={1521-3994},
	url={http://dx.doi.org/10.1002/asna.201512275},
	DOI={10.1002/asna.201512275},
	number={1–2},
	journal={Astronomische Nachrichten},
	publisher={Wiley},
	author={{Baldi}, R. D. and {Capetti}, A. and {Giovannini}, G.},
	year={2016},
	month={02}, pages={114--119} }

@article{FR_1974,
       author = {{Fanaroff}, B. L. and {Riley}, J. M.},
        title = {The morphology of extragalactic radio sources of high and low luminosity},
      journal = {\mnras},
         year = {1974},
        month = {05},
       volume = {167},
        pages = {31P--36P},
          doi = {10.1093/mnras/167.1.31P},
       adsurl = {https://ui.adsabs.harvard.edu/abs/1974MNRAS.167P..31F}
}

@inproceedings{Mancuso_2017,
	author = {{Mancuso}, C. and {Lapi}, A. and {Massardi}, M. and {Danese}, L.},
	title = {The Main Sequence of Star Forming Galaxies and AGNs at High-Z: Evidence for in situ Coevolution},
	booktitle = {Workshop sull'Astronomia Millimetrica in Italia},
	year = {2017},
	month = {11},
	pages = {2},
	doi = {10.5281/zenodo.1048809},
	adsurl = {https://ui.adsabs.harvard.edu/abs/2017ami..confE...2M}
}

@article{TW_2003,
	author = {{Terashima}, Y. and {Wilson}, A. S.},
	title = {Chandra Snapshot Observations of Low-Luminosity Active Galactic Nuclei with a Compact Radio Source},
	journal = {\apj},
	year = {2003},
	month = {01},
	volume = {583},
	number = {1},
	pages = {145--158},
	doi = {10.1086/345339},
	archivePrefix = {arXiv},
	eprint = {astro-ph/0209607},
	primaryClass = {astro-ph},
	adsurl = {https://ui.adsabs.harvard.edu/abs/2003ApJ...583..145T}
}

@article{Panessa_2007,
		author = {{Panessa}, F. and {Barcons}, X. and {Bassani}, L. and {Cappi}, M. and {Carrera}, F. and {Ho}, L. and {Pellegrini}, S.},
		title = {The X-ray and radio connection in low-luminosity active nuclei},
		DOI= {10.1051/0004-6361:20066943},
		url= {https://doi.org/10.1051/0004-6361:20066943},
		journal = {\aap},
		year = {2007},
		volume = {467},
		number = {2},
		pages = {519-527}}

@article{Laor_Behar_2008,
    author={{Laor}, A. and {Behar}, E.},
	title={On the origin of radio emission in radio-quiet quasars},
	volume={390},
	ISSN={1365-2966},
	url={http://dx.doi.org/10.1111/j.1365-2966.2008.13806.x},
	DOI={10.1111/j.1365-2966.2008.13806.x},
	number={2},
	journal={\mnras},
	publisher={Oxford University Press (OUP)},
	year={2008},
	month={10}, pages={847--862} }

@article{Padovani_2017,
	author = {{Padovani}, P.},
	title = {On the two main classes of active galactic nuclei},
	journal = {\nat},
	year = {2017},
	month = {08},
	volume = {1},
	pages = {0194},
	doi = {10.1038/s41550-017-0194},
	archivePrefix = {arXiv},
	eprint = {1707.08069},
	primaryClass = {astro-ph.GA},
	adsurl = {https://ui.adsabs.harvard.edu/abs/2017NatAs...1E.194P}
}

@article{Villar_Martin_2017,
		author = {{Villar-Mart{\'\i}n}, M. and {Emonts}, B. and {Cabrera Lavers}, A. and {Tadhunter}, C. and {Mukherjee}, D. and {Humphrey}, A. and {Rodr{\'\i}guez Zaur{\'\i}n}, J. and {Ramos Almeida}, C. and {P{\'e}rez Torres}, M. and {Bessiere}, P.},
		title = {Galaxy-wide radio-induced feedback in a radio-quiet quasar},
		journal = {\mnras},
		year = {2017},
		month = {12},
		volume = {472},
		number = {4},
		pages = {4659--4678},
		doi = {10.1093/mnras/stx2209},
		adsurl = {https://ui.adsabs.harvard.edu/abs/2017MNRAS.472.4659V}
	}

@article{Condon_1992,
	author = {{Condon}, J. J.},
	title = {Radio emission from normal galaxies.},
	journal = {\araa},
	year = {1992},
	month = {01},
	volume = {30},
	pages = {575--611},
	doi = {10.1146/annurev.aa.30.090192.003043},
	adsurl = {https://ui.adsabs.harvard.edu/abs/1992ARAA&AA..30..575C}
}

@article{Cluver_2017,
	doi = {10.3847/1538-4357/aa92c7},
	adsurl = {https://dx.doi.org/10.3847/1538-4357/aa92c7},
	year = {2017},
	month = {11},
	publisher = {American Astronomical Society},
	volume = {850},
	number = {1},
	pages = {68},
	author = {{Cluver}, M. E. and {Jarrett}, T. H. and {Dale}, D. A. and {Smith}, J.-D. T. and {August}, T. and {Brown}, M. J. I.},
	title = {Calibrating Star Formation in WISE Using Total Infrared Luminosity},
	journal = {\apj}
}

@article{Panessa_Giroletti_2013,
		author = {{Panessa}, F. and {Giroletti}, M.},
		title = {Sub-parsec radio cores in nearby Seyfert galaxies},
		journal = {\mnras},
		year = {2013},
		month = {06},
		volume = {432},
		number = {2},
		pages = {1138--1143},
		doi = {10.1093/mnras/stt547},
		adsurl = {https://ui.adsabs.harvard.edu/abs/2013MNRAS.432.1138P}
	}

@article{Baldi_2022,
	author = {{Baldi}, R. D. and {Laor}, A. and {Behar}, E. and {Horesh}, A. and  {Panessa}, F. and {McHardy}, I.  and {Kimball}, A.},
	title = {The PG-RQS survey. Building the radio spectral distribution of radio-quiet quasars. I. The 45-GHz data},
	journal = {\mnras},
	volume = {510},
	number = {1},
	pages = {1043-1058},
	year = {2022},
	month = {11},
	issn = {0035-8711},
	doi = {10.1093/mnras/stab3445},
	adsurl = {https://doi.org/10.1093/mnras/stab3445}}

@article{Zakamska_2016,
	author = {{Zakamska}, N. L. and {Lampayan}, K. and {Petric}, A. and {Dicken}, D. and {Greene}, J. E. and {Heckman}, T. M. and {Hickox}, R. C. and {Ho},  L. C. and {Krolik},  J. H. and {Nesvadba}, N. P. H. and {Strauss}, M. A. and {Geach}, J. E. and {Oguri}, M. and {Strateva}, I. V.},
	title = {Star formation in quasar hosts and the origin of radio emission in radio-quiet quasars},
	journal = {\mnras},
	year = {2016},
	month = {02},
	volume = {455},
	number = {4},
	pages = {4191--4211},
	doi = {10.1093/mnras/stv2571},
	archivePrefix = {arXiv},
	eprint = {1511.00013},
	primaryClass = {astro-ph.GA},
	adsurl = {https://ui.adsabs.harvard.edu/abs/2016MNRAS.455.4191Z}
}

@article{Zakamska_Greene_2014,
	author = {{Zakamska}, N. L. and {Greene}, J. E.},
	title = {Quasar feedback and the origin of radio emission in radio-quiet quasars},
	journal = {\mnras},
	volume = {442},
	number = {1},
	pages = {784--804},
	year = {2014},
	month = {06},
	issn = {0035-8711},
	doi = {10.1093/mnras/stu842}}

@article{Richards_2021,
	author = {{Richards}, Gordon T. and {McCaffrey}, Trevor V. and {Kimball}, Amy and {Rankine}, Amy L. and {Matthews}, James H. and {Hewett}, Paul C. and {Rivera}, Angelica B.},
	title = {Probing the Wind Component of Radio Emission in Luminous High-redshift Quasars},
	journal = {\aj},
	year = {2021},
	month = {12},
	volume = {162},
	number = {6},
	pages = {270},
	doi = {10.3847/1538-3881/ac283b},
	archivePrefix = {arXiv},
	eprint = {2106.07783},
	primaryClass = {astro-ph.GA},
	adsurl = {https://ui.adsabs.harvard.edu/abs/2021AJ....162..270R}
}

@article{Raginski_Laor_2016,
	author = {{Raginski}, I.  and {Laor}, A. },
	title = {AGN coronal emission models – I. The predicted radio emission},
	journal = {\mnras},
	volume = {459},
	number = {2},
	pages = {2082--2096},
	year = {2016},
	month = {04},
	issn = {0035-8711},
	doi = {10.1093/mnras/stw772},
	adsurl = {https://doi.org/10.1093/mnras/stw772},
	eprint = {https://academic.oup.com/mnras/article-pdf/459/2/2082/8041696/stw772.pdf}
}

@article{Patil_2020,
	doi = {10.3847/1538-4357/ab9011},
	adsurl = {https://dx.doi.org/10.3847/1538-4357/ab9011},
	year = {2020},
	month = {06},
	publisher = {The American Astronomical Society},
	volume = {896},
	number = {1},
	pages = {18},
	author = {{Patil}, P. and {Nyland}, K. and {Whittle}, M. and {Lonsdale}, C. and {Lacy}, M. and {Lonsdale}, C. and {Mukherjee}, D. and {Trapp}, A. C. and {Kimball}, A. E. and {Lanz}, L. and  {J. Wilkes}, B. and {Blain}, A. and {Harwood}, J. J. and {Efstathiou}, A. and {Vlahakis}, C.},
	title = {High-resolution VLA Imaging of Obscured Quasars: Young Radio Jets Caught in a Dense ISM},
	journal = {\apj}}

@article{Guedel_1993,
	author = {{Guedel}, M. and {Benz}, A. O.},
	title = {{X-Ray/Microwave Relation of Different Types of Active Stars}},
	journal = {\apjl},
	year = {1993},
	month = {03},
	volume = {405},
	pages = {L63},
	doi = {10.1086/186766},
	adsurl = {https://ui.adsabs.harvard.edu/abs/1993ApJ...405L..63G}
}

@article{Chen_2024,
       author = {{Chen}, Sina and {Laor}, Ari and {Behar}, Ehud and {Baldi}, Ranieri D. and {Gelfand}, Joseph D. and {Kimball}, Amy E. and {McHardy}, Ian M. and {Orosz}, Gabor and {Paragi}, Zsolt},
        title = "{Windy or Not: Radio Parsec-scale Evidence for a Broad-line Region Wind in Radio-quiet Quasars}",
      journal = {\apj},
         year = 2024,
        month = nov,
       volume = {975},
       number = {1},
        pages = {35},
          doi = {10.3847/1538-4357/ad74fc},
archivePrefix = {arXiv},
       eprint = {2408.15934},
 primaryClass = {astro-ph.GA},
       adsurl = {https://ui.adsabs.harvard.edu/abs/2024ApJ...975...35C}
}

@article{Shimwell_2026,
       author = {{Shimwell}, T.~W. and {Hardcastle}, M.~J. and {Tasse}, C. and {Drabent}, A. and {Botteon}, A. and {Williams}, W.~L. and {Best}, P.~N. and {R{\"o}ttgering}, H.~J.~A. and {Br{\"u}ggen}, M. and {Brunetti}, G. and {Callingham}, J.~R. and {Chy{\.z}y}, K.~T. and {Conway}, J.~E. and {De Gasperin}, F. and {Haverkorn}, M. and {Horellou}, C. and {Jackson}, N. and {Miley}, G.~K. and {Morabito}, L.~K. and {Morganti}, R. and {O'Sullivan}, S.~P. and {Schwarz}, D.~J. and {Smith}, D.~J.~B. and {van Weeren}, R.~J. and {Vedantham}, H.~K. and {White}, G.~J. and {Ahmadi}, A. and {Alegre}, L. and {Arias}, M. and {Asabere}, B. and {Bahr-Kalus}, B. and {Barkus}, B. and {Bilicki}, M. and {B{\"o}hme}, L. and {Brentjens}, M. and {Brienza}, M. and {Bomans}, D.~J. and {Bonafede}, A. and {Bonato}, M. and {Bonnassieux}, E. and {Boxelaar}, J.~M. and {Camera}, S. and {Cassano}, R. and {Chilufya}, J. and {Cianfaglione}, M. and {Croston}, J.~H. and {Cuciti}, V. and {Dabhade}, P. and {De Rubeis}, E. and {de Jong}, J.~M.~G.~H.~J. and {Dallacasa}, D. and {Dettmar}, R.~J. and {Duncan}, K.~J. and {Di Gennaro}, G. and {Edler}, H.~W. and {Groeneveld}, C. and {G{\"u}rkan}, G. and {Hajduk}, M. and {Hale}, C.~L. and {Heesen}, V. and {Hoang}, D.~N. and {Hoeft}, M. and {Holties}, H. and {Horton}, M.~A. and {Iacobelli}, M. and {Jamrozy}, M. and {Jarvis}, M.~J. and {Jelic}, V. and {Kadler}, M. and {Kondapally}, R. and {Kunert-Bajraszewska}, M. and {Loose}, M. and {Magliocchetti}, M. and {Ma{\l}ek}, K. and {Manzano}, C. and {McKean}, J.~P. and {Mevius}, M. and {Mingo}, B. and {Miskolczi}, A. and {Misra}, A. and {Mold{\'o}n}, J. and {Nair}, D.~G. and {Nakoneczny}, S.~J. and {Orru}, E. and {Pashapour-Ahmadabadi}, M. and {Pasini}, T. and {Petley}, J. and {Pierce}, J.~C.~S. and {Prandoni}, I. and {Rafferty}, D. and {Rajpurohit}, K. and {Riseley}, C.~J. and {Roberts}, I.~D. and {Sethi}, S. and {Shulevski}, A. and {Stein}, M. and {Stuardi}, C. and {Sweijen}, F. and {ter Veen}, S. and {Timmerman}, R. and {Vaccari}, M. and {Wijnholds}, S.},
         title = "{The LOFAR Two-metre Sky Survey: VII. Third Data Release}",
      journal = {\aap},
         year = 2026,
        month = mar,
       volume = {707},
        pages = {A198},
          doi = {10.1051/0004-6361/202557749},
archivePrefix = {arXiv},
       eprint = {2602.15949},
 primaryClass = {astro-ph.GA},
       adsurl = {https://ui.adsabs.harvard.edu/abs/2026A&A...707A.198S}
}

@article{Chen_2023,
       author = {{Chen}, Sina and {Laor}, Ari and {Behar}, Ehud and {Baldi}, Ranieri D. and {Gelfand}, Joseph D.},
        title = "{The radio emission in radio-quiet quasars: the VLBA perspective}",
      journal = {\mnras},
         year = 2023,
        month = oct,
       volume = {525},
       number = {1},
        pages = {164-182},
          doi = {10.1093/mnras/stad2289},
archivePrefix = {arXiv},
       eprint = {2307.13599},
 primaryClass = {astro-ph.GA},
       adsurl = {https://ui.adsabs.harvard.edu/abs/2023MNRAS.525..164C}
}

@article{Karwin_2023,
       author = {{Karwin}, Chris and {Ajello}, Marco and {Diesing}, Rebecca and {Caprioli}, Damiano and {Chartas}, George},
        title = "{Gamma rays from Fast Black-Hole Winds}",
    booktitle = {AAS/High Energy Astrophysics Division},
         year = 2023,
       journal = {BAAS},
       volume = {55},
        month = sep,
        pages = {300.03},
       adsurl = {https://ui.adsabs.harvard.edu/abs/2023HEAD...2030003K}
}

@article{Falcke_1996,
       author = {{Falcke}, Heino and {Sherwood}, William and {Patnaik}, Alok R.},
        title = "{The Nature of Radio-intermediate Quasars: What Is Radio-loud and What Is Radio-quiet?}",
      journal = {\apj},
         year = 1996,
        month = nov,
       volume = {471},
        pages = {106},
          doi = {10.1086/177956},
archivePrefix = {arXiv},
       eprint = {astro-ph/9605165},
 primaryClass = {astro-ph},
       adsurl = {https://ui.adsabs.harvard.edu/abs/1996ApJ...471..106F}
}

@article{Yuan_2008,
       author = {{Yuan}, W. and {Zhou}, H.~Y. and {Komossa}, S. and {Dong}, X.~B. and {Wang}, T.~G. and {Lu}, H.~L. and {Bai}, J.~M.},
        title = "{A Population of Radio-Loud Narrow-Line Seyfert 1 Galaxies with Blazar-Like Properties?}",
      journal = {\apj},
         year = 2008,
        month = oct,
       volume = {685},
       number = {2},
        pages = {801-827},
          doi = {10.1086/591046},
archivePrefix = {arXiv},
       eprint = {0806.3755},
 primaryClass = {astro-ph},
       adsurl = {https://ui.adsabs.harvard.edu/abs/2008ApJ...685..801Y}
}

@article{Ehlert_2025,
       author = {{Ehlert}, Domenik and {Oikonomou}, Foteini and {Peretti}, Enrico},
        title = "{Ultra-high-energy cosmic rays from ultra-fast outflows of active galactic nuclei}",
      journal = {\mnras},
         year = 2025,
        month = may,
       volume = {539},
       number = {3},
        pages = {2435-2462},
          doi = {10.1093/mnras/staf457},
archivePrefix = {arXiv},
       eprint = {2411.05667},
 primaryClass = {astro-ph.HE},
       adsurl = {https://ui.adsabs.harvard.edu/abs/2025MNRAS.539.2435E}
}

@article{Ulivi_2024,
	author = {{Ulivi}, L. and {Venturi, G.} and {Cresci, G.} and {Marconi, A.} and {Marconcini, C.} and {Amiri, A.} and {Belfiore, F.} and {Bertola, E.} and {Carniani, S.} and {D’Amato, Q.} and {Di Teodoro, E.} and {Ginolfi, M.} and {Girdhar, A.} and {Harrison, C.} and {Maiolino, R.} and {Mannucci, F.} and {Mingozzi, M.} and {Perna, M.} and {Scialpi, M.} and {Tomicic, N.} and {Tozzi, G.} and {Treister, E.}},
	title = {Feedback and ionized gas outflows in four low-radio power AGN at z ~ 0.15},
	doi= {10.1051/0004-6361/202347436},
	url= {https://doi.org/10.1051/0004-6361/202347436},
	journal = {\aap},
	year = {2024},
	volume = {685},
	pages = {A122},
}

@article{Ricci_2023,
       author = {{Ricci}, Claudio and {Chang}, Chin-Shin and {Kawamuro}, Taiki and {Privon}, George C. and {Mushotzky}, Richard and {Trakhtenbrot}, Benny and {Laor}, Ari and {Koss}, Michael J. and {Smith}, Krista L. and {Gupta}, Kriti K. and {Dimopoulos}, Georgios and {Aalto}, Susanne and {Ros}, Eduardo},
        title = "{A Tight Correlation between Millimeter and X-Ray Emission in Accreting Massive Black Holes from <100 mas Resolution ALMA Observations}",
      journal = {\apjl},
         year = 2023,
        month = aug,
       volume = {952},
       number = {2},
        pages = {L28},
          doi = {10.3847/2041-8213/acda27},
archivePrefix = {arXiv},
       eprint = {2306.04679},
 primaryClass = {astro-ph.HE},
       adsurl = {https://ui.adsabs.harvard.edu/abs/2023ApJ...952L..28R}
}

@inproceedings{Briggs_1995,
       author = {{Briggs}, D.~S.},
        title = "{High Fidelity Interferometric Imaging: Robust Weighting and NNLS Deconvolution}",
    booktitle = {American Astronomical Society Meeting Abstracts},
         year = 1995,
       series = {American Astronomical Society Meeting Abstracts},
       volume = {187},
        month = dec,
        pages = {112.02},
       adsurl = {https://ui.adsabs.harvard.edu/abs/1995AAS...18711202B}
}

@article{Perley_Butler_2013,
       author = {{Perley}, R.~A. and {Butler}, B.~J.},
        title = "{An Accurate Flux Density Scale from 1 to 50 GHz}",
      journal = {\apjs},
         year = 2013,
        month = feb,
       volume = {204},
       number = {2},
        pages = {19},
          doi = {10.1088/0067-0049/204/2/19},
archivePrefix = {arXiv},
       eprint = {1211.1300},
 primaryClass = {astro-ph.IM},
       adsurl = {https://ui.adsabs.harvard.edu/abs/2013ApJS..204...19P}
}

@article{del_Palacio_2025,
   title={Millimeter emission from supermassive black hole coronae},
   volume={701},
   ISSN={1432-0746},
   url={http://dx.doi.org/10.1051/0004-6361/202554936},
   DOI={10.1051/0004-6361/202554936},
   journal={\aap},
   publisher={EDP Sciences},
   author={del Palacio, S. and Yang, C. and Aalto, S. and Ricci, C. and Lankhaar, B. and König, S. and Becker Tjus, J. and Magno, M. and Smith, K. L. and Yang, J. and Barcos-Muñoz, L. and Combes, F. and Linden, S. and Henkel, C. and Mangum, J. G. and Martín, S. and Olander, G. and Privon, G. and Wethers, C. and Baczko, A.-K. and Beswick, R. J. and García-Bernete, I. and García-Burillo, S. and González-Alfonso, E. and Gorski, M. and Imanishi, M. and Izumi, T. and Muller, S. and Nishimura, Y. and Pereira-Santaella, M. and van der Werf, P. P.},
   year={2025},
   month=sep, pages={A41} }

@article{Dabhade_2020,
       author = {{Dabhade}, P. and {R{\"o}ttgering}, H.~J.~A. and {Bagchi}, J. and {Shimwell}, T.~W. and {Hardcastle}, M.~J. and {Sankhyayan}, S. and {Morganti}, R. and {Jamrozy}, M. and {Shulevski}, A. and {Duncan}, K.~J.},
        title = "{Giant radio galaxies in the LOFAR Two-metre Sky Survey. I. Radio and environmental properties}",
      journal = {\aap},
         year = 2020,
        month = mar,
       volume = {635},
        pages = {A5},
          doi = {10.1051/0004-6361/201935589},
archivePrefix = {arXiv},
       eprint = {1904.00409},
 primaryClass = {astro-ph.GA},
       adsurl = {https://ui.adsabs.harvard.edu/abs/2020A&A...635A...5D}
}

@article{Kuzmicz_2021,
       author = {{Ku{\'z}micz}, Agnieszka and {Jamrozy}, Marek},
        title = "{Giant Radio Quasars: Sample and Basic Properties}",
      journal = {\apjs},
         year = 2021,
        month = mar,
       volume = {253},
       number = {1},
        pages = {25},
          doi = {10.3847/1538-4365/abd483},
archivePrefix = {arXiv},
       eprint = {2012.08857},
 primaryClass = {astro-ph.GA},
       adsurl = {https://ui.adsabs.harvard.edu/abs/2021ApJS..253...25K}
}

@article{Sabater_2019,
       author = {{Sabater}, J. and {Best}, P.~N. and {Hardcastle}, M.~J. and {Shimwell}, T.~W. and {Tasse}, C. and {Williams}, W.~L. and {Br{\"u}ggen}, M. and {Cochrane}, R.~K. and {Croston}, J.~H. and {de Gasperin}, F. and {Duncan}, K.~J. and {G{\"u}rkan}, G. and {Mechev}, A.~P. and {Morabito}, L.~K. and {Prandoni}, I. and {R{\"o}ttgering}, H.~J.~A. and {Smith}, D.~J.~B. and {Harwood}, J.~J. and {Mingo}, B. and {Mooney}, S. and {Saxena}, A.},
        title = "{The LoTSS view of radio AGN in the local Universe. The most massive galaxies are always switched on}",
      journal = {\aap},
         year = 2019,
        month = feb,
       volume = {622},
        pages = {A17},
          doi = {10.1051/0004-6361/201833883},
archivePrefix = {arXiv},
       eprint = {1811.05528},
 primaryClass = {astro-ph.GA},
       adsurl = {https://ui.adsabs.harvard.edu/abs/2019A&A...622A..17S}
}

@article{Tombesi_2014,
       author = {{Tombesi}, F. and {Tazaki}, F. and {Mushotzky}, R.~F. and {Ueda}, Y. and {Cappi}, M. and {Gofford}, J. and {Reeves}, J.~N. and {Guainazzi}, M.},
        title = "{Ultrafast outflows in radio-loud active galactic nuclei}",
      journal = {\mnras},
         year = 2014,
        month = sep,
       volume = {443},
       number = {3},
        pages = {2154-2182},
          doi = {10.1093/mnras/stu1297},
archivePrefix = {arXiv},
       eprint = {1406.7252},
 primaryClass = {astro-ph.HE},
       adsurl = {https://ui.adsabs.harvard.edu/abs/2014MNRAS.443.2154T}
}

@article{Giroletti_2017,
       author = {{Giroletti}, M. and {Panessa}, F. and {Longinotti}, A.~L. and {Krongold}, Y. and {Guainazzi}, M. and {Costantini}, E. and {Santos-Lleo}, M.},
        title = "{Coexistence of a non-thermal jet and a complex ultra-fast X-ray outflow in a moderately luminous AGN}",
      journal = {\aap},
         year = 2017,
        month = apr,
       volume = {600},
        pages = {A87},
          doi = {10.1051/0004-6361/201630161},
archivePrefix = {arXiv},
       eprint = {1701.07298},
 primaryClass = {astro-ph.GA},
       adsurl = {https://ui.adsabs.harvard.edu/abs/2017A&A...600A..87G}
}

@article{Morganti_2017,
       author = {{Morganti}, Raffaella},
        title = "{The many routes to AGN feedback}",
      journal = {Front. Astron. Space Sci.},
         year = 2017,
        month = nov,
       volume = {4},
        pages = {42},
          doi = {10.3389/fspas.2017.00042},
archivePrefix = {arXiv},
       eprint = {1712.05301},
 primaryClass = {astro-ph.GA},
       adsurl = {https://ui.adsabs.harvard.edu/abs/2017FrASS...4...42M}
}

@article{Mestici_2024,
       author = {{Mestici}, S. and {Tombesi}, F. and {Gaspari}, M. and {Piconcelli}, E. and {Panessa}, F.},
        title = "{Unified properties of supermassive black hole winds in radio-quiet and radio-loud AGN}",
      journal = {\mnras},
         year = 2024,
        month = aug,
       volume = {532},
       number = {3},
        pages = {3036-3055},
          doi = {10.1093/mnras/stae1617},
archivePrefix = {arXiv},
       eprint = {2407.00393},
 primaryClass = {astro-ph.HE},
       adsurl = {https://ui.adsabs.harvard.edu/abs/2024MNRAS.532.3036M}
}

@article{Yamada_2024,
       author = {{Yamada}, Satoshi and {Kawamuro}, Taiki and {Mizumoto}, Misaki and {Ricci}, Claudio and {Ogawa}, Shoji and {Noda}, Hirofumi and {Ueda}, Yoshihiro and {Enoto}, Teruaki and {Kokubo}, Mitsuru and {Minezaki}, Takeo and {Sameshima}, Hiroaki and {Horiuchi}, Takashi and {Mizukoshi}, Shoichiro},
        title = "{X-Ray Winds in Nearby-to-distant Galaxies (X-WING). I. Legacy Surveys of Galaxies with Ultrafast Outflows and Warm Absorbers in z {\ensuremath{\sim}} 0─4}",
      journal = {\apjs},
         year = 2024,
        month = sep,
       volume = {274},
       number = {1},
        pages = {8},
          doi = {10.3847/1538-4365/ad5961},
archivePrefix = {arXiv},
       eprint = {2405.02391},
 primaryClass = {astro-ph.HE},
       adsurl = {https://ui.adsabs.harvard.edu/abs/2024ApJS..274....8Y}
}

@article{Panessa_2022,
    author={{Panessa}, F. and {Chiaraluce}, E. and {Bruni}, G. and {Dallacasa}, D. and {Laor}, A. and {Baldi}, R. D. and {Behar}, E. and {McHardy}, I. and {Tombesi}, F. and {Vagnetti}, F.},
	title={Hard-X-ray-selected active galactic nuclei - II. Spectral energy distributions in the 5–45 GHz domain},
	volume={515},
	ISSN={1365-2966},
	url={http://dx.doi.org/10.1093/mnras/stac1745},
	DOI={10.1093/mnras/stac1745},
	number={1},
	journal={\mnras},
	publisher={Oxford University Press (OUP)},
	year={2022},
	month={06}, 
	pages={473--490} 
	}

@article{Nims_2015,
	author = {{Nims}, J. and {Quataert}, E. and {Faucher-Gigu{\`e}re}, C. A.},
	title = {Observational signatures of galactic winds powered by active galactic nuclei},
	journal = {\mnras},
	year = {2015},
	month = {03},
	volume = {447},
	number = {4},
	pages = {3612--3622},
	doi = {10.1093/mnras/stu2648},
	archivePrefix = {arXiv},
	eprint = {1408.5141},
	primaryClass = {astro-ph.GA},
	adsurl = {https://ui.adsabs.harvard.edu/abs/2015MNRAS.447.3612N}
}

@article{FGQ_2012,
	author = {{Faucher-Gigu{\`e}re}, C. A. and {Quataert}, E.},
	title = {The physics of galactic winds driven by active galactic nuclei},
	journal = {\mnras},
	year = {2012},
	month = {09},
	volume = {425},
	number = {1},
	pages = {605--622},
	doi = {10.1111/j.1365-2966.2012.21512.x},
	archivePrefix = {arXiv},
	eprint = {1204.2547},
	primaryClass = {astro-ph.CO},
	adsurl = {https://ui.adsabs.harvard.edu/abs/2012MNRAS.425..605F}
}

@article{LaFranca_2010,
       author = {{La Franca}, F. and {Melini}, G. and {Fiore}, F.},
        title = "{Tools for Computing the AGN Feedback: Radio-loudness Distribution and the Kinetic Luminosity Function}",
      journal = {\apj},
         year = 2010,
        month = jul,
       volume = {718},
       number = {1},
        pages = {368-379},
          doi = {10.1088/0004-637X/718/1/368},
archivePrefix = {arXiv},
       eprint = {1006.1247},
 primaryClass = {astro-ph.CO},
       adsurl = {https://ui.adsabs.harvard.edu/abs/2010ApJ...718..368L}
}

@article{White_2017,
       author = {{White}, Sarah V. and {Jarvis}, Matt J. and {Kalfountzou}, Eleni and {Hardcastle}, Martin J. and {Verma}, Aprajita and {Cao Orjales}, Jos{\'e} M. and {Stevens}, Jason},
        title = "{Evidence that the AGN dominates the radio emission in z {\ensuremath{\sim}} 1 radio-quiet quasars}",
      journal = {\mnras},
         year = 2017,
        month = jun,
       volume = {468},
       number = {1},
        pages = {217-238},
          doi = {10.1093/mnras/stx284},
archivePrefix = {arXiv},
       eprint = {1702.00904},
 primaryClass = {astro-ph.GA},
       adsurl = {https://ui.adsabs.harvard.edu/abs/2017MNRAS.468..217W}
}

@article{Kawamuro_2022,
       author = {{Kawamuro}, Taiki and {Ricci}, Claudio and {Imanishi}, Masatoshi and {Mushotzky}, Richard F. and {Izumi}, Takuma and {Ricci}, Federica and {Bauer}, Franz E. and {Koss}, Michael J. and {Trakhtenbrot}, Benny and {Ichikawa}, Kohei and {Rojas}, Alejandra F. and {Smith}, Krista Lynne and {Shimizu}, Taro and {Oh}, Kyuseok and {den Brok}, Jakob S. and {Baba}, Shunsuke and {Balokovi{\'c}}, Mislav and {Chang}, Chin-Shin and {Kakkad}, Darshan and {Pfeifle}, Ryan W. and {Privon}, George C. and {Temple}, Matthew J. and {Ueda}, Yoshihiro and {Harrison}, Fiona and {Powell}, Meredith C. and {Stern}, Daniel and {Urry}, Meg and {Sanders}, David B.},
        title = "{BASS XXXII: Studying the Nuclear Millimeter-wave Continuum Emission of AGNs with ALMA at Scales {\ensuremath{\lesssim}}100-200 pc}",
      journal = {\apj},
         year = 2022,
        month = oct,
       volume = {938},
       number = {1},
        pages = {87},
          doi = {10.3847/1538-4357/ac8794},
archivePrefix = {arXiv},
       eprint = {2208.03880},
 primaryClass = {astro-ph.GA},
       adsurl = {https://ui.adsabs.harvard.edu/abs/2022ApJ...938...87K}
}

@article{Behar_2015,
       author = {{Behar}, Ehud and {Baldi}, Ranieri D. and {Laor}, Ari and {Horesh}, Assaf and {Stevens}, Jamie and {Tzioumis}, Tasso},
        title = "{Discovery of millimetre-wave excess emission in radio-quiet active galactic nuclei}",
      journal = {\mnras},
         year = 2015,
        month = jul,
       volume = {451},
       number = {1},
        pages = {517-526},
          doi = {10.1093/mnras/stv988},
archivePrefix = {arXiv},
       eprint = {1504.01226},
 primaryClass = {astro-ph.GA},
       adsurl = {https://ui.adsabs.harvard.edu/abs/2015MNRAS.451..517B}
}

@article{Wang_2023,
       author = {{Wang}, Ailing and {An}, Tao and {Zhang}, Yingkang and {Cheng}, Xiaopeng and {Ho}, Luis C. and {Kellermann}, Kenneth I. and {Baan}, Willem A.},
        title = "{VLBI Observations of a sample of Palomar-Green quasars II: characterizing the parsec-scale radio emission}",
      journal = {\mnras},
         year = 2023,
        month = nov,
       volume = {525},
       number = {4},
        pages = {6064-6083},
          doi = {10.1093/mnras/stad2651},
archivePrefix = {arXiv},
       eprint = {2308.16780},
 primaryClass = {astro-ph.GA},
       adsurl = {https://ui.adsabs.harvard.edu/abs/2023MNRAS.525.6064W}
}

@article{Zubovas_2020,
       author = {{Zubovas}, Kastytis and {Nardini}, Emanuele},
        title = "{Intermittent AGN episodes drive outflows with a large spread of observable loading factors}",
      journal = {\mnras},
         year = 2020,
        month = nov,
       volume = {498},
       number = {3},
        pages = {3633-3647},
          doi = {10.1093/mnras/staa2652},
archivePrefix = {arXiv},
       eprint = {2008.12492},
 primaryClass = {astro-ph.GA},
       adsurl = {https://ui.adsabs.harvard.edu/abs/2020MNRAS.498.3633Z}
}

@article{Meenakshi_2024,
       author = {{Meenakshi}, Moun and {Mukherjee}, Dipanjan and {Bodo}, Gianluigi and {Rossi}, Paola and {Harrison}, Chris M.},
        title = "{A comparative study of radio signatures from winds and jets: modelling synchrotron emission and polarization}",
      journal = {\mnras},
         year = 2024,
        month = sep,
       volume = {533},
       number = {2},
        pages = {2213-2231},
          doi = {10.1093/mnras/stae1890},
archivePrefix = {arXiv},
       eprint = {2408.00099},
 primaryClass = {astro-ph.HE},
       adsurl = {https://ui.adsabs.harvard.edu/abs/2024MNRAS.533.2213M}
}

@article{Baldi_2018b,
       author = {{Baldi}, R.~D. and {Williams}, D.~R.~A. and {McHardy}, I.~M. and {Beswick}, R.~J. and {Argo}, M.~K. and {Dullo}, B.~T. and {Knapen}, J.~H. and {Brinks}, E. and {Muxlow}, T.~W.~B. and {Aalto}, S. and {Alberdi}, A. and {Bendo}, G.~J. and {Corbel}, S. and {Evans}, R. and {Fenech}, D.~M. and {Green}, D.~A. and {Kl{\"o}ckner}, H.-R. and {K{\"o}rding}, E. and {Kharb}, P. and {Maccarone}, T.~J. and {Mart{\'\i}-Vidal}, I. and {Mundell}, C.~G. and {Panessa}, F. and {Peck}, A.~B. and {P{\'e}rez-Torres}, M.~A. and {Saikia}, D.~J. and {Saikia}, P. and {Shankar}, F. and {Spencer}, R.~E. and {Stevens}, I.~R. and {Uttley}, P. and {Westcott}, J.},
        title = "{LeMMINGs - I. The eMERLIN legacy survey of nearby galaxies. 1.5-GHz parsec-scale radio structures and cores}",
      journal = {\mnras},
         year = 2018,
        month = may,
       volume = {476},
       number = {3},
        pages = {3478-3522},
          doi = {10.1093/mnras/sty342},
archivePrefix = {arXiv},
       eprint = {1802.02162},
 primaryClass = {astro-ph.GA},
       adsurl = {https://ui.adsabs.harvard.edu/abs/2018MNRAS.476.3478B}
}

@article{Baldi_2019,
       author = {{Baldi}, Ranieri D. and {Capetti}, Alessandro and {Giovannini}, Gabriele},
        title = "{High-resolution VLA observations of FR0 radio galaxies: the properties and nature of compact radio sources}",
      journal = {\mnras},
         year = 2019,
        month = jan,
       volume = {482},
       number = {2},
        pages = {2294-2304},
          doi = {10.1093/mnras/sty2703},
archivePrefix = {arXiv},
       eprint = {1810.01894},
 primaryClass = {astro-ph.GA},
       adsurl = {https://ui.adsabs.harvard.edu/abs/2019MNRAS.482.2294B}
}

@article{Jarvis_2021,
       author = {{Jarvis}, M.~E. and {Harrison}, C.~M. and {Mainieri}, V. and {Alexander}, D.~M. and {Arrigoni Battaia}, F. and {Calistro Rivera}, G. and {Circosta}, C. and {Costa}, T. and {De Breuck}, C. and {Edge}, A.~C. and {Girdhar}, A. and {Kakkad}, D. and {Kharb}, P. and {Lansbury}, G.~B. and {Molyneux}, S.~J. and {Mukherjee}, D. and {Mullaney}, J.~R. and {Farina}, E.~P. and {Silpa}, S. and {Thomson}, A.~P. and {Ward}, S.~R.},
        title = "{The quasar feedback survey: discovering hidden Radio-AGN and their connection to the host galaxy ionized gas}",
      journal = {\mnras},
         year = 2021,
        month = may,
       volume = {503},
       number = {2},
        pages = {1780-1797},
          doi = {10.1093/mnras/stab549},
archivePrefix = {arXiv},
       eprint = {2103.00014},
 primaryClass = {astro-ph.GA},
       adsurl = {https://ui.adsabs.harvard.edu/abs/2021MNRAS.503.1780J}
}

@article{AAnB_2012,
       author = {{An}, Tao and {Baan}, Willem A.},
        title = "{The Dynamic Evolution of Young Extragalactic Radio Sources}",
      journal = {\apj},
         year = 2012,
        month = nov,
       volume = {760},
       number = {1},
        pages = {77},
          doi = {10.1088/0004-637X/760/1/77},
archivePrefix = {arXiv},
       eprint = {1211.1760},
 primaryClass = {astro-ph.CO},
       adsurl = {https://ui.adsabs.harvard.edu/abs/2012ApJ...760...77A}
}

@article{JG_2019,
       author = {{Jimenez-Gallardo}, A. and {Massaro}, F. and {Capetti}, A. and {Prieto}, M.~A. and {Paggi}, A. and {Baldi}, R.~D. and {Grossova}, R. and {Ostorero}, L. and {Siemiginowska}, A. and {Viada}, S.},
        title = "{COMP2CAT: hunting compact double radio sources in the local Universe}",
      journal = {\aap},
         year = 2019,
        month = jul,
       volume = {627},
        pages = {A108},
          doi = {10.1051/0004-6361/201935104},
archivePrefix = {arXiv},
       eprint = {1905.02212},
 primaryClass = {astro-ph.HE},
       adsurl = {https://ui.adsabs.harvard.edu/abs/2019A&A...627A.108J}
}

@article{Liu_2007,
       author = {{Liu}, X. and {Cui}, L. and {Luo}, W.-F. and {Shi}, W.-Z. and {Song}, H.-G.},
        title = "{VLBI observations of nineteen GHz-peaked-spectrum radio sources at 1.6 GHz}",
      journal = {\aap},
         year = 2007,
        month = jul,
       volume = {470},
       number = {1},
        pages = {97-104},
          doi = {10.1051/0004-6361:20077265},
archivePrefix = {arXiv},
       eprint = {0704.0310},
 primaryClass = {astro-ph},
       adsurl = {https://ui.adsabs.harvard.edu/abs/2007A&A...470...97L}
}

@article{Parma_1987,
       author = {{Fanti}, C. and {Fanti}, R. and {de Ruiter}, H.~R. and {Parma}, P.},
        title = "{VLA observations of low luminosity radio galaxies. IV. The B2 sample revisited.}",
      journal = {\aaps},
         year = 1987,
        month = apr,
       volume = {69},
        pages = {57-76},
       adsurl = {https://ui.adsabs.harvard.edu/abs/1987A&AS...69...57F}
}

@article{Bischetti_2019,
       author = {{Bischetti}, M. and {Piconcelli}, E. and {Feruglio}, C. and {Fiore}, F. and {Carniani}, S. and {Brusa}, M. and {Cicone}, C. and {Vignali}, C. and {Bongiorno}, A. and {Cresci}, G. and {Mainieri}, V. and {Maiolino}, R. and {Marconi}, A. and {Nardini}, E. and {Zappacosta}, L.},
        title = "{The gentle monster PDS 456. Kiloparsec-scale molecular outflow and its implications for QSO feedback}",
      journal = {\aap},
         year = 2019,
        month = aug,
       volume = {628},
        pages = {A118},
          doi = {10.1051/0004-6361/201935524},
archivePrefix = {arXiv},
       eprint = {1903.10528},
 primaryClass = {astro-ph.GA},
       adsurl = {https://ui.adsabs.harvard.edu/abs/2019A&A...628A.118B}
}

@article{Tozzi_2021,
       author = {{Tozzi}, G. and {Cresci}, G. and {Marasco}, A. and {Nardini}, E. and {Marconi}, A. and {Mannucci}, F. and {Chartas}, G. and {Rizzo}, F. and {Amiri}, A. and {Brusa}, M. and {Comastri}, A. and {Dadina}, M. and {Lanzuisi}, G. and {Mainieri}, V. and {Mingozzi}, M. and {Perna}, M. and {Venturi}, G. and {Vignali}, C.},
        title = "{Connecting X-ray nuclear winds with galaxy-scale ionised outflows in two z {\ensuremath{\sim}} 1.5 lensed quasars}",
      journal = {\aap},
         year = 2021,
        month = apr,
       volume = {648},
        pages = {A99},
          doi = {10.1051/0004-6361/202040190},
archivePrefix = {arXiv},
       eprint = {2102.07789},
 primaryClass = {astro-ph.GA},
       adsurl = {https://ui.adsabs.harvard.edu/abs/2021A&A...648A..99T}
}

@article{Bonanomi_2023,
       author = {{Bonanomi}, Francesca and {Cicone}, Claudia and {Severgnini}, Paola and {Braito}, Valentina and {Vignali}, Cristian and {Reeves}, James N. and {Sirressi}, Mattia and {Montoya Arroyave}, Isabel and {Della Ceca}, Roberto and {Ballo}, Lucia and {Dotti}, Massimo},
        title = "{Another X-ray UFO without a momentum-boosted molecular outflow. ALMA CO(1-0) observations of the galaxy pair IRAS 05054+1718}",
      journal = {\aap},
         year = 2023,
        month = may,
       volume = {673},
        pages = {A46},
          doi = {10.1051/0004-6361/202245630},
archivePrefix = {arXiv},
       eprint = {2303.00770},
 primaryClass = {astro-ph.GA},
       adsurl = {https://ui.adsabs.harvard.edu/abs/2023A&A...673A..46B}
}

@article{Cheng_2018,
       author = {{Cheng}, X.-P. and {An}, T.},
        title = "{Parsec-scale Radio Structure of 14 Fanaroff-Riley Type 0 Radio Galaxies}",
      journal = {\apj},
         year = 2018,
        month = aug,
       volume = {863},
       number = {2},
        pages = {155},
          doi = {10.3847/1538-4357/aad22c},
archivePrefix = {arXiv},
       eprint = {1807.02505},
 primaryClass = {astro-ph.HE},
       adsurl = {https://ui.adsabs.harvard.edu/abs/2018ApJ...863..155C}
}

@article{KB_2010,
       author = {{Kunert-Bajraszewska}, M. and {Gawro{\'n}ski}, M.~P. and {Labiano}, A. and {Siemiginowska}, A.},
        title = "{A survey of low-luminosity compact sources and its implication for the evolution of radio-loud active galactic nuclei - I. Radio data}",
      journal = {\mnras},
         year = 2010,
        month = nov,
       volume = {408},
       number = {4},
        pages = {2261-2278},
          doi = {10.1111/j.1365-2966.2010.17271.x},
archivePrefix = {arXiv},
       eprint = {1009.5235},
 primaryClass = {astro-ph.CO},
       adsurl = {https://ui.adsabs.harvard.edu/abs/2010MNRAS.408.2261K}
}

@article{Zubovas_King_2012,
       author = {{Zubovas}, Kastytis and {King}, Andrew},
        title = "{Clearing Out a Galaxy}",
      journal = {\apjl},
         year = 2012,
        month = feb,
       volume = {745},
       number = {2},
        pages = {L34},
          doi = {10.1088/2041-8205/745/2/L34},
archivePrefix = {arXiv},
       eprint = {1201.0866},
 primaryClass = {astro-ph.GA},
       adsurl = {https://ui.adsabs.harvard.edu/abs/2012ApJ...745L..34Z}
}

@article{Lao_2024,
       author = {{Lao}, Bao-Qiang and {Yang}, Xiao-Long and {Jaiswal}, Sumit and {Mohan}, Prashanth and {Sun}, Xiao-Hui and {Qin}, Sheng-Li and {Zhao}, Ru-Shuang},
        title = "{A Machine Learning Made Catalog of FR-II Radio Galaxies from the FIRST Survey}",
      journal = {RAA},
         year = 2024,
        month = mar,
       volume = {24},
       number = {3},
        pages = {035021},
          doi = {10.1088/1674-4527/ad204f},
archivePrefix = {arXiv},
       eprint = {2401.08048},
 primaryClass = {astro-ph.GA},
       adsurl = {https://ui.adsabs.harvard.edu/abs/2024RAA....24c5021L}
}

@article{Cicone_2018,
       author = {{Cicone}, Claudia and {Brusa}, Marcella and {Ramos Almeida}, Cristina and {Cresci}, Giovanni and {Husemann}, Bernd and {Mainieri}, Vincenzo},
        title = "{The largely unconstrained multiphase nature of outflows in AGN host galaxies}",
      journal = {Nature Astronomy},
         year = 2018,
        month = feb,
       volume = {2},
        pages = {176-178},
          doi = {10.1038/s41550-018-0406-3},
archivePrefix = {arXiv},
       eprint = {1802.10308},
 primaryClass = {astro-ph.GA},
       adsurl = {https://ui.adsabs.harvard.edu/abs/2018NatAs...2..176C}
}

@article{King_2015,
    author={{King}, A. R. and {Pounds}, K.},
	title={Powerful Outflows and Feedback from Active Galactic Nuclei},
	volume={53},
	ISSN={1545-4282},
	url={http://dx.doi.org/10.1146/annurev-astro-082214-122316},
	DOI={10.1146/annurev-astro-082214-122316},
	number={1},
	journal={\araa},
	publisher={Annual Reviews},
	year={2015},
	month={08}, 
	pages={115--154} }

@article{Barvainis_1996,
       author = {{Barvainis}, Richard and {Lonsdale}, Colin and {Antonucci}, Robert},
        title = "{Radio Spectra of Radio Quiet Quasars}",
      journal = {\aj},
         year = 1996,
        month = apr,
       volume = {111},
        pages = {1431},
          doi = {10.1086/117888},
       adsurl = {https://ui.adsabs.harvard.edu/abs/1996AJ....111.1431B}
}

@article{Fiore_2017,
	title={AGN wind scaling relations and the co-evolution of black  holes and galaxies},
	volume={601},
	ISSN={1432-0746},
	url={http://dx.doi.org/10.1051/0004-6361/201629478},
	DOI={10.1051/0004-6361/201629478},
	journal={\aap},
	publisher={EDP Sciences},
	author={{Fiore}, F. and {Feruglio}, C. and {Shankar}, F. and {Bischetti}, M. and {Bongiorno}, A. and {Brusa}, M. and {Carniani}, S. and {Cicone}, C. and {Duras}, F. and {Lamastra}, A. and {Mainieri}, V. and {Marconi}, A. and {Menci}, N. and {Maiolino}, R. and {Piconcelli}, E. and {Vietri}, G. and {Zappacosta}, L.},
	year={2017},
	month={05},
	pages={A143} }

@article{Ciotti_2010,
	doi = {10.1088/0004-637X/717/2/708},
	adsurl = {https://dx.doi.org/10.1088/0004-637X/717/2/708},
	year = {2010},
	month = {06},
	publisher = {The American Astronomical Society},
	volume = {717},
	number = {2},
	pages = {708},
	author = {{Ciotti}, L. and {Ostriker}, J. P. and {Proga}, D.},
	title = {FEEDBACK FROM CENTRAL BLACK HOLES IN ELLIPTICAL GALAXIES. III. MODELS WITH BOTH RADIATIVE AND MECHANICAL FEEDBACK},
	journal = {\apj}
}

@article{Fabian_2012,
	title={Observational Evidence of Active Galactic Nuclei Feedback},
	volume={50},
	ISSN={1545-4282},
	url={http://dx.doi.org/10.1146/annurev-astro-081811-125521},
	DOI={10.1146/annurev-astro-081811-125521},
	number={1},
	journal={\araa},
	publisher={Annual Reviews},
	author={Fabian, A. C.},
	year={2012},
	month={09}, 
	pages={455--489} }

@article{Crenshaw_2003,
       author = {{Crenshaw}, D. Michael and {Kraemer}, Steven B. and {George}, Ian M.},
        title = "{Mass Loss from the Nuclei of Active Galaxies}",
      journal = {\araa},
         year = 2003,
        month = jan,
       volume = {41},
        pages = {117-167},
          doi = {10.1146/annurev.astro.41.082801.100328},
       adsurl = {https://ui.adsabs.harvard.edu/abs/2003ARA&A..41..117C}
}

@article{King_2005,
	author={{King}, A. R. and {Lubow}, S. H. and {Ogilvie}, G. I. and {Pringle}, J. E.},
    title={Aligning spinning black holes and accretion discs},
	volume={363},
	ISSN={1365-2966},
	url={http://dx.doi.org/10.1111/j.1365-2966.2005.09378.x},
	DOI={10.1111/j.1365-2966.2005.09378.x},
	number={1},
	journal={\mnras},
	publisher={Oxford University Press (OUP)},
	year={2005},
	month={10}, pages={49--56} }

@article{Cattaneo_1999,
       author = {{Cattaneo}, Andrea and {Haehnelt}, Martin G. and {Rees}, Martin J.},
        title = "{The distribution of supermassive black holes in the nuclei of nearby galaxies}",
      journal = {\mnras},
         year = 1999,
        month = sep,
       volume = {308},
       number = {1},
        pages = {77-81},
          doi = {10.1046/j.1365-8711.1999.02693.x},
archivePrefix = {arXiv},
       eprint = {astro-ph/9902223},
 primaryClass = {astro-ph},
       adsurl = {https://ui.adsabs.harvard.edu/abs/1999MNRAS.308...77C}
}

@article{Kauffman_2000,
       author = {{Kauffmann}, Guinevere and {Haehnelt}, Martin},
        title = "{A unified model for the evolution of galaxies and quasars}",
      journal = {\mnras},
         year = 2000,
        month = jan,
       volume = {311},
       number = {3},
        pages = {576-588},
          doi = {10.1046/j.1365-8711.2000.03077.x},
archivePrefix = {arXiv},
       eprint = {astro-ph/9906493},
 primaryClass = {astro-ph},
       adsurl = {https://ui.adsabs.harvard.edu/abs/2000MNRAS.311..576K}
}

@article{Pounds_2014,
       author = {{Pounds}, Ken},
        title = "{X-Ray Observations of Powerful AGN Outflows. Implications for Feedback}",
      journal = {\ssr},
         year = 2014,
        month = sep,
       volume = {183},
       number = {1-4},
        pages = {339-351},
          doi = {10.1007/s11214-013-0008-4},
archivePrefix = {arXiv},
       eprint = {1307.6700},
 primaryClass = {astro-ph.HE},
       adsurl = {https://ui.adsabs.harvard.edu/abs/2014SSRv..183..339P}
}

@article{Morganti_2016,
	title={Another piece of the puzzle: The fast H I outflow in Mrk 231},
	volume={593},
	ISSN={1432-0746},
	url={http://dx.doi.org/10.1051/0004-6361/201628978},
	DOI={10.1051/0004-6361/201628978},
	journal={\aap},
	publisher={EDP Sciences},
	author={{Morganti}, R. and {Veilleux}, S. and {Oosterloo}, T. and {Teng}, S. H. and {Rupke}, D.},
	year={2016},
	month={09}, 
    pages={A30}}

@article{KormendyHo_2013,
       author = {{Kormendy}, John and {Ho}, Luis C.},
        title = "{Coevolution (Or Not) of Supermassive Black Holes and Host Galaxies}",
      journal = {\araa},
         year = 2013,
        month = aug,
       volume = {51},
       number = {1},
        pages = {511-653},
          doi = {10.1146/annurev-astro-082708-101811},
archivePrefix = {arXiv},
       eprint = {1304.7762},
 primaryClass = {astro-ph.CO},
       adsurl = {https://ui.adsabs.harvard.edu/abs/2013ARA&A..51..511K}
}

@article{Panessa_2019,
       author = {{Panessa}, Francesca and {Baldi}, Ranieri Diego and {Laor}, Ari and {Padovani}, Paolo and {Behar}, Ehud and {McHardy}, Ian},
        title = "{The origin of radio emission from radio-quiet active galactic nuclei}",
      journal = {Nature Astronomy},
         year = 2019,
        month = apr,
       volume = {3},
        pages = {387-396},
          doi = {10.1038/s41550-019-0765-4},
archivePrefix = {arXiv},
       eprint = {1902.05917},
 primaryClass = {astro-ph.GA},
       adsurl = {https://ui.adsabs.harvard.edu/abs/2019NatAs...3..387P}
}

@article{Alexander_2025,
       author = {{Alexander}, D.~M. and {Hickox}, R.~C. and {Aird}, J. and {Combes}, F. and {Costa}, T. and {Habouzit}, M. and {Harrison}, C.~M. and {Leng}, R.~I. and {Morabito}, L.~K. and {Uckelman}, S.~L. and {Vickers}, P.},
         title = "{What drives the growth of black holes: A decade of progress}",
      journal = {\nar},
         year = 2025,
        month = dec,
       volume = {101},
        pages = {101733},
          doi = {10.1016/j.newar.2025.101733},
archivePrefix = {arXiv},
       eprint = {2506.19166},
 primaryClass = {astro-ph.GA},
       adsurl = {https://ui.adsabs.harvard.edu/abs/2025NewAR.10101733A}
}

@article{Mauch_2007,
       author = {{Mauch}, Tom and {Sadler}, Elaine M.},
        title = "{Radio sources in the 6dFGS: local luminosity functions at 1.4 GHz for star-forming galaxies and radio-loud AGN}",
      journal = {\mnras},
         year = 2007,
        month = mar,
       volume = {375},
       number = {3},
        pages = {931-950},
          doi = {10.1111/j.1365-2966.2006.11353.x},
archivePrefix = {arXiv},
       eprint = {astro-ph/0612018},
 primaryClass = {astro-ph},
       adsurl ={https://ui.adsabs.harvard.edu/abs/2007MNRAS.375..931M}
}

@inproceedings{Venturi_2021,
		author = {{Venturi}, G. and {Marconi}, A. and {Mingozzi}, M. and {Cresci}, G. and {Carniani}, S. and {Mannucci}, F.},
		title = {Turbulence/outflows perpendicular to low-power jets in Seyfert galaxies},
		booktitle = {Galaxy Evolution and Feedback across Different Environments},
		year = {2021},
		editor = {{Storchi Bergmann}, Thaisa and {Forman}, William and {Overzier}, Roderik and {Riffel}, Rog{\'e}rio},
		series = {IAU Symposium},
		volume = {359},
		month = {01},
		pages = {464--466},
		doi = {10.1017/S1743921320002197},
		adsurl = {https://ui.adsabs.harvard.edu/abs/2021IAUS..359..464V}
	}

@article{Speranza_2022,
       author = {{Speranza}, G. and {Ramos Almeida}, C. and {Acosta-Pulido}, J.~A. and {Riffel}, R.~A. and {Tadhunter}, C. and {Pierce}, J.~C.~S. and {Rodr{\'\i}guez-Ardila}, A. and {Coloma Puga}, M. and {Brusa}, M. and {Musiimenta}, B. and {Alexander}, D.~M. and {Lapi}, A. and {Shankar}, F. and {Villforth}, C.},
        title = "{Warm molecular and ionized gas kinematics in the type-2 quasar J0945+1737}",
      journal = {\aap},
         year = 2022,
        month = sep,
       volume = {665},
        pages = {A55},
          doi = {10.1051/0004-6361/202243585},
archivePrefix = {arXiv},
       eprint = {2206.15347},
 primaryClass = {astro-ph.GA},
       adsurl = {https://ui.adsabs.harvard.edu/abs/2022A&A...665A..55S}
}

@inproceedings{Murthy_2022,
       author = {{Murthy}, Suma and {Morganti}, Raffaella and {Wagner}, Alexander and {Oosterloo}, Tom and {Guillard}, Pierre and {Mukherjee}, Dipanjan and {Bicknell}, Geoffrey},
        title = "{Feedback from low-luminosity radio AGN: cold gas removal in B2 0258+35}",
    booktitle = {EAS2022, European Astronomical Society Annual Meeting},
         year = 2022,
        month = jul,
        pages = {874},
       adsurl = {https://ui.adsabs.harvard.edu/abs/2022eas..conf..874M}
}

@article{Mukherjee_2025,
       author = {{Mukherjee}, Dipanjan},
        title = "{Jet-Feedback on kpc scales: a review}",
      journal = {Galaxies},
         year = 2025,
        month = sep,
       volume = {13},
       number = {5},
        pages = {102},
          doi = {10.3390/galaxies13050102},
archivePrefix = {arXiv},
       eprint = {2506.03888},
 primaryClass = {astro-ph.GA},
       adsurl = {https://ui.adsabs.harvard.edu/abs/2025Galax..13..102M}
}

@article{Jarvis_2019,
       author = {{Jarvis}, M. E. and {Harrison}, C. M. and {Thomson}, A. P. and {Circosta}, C. and {Mainieri}, V. and {Alexander}, D. M. and {Edge}, A. C. and {Lansbury}, G. B. and {Molyneux}, S. J. and {Mullaney}, J. R.},
        title = {Prevalence of radio jets associated with galactic outflows and feedback from quasars},
      journal = {\mnras},
         year = {2019},
        month = {05},
       volume = {485},
       number = {2},
        pages = {2710--2730},
          doi = {10.1093/mnras/stz556},
archivePrefix = {arXiv},
       eprint = {1902.07727},
 primaryClass = {astro-ph.GA},
       adsurl = {https://ui.adsabs.harvard.edu/abs/2019MNRAS.485.2710J}
}

@article{Bicknell_1994,
       author = {{Bicknell}, Geoffrey V.},
        title = "{Extragalactic radio sources and the role of relativistic jets.}",
      journal = {Australian Journal of Physics},
         year = 1994,
        month = jan,
       volume = {47},
        pages = {669-680},
          doi = {10.1071/PH940669},
       adsurl = {https://ui.adsabs.harvard.edu/abs/1994AuJPh..47..669B}
}

@article{Tombesi_2012,
       author = {{Tombesi}, F. and {Cappi}, M. and {Reeves}, J.~N. and {Braito}, V.},
        title = "{Evidence for ultrafast outflows in radio-quiet AGNs - III. Location and energetics}",
      journal = {\mnras},
         year = 2012,
        month = may,
       volume = {422},
       number = {1},
        pages = {L1-L5},
          doi = {10.1111/j.1745-3933.2012.01221.x},
archivePrefix = {arXiv},
       eprint = {1201.1897},
 primaryClass = {astro-ph.HE},
       adsurl = {https://ui.adsabs.harvard.edu/abs/2012MNRAS.422L...1T}
}

@article{Tombesi_2015,
       author = {{Tombesi}, F. and {Mel{\'e}ndez}, M. and {Veilleux}, S. and {Reeves}, J.~N. and {Gonz{\'a}lez-Alfonso}, E. and {Reynolds}, C.~S.},
        title = "{Wind from the black-hole accretion disk driving a molecular outflow in an active galaxy}",
      journal = {\nat},
         year = 2015,
        month = mar,
       volume = {519},
       number = {7544},
        pages = {436-438},
          doi = {10.1038/nature14261},
archivePrefix = {arXiv},
       eprint = {1501.07664},
 primaryClass = {astro-ph.HE},
       adsurl = {https://ui.adsabs.harvard.edu/abs/2015Natur.519..436T}
}
%
\clearpage

\begin{appendix}
\begin{table*}[tp]
 \section{Tables}\label{tabs}
 \caption{Imaging parameters.}
 \label{tab:image_params}
 \centering
 \footnotesize
 \begin{tabular}{l c l c c c l}
 \hline\hline
 Target Name & Freq. & Restoring Beam & $\langle \sigma_\mathrm{rms} \rangle$ & Dyn. Range & Robust & UV Taper\\
 & (GHz) & (arcsec, arcsec, deg) & (mJy beam$^{-1}$) & & & (b/$\lambda$) \\
 (1) & (2) & (3) & (4) & (5)& (6) & (7)\\
 \hline
 PG0804+761 & 1.5 & 5.28, 3.13, 58.56 & 0.025 & 70 & 0.5 & / \\
 & 6.0$^{\dagger}$ & 5.29, 2.63, 56.55 & 0.019 & 42 & 2 & / \\
 PG0947+396 & 1.5 & 5.28, 3.34, -87.17 & 0.019 & 9 & 0.5 & / \\
 & 6.0$^{\dagger}$ & 5.52, 3.47, -86.34 & 0.014 & 10 & 2 & / \\
 2MASXJ1051+35$^{*}$ & 1.5 & 5.33, 3.49, -81.18 & 0.018 & 633 & 0.5 & / \\
 & 6.0$^{*}$ & 5.65, 3.71, -81.37 & 0.022 & 182 & 2 & / \\
 PG1114+445 & 1.5 & 4.78, 3.57, 86.45 & 0.017 & 29 & 0.5 & / \\
 & 6.0$^{\dagger}$ & 4.94, 3.56, 81.79 & 0.019 & 10 & 2 & / \\
 PG1202+281 & 1.5 & 4.56, 3.73, 75.17 & 0.018 & 73 & 0.5 & / \\
 & 6.0 & 4.76, 4.00, 74.25& 0.011 & 57 & 2 & / \\
 LBQS1338-0038$^{*}$ & 1.5 & 5.75, 3.71, -39.67 & 0.016 & 340 & 0.5 & / \\
 & 6.0$^{\dagger}$ & 5.99, 3.85, -39.53 & 0.068 & 25 & 2 & / \\
 2MASXJ1653+23$^{\dagger}$ & 1.5 & 3.56, 3.08, 78.78 & 0.021 & 290 & 0.5 & / \\
 & 6.0 & 3.71, 3.24, 78.36 & 0.018 & 82 & 2 & / \\
 \hline
 PG1307+085 & 1.5 & 4.22, 4.00, -29.88 & 0.025 & 24 & 0.5 & / \\
 & 6.0 & 1.15, 1.06, -48.83 & 0.012 & 23 & 0.5 & / \\
 PG1352+183 & 1.5 & 3.94, 3.72, 81.14 & 0.031 & 5 & 0.5 & / \\
 & 6.0 & 1.09, 1.03, 83.09 & 0.013 & 8 & 0.5 & / \\
 PG0052+251 & 1.5 & 3.70, 3.54, 6.26 & 0.018 & 75 & 0.5 & / \\
 & 6.0$^{\dagger}$ & 3.93, 3.73, 2.45 & 0.039 & 15 & 2 & / \\
 2MASXJ0220-07$^{*}$ & 1.5 & 3.72, 2.95, -11.75 & 0.021 & 72 & 0.5 & / \\
 & 6.0$^{\dagger}$ & 4.12, 3.17, -11.40 & 0.014 & 86 & 2 & / \\
 WISEJ0537-02$^{*}$ & 1.5 & 4.89, 3.64, -32.96 & 0.023 & 20 & 0.5 & / \\
 & 6.0 & 1.48, 0.95, -31.29 & 0.012 & 17 & 0.5 & / \\
 PG0953+414 & 1.5 & 4.70, 3.53, 88.78 & 0.015 & 23 & 0.5 & / \\
 & 6.0$^{\dagger}$ & 4.89, 3.65, -89.02 & 0.012 & 17 & 2 & / \\
 2MASXJ1402+26 & 1.5 & 6.05, 3.74, -69.52 & 0.018 & 23 & 0.5 & / \\
 & 6.0 & 1.70, 1.00, -70.95 & 0.013 & 14 & 0.5 & / \\
 PG1402+261 & 1.5 & 3.57, 2.60, 85.26 & 0.020 & 53 & 0.5 & / \\
 & 6.0 & 1.35, 1.00, -76.51 & 0.012 & 33 & 0.5 & / \\
 PG1416-129 & 1.5 & 7.81, 3.64, -34.05 & 0.023 & 85 & 0.5 & / \\
 & 6.0$^{\dagger}$ & 8.09, 3.76, -34.14 & 0.029 & 36 & 2 & / \\
 PG1425+267$^{*}$ & 1.5 & 3.16, 2.61, 77.13 & 0.022 & 1327 & -0.5 & $b_\mathrm{min}\sim 2585$ \\
 & 6.0$^{* \dagger}$ & 3.44, 2.91, 83.78 & 0.014 & 1950 & 2 & / \\
 PG1427+480 & 1.5 & 4.05, 3.60, 46.64 & 0.015 & 15 & 0.5 & / \\
 & 6.0 & 1.11, 0.92, 46.61 & 0.010 & 14 & 0.5 & / \\
 PG1435-067$^{*}$ & 1.5 & 5.99, 5.99, 0.0 & 0.028 & 19 & 0.5 & / \\
 & 6.0 & 1.57, 0.97, -30.61 & 0.019 & 7 & 0.5 & / \\
 SDSSJ1444+06$^{*}$ & 1.5 & 4.13, 3.48, -24.60 & 0.017 & 24 & 0.5 & / \\
 & 6.0 & 1.18, 0.96, -11.17 & 0.014 & 15 & 0.5 & / \\
 HB891529+050$^{*}$ & 1.5 & 4.40, 4.40, 0.0 & 0.018 & 555 & 0.5 & / \\
 & 6.0$^{* \dagger}$ & 6.79, 4.56, -6.49 & 0.079 & 40 & 2 & / \\
 PG1626+554 & 1.5 & 3.58, 2.81, -50.3 & 0.024 & 12 & 0.5 & / \\
 & 6.0 & 1.10, 0.87, -43.72 & 0.010 & 19 & 0.5 & / \\
 \hline
 \end{tabular}
 \tablefoot{Column 1: target name. Column 2: central observing frequency. Column 3: restoring beam major axis, minor axis and position angle. Column 4: mean rms noise. Column 5: image dynamic range. Column 6: robust Briggs weighting. Column 7: minimum baseline. Below the reported value the baselines have been removed.\\
 (*) Images obtained with WSClean.\\
 ($\dagger$) Images that have been smoothed. The UFO subsample is shown before the horizontal line.}
\end{table*}

\begin{table*}[tp]
\section{target by target notes}\label{A:SEDs}
\begin{minipage}{\textwidth}
In the following we report a summary of the main properties derived for each target. For each target we also show:

\vspace{2mm}
- Left: Optical image (SDSS u filter) with $\sigma_{rms}\cdot$[-3,3,6,12,24,48,96] contours of the LoTSS (145 MHz, green), and proprietary ones (1.5 GHz, white and 6 GHz, yellow).

\vspace{2mm}
- Middle: Radio map. Colours at 6 GHz with $\sigma_{rms}\cdot [-3,3,6,12,24,48,96]$ contours at 1.5 GHz (white) and 145 MHz (green). The beams are shown at the bottom left of each, so that it is possible to appreciate the different resolutions. The black cross marks the optical position of the AGN.

\vspace{2mm}
- Right: SED. The points are the integrated flux densities extracted from the proprietary data and LoTSS (blue filled dots), those of TGSS, RACS and FIRST and VLASS (grey dots). When different epochs of VLASS are shown, they are plotted with xs in different shades of grey. 
\end{minipage}
\end{table*}
\vspace{1cm}
\begin{figure*}[!ht]
		\centering
		\includegraphics[width=0.31\textwidth]{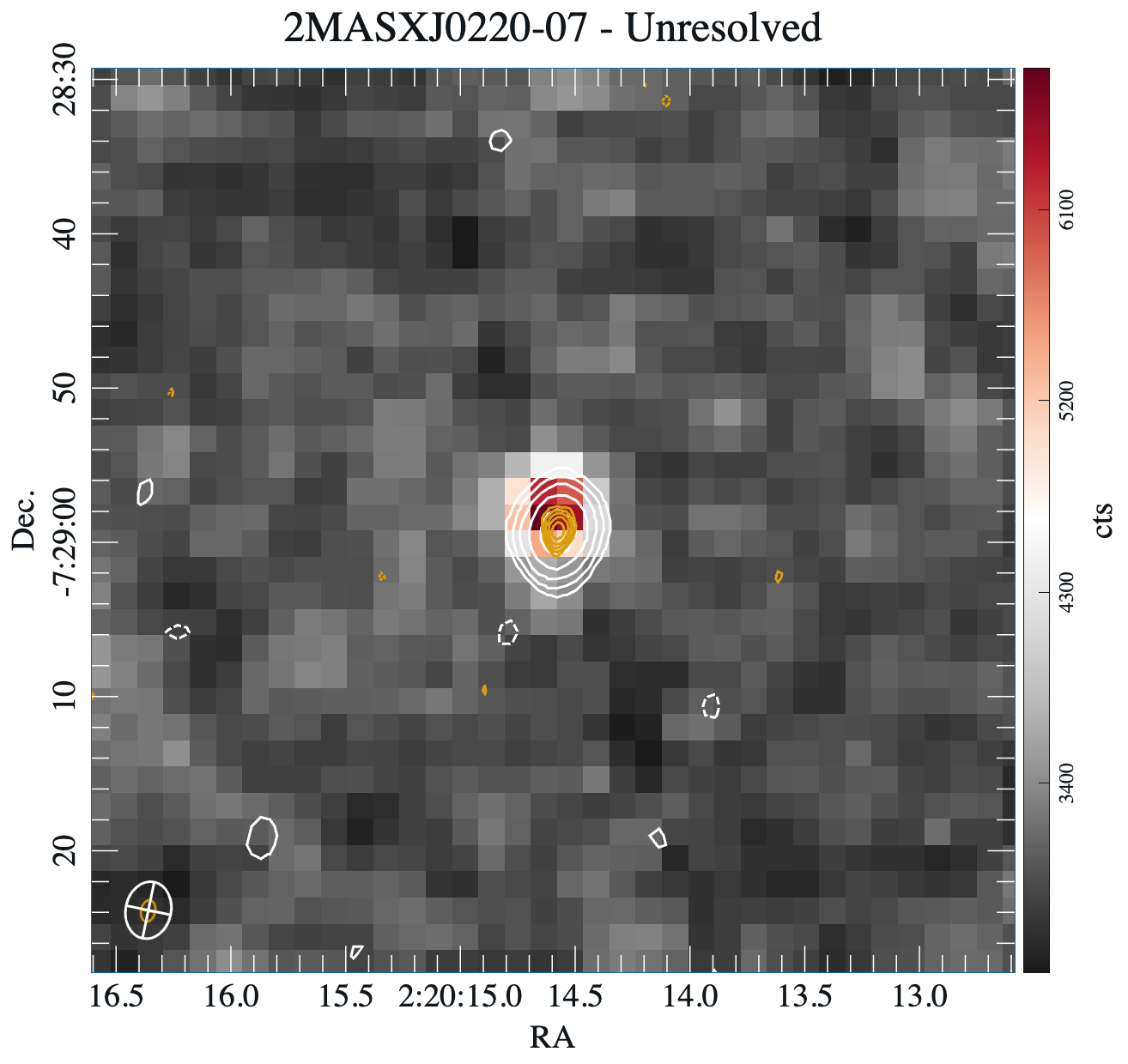}
\hfill
 \includegraphics[width=0.31\textwidth]{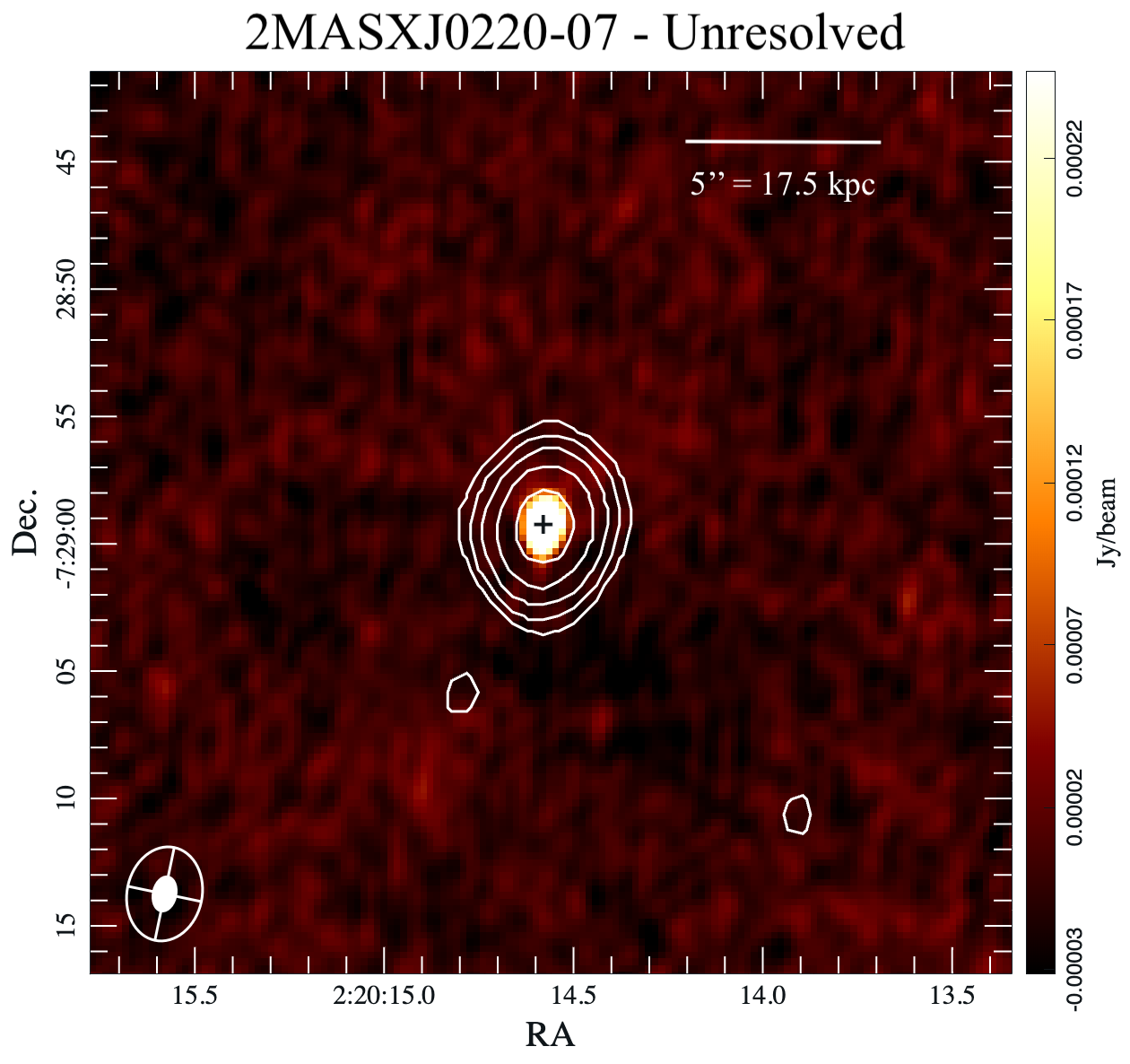}
\hfill
			\includegraphics[width=0.36\textwidth]{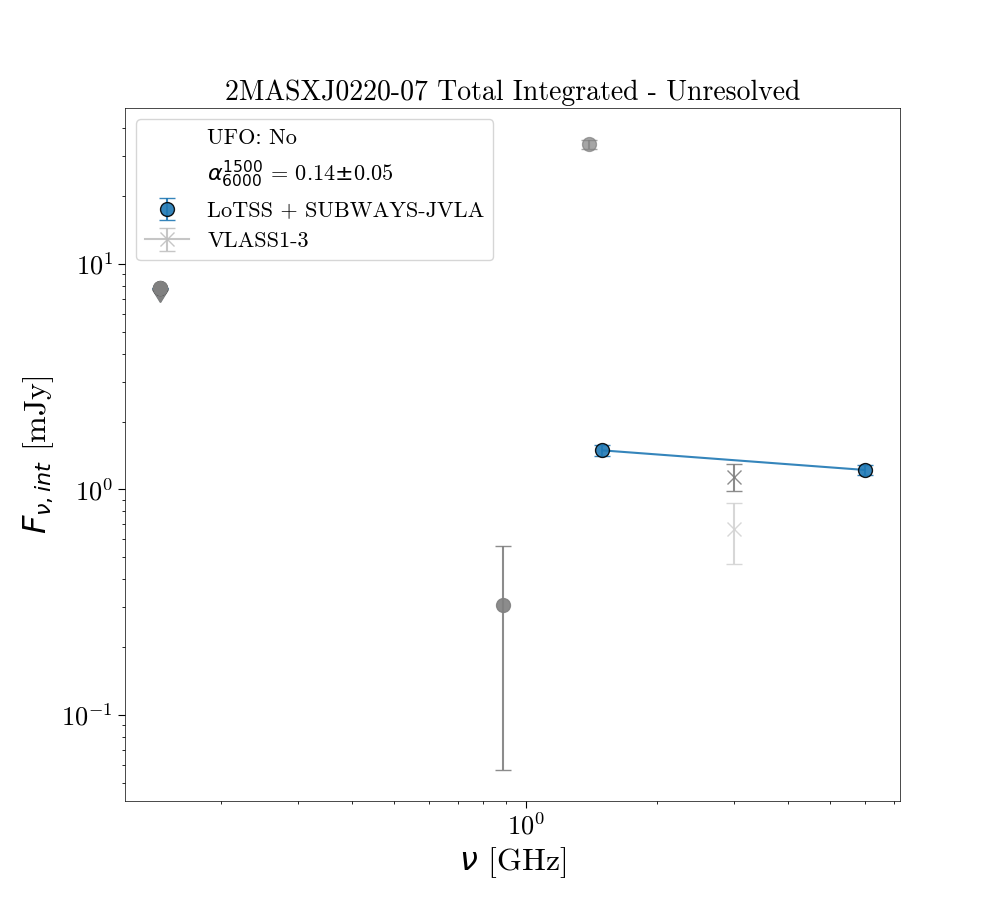}
\caption{2MASXJ0220-07. The linear size is <3.66 kpc. The SED is very flat at GHz frequencies, and the offset of RACS, FIRST and VLASS flux densities with respect to the computed slopes suggests strong variability. No UFO is detected. The SFR predicted with radio emission is consistent with the IR deduced one within $3\sigma$. The target exceeds the 3$\sigma$ uncertainty around the Güdel–Benz relation. According to the spectral shape, and given the offset with respect to the Güdel–Benz relation, we conclude the radio emission is most likely due to an unresolved compact jet.}
	\label{sed:2MASXJ0220}
\end{figure*}	

\begin{figure*}[hp!]
		\centering
		\includegraphics[width=0.31\textwidth]{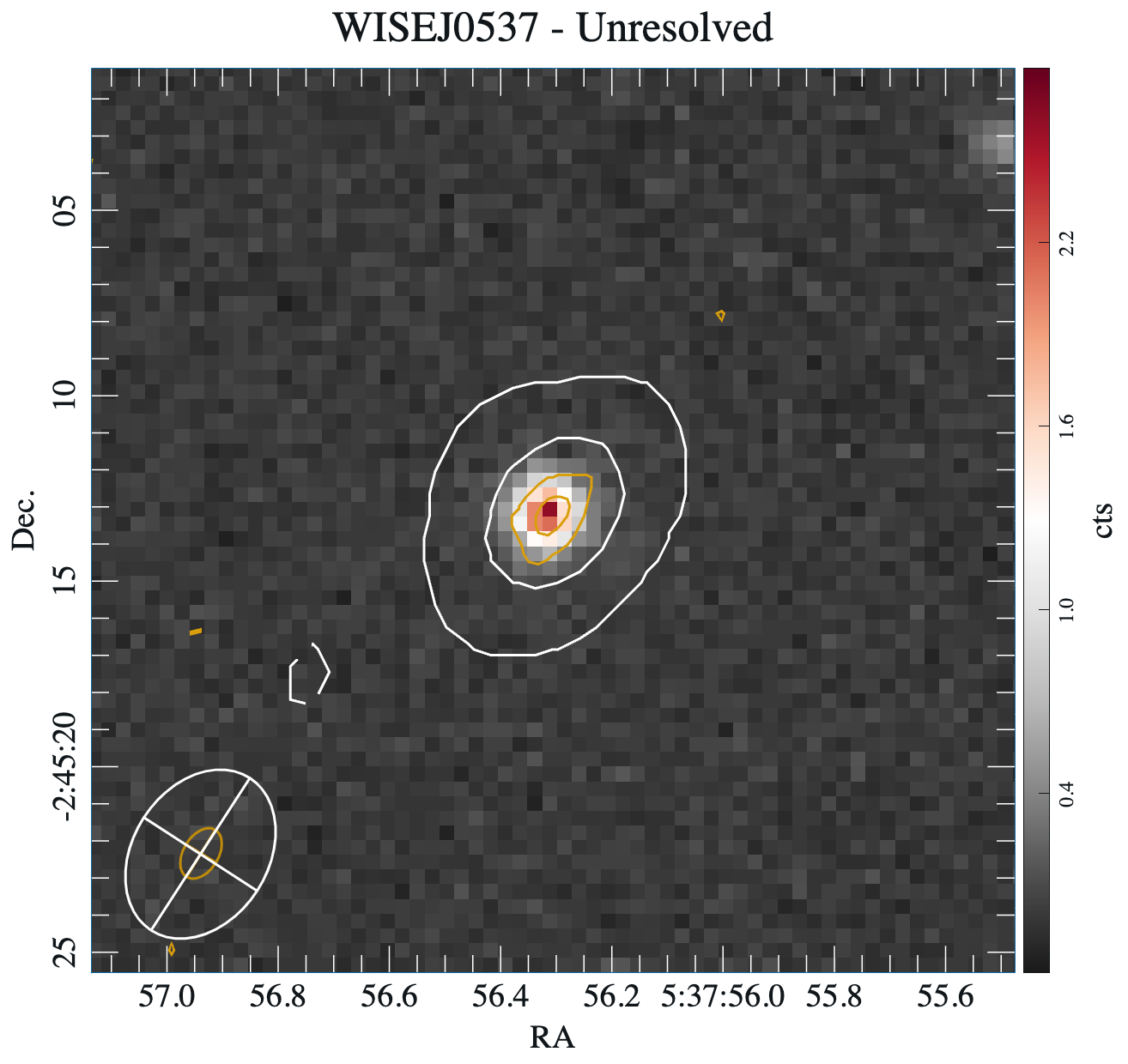}
 \hfill
 \includegraphics[width=0.31\textwidth]{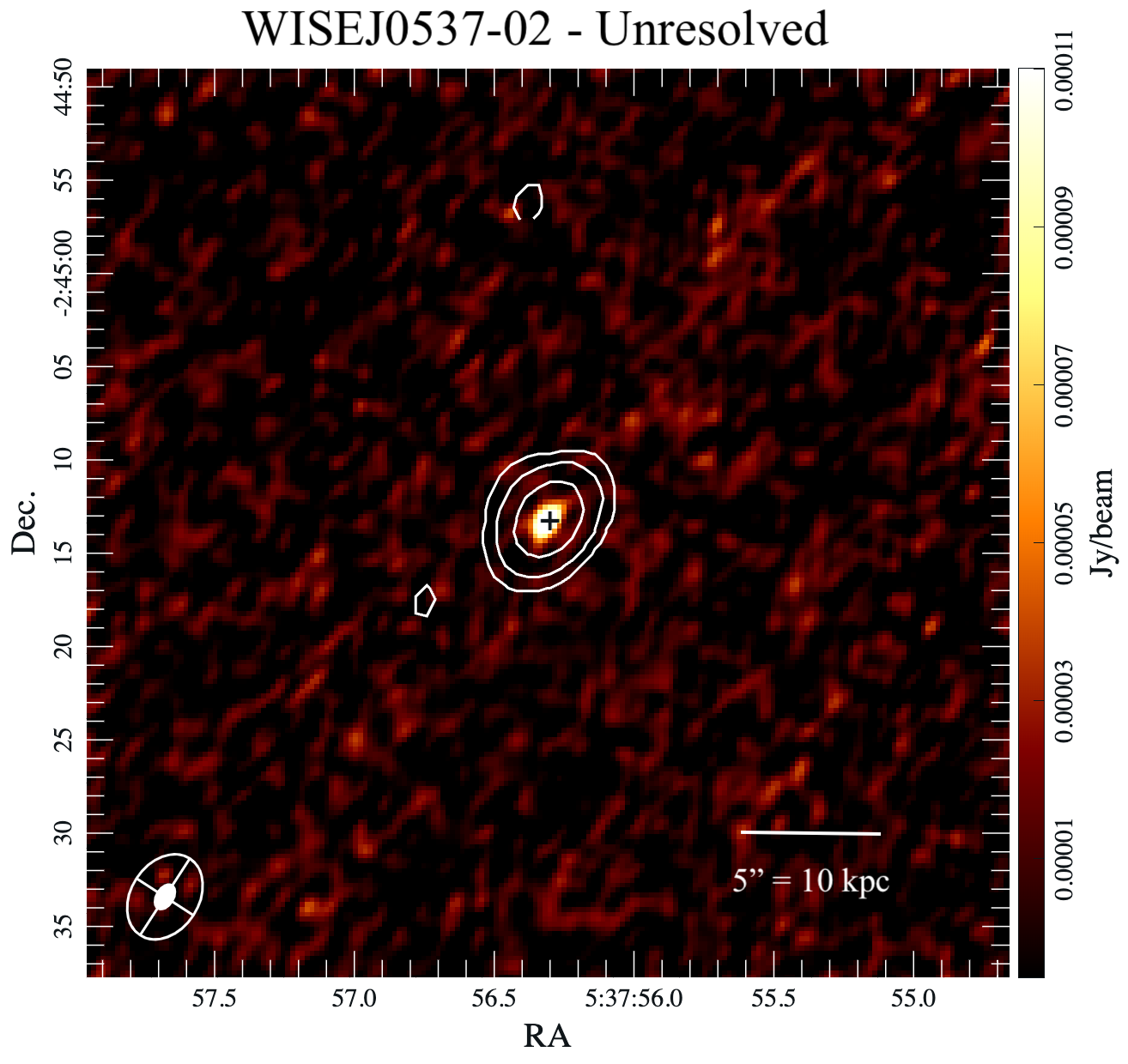}
	\hfill
		\includegraphics[width=0.36\textwidth]{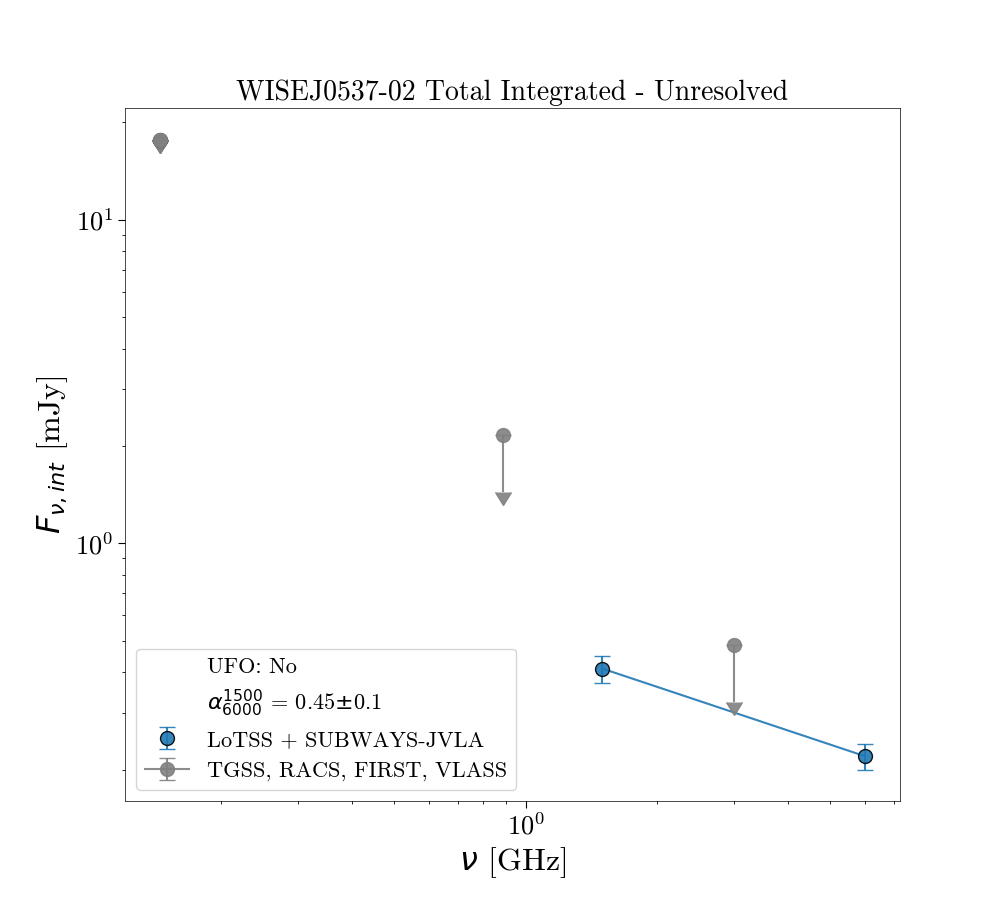}
	\caption{WISEJ0537-02. The linear size is <1.8 kpc. The SED is steep at GHz, within the uncertainties. No UFO detected. The SFR predicted with radio emission is consistent with the IR deduced one within $2\sigma$. The target follows the Güdel–Benz relation within $1\sigma$ uncertainty. According to the spectral shape, we conclude the radio emission is most likely due to an unresolved compact jet, however some contribution from SF cannot be excluded.}
	\label{sed:WISEJ0537-02}
\end{figure*}	

\begin{figure*}[hp!]
		\centering
			\includegraphics[width=0.31\textwidth]{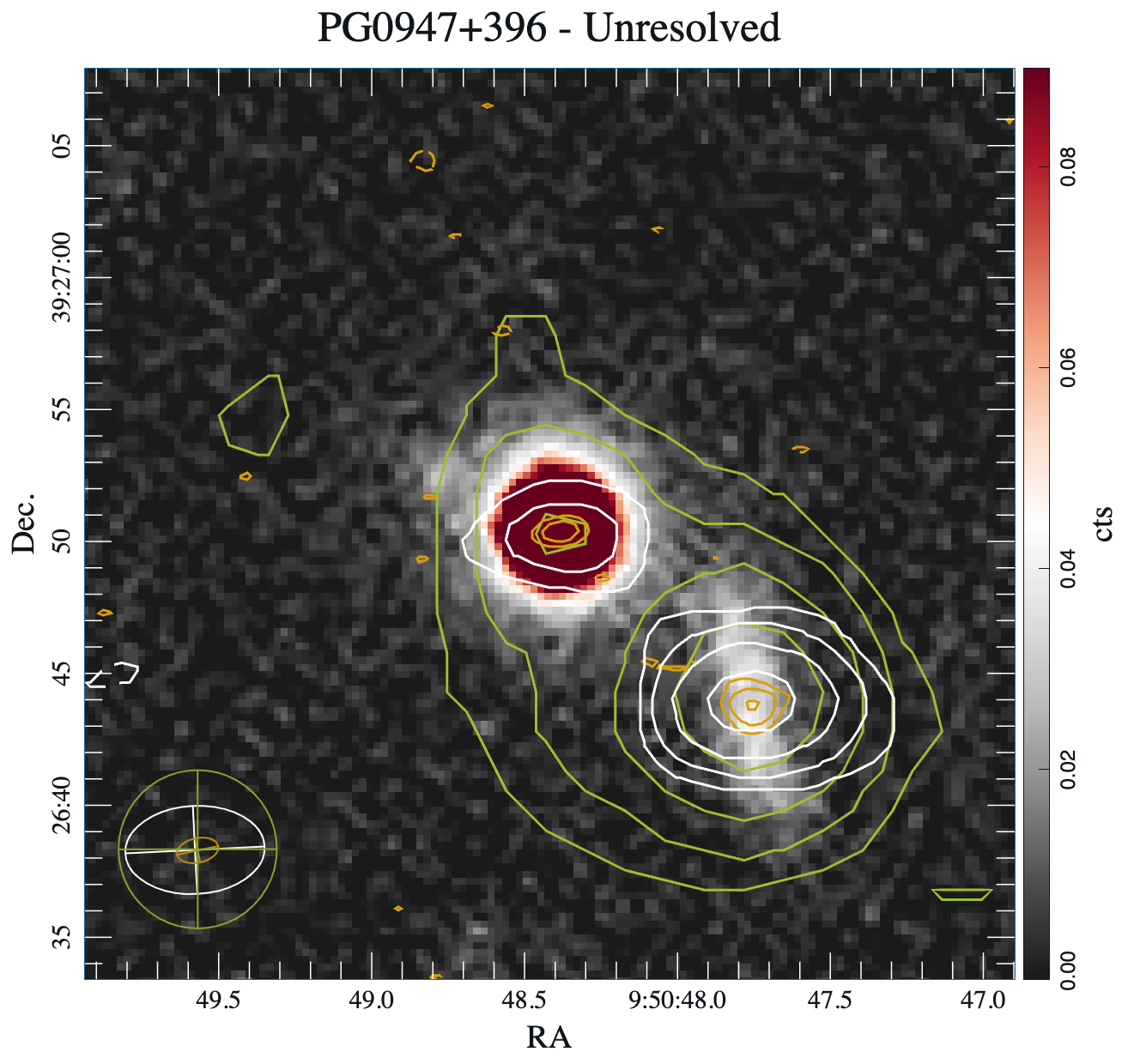}
\hfill
 \includegraphics[width=0.31\textwidth]{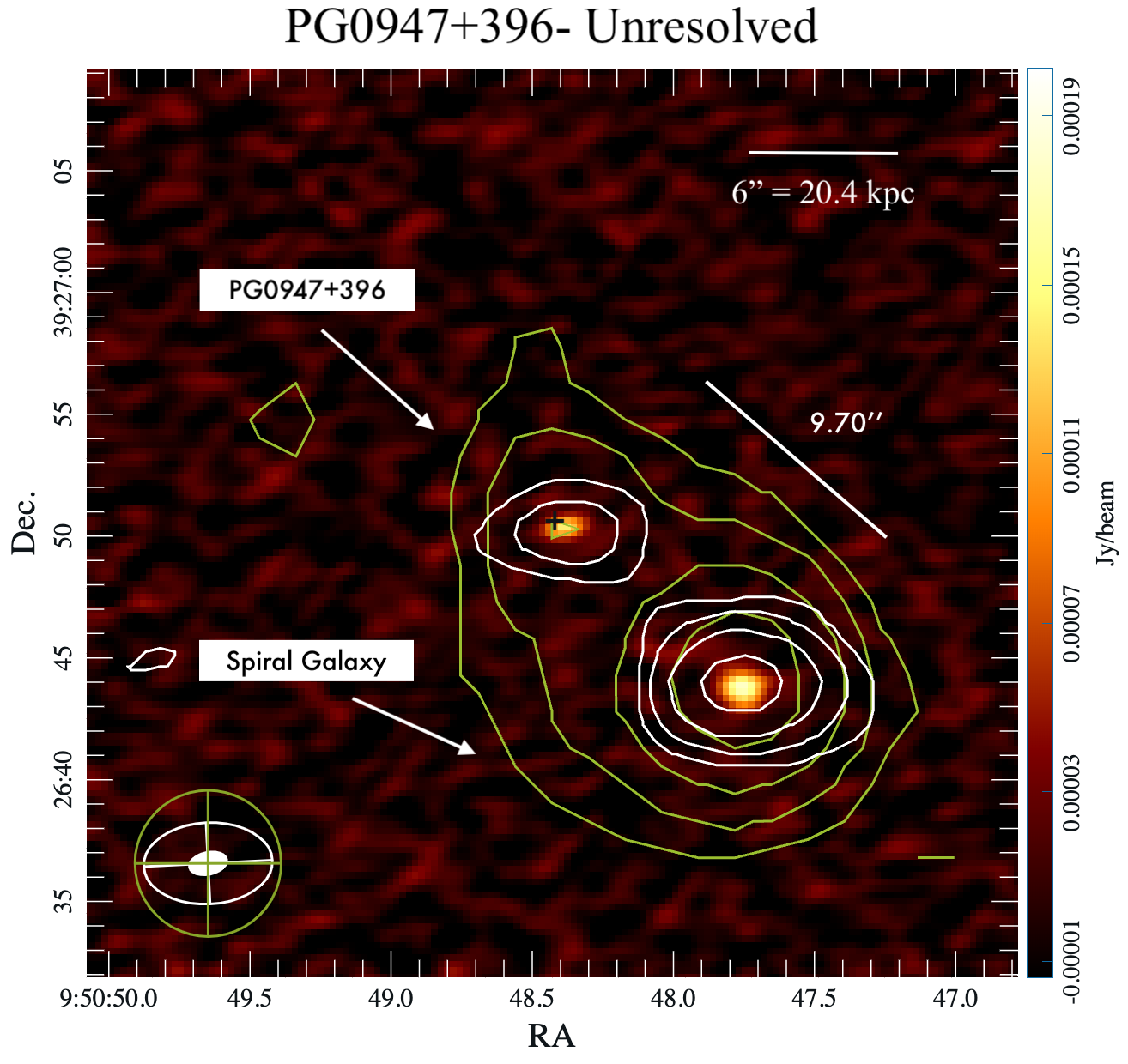}
\hfill
		\includegraphics[width=0.36\textwidth]{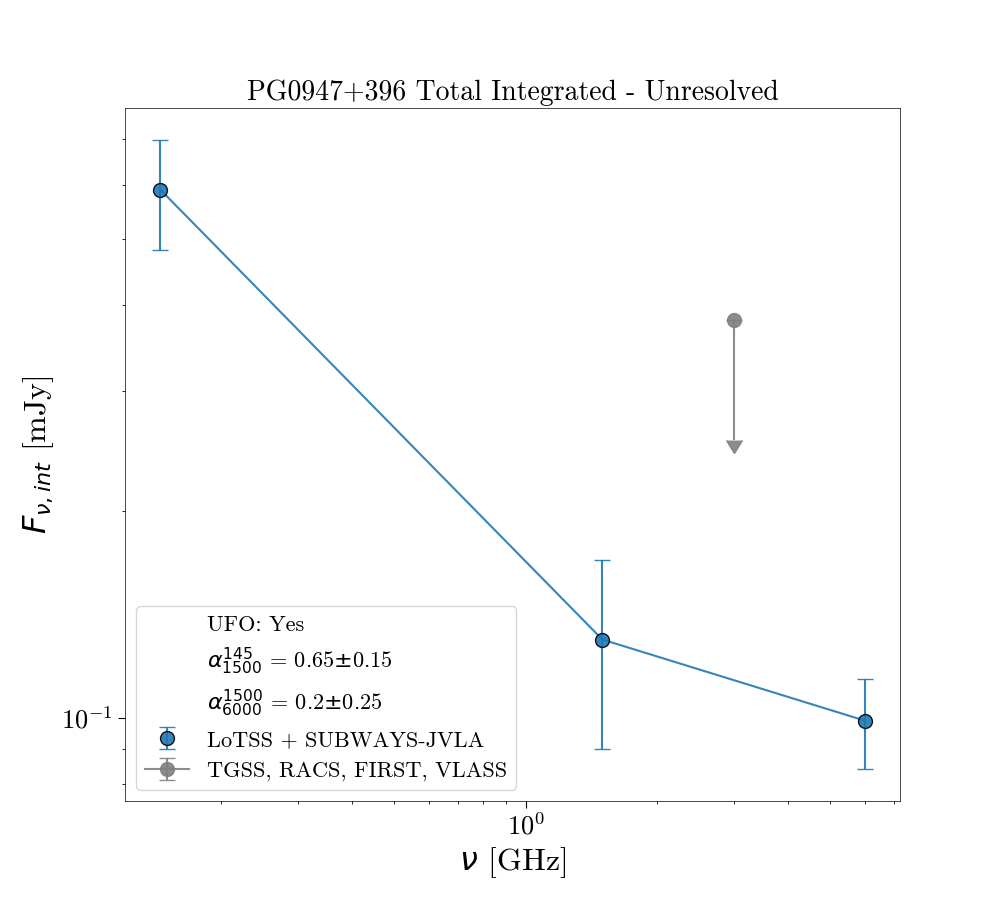}
	\caption{PG0947+396. The linear size is < 4.28 kpc. The SED is flat at GHz frequencies but it can be considered steep at all frequencies within uncertainties. The MHz-frequency slope may be not realistic due to technical difficulties in excluding the spiral galaxy south west of the target. A UFO is detected. The SFR predicted with radio emission is consistent with the IR deduced one within $2\sigma$. The target follows the Güdel–Benz relation within $1\sigma$ uncertainty. Estimates done with the wind model as in \citet{Nims_2015} support the wind scenario. This object is particularly ambiguous, however the most constraining results favour either the wind scenario (if the SED is considered steep) or the corona one (if the SED is considered flat at GHz frequencies).}
	\label{sed:PG0947+396}
\end{figure*}

\begin{figure*}[hp!]
		\centering
			\includegraphics[width=0.31\textwidth]{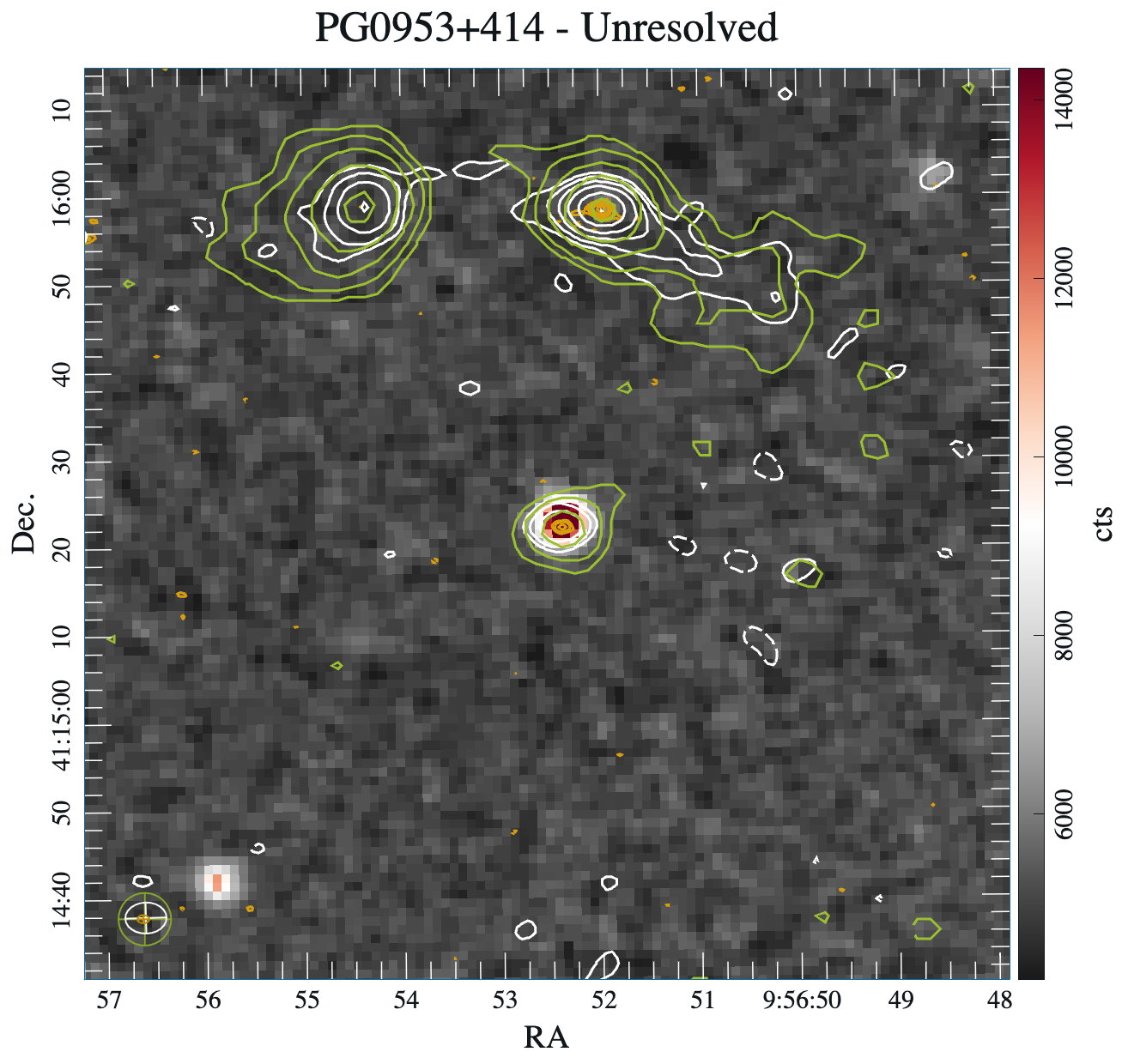}
\hfill
 \includegraphics[width=0.31\textwidth]{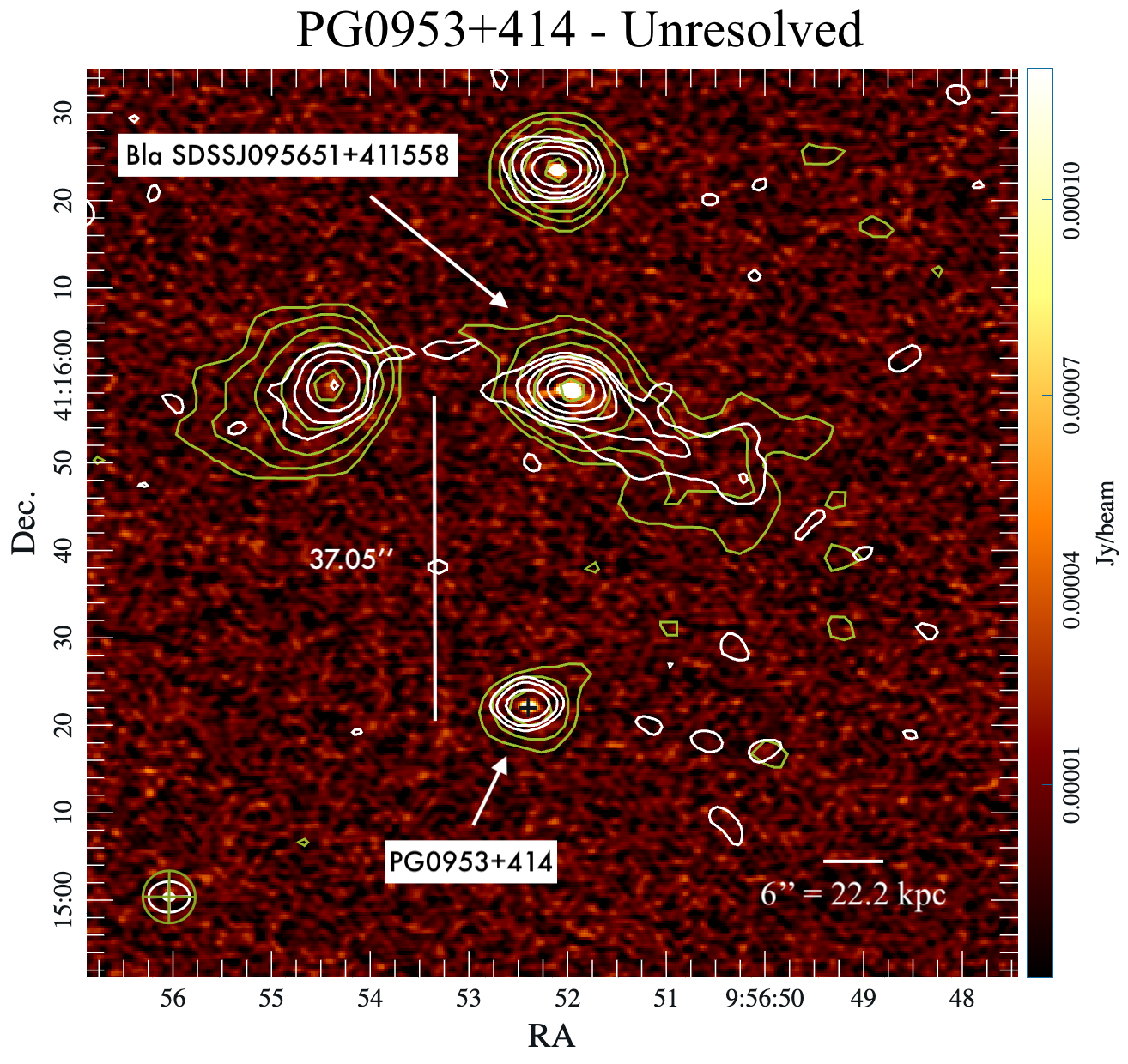}
\hfill
		\includegraphics[width=0.36\textwidth]{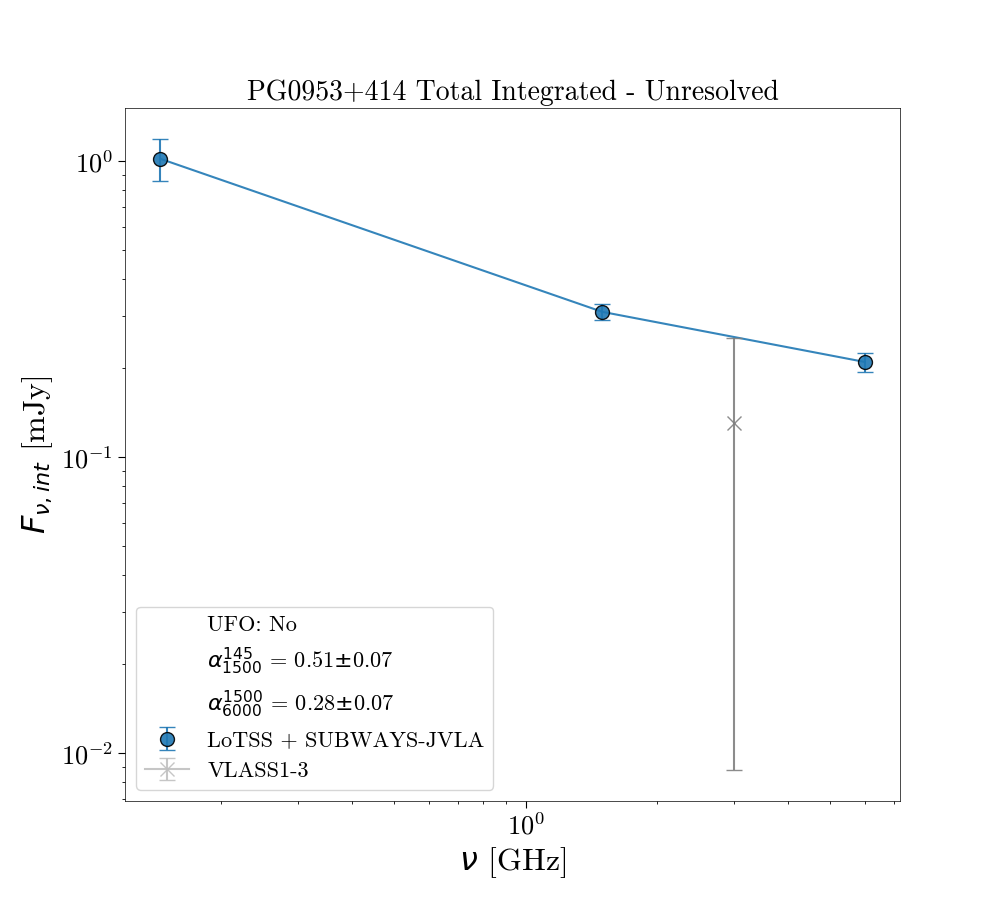}
	\caption{PG0953+414. The linear size is < 3.85 kpc. The SED is flat at GHz frequencies but steeper at MHz ones. No UFO detected. The SFR predicted with radio emission is consistent with the IR deduced one within $1\sigma$. The target follows the Güdel–Benz relation within $1\sigma$ uncertainty. The radio emission is most likely the result of a superposition of processes and it is difficult to disentangle the dominant one. The GHz spectral index is flat but not as flat as we expect for a corona dominated emission, therefore an unresolved compact jet may explain the the SED curvature.}
	\label{sed:PG0953+414}
\end{figure*}	

\begin{figure*}[hp!]
		\centering
		\includegraphics[width=0.31\textwidth]{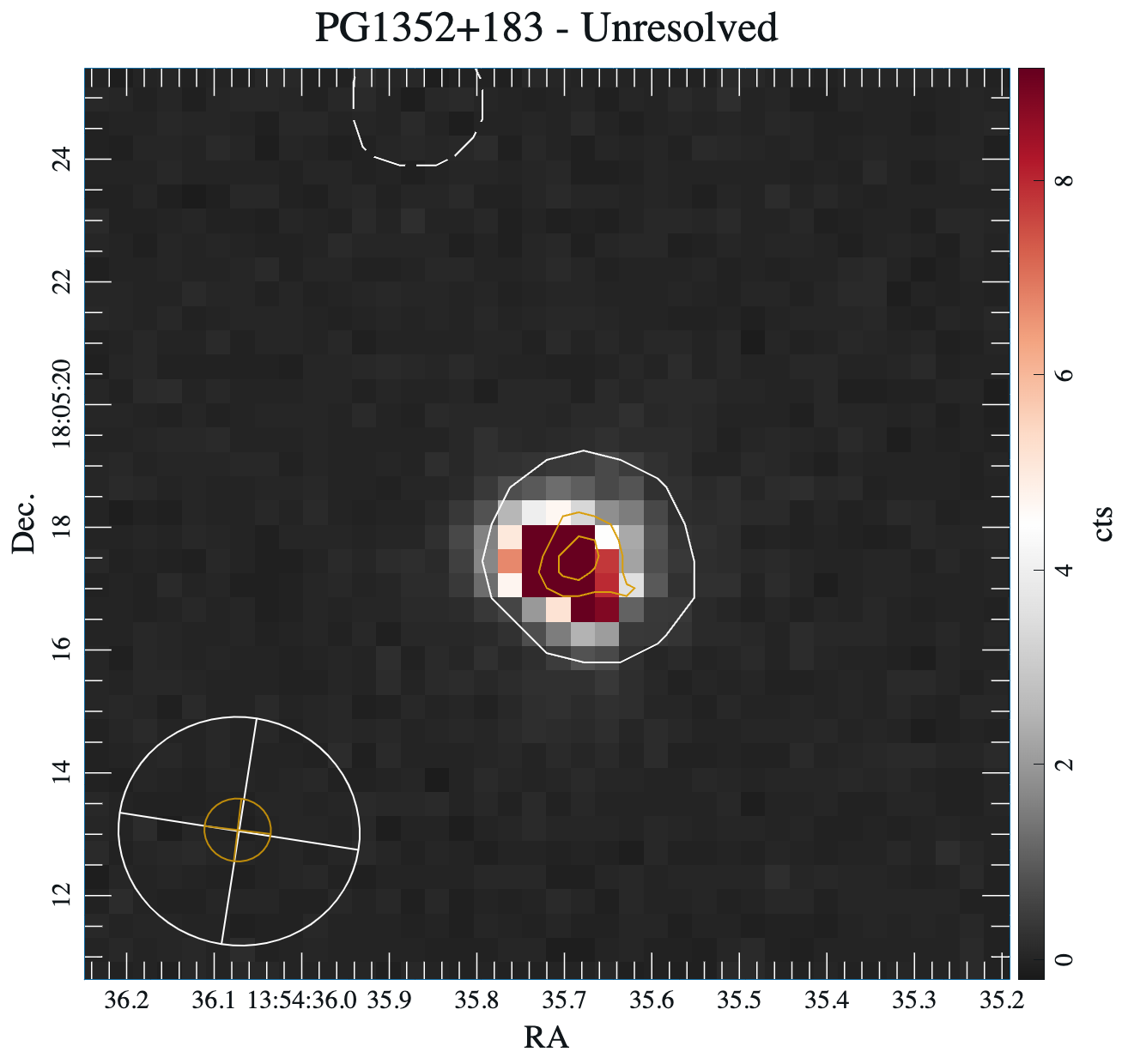}
\hfill
 \includegraphics[width=0.31\textwidth]{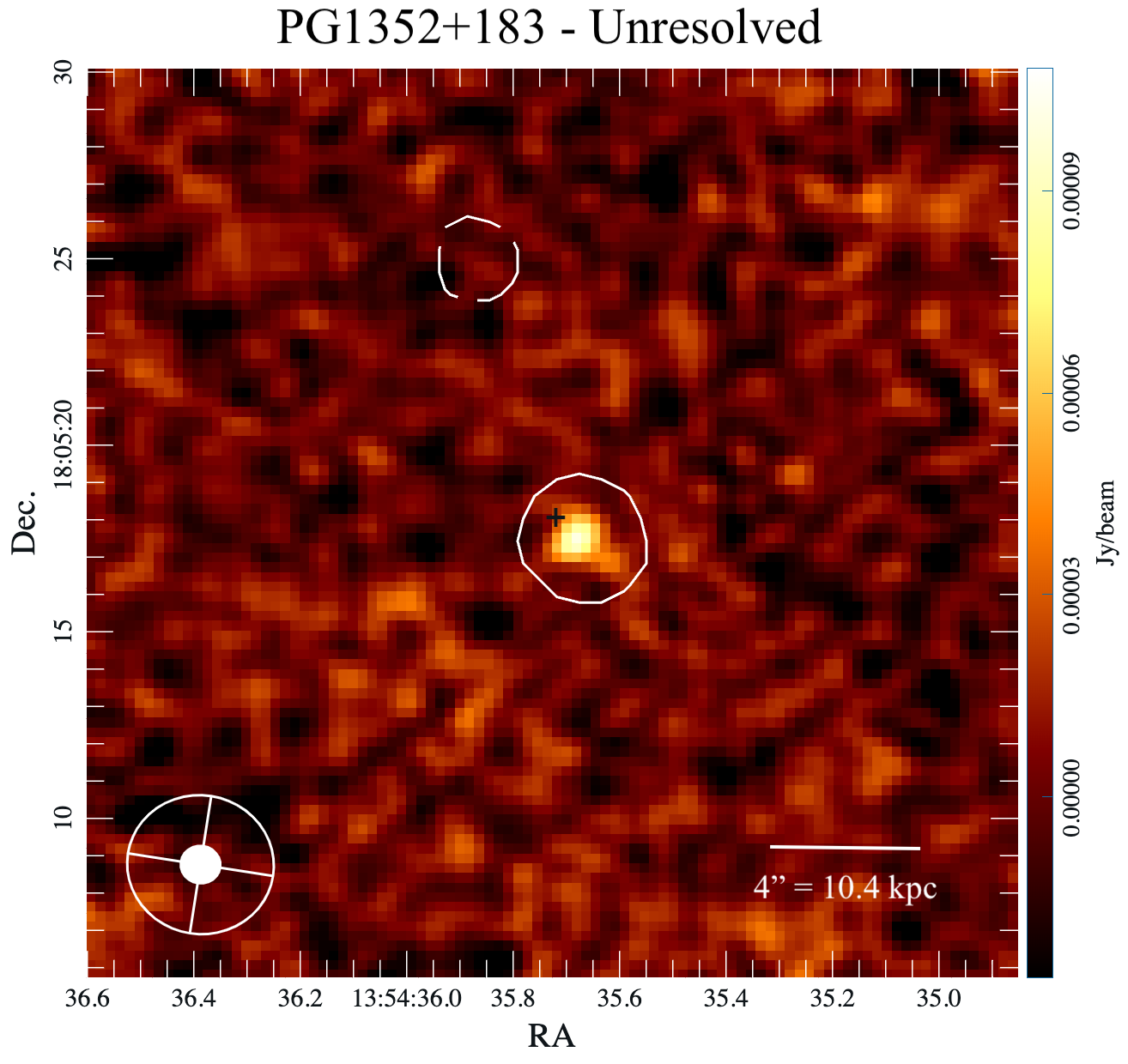}
\hfill
		\includegraphics[width=0.36\textwidth]{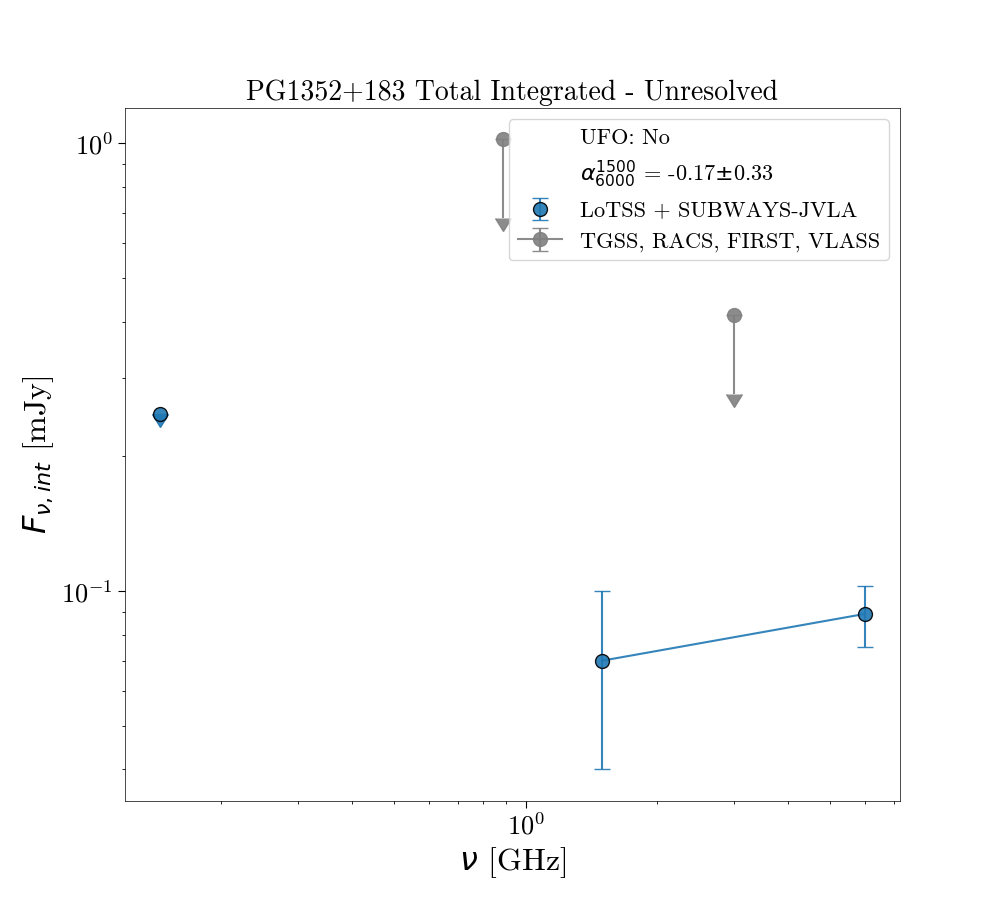}
	\caption{PG1352+183. The linear size is < 2.44 kpc. The SED is flat at GHz frequencies (can be also inverted). No UFO is detected. The SFR predicted with radio emission is consistent with the IR deduced one within $1\sigma$. The target follows the Güdel–Benz relation within $1\sigma$ uncertainty. According to the SED shape, the radio emission could be fully explained with coronal emission, however an unresolved jet cannot be ruled-oud.}
	\label{sed:PG1352+183}
\end{figure*}	

\begin{figure*}[hp!]
		\centering
		\includegraphics[width=0.31\textwidth]{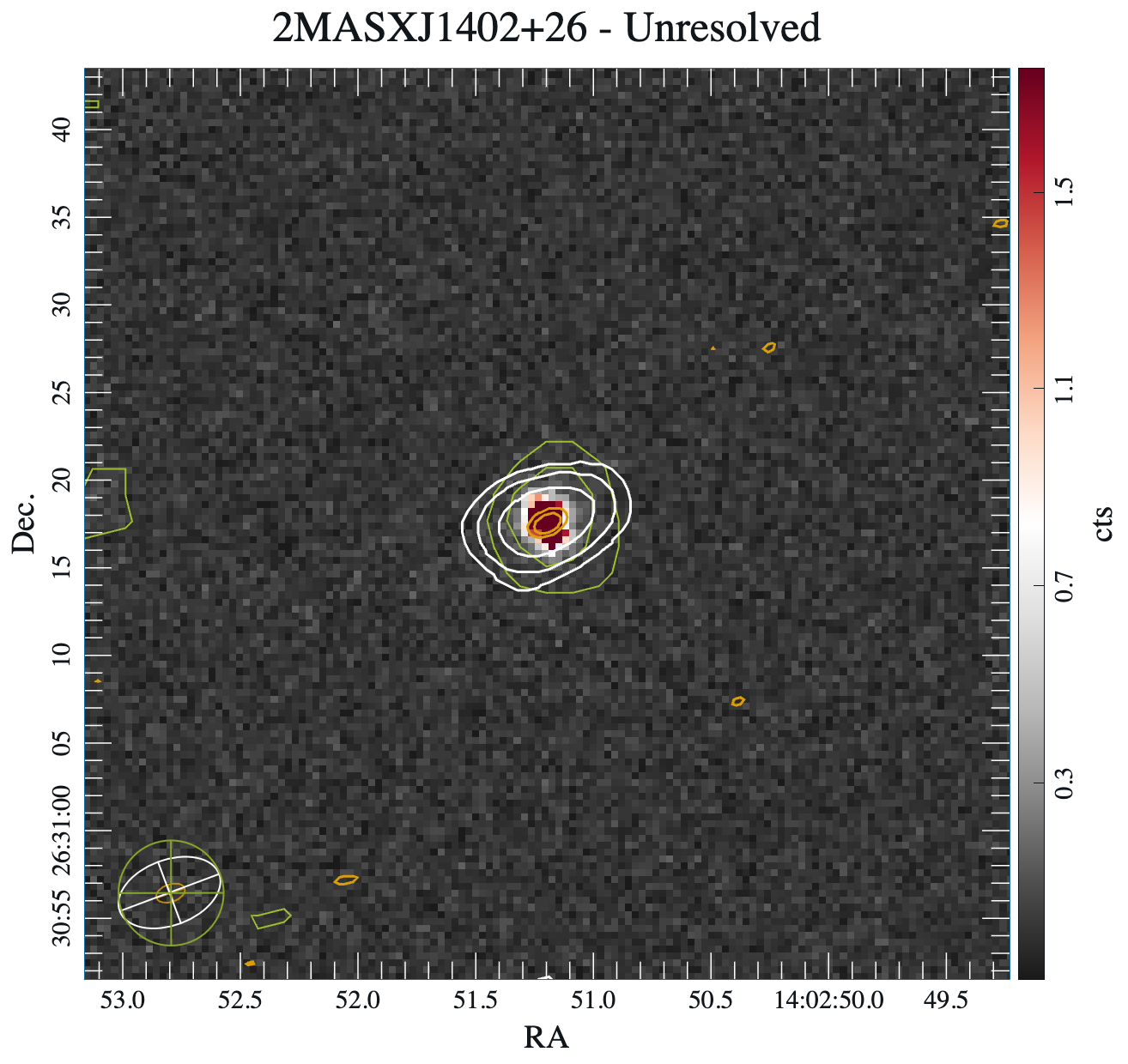}
\hfill
 \includegraphics[width=0.31\textwidth]{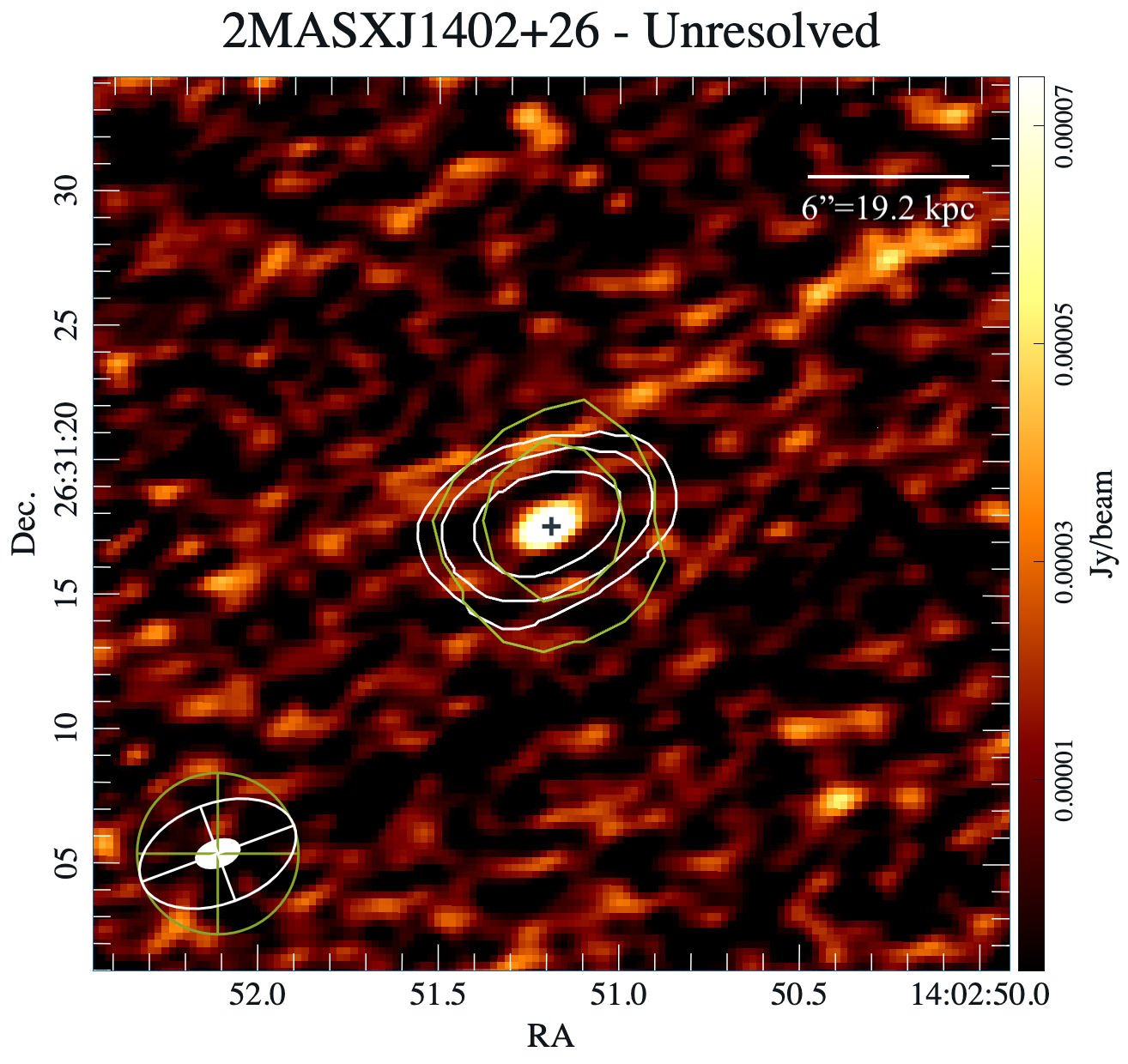}
\hfill
		\includegraphics[width=0.36\textwidth]{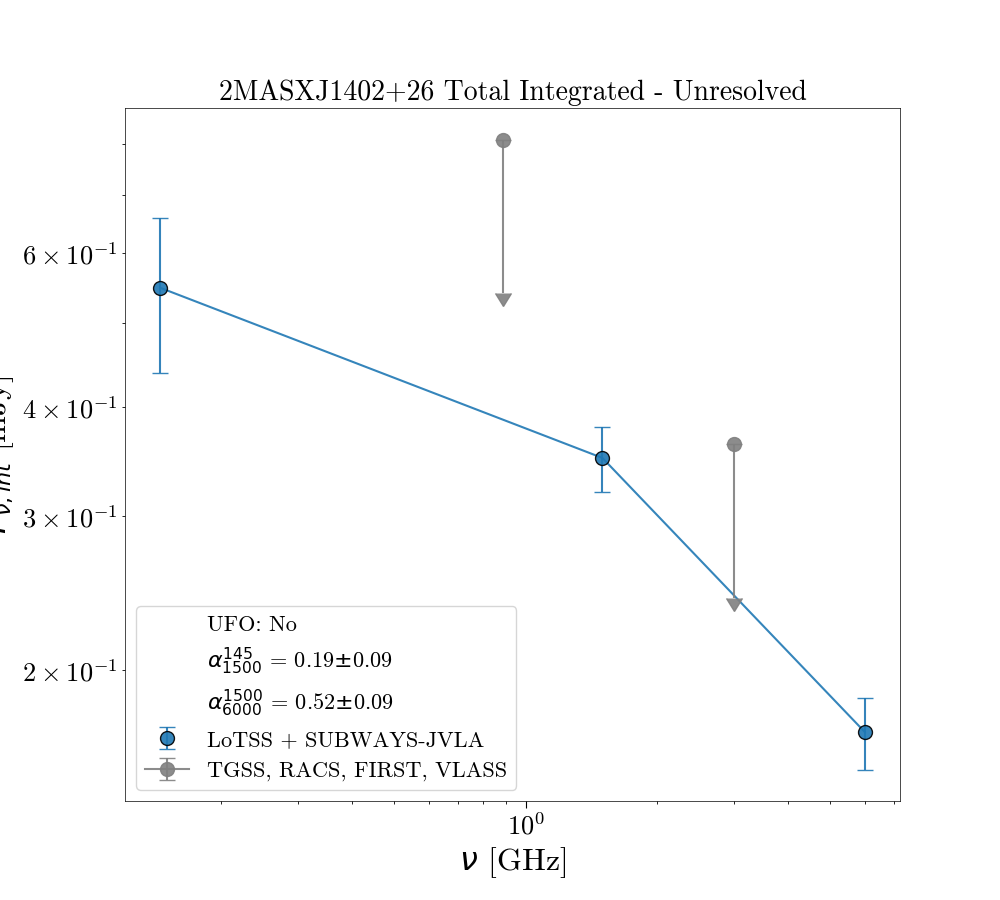}

	\caption{2MASXJ1402+26. The linear size is < 4.42 kpc. The SED is flat then steep, with an optically thin component becoming dominant at GHz frequencies. No UFO is detected. The SFR predicted with radio emission is compatible with the IR deduced one within $1\sigma$. The target follows the Güdel–Benz relation within $1\sigma$ uncertainty. The spectral shape suggests radio emission at GHz frequencies could be fully explained with SF, but the flattening toward lower ones suggests some self-absorption may be present.}
	\label{sed:2MASXJ1402+26}
\end{figure*}	

\begin{figure*}[hp!]
		\centering
		\includegraphics[width=0.31\textwidth]{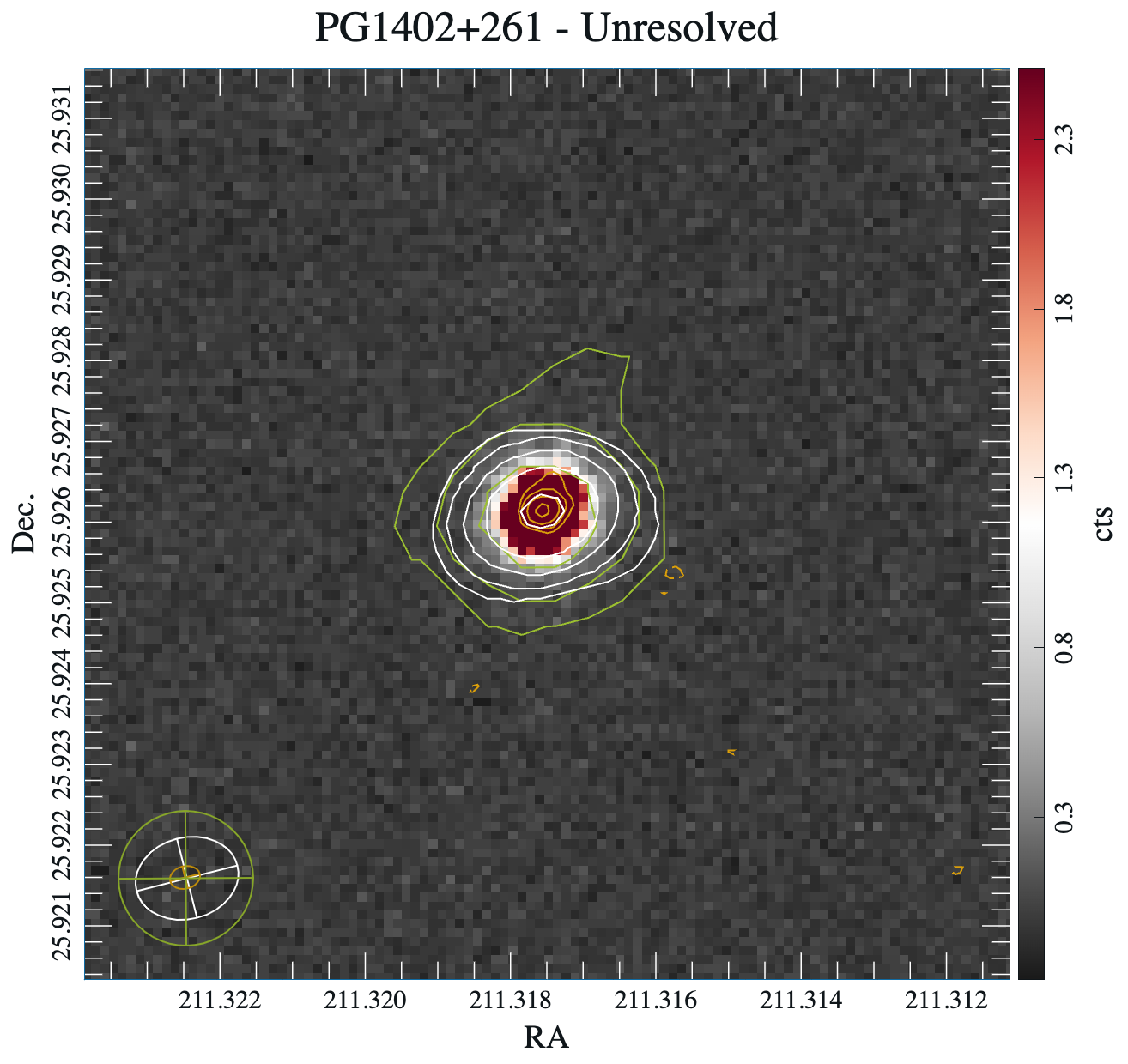}
\hfill
 \includegraphics[width=0.31\textwidth]{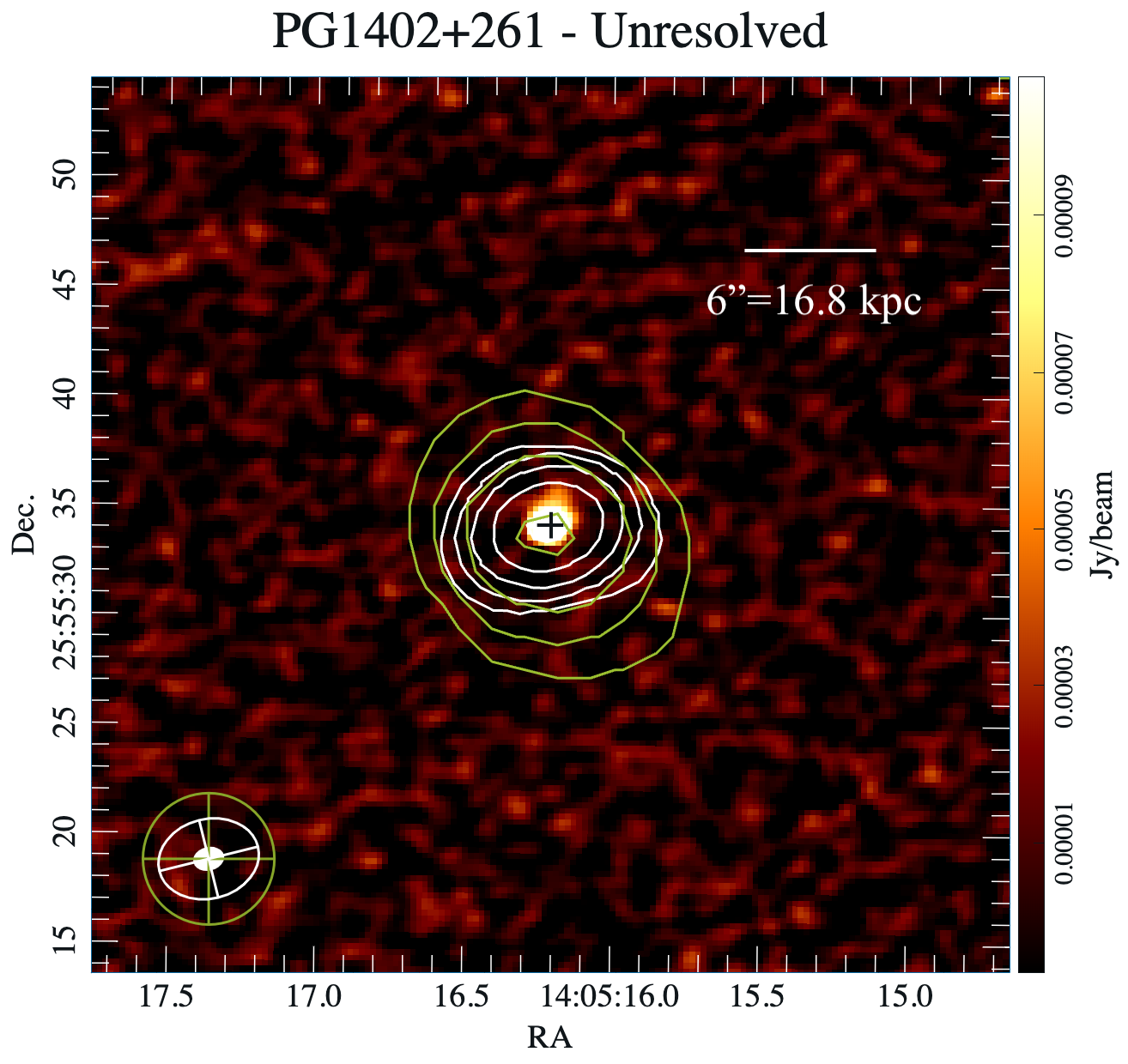}
\hfill
		\includegraphics[width=0.36\textwidth]{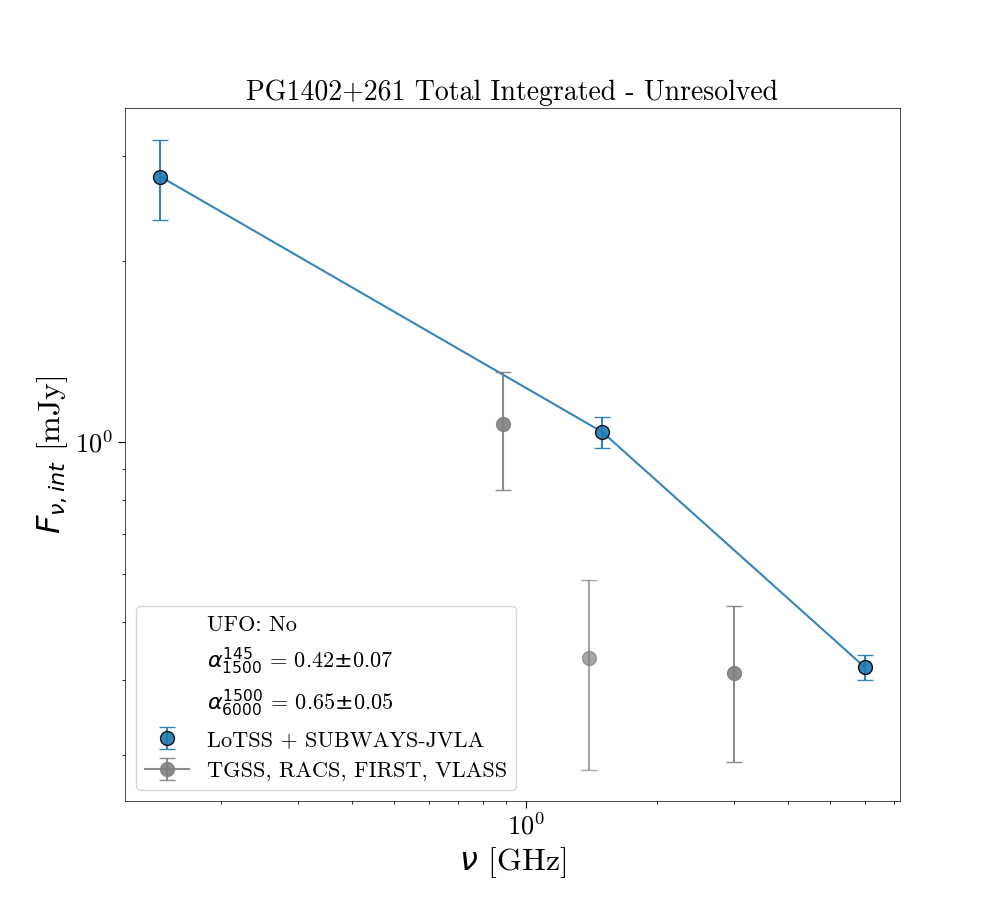}

	\caption{PG1402+26. The linear size is < 7.25 kpc. The SED is flat then steep, with an optically thin component becoming dominant at GHz frequencies. Variability is suggested by the location of FIRST and VLASS flux density. No UFO is detected. The SFR predicted with radio emission is compatible with the IR deduced one within $1\sigma$. The target follows the Güdel–Benz relation within $1\sigma$ uncertainty. Radio emission could be fully explained with SF at higher frequencies, but the joint presence of variability and the break in the spectrum may point to a different scenario, as a jet base.}
	\label{sed:PG1402+261}
\end{figure*}	

\begin{figure*}[hp!]
		\centering
		\includegraphics[width=0.31\textwidth]{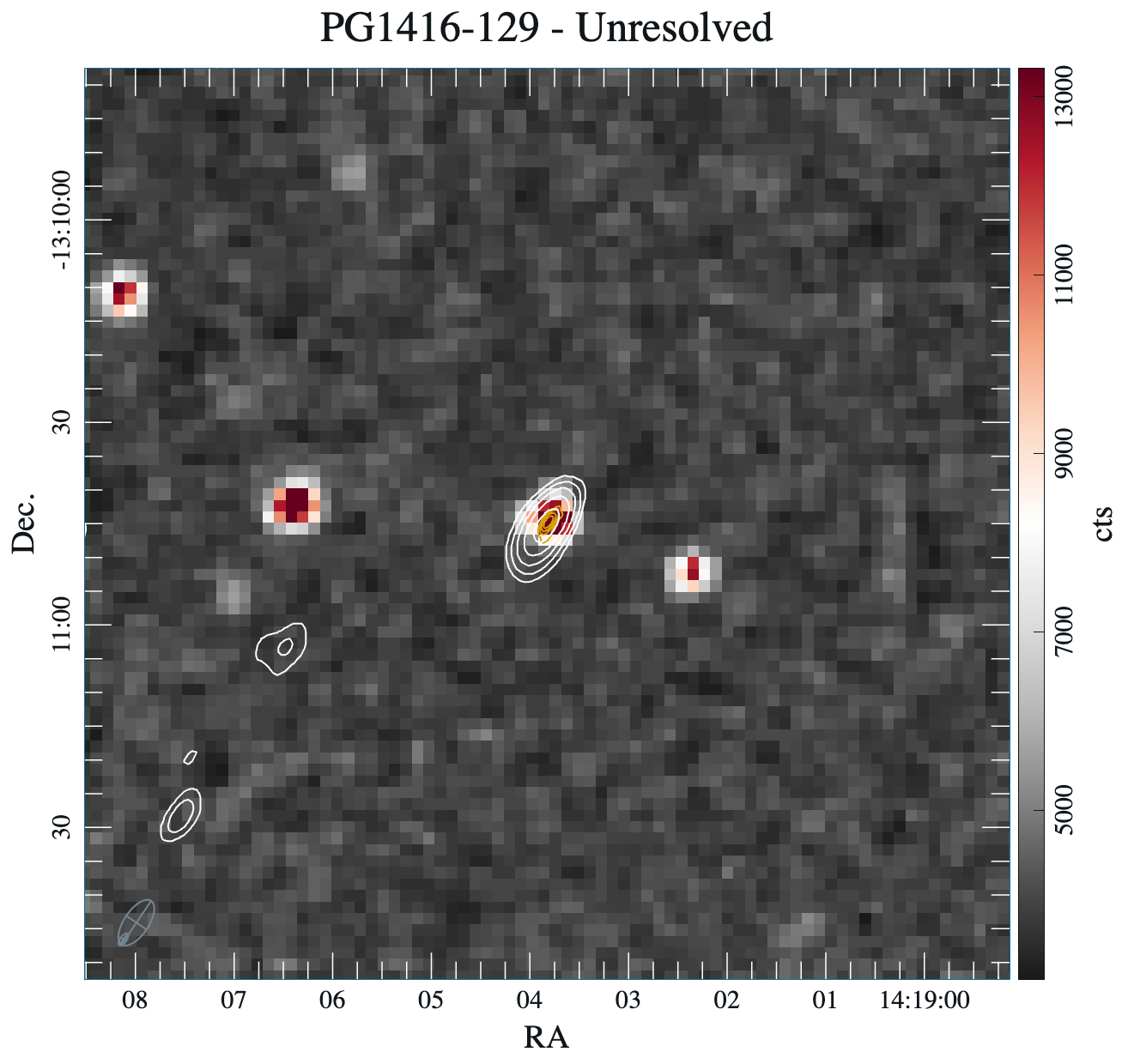}
\hfill
 \includegraphics[width=0.31\textwidth]{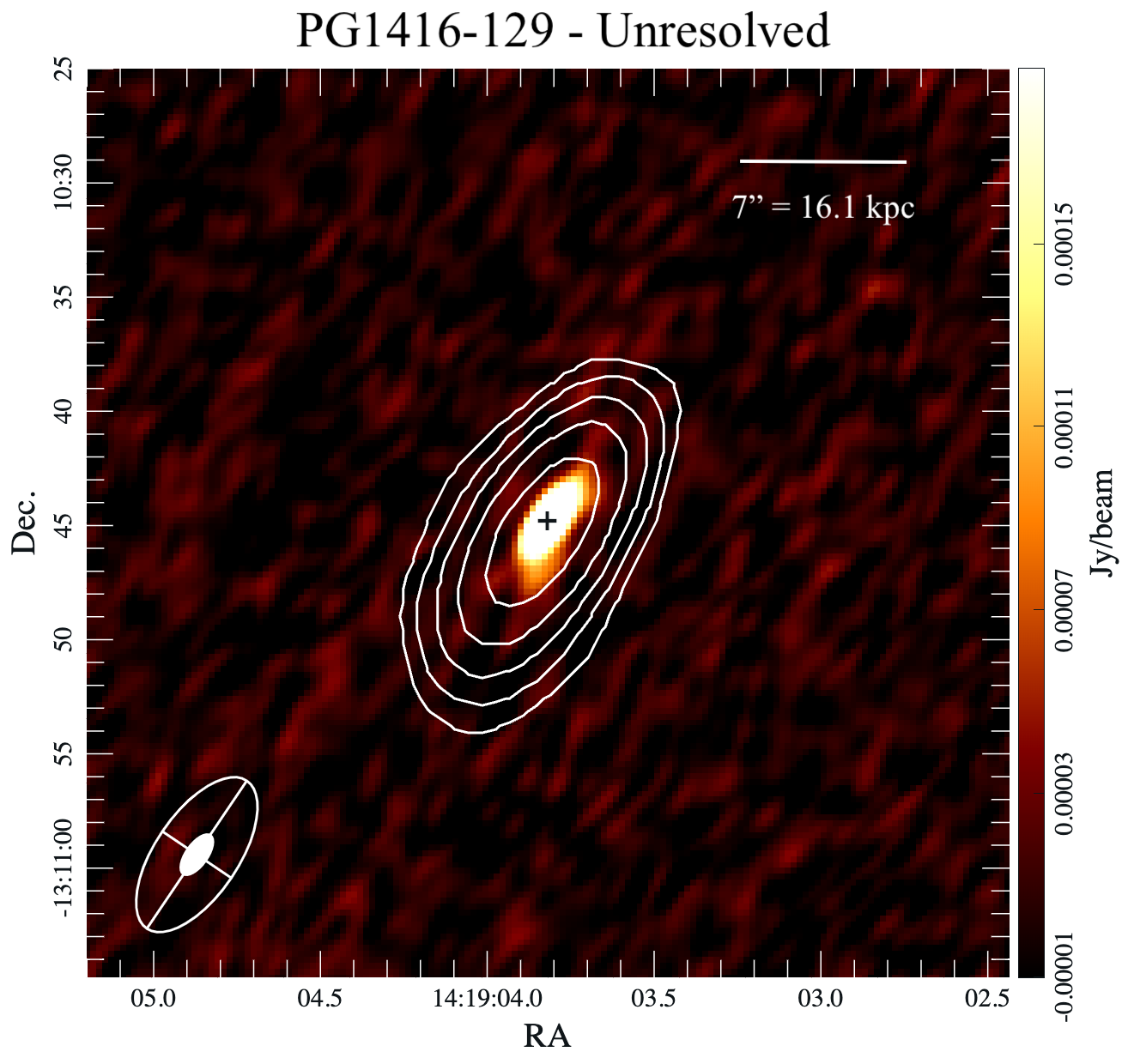}
\hfill
		\includegraphics[width=0.36\textwidth]{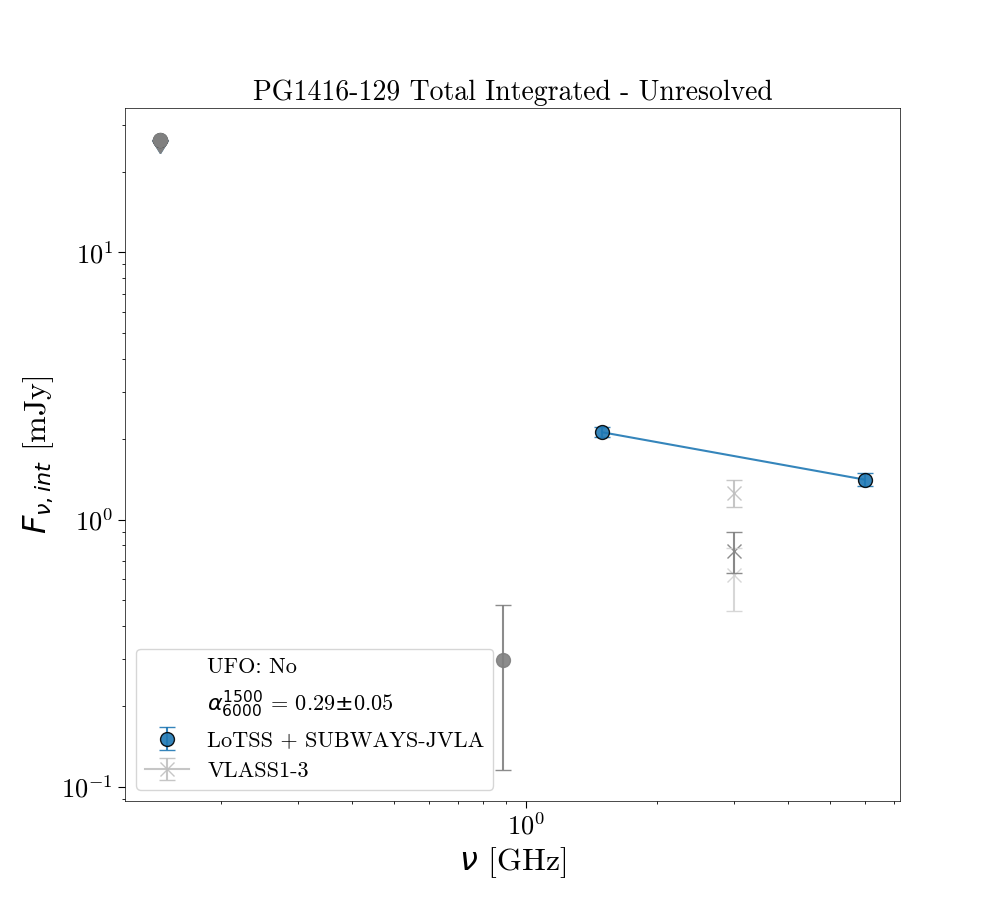}
	
	\caption{PG1416-129. The linear size is < 2 kpc. The SED is flat at GHz frequencies and shows signs of variability. No UFO detected. The SFR predicted with radio emission is a factor 34 higher than the IR deduced one (out of the $3\sigma$ uncertainty). The target follows the Güdel–Benz relation within $1\sigma$ uncertainty. The SED is not flat enough to be dominated by the corona, therefore an unresolved jet cannot be ruled-out. However \citet{Chen_2024}, with EVN observations, detect a compact core (linear size < 10 pc). The authors also report (from \citealt{Barvainis_1996}) a C-band flux density a factor $\sim0.42$ smaller than the one presented in this work, then attribute the dominant radio origin to the corona. We can infer our observations collect the superposition of more extended emission to the one of a pure compact corona.}
	\label{sed:PG1416-129}
\end{figure*}

\begin{figure*}[hp!]
\centering
	\includegraphics[width=0.31\textwidth]{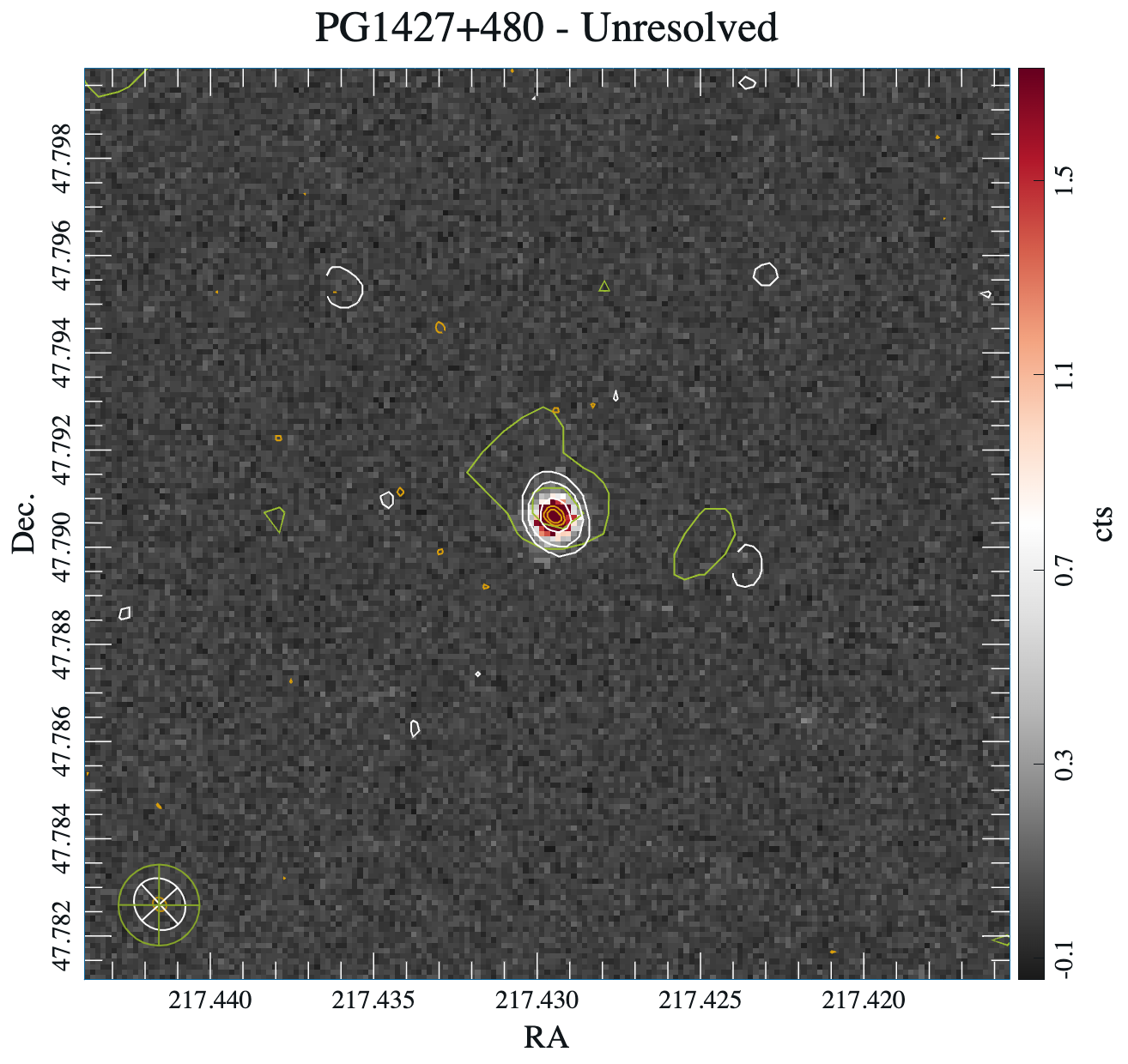}
	\hfill
	\includegraphics[width=0.31\textwidth]{IMG_new/PG1427+480_overlays_redNEW2.png}
\hfill
 \includegraphics[width=0.36\textwidth]{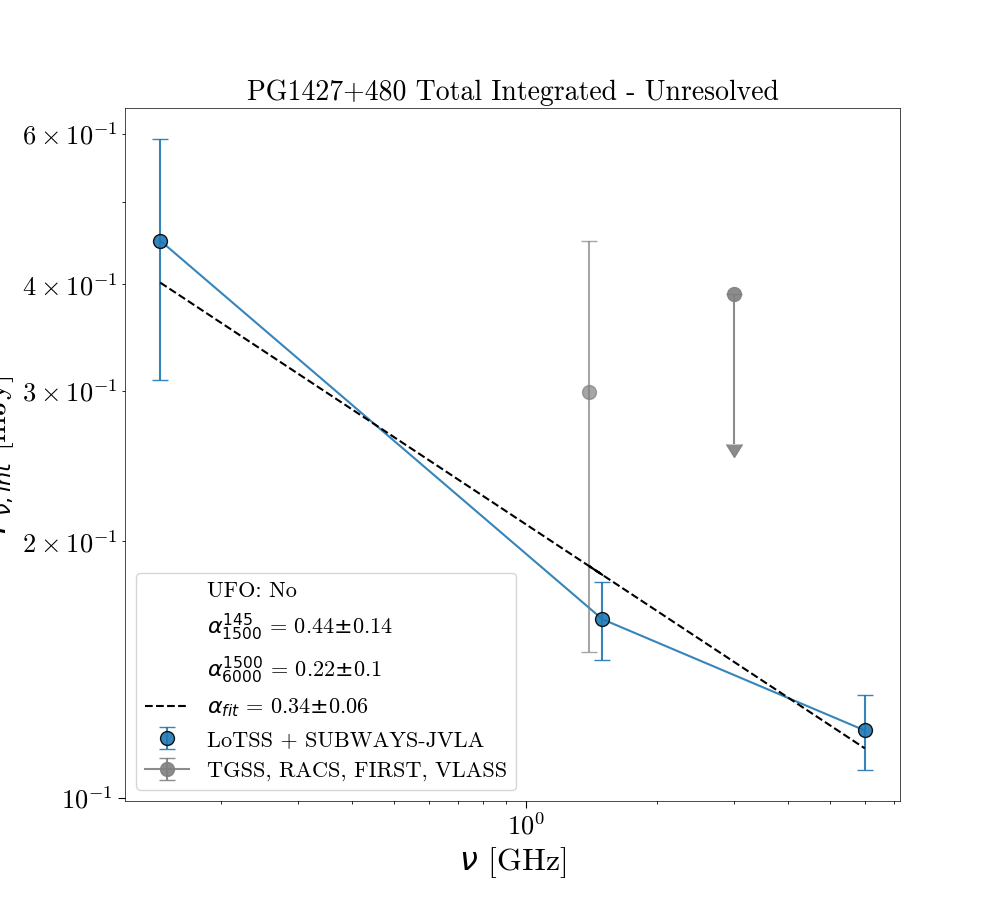}

	\caption{PG1427+480. The linear size is < 3.02 kpc. The SED is flat. No UFO is detected. The SFR predicted with radio emission is compatible with the IR deduced one within $1\sigma$. The target follows the Güdel–Benz relation within $1\sigma$ uncertainty. The target follows the Güdel–Benz relation within $1\sigma$ uncertainty. The MHz detection of a faint feature on the north-east side of the core deserves further investigations. Maybe a young jet could be the responsible for the SED shape, while the extended feature belongs to an older episode.}
	\label{sed:PG1427+480}
\end{figure*}

\begin{figure*}[hp!]
		\centering
	\includegraphics[width=0.31\textwidth]{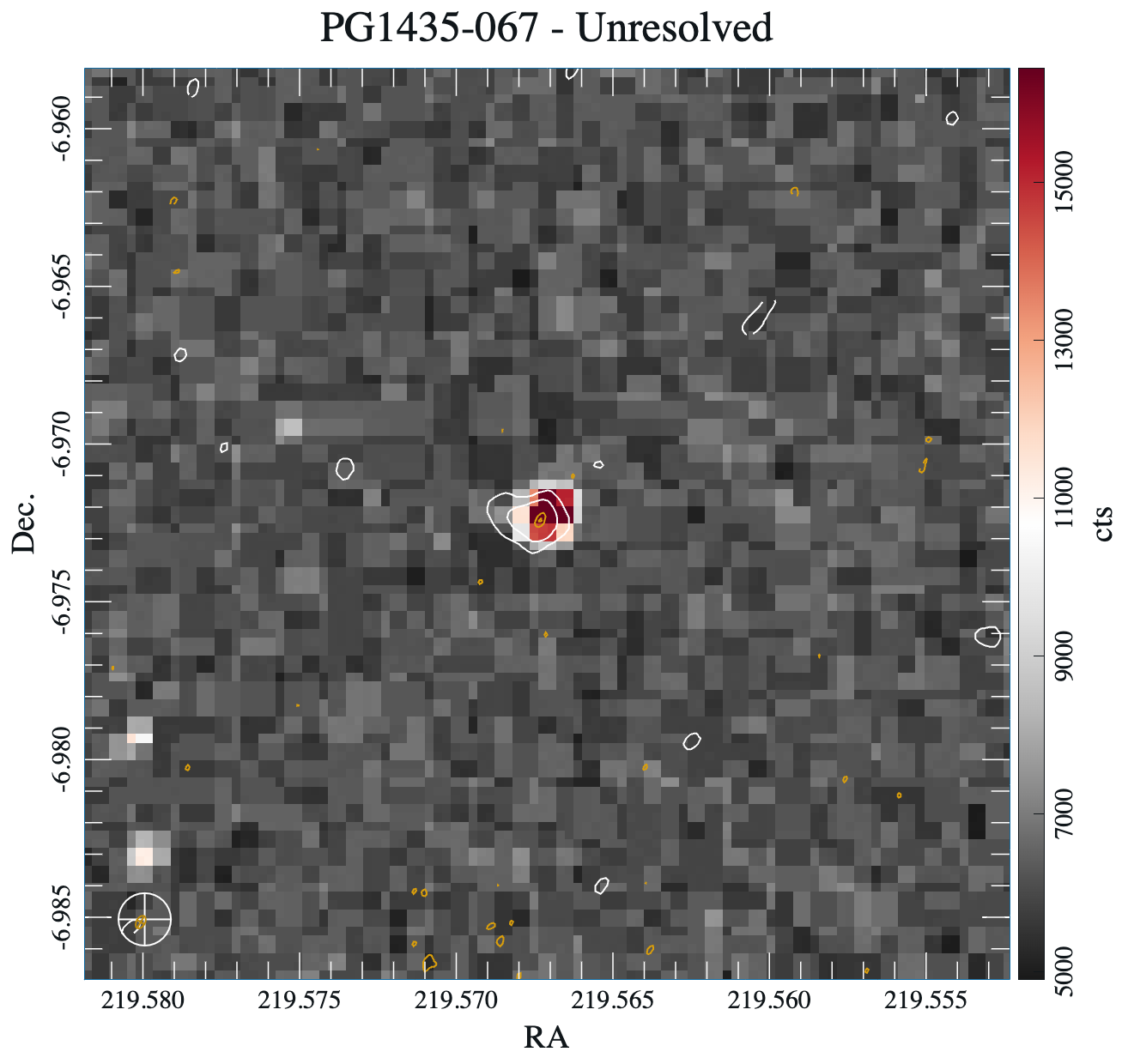}
\hfill
 {\includegraphics[width=0.31\textwidth]{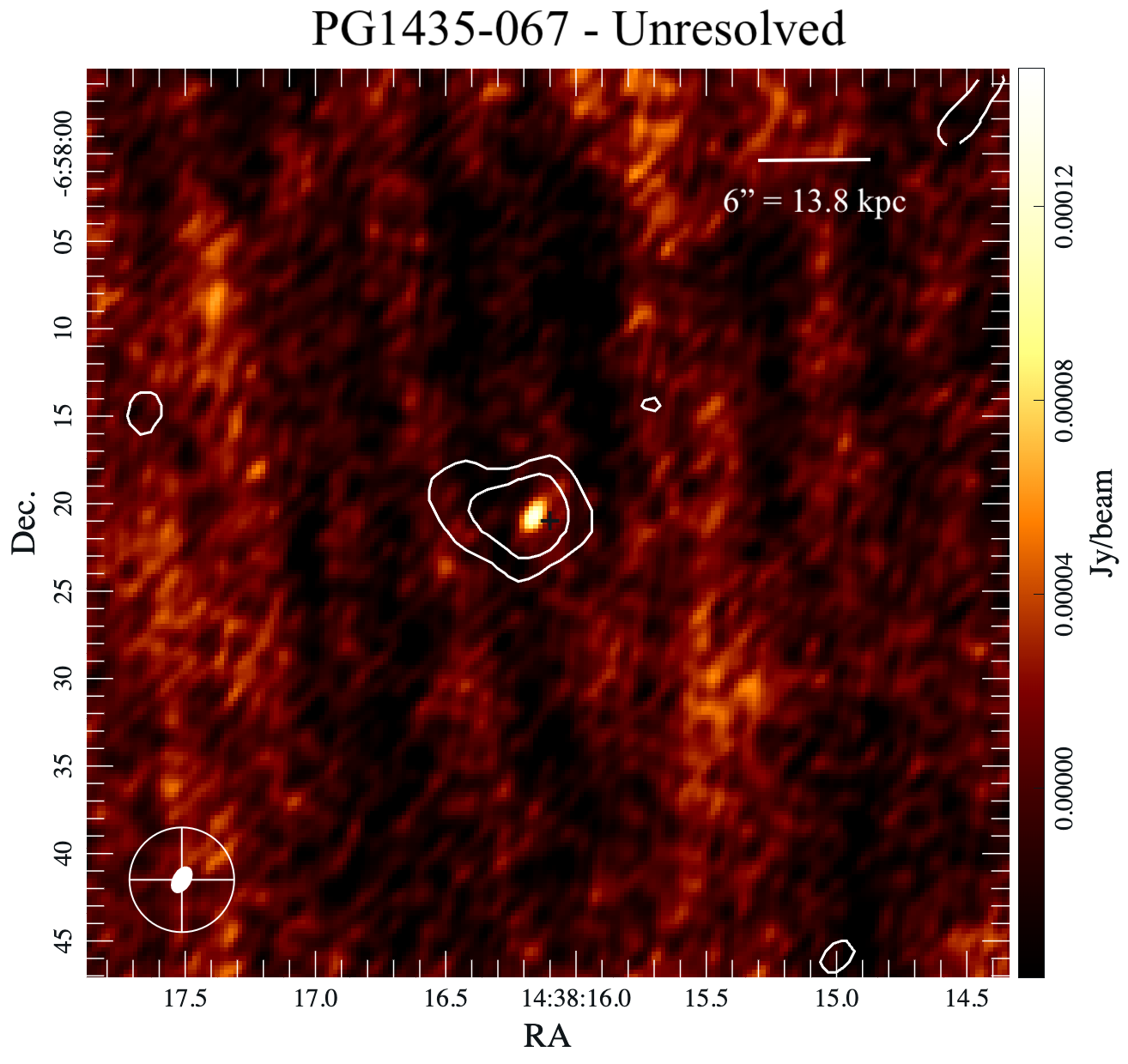}}
\hfill
	\includegraphics[width=0.36\textwidth]{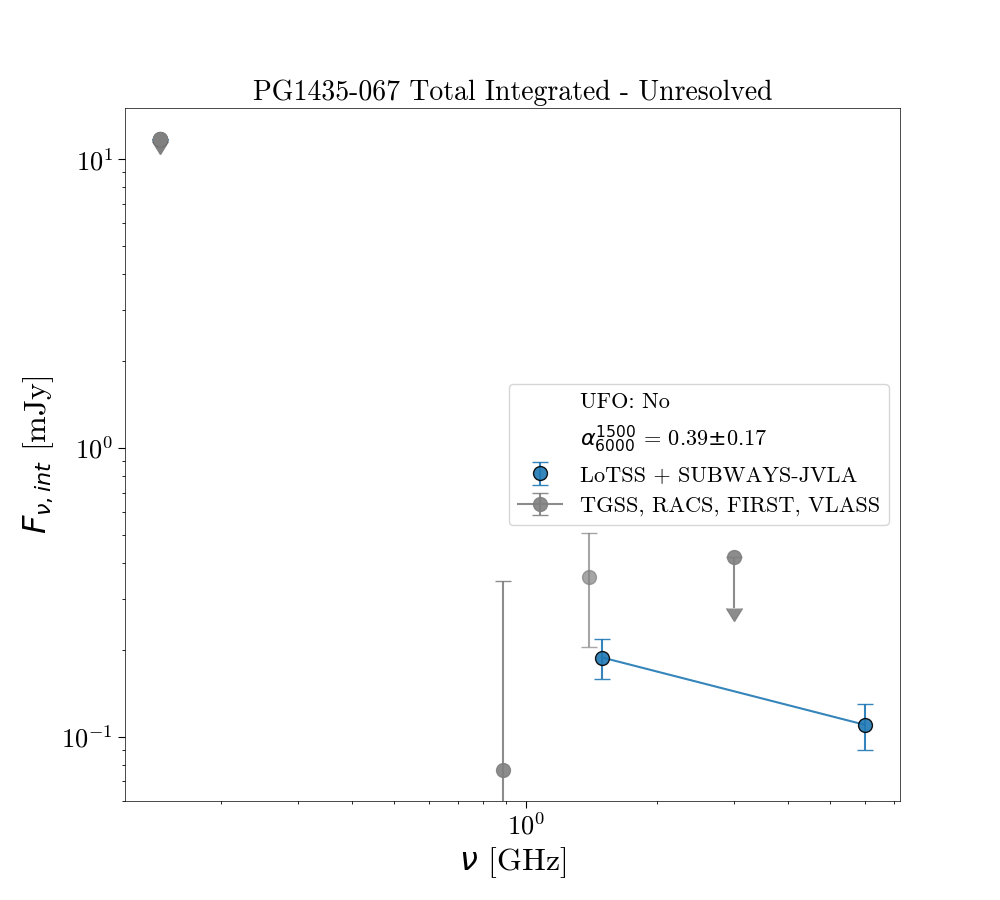}

	\caption{PG1435-067. The linear size is < 2.83 kpc. The SED is flat at GHz frequencies. No UFO is detected. The SFR predicted with radio emission is compatible with the IR deduced one within $1\sigma$. The target follows the Güdel–Benz relation within $1\sigma$ uncertainty. While the SED is not flat enough to favour the corona as main source of radio emission, the spectrum could be dominated by an unresolved jet.}
	\label{sed:PG1435-067}
\end{figure*}	

\begin{figure*}[hp!]
		\centering
		\includegraphics[width=0.31\textwidth]{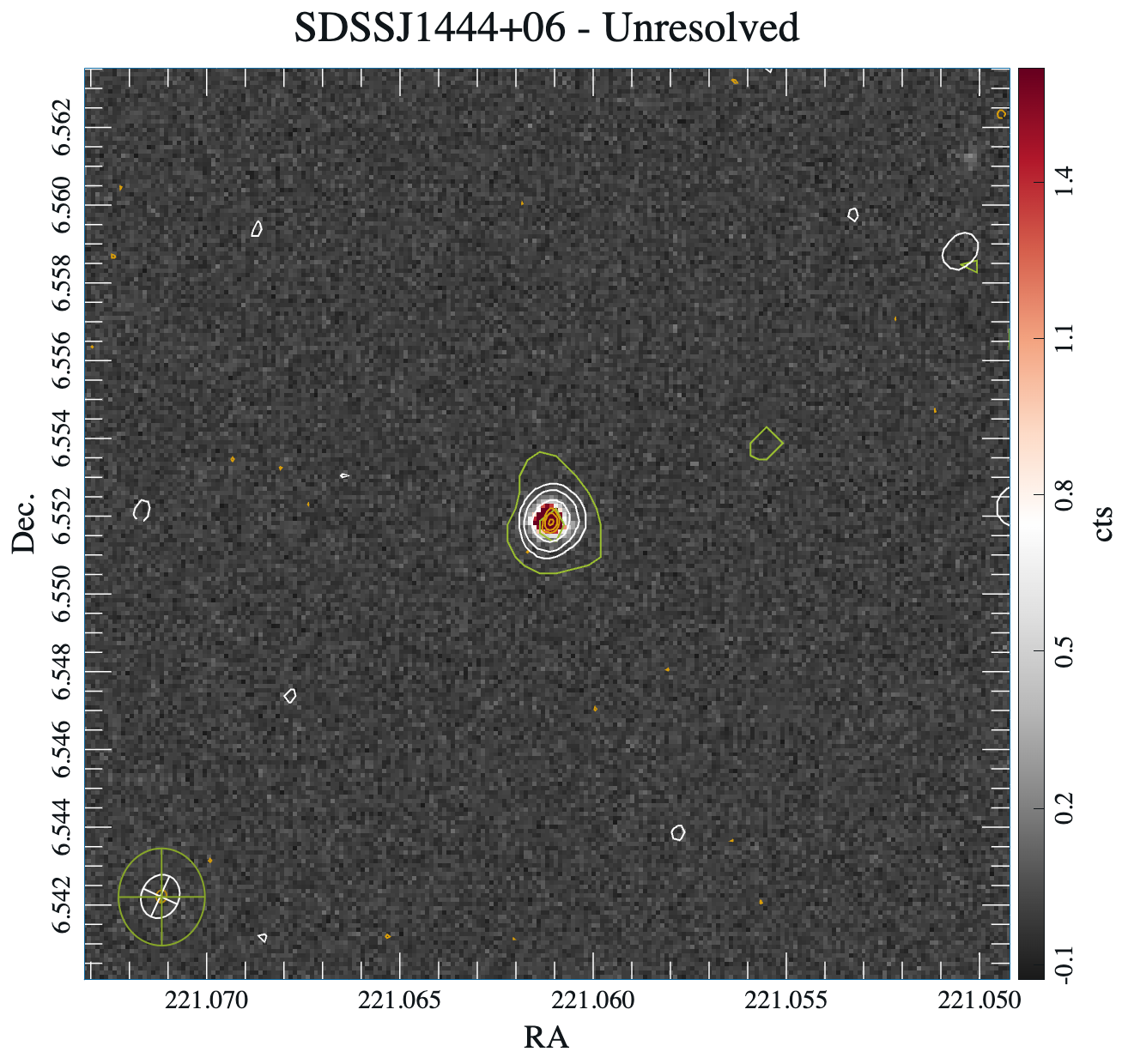}
 \hfill
 \includegraphics[width=0.31\textwidth]{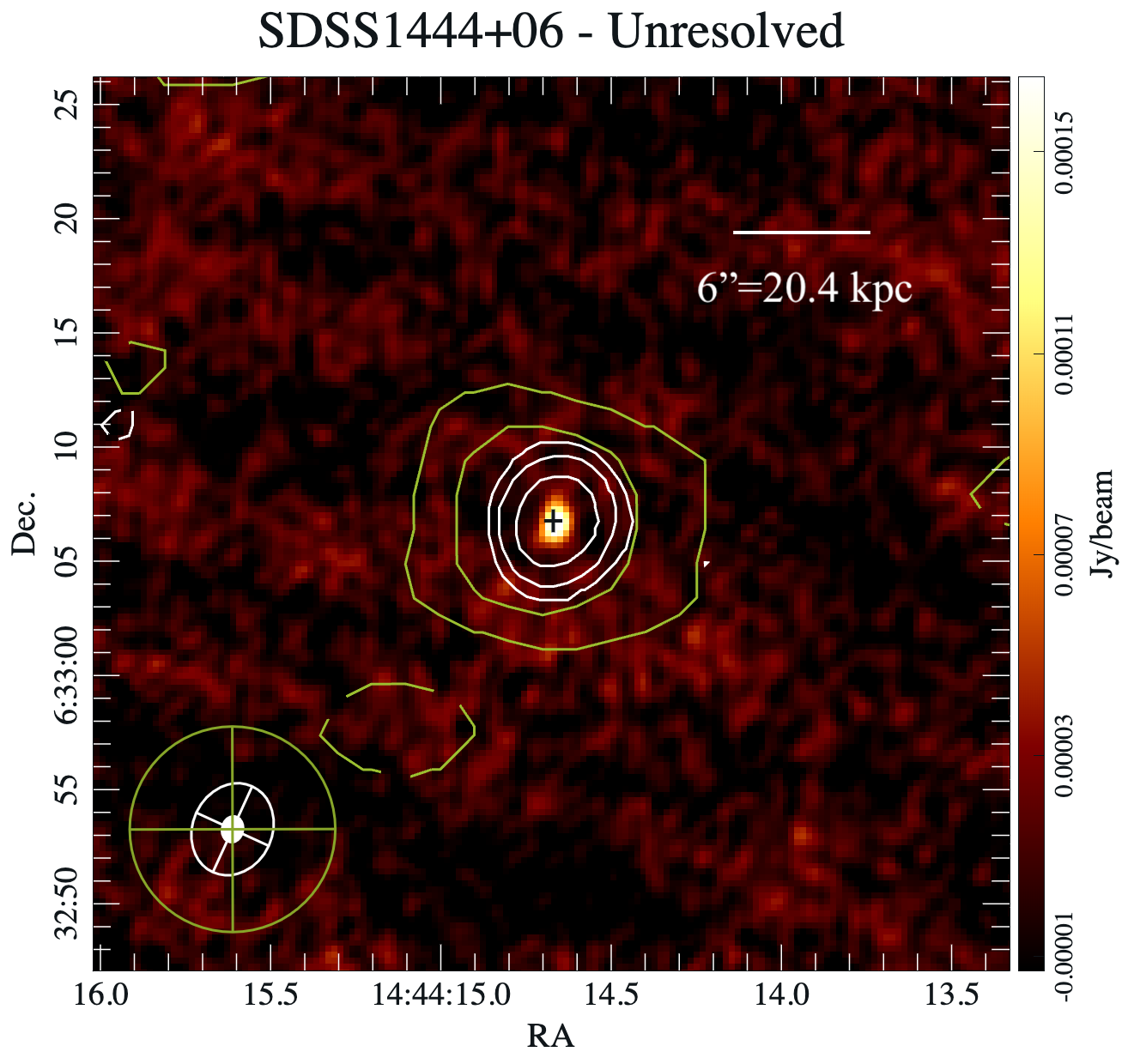}
	\hfill
	 \includegraphics[width=0.36\textwidth]{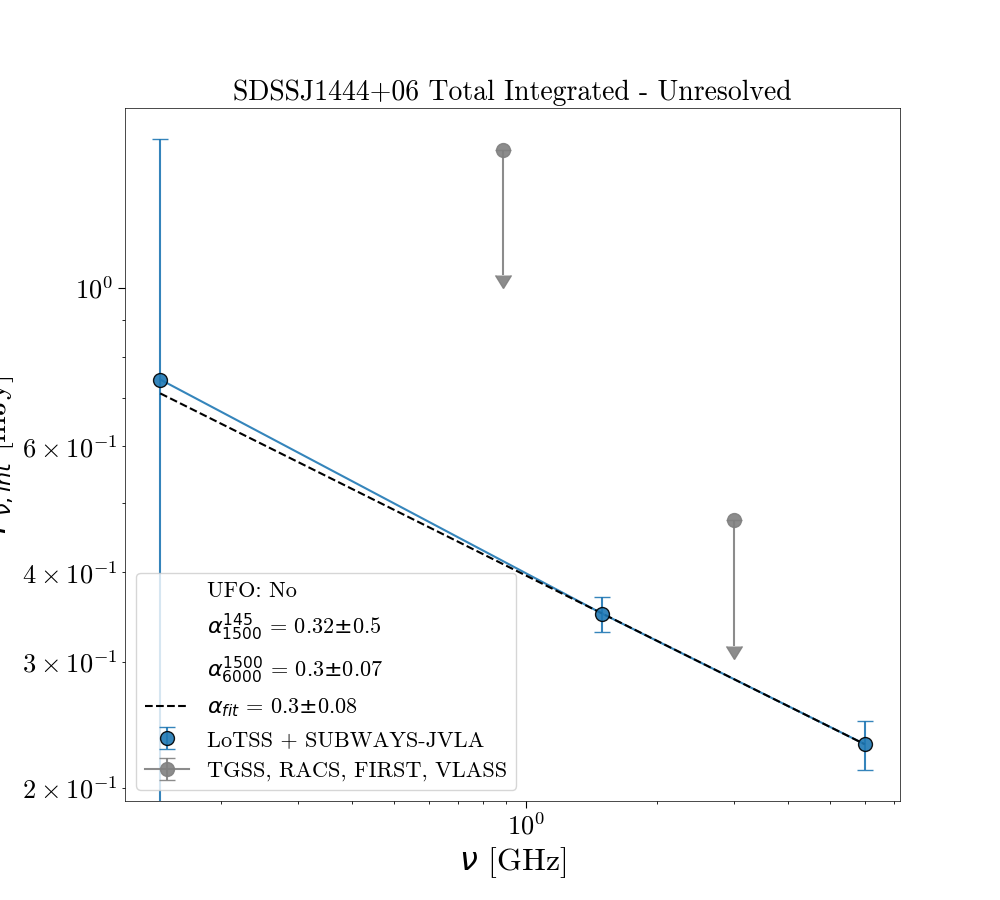}

	\caption{SDSSJ1444+06. The linear size is < 2.96 kpc. The SED is flat in both regimes. No UFO is detected. The SFR predicted with radio emission is compatible with the IR deduced one within $3\sigma$. The target follows the Güdel–Benz relation within $1\sigma$ uncertainty. The spectral slope is too steep for the radio emission to be dominated by the corona. An unresolved jet is the most likely the dominant radio origin.}
	\label{sed:SDSSJ1444+06}
\end{figure*}

\begin{figure*}[hp!]
\centering
 \includegraphics[width=0.31\textwidth]{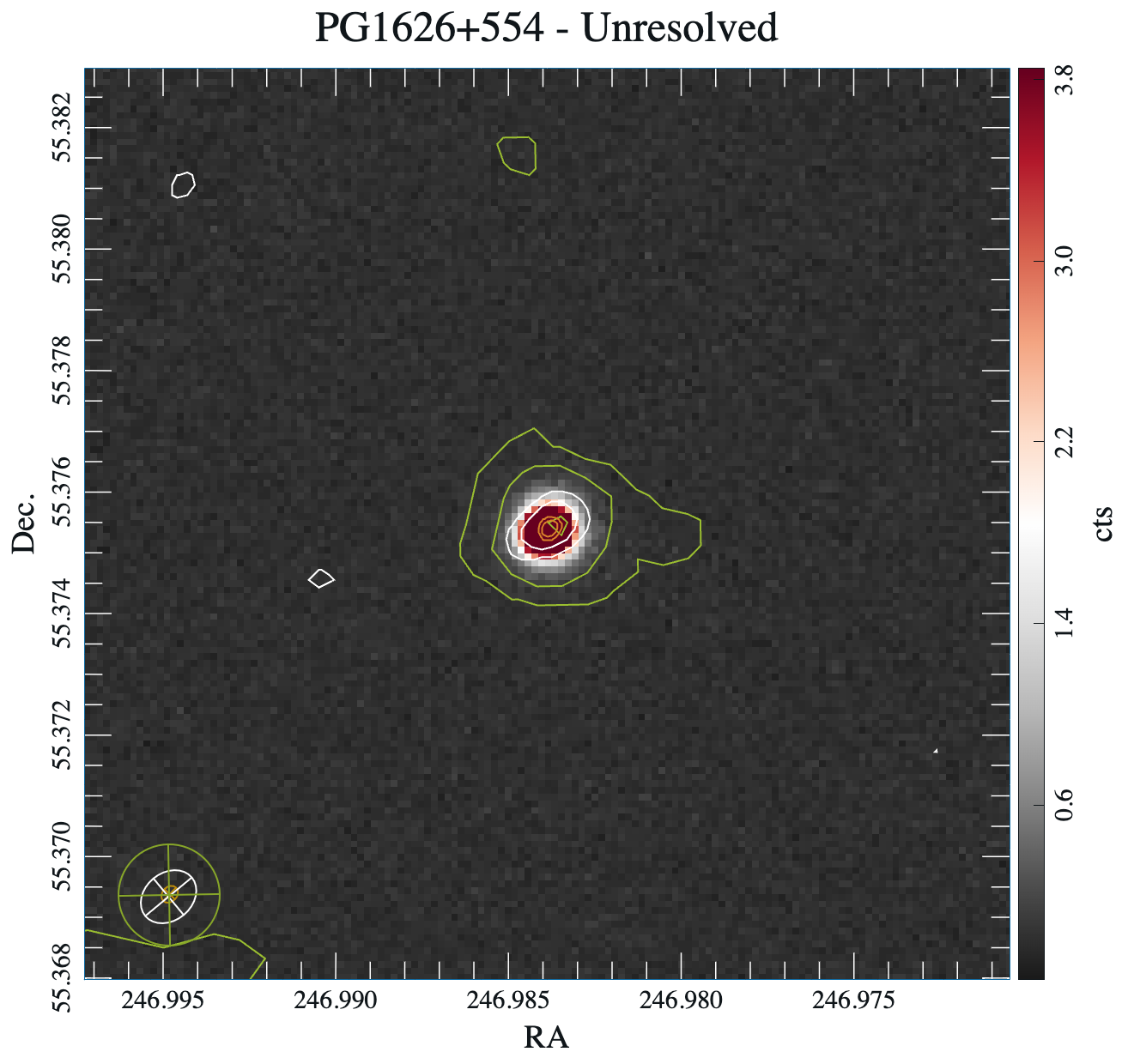}
\hfill
 \includegraphics[width=0.31\textwidth]{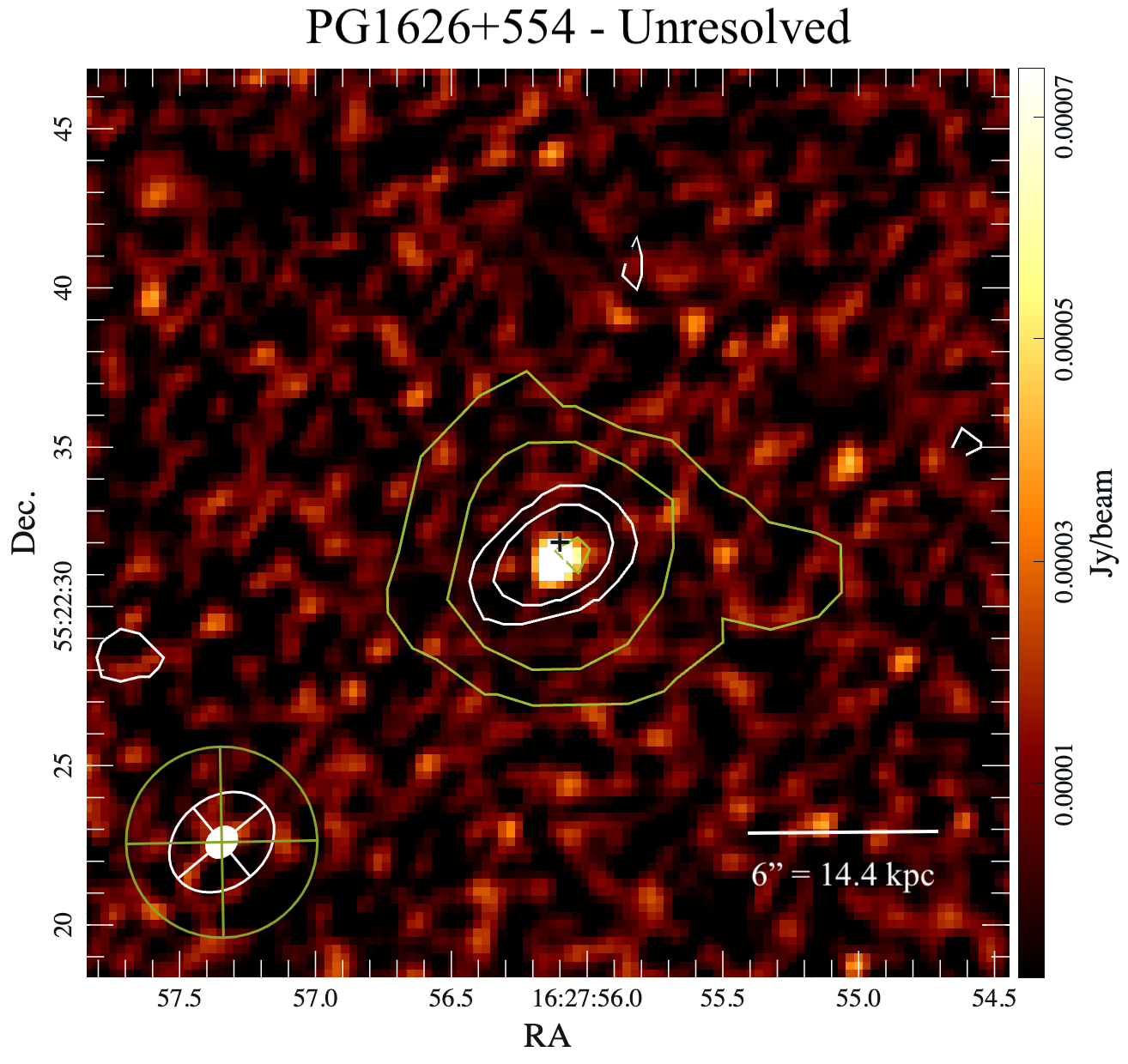}
\hfill
	\includegraphics[width=0.36\textwidth]{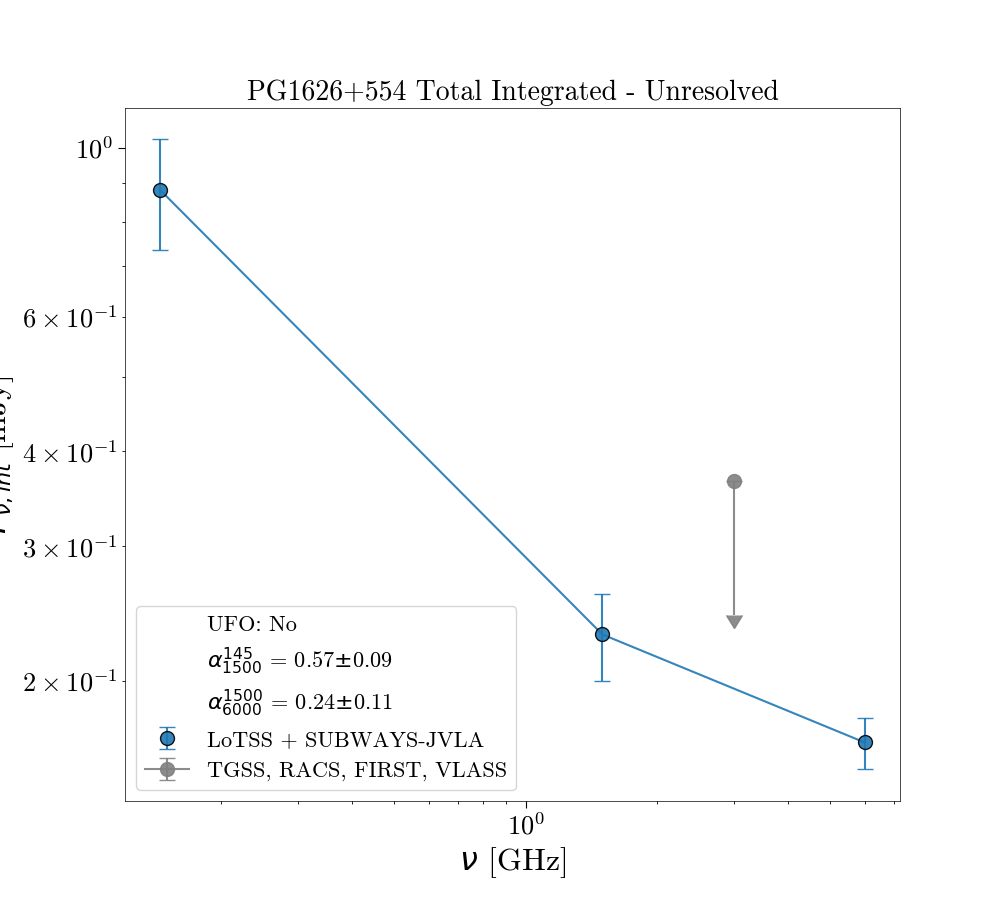}

	\caption{PG1626+554.The linear size is < 2.09 kpc. The SED is steep then flat, with an optically thick component becoming dominant at GHz frequencies. The steep MHz-frequency spectrum may be due to some negative flux density contours at 1.5 GHz. No UFO detected. The SFR predicted with radio emission is compatible with the IR deduced one within $3\sigma$. The target follows the Güdel–Benz relation within $1\sigma$ uncertainty. The SED is too steep to be dominated by coronal emission. Single mechanism at the origin of radio is difficult to constrain, however an unresolved jet may explain the SED curvature.}
	\label{sed:PG1626+554}
\end{figure*}	


\begin{figure*}[hp!]
		\centering
		\includegraphics[width=0.31\textwidth]{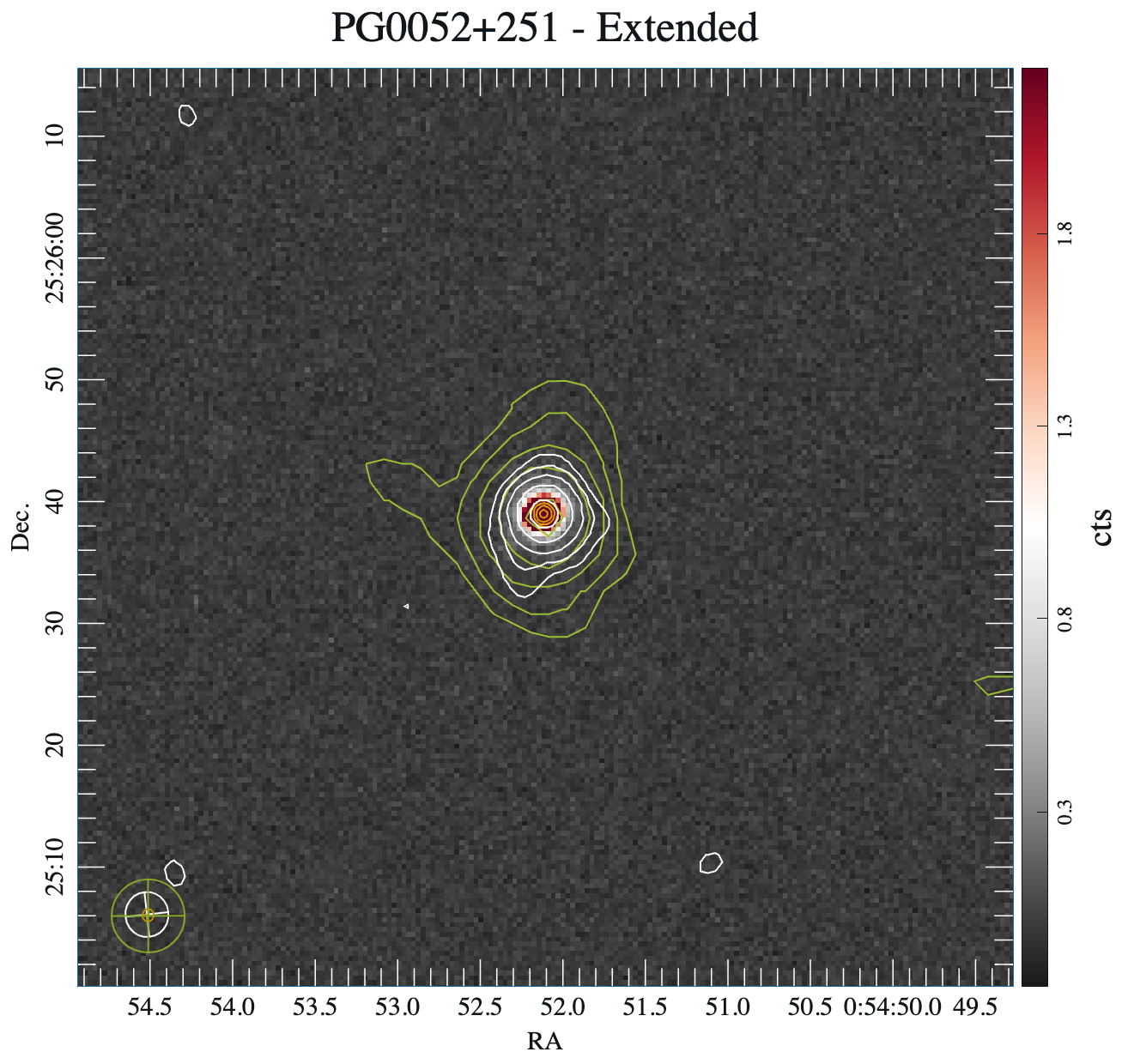}
 \hfill
 \includegraphics[width=0.31\textwidth]{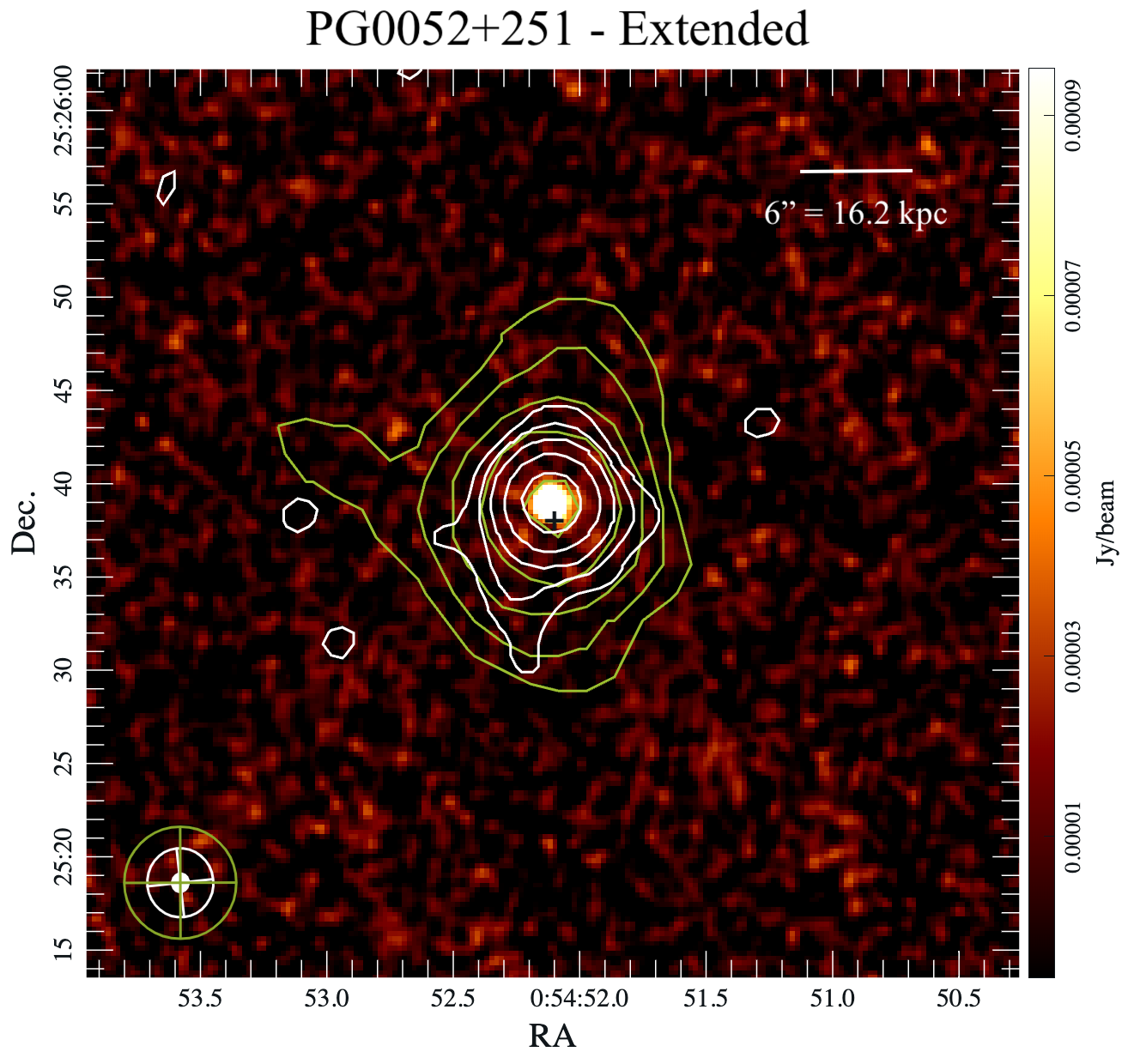}
	\hfill
	 \includegraphics[width=0.36\textwidth]{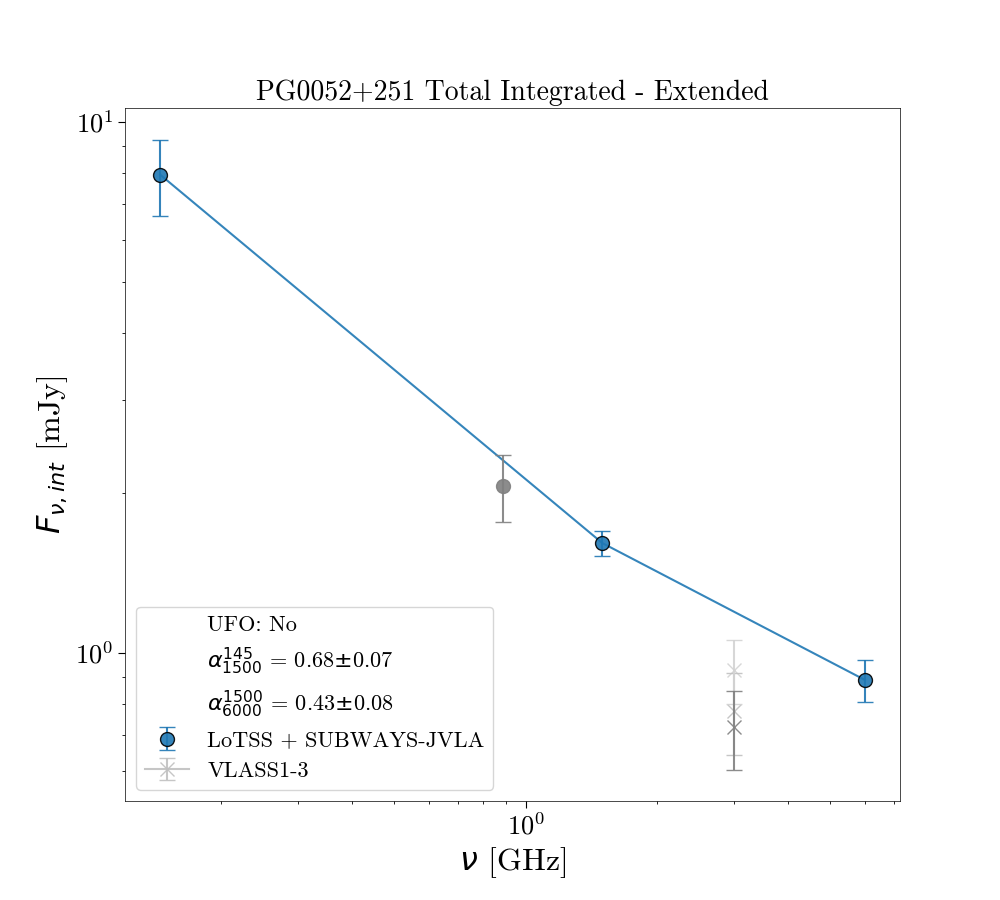}
	
	\caption{PG0052+251.The linear size is 11.4 kpc. The SED is steep at all frequencies within uncertainties. No UFO detected. The SFR predicted with radio emission is compatible with the IR deduced one within $2\sigma$. The target follows the Güdel–Benz relation within $1\sigma$ uncertainty. The north-south extension, together with the steep SED shape, suggests a symmetric outflow, either in the form of a wind or a jet. However \citet{Chen_2023} report (VLA in A configuration) a C-band flux density a factor $\sim0.57$ smaller than the one presented in this work. The authors also detect a VLBA compact core, then attribute the dominant radio origin to the corona. We can infer our observations collect the superposition of extended emission to the one of a compact corona.}
	\label{sed:PG0052+251}
\end{figure*}	

\begin{figure*}[hp!]
		\centering
		\includegraphics[width=0.31\textwidth]{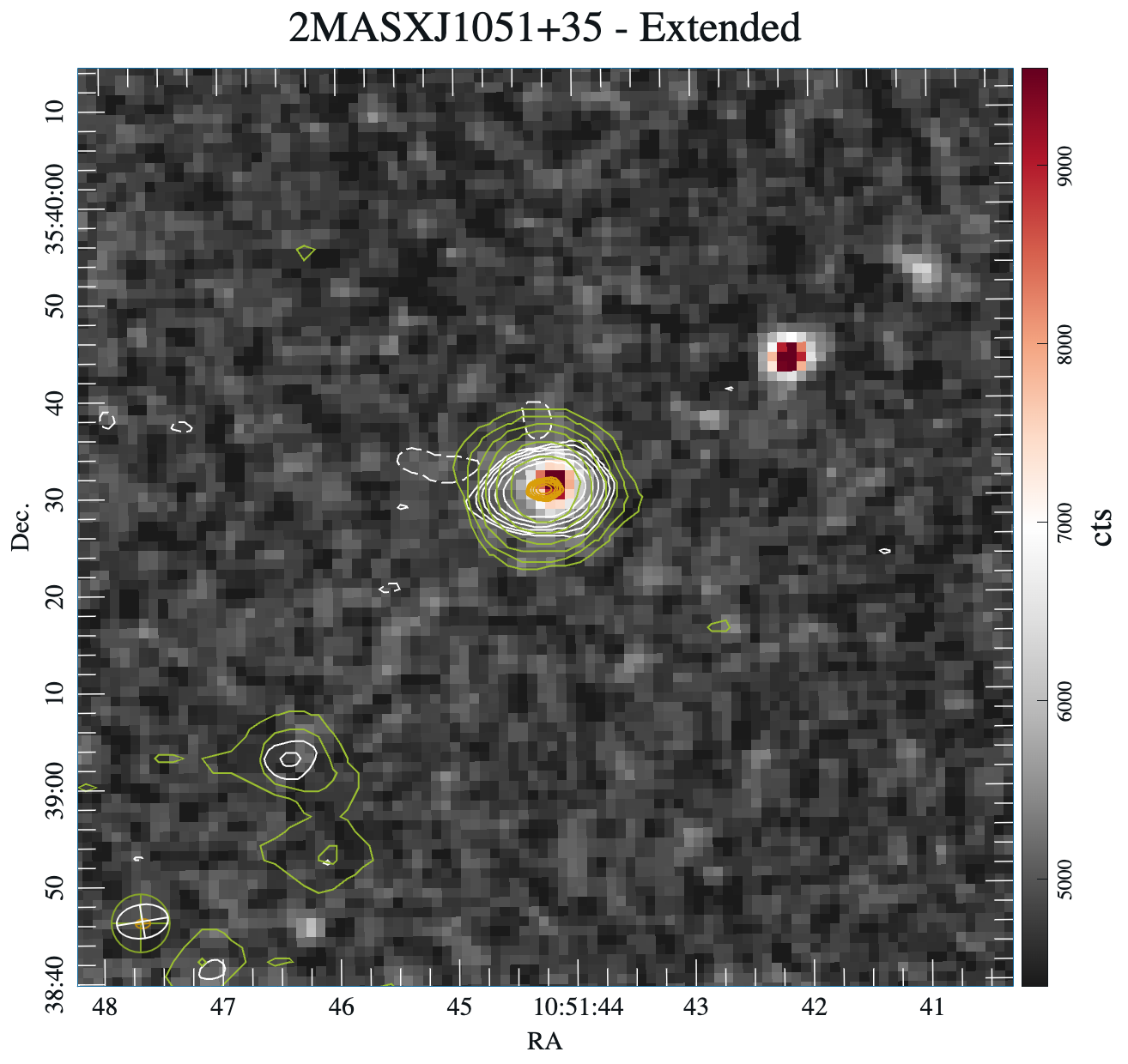}
 \hfill
 \includegraphics[width=0.31\textwidth]{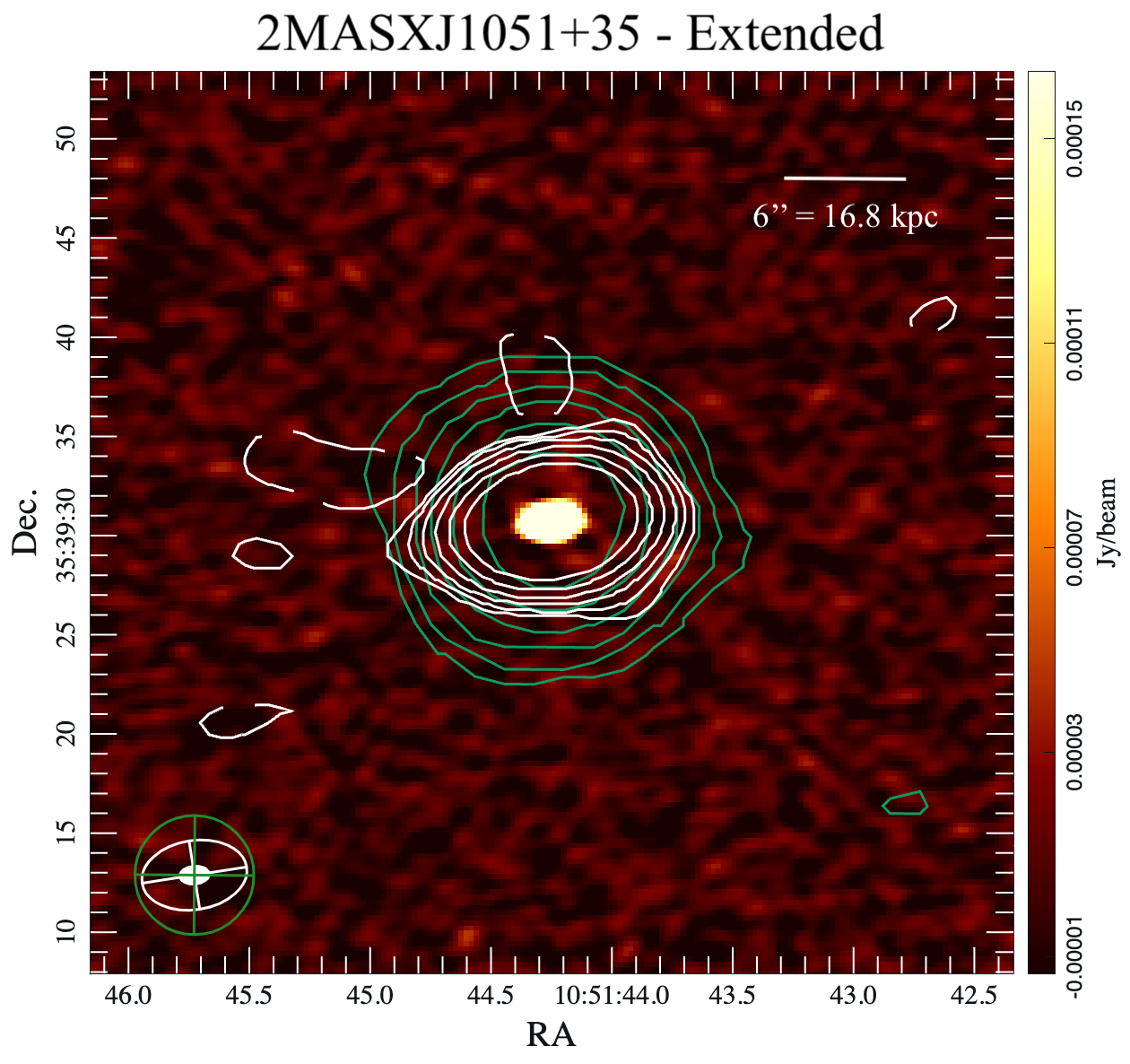}
	\hfill
	 \includegraphics[width=0.36\textwidth]{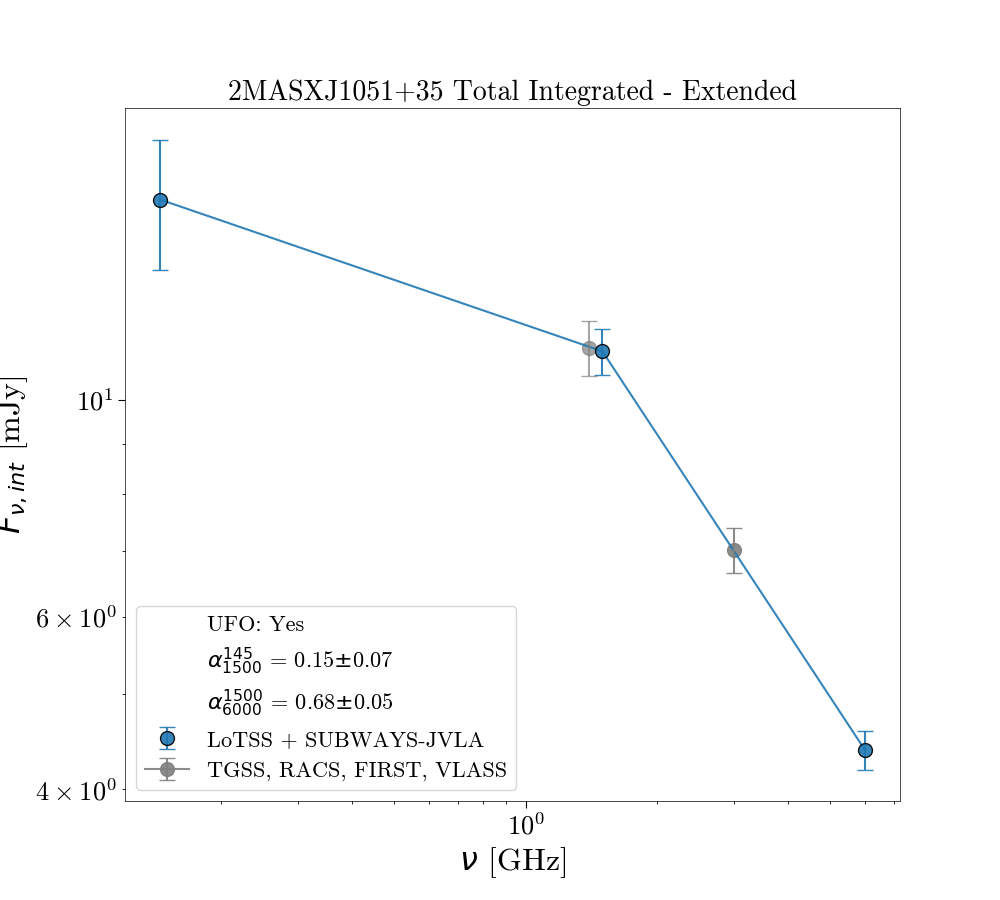}
	
	\caption{2MASXJ1051+35. the linear size is 8.7 kpc. The SED is flat then steep, with an optically thin component becoming dominant at GHz frequencies. A UFO is detected. The SFR predicted with radio emission is a factor 32 higher than the IR deduced one. The target exceeds the $3\sigma$ uncertainty region around the Güdel–Benz relation. Estimates done with the wind model as in \citet{Nims_2015} favour the wind scenario, indeed the extension, together with the synchrotron break, suggests either a wind or a compact jet.}
	\label{sed:2MASXJ1051+35}
\end{figure*}	

\begin{figure*}[hp!]
		\centering
		\includegraphics[width=0.31\textwidth]{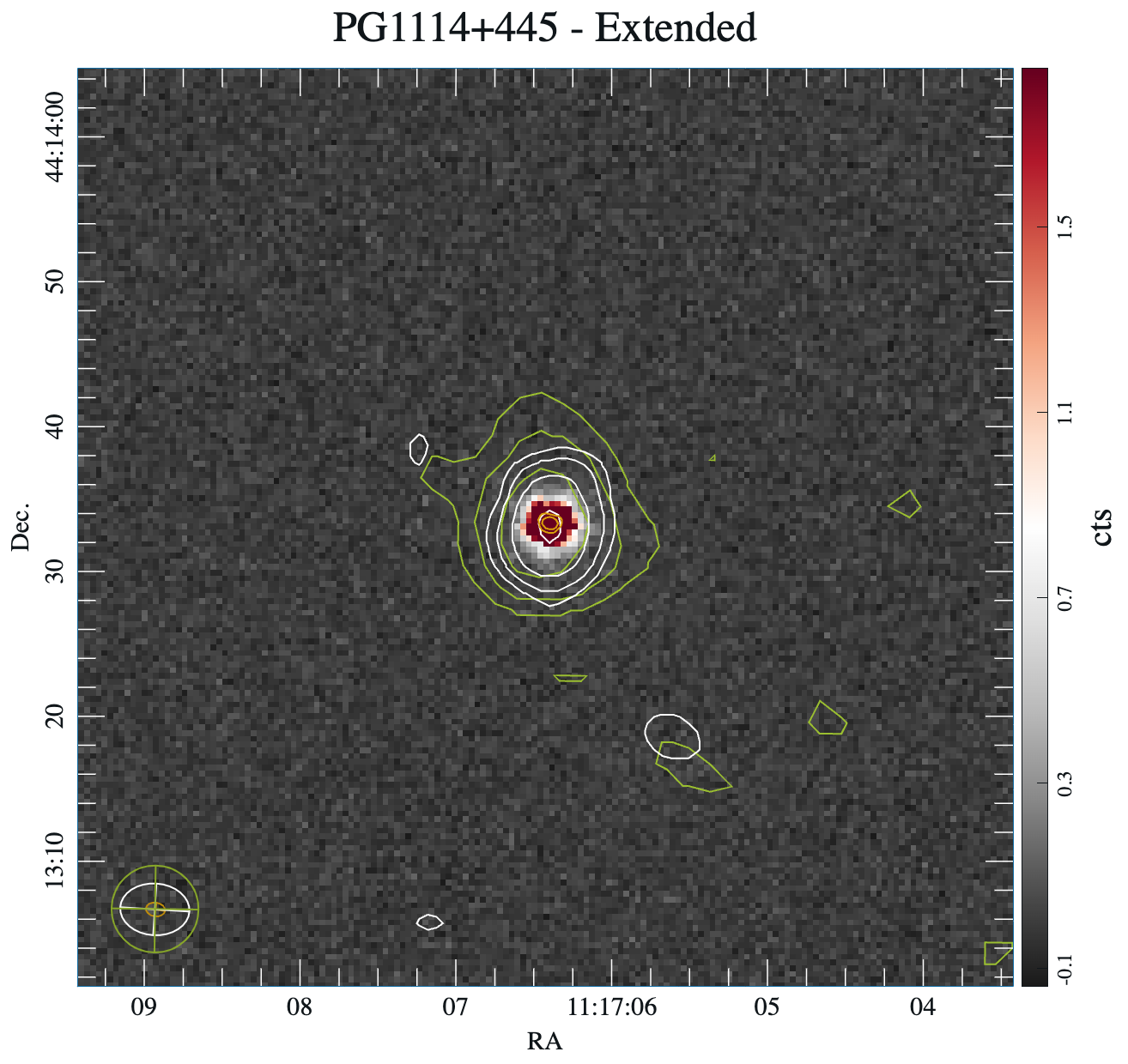}
 \hfill
 \includegraphics[width=0.31\textwidth]{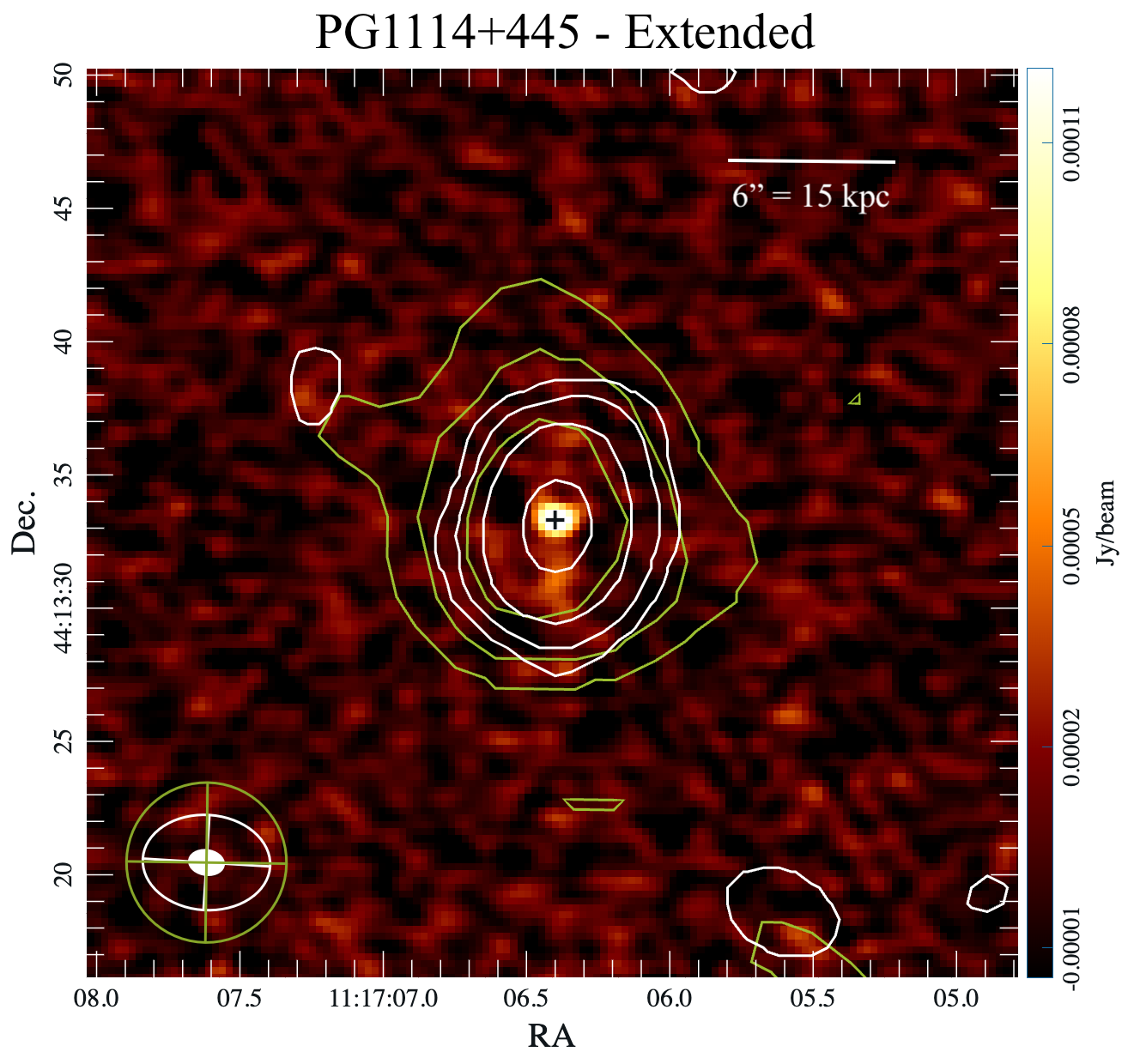}
	\hfill
	 \includegraphics[width=0.36\textwidth]{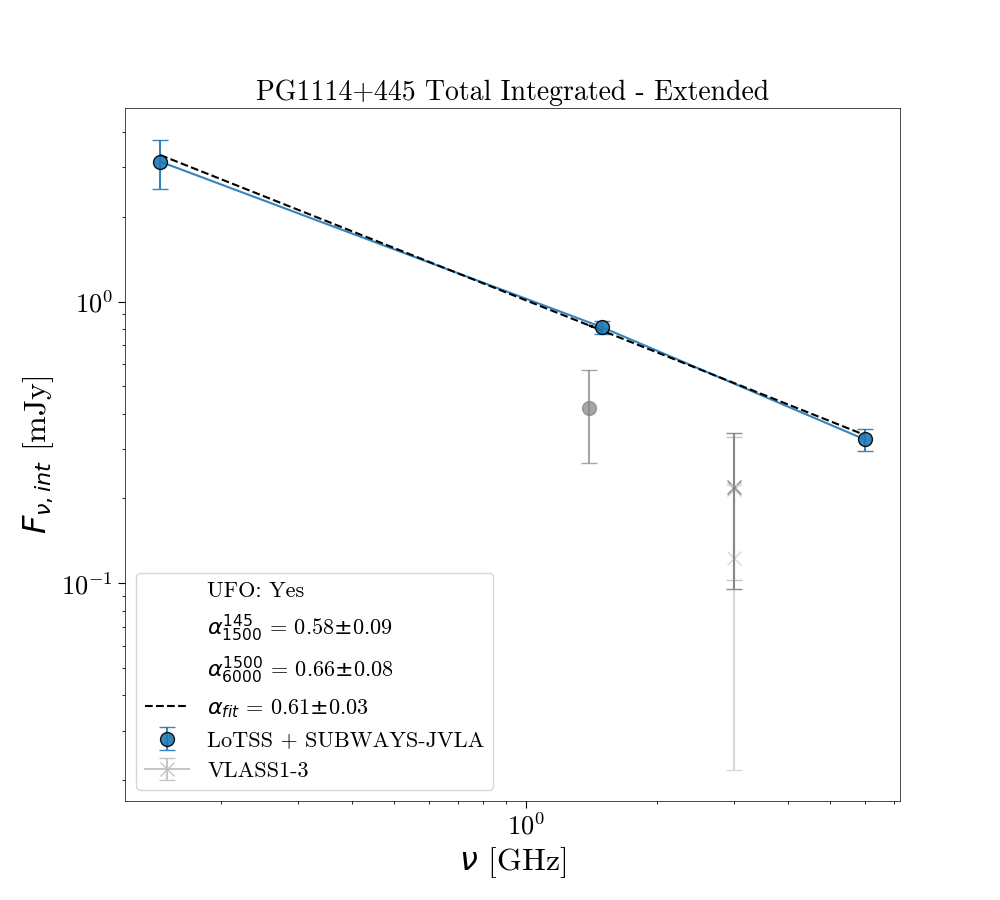}
	
	\caption{PG1114+445.The linear size is 22.35 kpc. The SED is steep dominated by an optically thin component. The UFO is detected. The SFR predicted with radio emission consistent with the IR deduced one within $3\sigma$. The target follows the Güdel–Benz relation within $1\sigma$. Estimates done with the wind model as in \citet{Nims_2015} favour the wind scenario. The extended shape and the UFO detection suggest radio emission could be related to a wind or a jet.}
	\label{sed:PG1114+445}
\end{figure*}	

\begin{figure*}[hp!]
		\centering
			\includegraphics[width=0.31\textwidth]{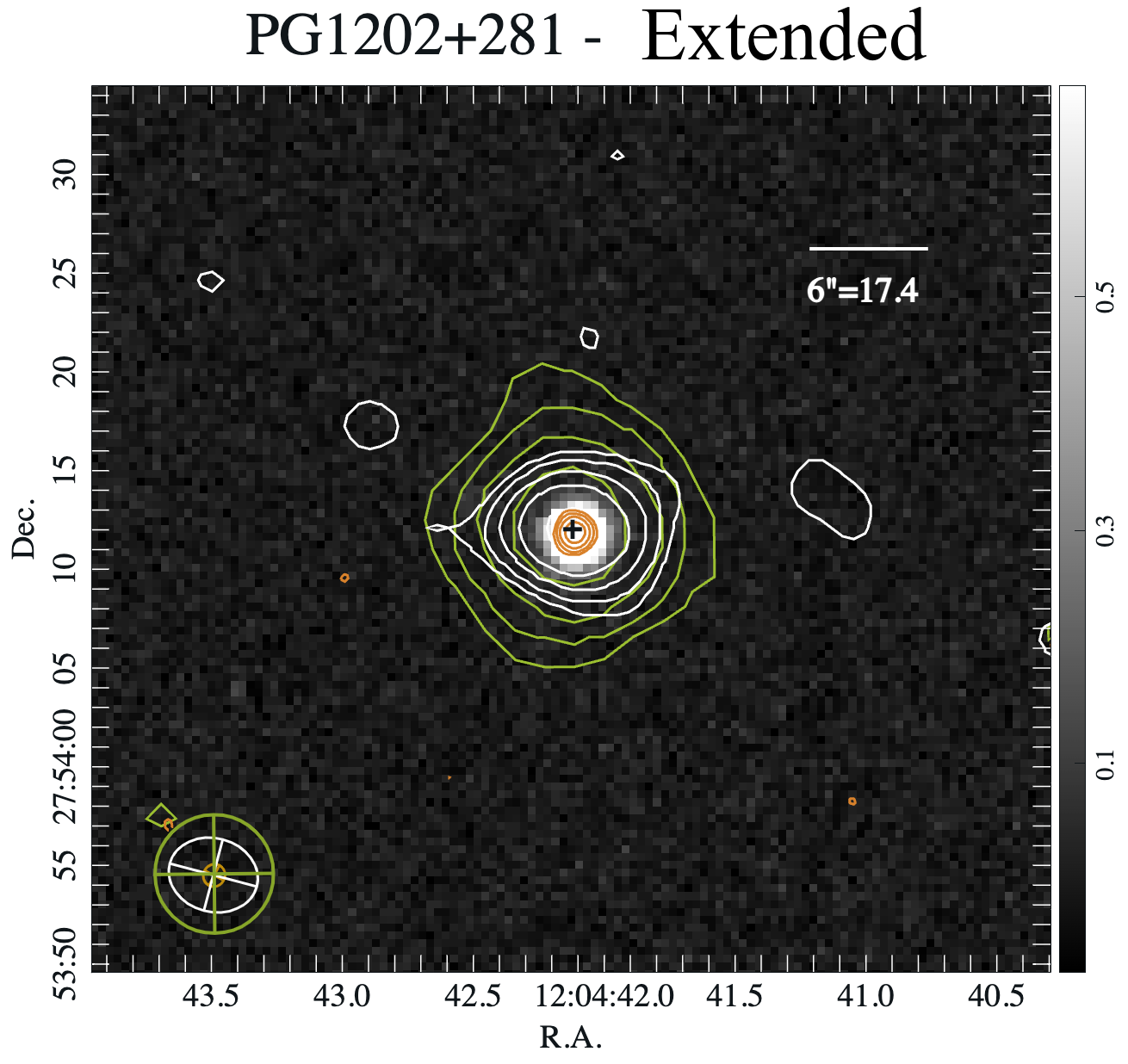}
\hfill
 \includegraphics[width=0.31\textwidth]{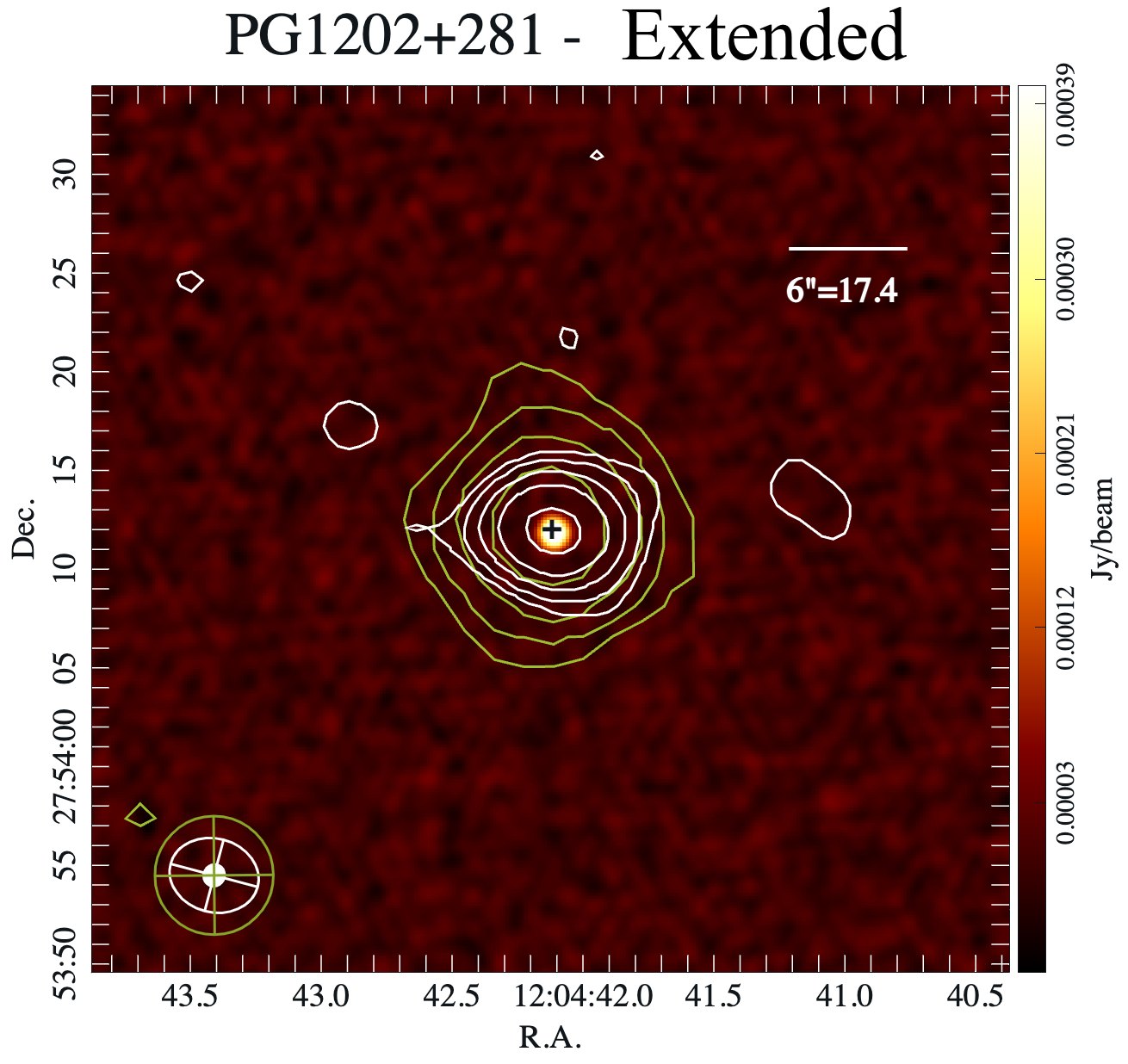}
\hfill
		\includegraphics[width=0.36\textwidth]{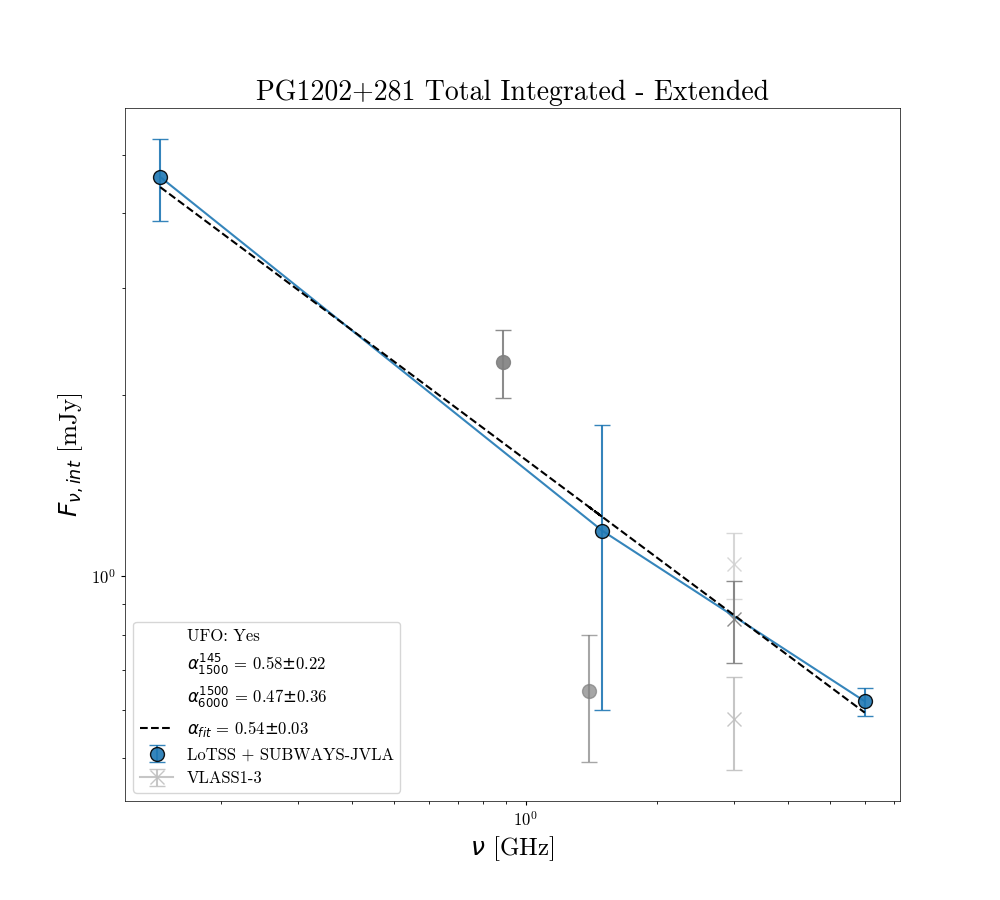}
	\caption{PG1202+281. The linear size is 8.7 kpc. The integrated SED is steep dominated by an optically thin component. A UFO is detected. The SFR predicted with radio emission is consistent with the IR deduced one within $1/2\sigma$. The target follows the Güdel–Benz relation within $1/2\sigma$ uncertainty. Estimates done with the wind model as in \citet{Nims_2015} favour the wind scenario. The bulk of radio emission is probably related to the presence of outflows, even if it remains difficult to distinguish between AGN driven winds and a low-power jet.}
	\label{sed:PG1202+281}
\end{figure*}

\begin{figure*}[hp!]
		\centering
		\includegraphics[width=0.31\textwidth]{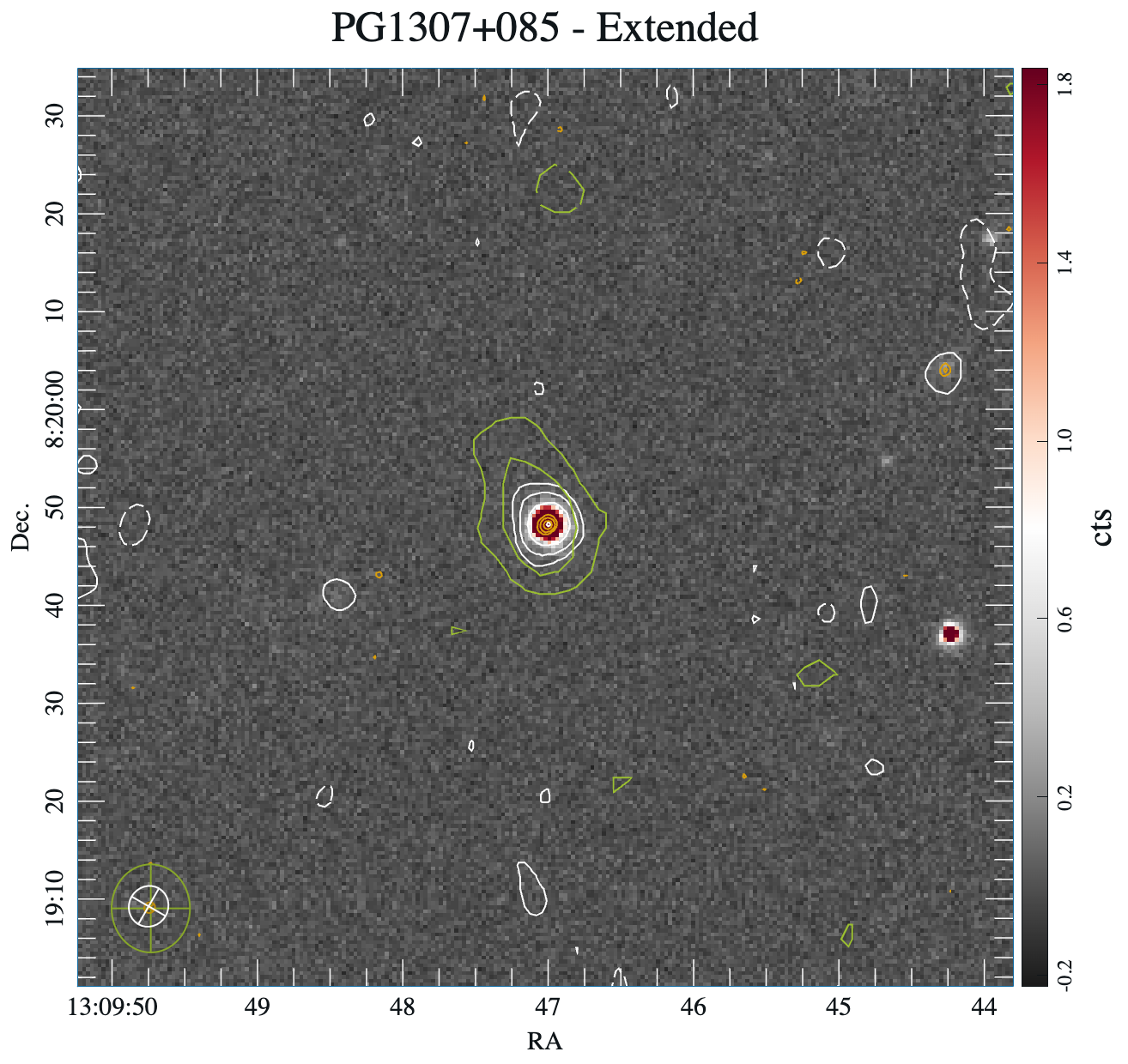}
 \hfill
 \includegraphics[width=0.31\textwidth]{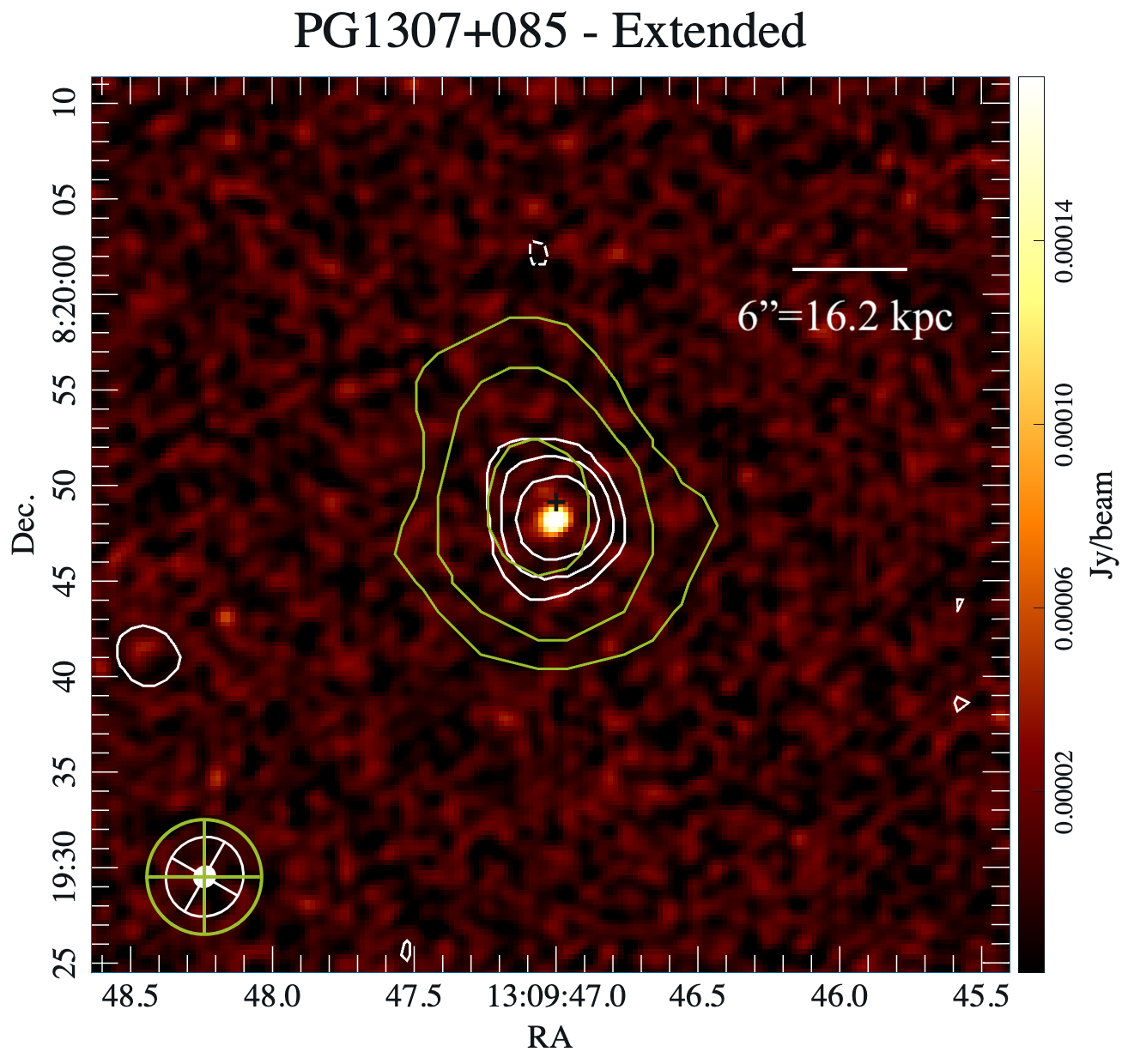}
	\hfill
	 \includegraphics[width=0.36\textwidth]{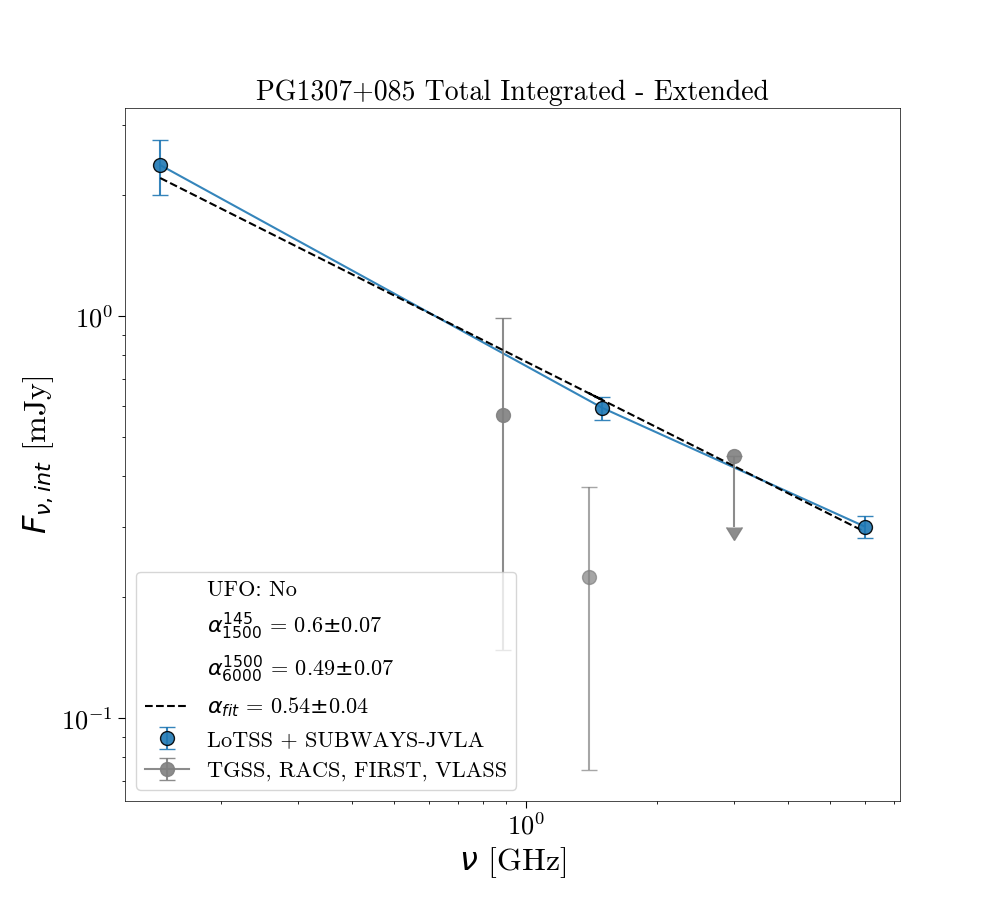}
	
	\caption{PG1307+085. The linear size is 35 kpc. The SED is steep within the uncertainties. No UFO is detected. The SFR predicted with radio emission is consistent with the IR deduced one withn $2\sigma$. The offset of FIRST flux density of a factor 0.37 with respect to the proprietary data suggests variability. The target follows the Güdel–Benz relation within $1\sigma$ uncertainty. The extension together with the hints of variability suggest contribution from a jet.}
	\label{sed:PG1307+085}
\end{figure*}	

\begin{figure*}[hp!]
		\centering
 \includegraphics[width=0.31\textwidth]{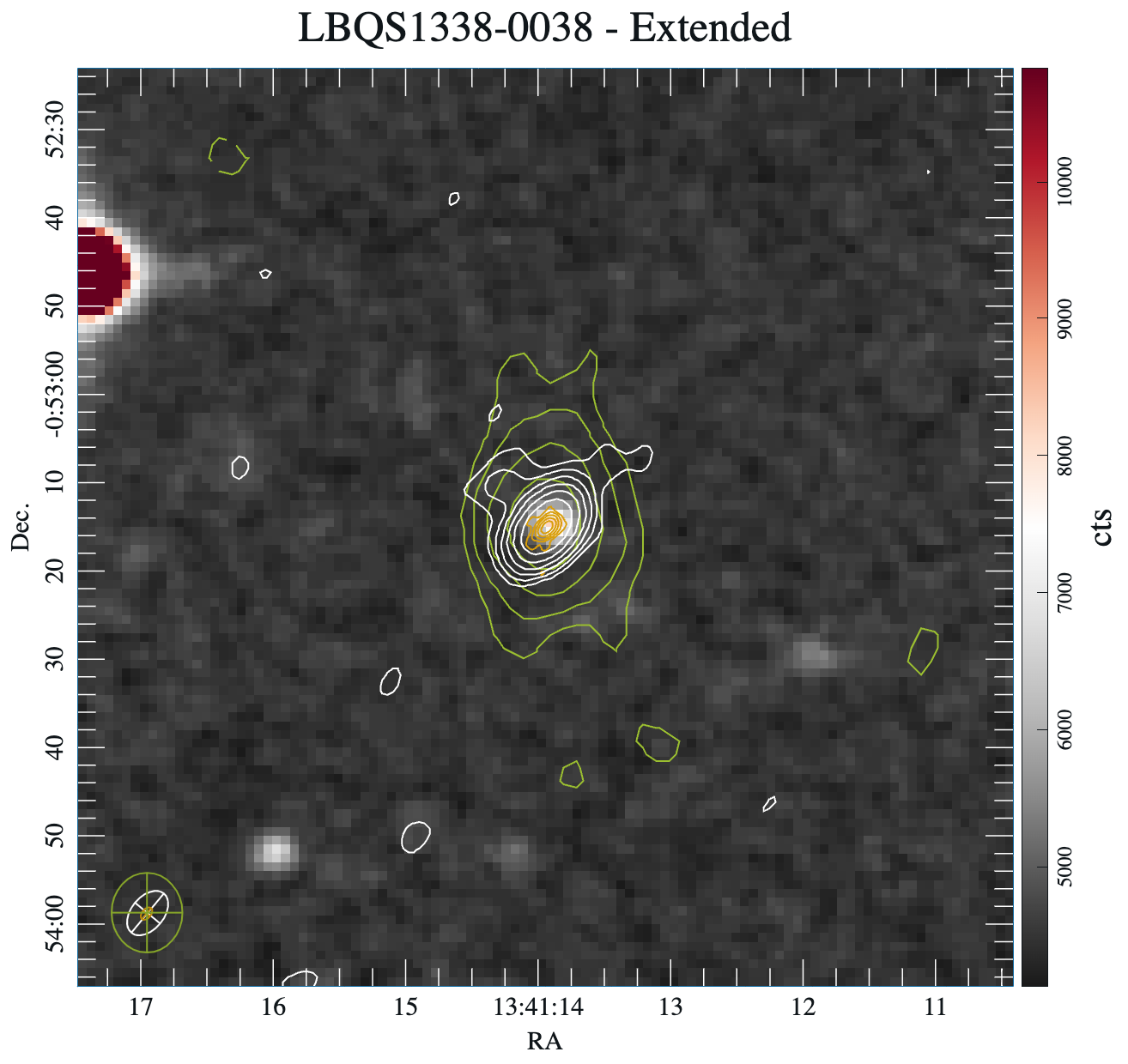}
 \hfill
 \includegraphics[width=0.31\textwidth]{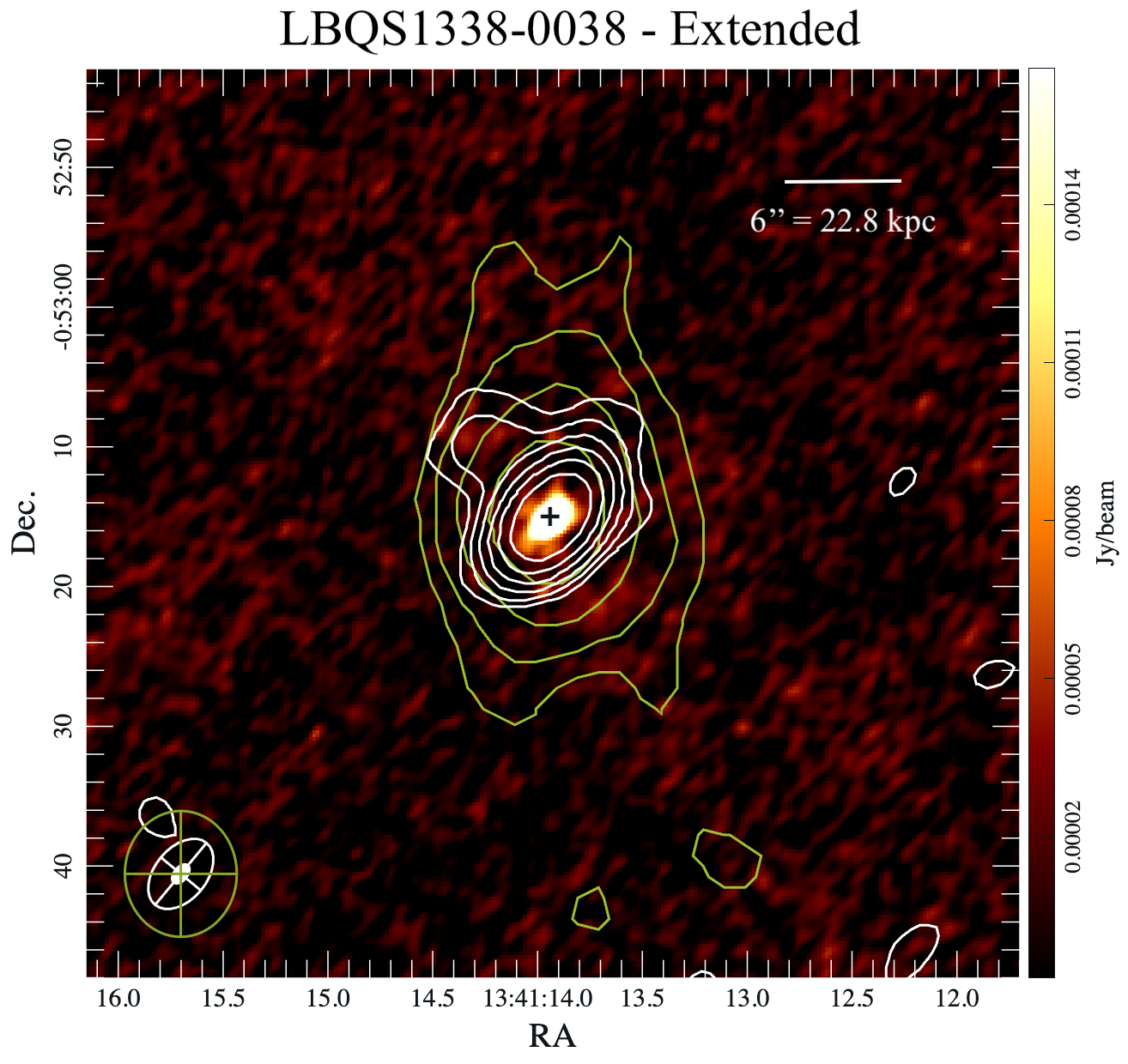}
 \hfill
 \includegraphics[width=0.36\textwidth]{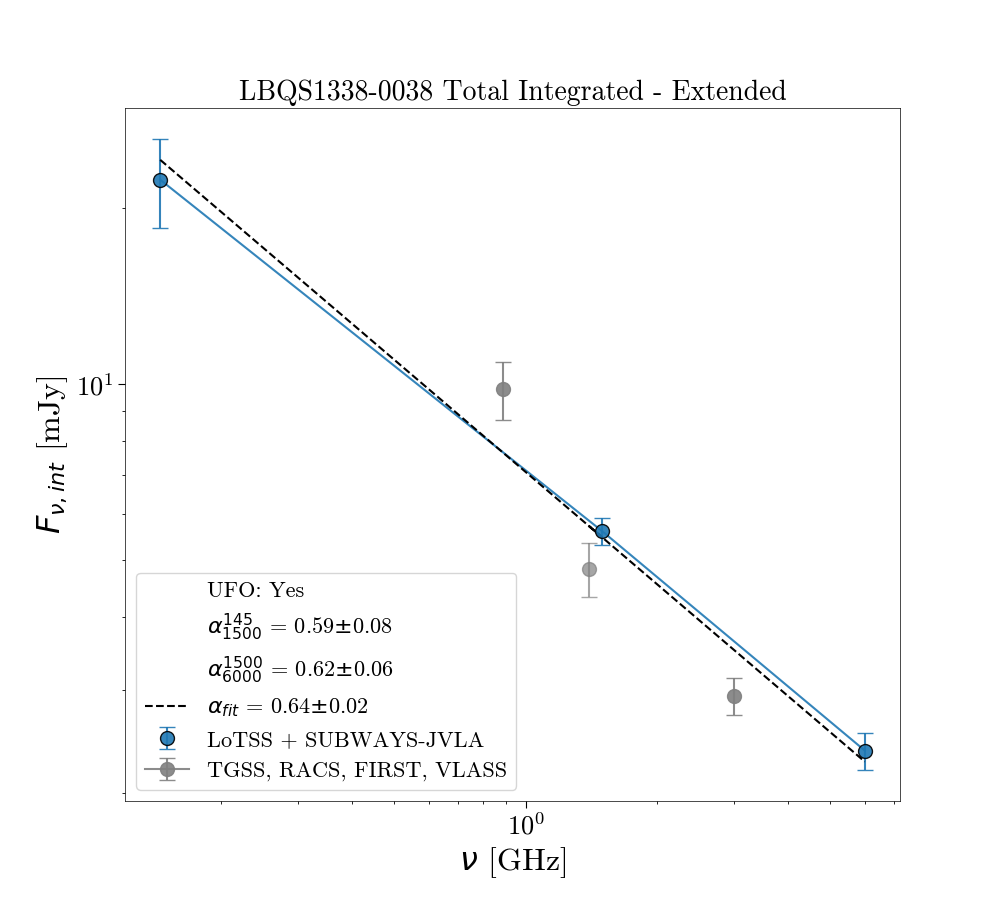}
	
	\caption{LBQS1338-0038.The linear size is 26.1 kpc. The SED is steep, dominated by an optically thin extended component. The UFO is detected. The SFR predicted with radio emission is consistent with the IR deduced one within $3\sigma$. The offset of RACS, and VLASS flux densities with respect to the computed slope suggests some variability. The target exceeds the $3\sigma$ uncertainty region around the Güdel–Benz relation. Estimates done with the wind model as in \citet{Nims_2015} favour the wind scenario, therefore the radio emission may be dominated by wind shocks, but a jet scenario cannot be ruled-out.}
	\label{sed:LBQS1338-0038}
\end{figure*}	

\begin{figure*}[hp!]
		\centering
	 \includegraphics[width=0.31\textwidth]{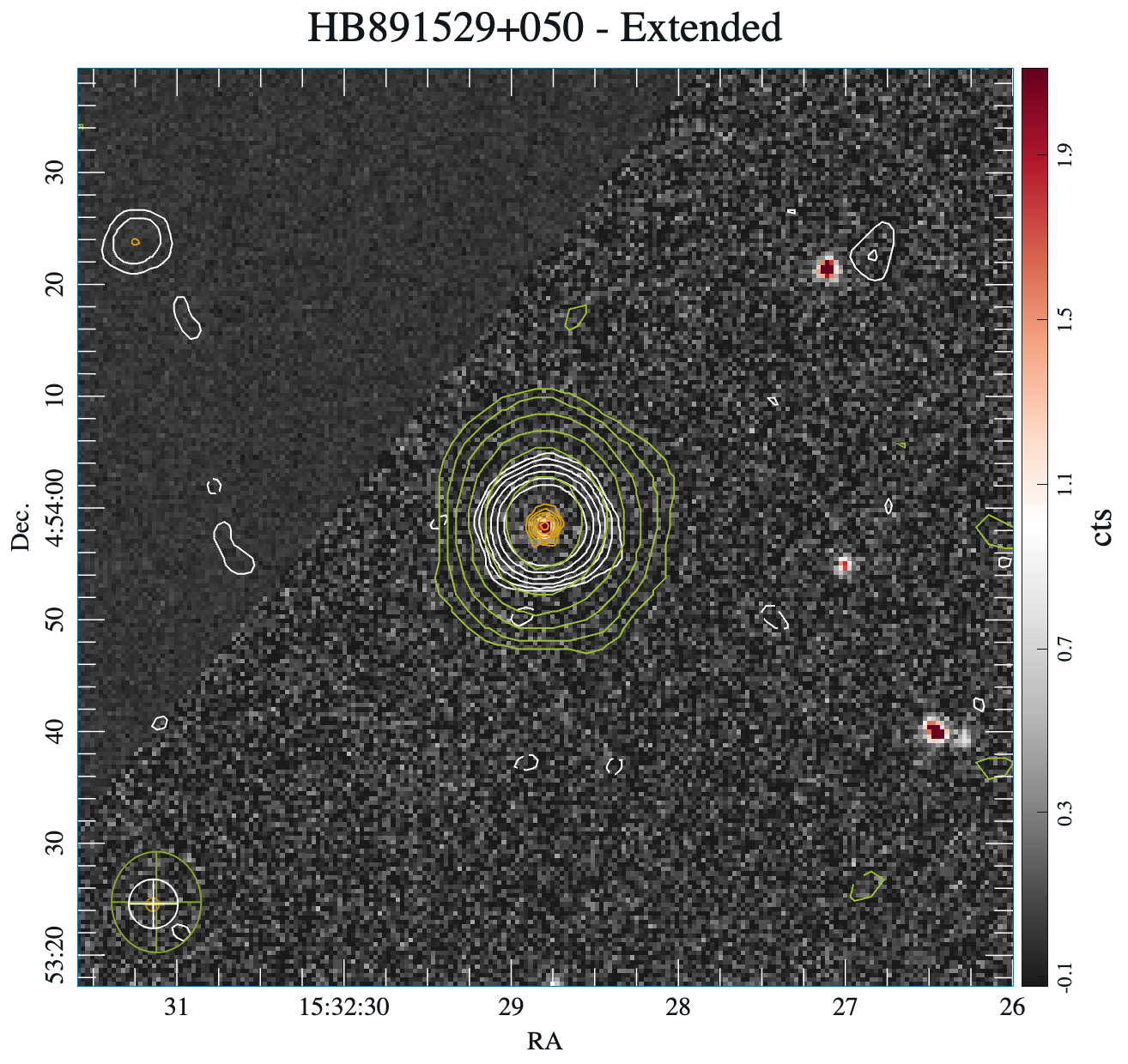}
 \hfill
 \includegraphics[width=0.31\textwidth]{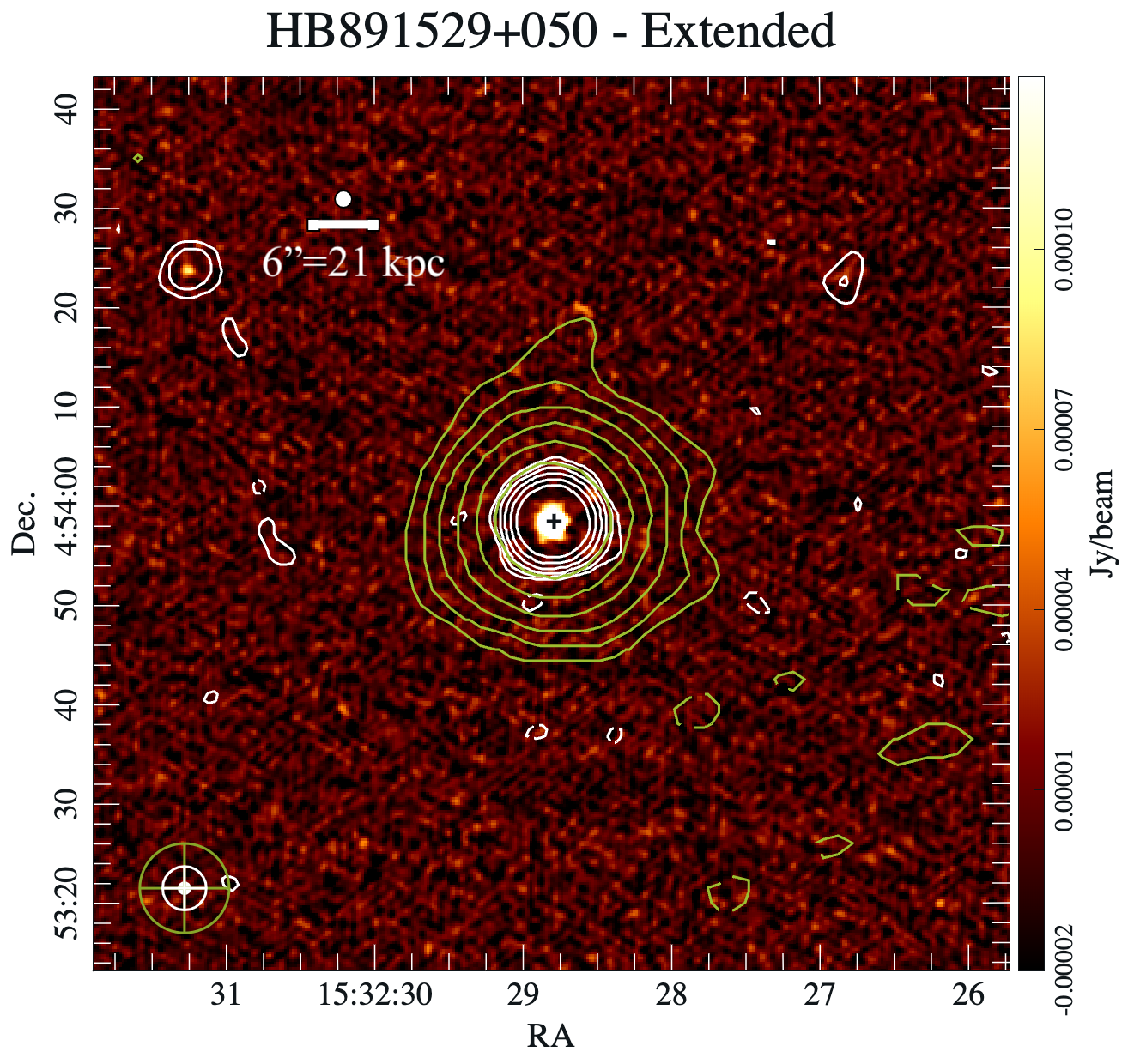}
	\hfill
 \includegraphics[width=0.36\textwidth]{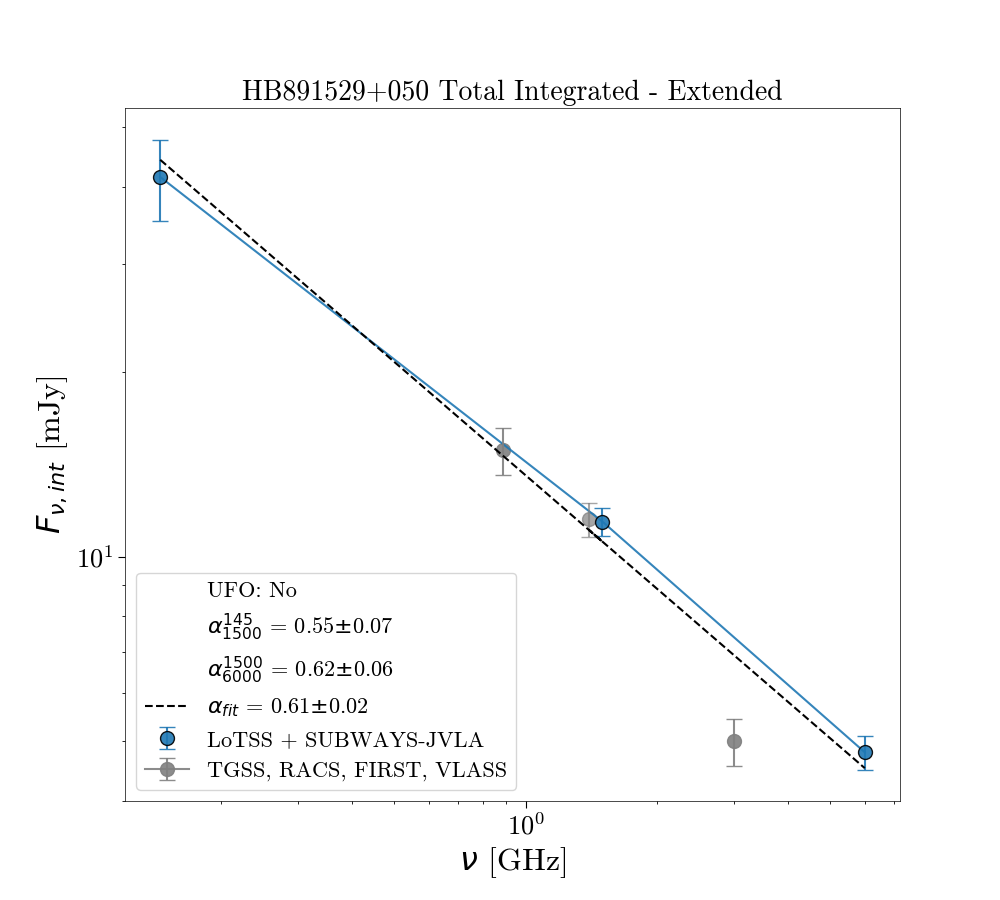}
	
	\caption{HB891529+050. The linear size is 14.35 kpc. The SED is steep dominated by an optically thin component. The offset of VLASS flux density with respect to the predicted slope may be hint of variability, even though all the other points are consistent with expectations. No UFO is detected. The SFR predicted with radio emission is a factor 9 higher than the IR deduced one. The target does exceeds the $3\sigma$ uncertainty region around the Güdel–Benz relation. The extension suggests the presence of an outflow, maybe an unresolved jet.}
	\label{sed:HB891529+050}
\end{figure*}	


\begin{figure*}[hp!]
		\centering
 \includegraphics[width=0.31\textwidth]{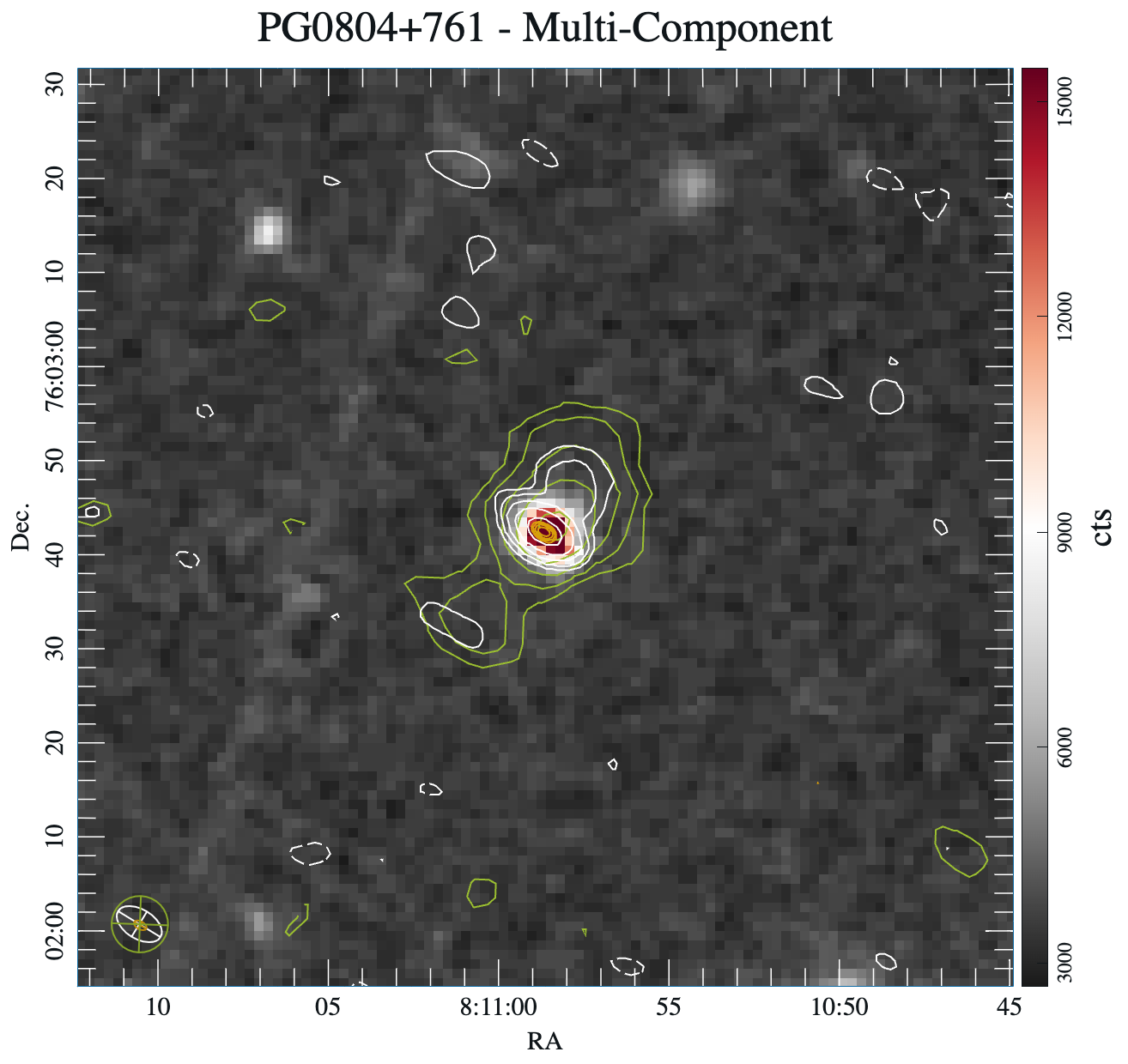}
 \hfill
 \includegraphics[width=0.31\textwidth]{IMG_new/PG0804+761_overlays_red2.png}
	\hfill
 \includegraphics[width=0.36\textwidth]{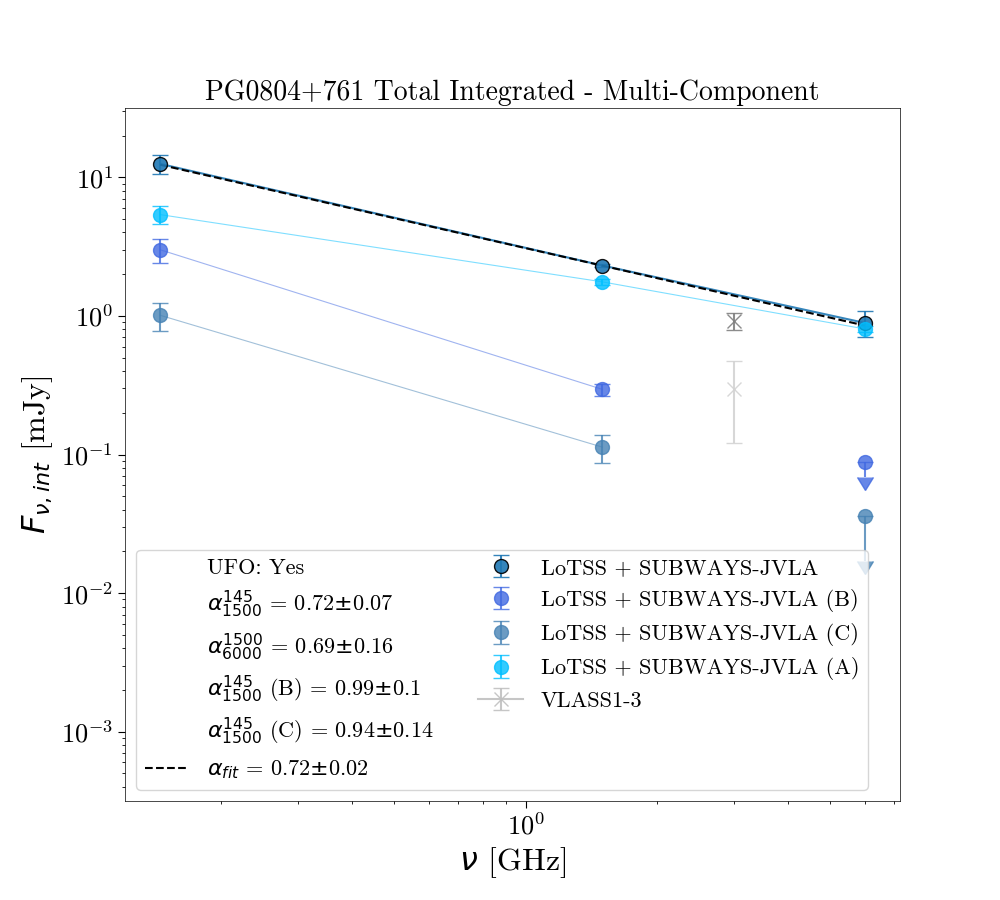}
	
	\caption{PG0804+761. The linear size is 41.8 kpc. The SED is steep, dominated by an optically thin extended component. The B,C extranuclear component have a steep spectrum, compatible with shock driven radio emission. The UFO is detected. The SFR predicted with radio emission, for the nuclear component (A), is a factor 25 higher than the IR deduced one. The core of the target follows the Güdel–Benz relation within $1\sigma$ uncertainty. Estimates done with the wind model as in \citet{Nims_2015} favour the wind scenario. Very similar to the already studied 2MASXJ1653+23. The radio emission is probably dominated by shocks from winds or jets.}
	\label{sed:PG0804+761}
\end{figure*}	

\begin{figure*}[hp!]
		\centering
 \includegraphics[width=0.31\textwidth]{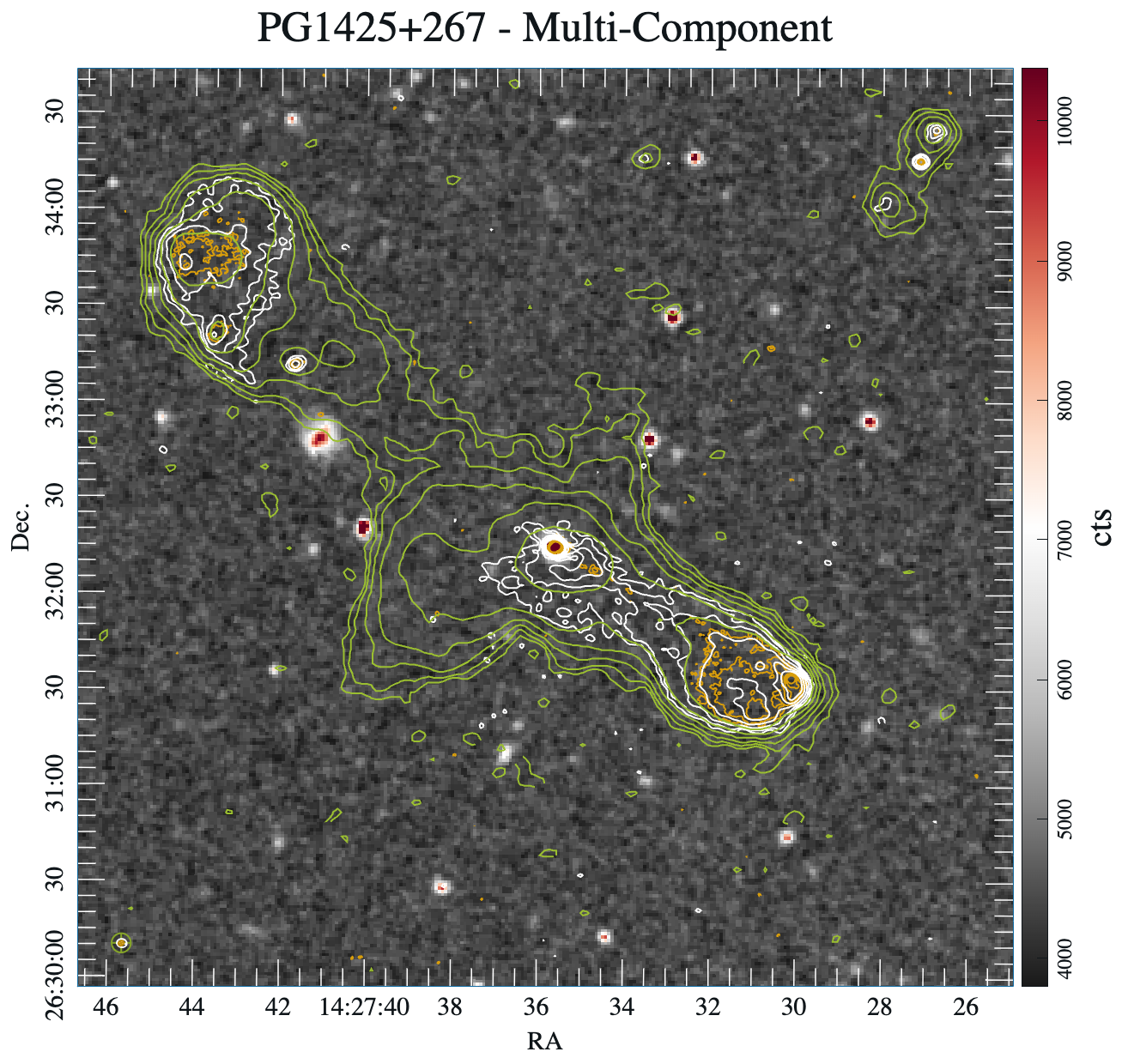}
 \hfill
 \includegraphics[width=0.31\textwidth]{IMG_new/PG1425+267_overlays_red2.png}
	\hfill
	 \includegraphics[width=0.36\textwidth]{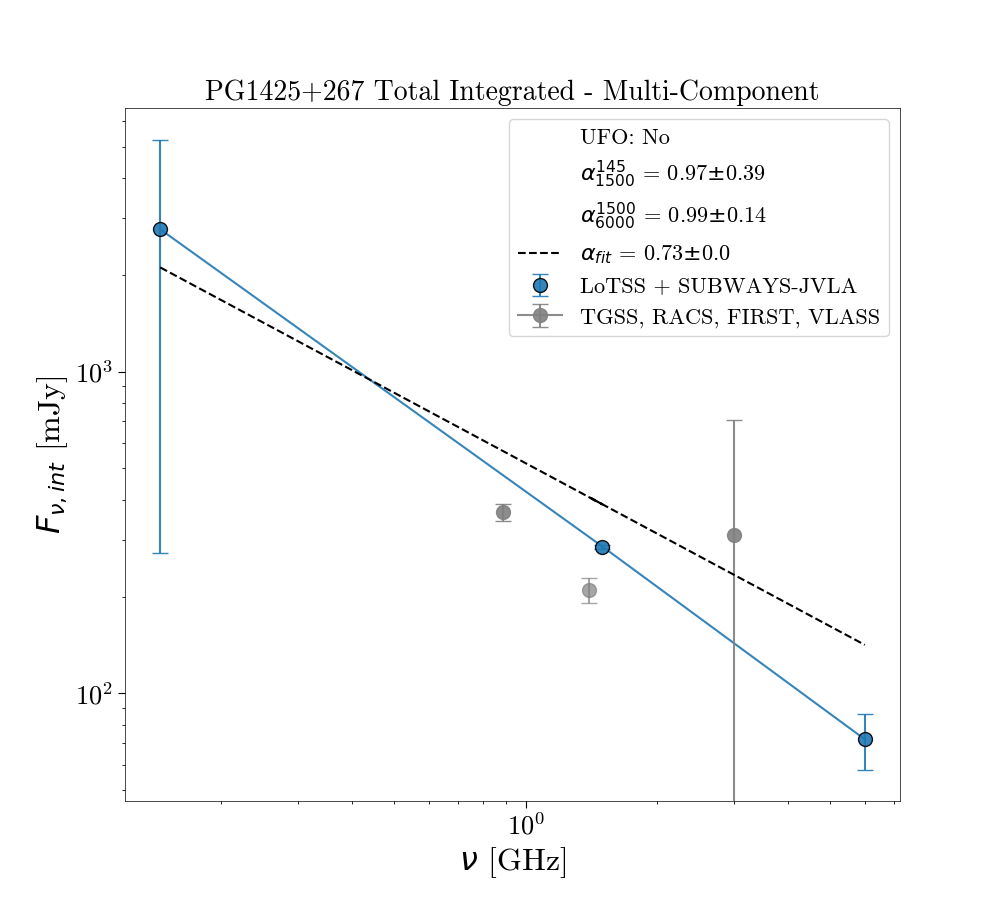}
	
	\caption{PG1425+267.The linear size is 1.2 Mpc. The SED is steep dominated by an optically thin component. The core SED is flat, as expected from a jet base. No UFO is detected. The SFR predicted with radio emission is a factor 76 higher than the IR deduced one. The core of the target exceeds the $3\sigma$ uncertainty region around the Güdel–Benz. The radio emission is unambiguously related to the presence of a relativistic jet.}
	\label{sed:PG1425+267}
\end{figure*}	

\begin{figure*}[hp!]
	\centering
	\includegraphics[width=0.31\textwidth]{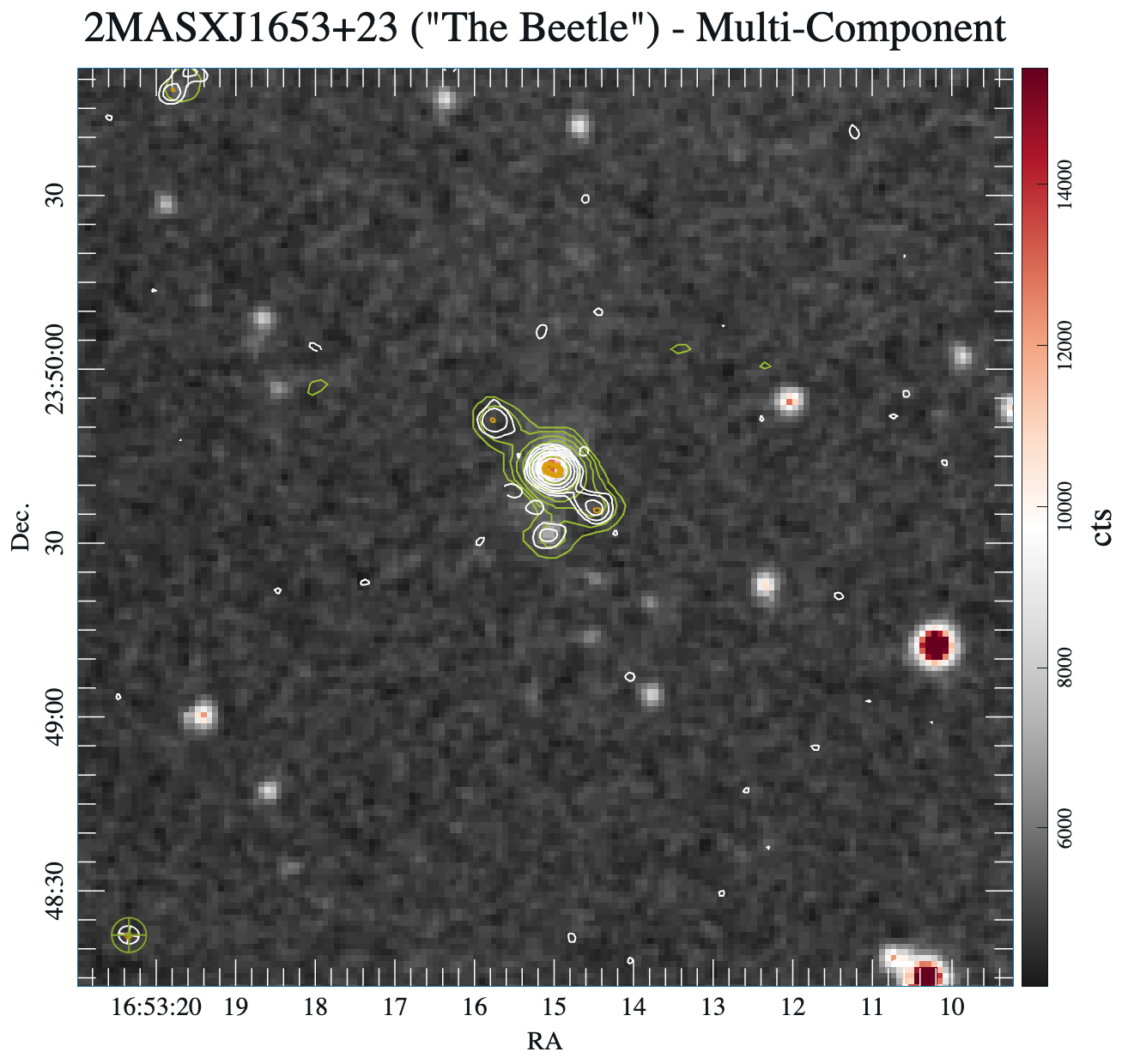}
\hfill
 \includegraphics[width=0.31\textwidth]{IMG_new/2MASXJ1653+23_overlays_red2.png}
\hfill
	\includegraphics[width=0.36\textwidth]{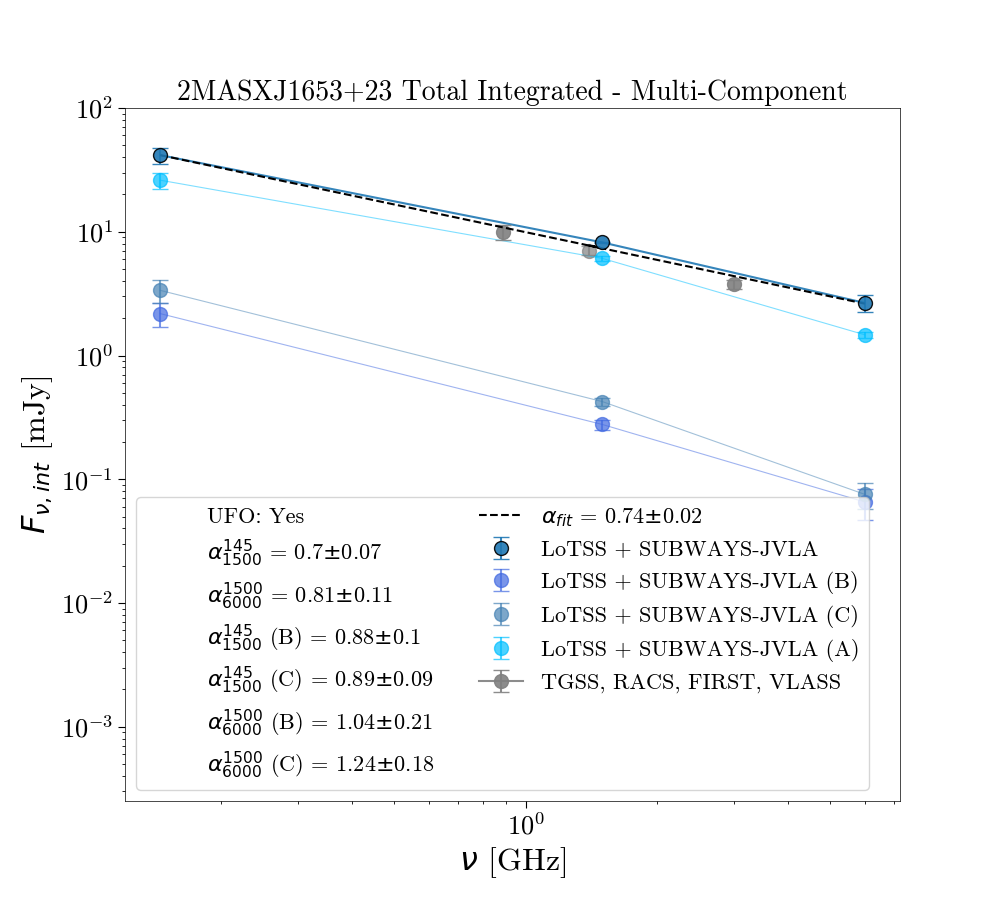}
	
	\caption{2MASXJ1653+23.The linear size is 45.6 kpc. The integrated SED is steep dominated by an optically thin component, but the core one shows a break. The B,C extranuclear component have a steep spectrum, compatible with shock driven radio emission. UFO detected. The SFR predicted with radio emission is comparable to the IR deduced one withun $1\sigma$. The core of the target exceeds the $3\sigma$ uncertainty region around the Güdel–Benz relation. Estimates done with the wind model as in \citet{Nims_2015} favour the wind scenario. The bulk of radio emission is unambiguously related to the presence of outflows, even if it remains difficult to distinguish between AGN driven winds and a low-power jet.}
	\label{sed:2MASXJ1653+23}
\end{figure*}

\end{appendix}

\end{document}